\documentclass[twocolumn,floatfix,tighten]{aastex62}

\newcommand{\mangafd}{10$^{-17}$ erg/s/cm$^2$/\AA}
\newcommand{\kms}{{km$~\!$s$^{-1}$}}
\newcommand{\halpha}{H$\alpha$}

\newcommand{\drp}{{\tt DRP}}	
\newcommand{\dap}{{\tt DAP}}	
\newcommand{\dapall}{{\tt DAPall}}
\newcommand{\drpall}{{\tt DRPall}}
\newcommand{\dapmaps}{{\tt MAPS}}
\newcommand{\dapcube}{{\tt LOGCUBE}}
\newcommand{\ppxf}{{\tt pPXF}}
\newcommand{\mileshc}{{\tt MILES-HC}}
\newcommand{\marvin}{{\tt Marvin}}
\newcommand{\mastar}{MaStar}

\newcommand{\mass}{\mathcal{M}}

\newcommand{\cindx}{{\mathcal I}_a^c}
\newcommand{\snrg}{S/N$_g$}

\newcommand{\reff}{R_e}

\newcommand{\sigo}{\sigma_{\rm obs}}
\newcommand{\sigs}{\sigma_\ast}
\newcommand{\sigg}{\sigma_{\rm g}}
\newcommand{\sigt}{\sigma_{\rm t}}
\newcommand{\dsigi}{\delta\sigma_{\rm inst}}
\newcommand{\sigi}{\sigma_{\rm inst}}

\definecolor{todo}{RGB}{200,0,0}

\received{}
\revised{}
\accepted{}
\submitjournal{AJ}

\shortauthors{Westfall et al.}
\shorttitle{SDSS-IV/MaNGA Data Analysis Pipeline: Overview}

\begin{document}

\title{ The Data Analysis Pipeline for the SDSS-IV MaNGA IFU Galaxy
Survey: Overview }

\correspondingauthor{Kyle B. Westfall}
\email{westfall@ucolick.org}

\author[0000-0003-1809-6920]{Kyle B. Westfall}
\affiliation{University of California Observatories, University of California, Santa Cruz, 1156 High St., Santa Cruz, CA 95064, USA}

\author[0000-0002-1283-8420]{Michele Cappellari}
\affiliation{Sub-department of Astrophysics, Department of Physics, University of Oxford, Denys Wilkinson Building, Keble Road, Oxford OX1 3RH, UK}

\author[0000-0002-3131-4374]{Matthew A. Bershady}
\affiliation{Department of Astronomy, University of Wisconsin-Madison, 475N. Charter St., Madison WI 53703, USA}
\affiliation{South African Astronomical Observatory, P.O. Box 9, Observatory 7935, Cape Town, South Africa}

\author[0000-0001-9742-3138]{Kevin Bundy}
\affiliation{University of California Observatories, University of California, Santa Cruz, 1156 High St., Santa Cruz, CA 95064, USA}

\author[0000-0002-2545-5752]{Francesco Belfiore}
\affiliation{European Southern Observatory, Karl-Schwarzchild-Str. 2, Garching bei M{\"u}nchen, 85748, Germany}
\affiliation{University of California Observatories, University of California, Santa Cruz, 1156 High St., Santa Cruz, CA 95064, USA}

\author{Xihan Ji}
\affiliation{Tsinghua Center of Astrophysics \& Department of Physics, Tsinghua University, Beijing 100084, China}

\author[0000-0002-9402-186X]{David R. Law}
\affiliation{Space Telescope Science Institute, 3700 San Martin Drive, Baltimore, MD 21218, USA}

\author{Adam Schaefer}
\affiliation{Department of Astronomy, University of Wisconsin-Madison, 475N. Charter St., Madison WI 53703, USA}

\author{Shravan Shetty}
\affiliation{Department of Astronomy, University of Wisconsin-Madison, 475N. Charter St., Madison WI 53703, USA}

\author[0000-0003-3097-5178]{Christy A. Tremonti}
\affiliation{Department of Astronomy, University of Wisconsin-Madison, 475N. Charter St., Madison WI 53703, USA}

\author[0000-0003-1025-1711]{Renbin Yan}
\affiliation{Department of Physics and Astronomy, University of Kentucky, 505 Rose Street, Lexington, KY 40506, USA}

\author[0000-0001-8085-5890]{Brett H. Andrews}
\affiliation{University of Pittsburgh, PITT PACC, Department of Physics and Astronomy, Pittsburgh, PA 15260, USA}

\author[0000-0002-8725-1069]{Joel R. Brownstein}
\affiliation{University of Utah, Department of Physics and Astronomy, 115 S. 1400 E., Salt Lake City, UT 84112, USA}

\author[0000-0002-4289-7923]{Brian Cherinka}
\affiliation{Center for Astrophysical Sciences, Department of Physics and Astronomy, Johns Hopkins University, 3400 North Charles Street, Baltimore, MD 21218, USA}
\affiliation{Space Telescope Science Institute, 3700 San Martin Drive, Baltimore, MD 21218, USA}

\author[0000-0001-7817-6995]{Lodovico Coccato}
\affiliation{European Southern Observatory, Karl-Schwarzchild-str., 2, Garching b. M\"unchen, 85748, Germany}

\author[0000-0002-7339-3170]{Niv Drory}
\affiliation{McDonald Observatory, The University of Texas at Austin, 1 University Station, Austin, TX 78712, USA}

\author[0000-0001-7711-3677]{Claudia Maraston}
\affiliation{Institute of Cosmology \& Gravitation, University of Portsmouth, Dennis Sciama Building, Portsmouth, PO1 3FX, UK}

\author[0000-0002-0621-6238]{Taniya Parikh}
\affiliation{Institute of Cosmology \& Gravitation, University of Portsmouth, Dennis Sciama Building, Portsmouth, PO1 3FX, UK}

\author[0000-0003-2405-7258]{Jos\'e R. S\'anchez-Gallego}
\affiliation{Department of Astronomy, University of Washington, Box 351580, Seattle, WA 98195, USA}

\author[0000-0002-6325-5671]{Daniel Thomas}
\affiliation{Institute of Cosmology \& Gravitation, University of Portsmouth, Dennis Sciama Building, Portsmouth, PO1 3FX, UK}

\author[0000-0002-5908-6852]{Anne-Marie Weijmans}
\affiliation{School of Physics and Astronomy, University of St Andrews, North Haugh, St Andrews KY16 9SS, UK}

\author[0000-0003-2405-7258]{Jorge Barrera-Ballesteros}
\affiliation{Center for Astrophysical Sciences, Department of Physics and Astronomy, Johns Hopkins University, 3400 North Charles Street, Baltimore, MD 21218, USA}

\author{Cheng Du}
\affiliation{Tsinghua Center of Astrophysics \& Department of Physics, Tsinghua University, Beijing 100084, China}

\author{Daniel Goddard}
\affiliation{Institute of Cosmology \& Gravitation, University of Portsmouth, Dennis Sciama Building, Portsmouth, PO1 3FX, UK}

\author[0000-0003-2778-002X]{Niu Li}
\affiliation{Tsinghua Center of Astrophysics \& Department of Physics, Tsinghua University, Beijing 100084, China}

\author[0000-0003-0846-9578]{Karen Masters}
\affiliation{Department of Physics and Astronomy, Haverford College, 370 Lancaster Ave, Haverford, PA 19041}

\author[0000-0002-9790-6313]{H\'ector Javier Ibarra Medel}
\affiliation{Instituto de Astronomıa, Universidad Nacional Aut\'onoma de M\'exico, A.P. 70-264, 04510, Mexico, D.F., M\'exico}

\author[0000-0001-6444-9307]{Sebasti\'an F. S\'anchez}
\affiliation{Instituto de Astronomıa, Universidad Nacional Aut\'onoma de M\'exico, A.P. 70-264, 04510, Mexico, D.F., M\'exico}

\author[0000-0002-1749-1892]{Meng Yang}
\affiliation{School of Physics and Astronomy, University of St Andrews, North Haugh, St Andrews KY16 9SS, UK}

\author{Zheng Zheng}
\affiliation{National Astronomical Observatories of China, Chinese Academy of Sciences, 20A Datun Road, Chaoyang District, Beijing 100012, China}
\affiliation{CAS Key Laboratory of FAST, NAOC, Chinese Academy of Sciences}

\author[0000-0002-8999-6814]{Shuang Zhou}
\affiliation{Tsinghua Center of Astrophysics \& Department of Physics, Tsinghua University, Beijing 100084, China}

\begin{abstract}

Mapping Nearby Galaxies at Apache Point Observatory (MaNGA) is
acquiring integral-field spectroscopy for the largest sample of
galaxies to date. By 2020, the MaNGA Survey --- one of three core
programs in the fourth-generation Sloan Digital Sky Survey (SDSS-IV)
--- will have observed a statistically representative sample of
10$^4$ galaxies in the local Universe ($z\lesssim0.15$). In addition
to a robust data-reduction pipeline (\drp), MaNGA has developed a
data-analysis pipeline (\dap) that provides higher-level data
products. To accompany the first public release of its code base and
data products, we provide an overview of the MaNGA \dap, including
its software design, workflow, measurement procedures and algorithms,
performance, and output data model. In conjunction with our companion
paper Belfiore et al., we also assess the \dap\ output provided for
4718 observations of 4648 unique galaxies in the recent SDSS Data
Release 15 (DR15). These analysis products focus on measurements that
are close to the data and require minimal model-based assumptions.
Namely, we provide stellar kinematics (velocity and velocity
dispersion), emission-line properties (kinematics, fluxes, and
equivalent widths), and spectral indices (e.g., D4000 and the Lick
indices). We find that the \dap\ provides robust measurements and
errors for the vast majority ($>$99\%) of analyzed spectra. We
summarize assessments of the precision and accuracy of our
measurements as a function of signal-to-noise, and provide specific
guidance to users regarding the limitations of the data. The MaNGA
\dap\ software is publicly available and we encourage community
involvement in its development.

\end{abstract}


\section{Introduction}
\label{sec:intro}

Publicly available data sets in accessible formats play an increasingly
important role in astronomy, broadening access and enabling analyses
that combine observations across wavelengths and telescopes.  The
original Sloan Digital Sky Survey \citep[SDSS,][]{2000AJ....120.1579Y}
initiated an ongoing commitment to publicly release raw and reduced data
that continues through the current generation, SDSS-IV
\citep{blanton17}.  

With the introduction of the MaNGA Survey \citep[Mapping Nearby Galaxies
at Apache Point Observatory,][]{2015ApJ...798....7B}, SDSS-IV data
releases have included ``higher dimensional'' sets of datacubes whose
production requires a sophisticated Data Reduction Pipeline
\citep[\drp]{2016AJ....152...83L}.  Over its six-year duration ending
July 2020, MaNGA is providing spatially resolved spectroscopy for 10,000
nearby galaxies selected with $\mass_* \gtrsim 10^9 \mass_{\odot}$ and
$\langle z\rangle \sim 0.03$ \citep{2017AJ....154...86W}.  MaNGA uses
specially designed fiber bundles \citep{2015AJ....149...77D} that feed
the BOSS spectrographs \citep{2013AJ....146...32S} on the 2.5-meter
Sloan Telescope \citep{2006AJ....131.2332G}.  Spanning 0.36--1.0 $\mu m$
at a resolution of $R \sim 2000$ and with excellent flux calibration
\citep{2016AJ....151....8Y}, MaNGA executes approximately three-hour
long dithered integrations \citep{2015AJ....150...19L} to reach signal-to-noise
requirements for galaxies observed to approximately 1.5 effective
(half-light) radii, $R_e$, (two-thirds of the sample) and 2.5 $R_e$
(one-third of the sample) \citep{2016AJ....152..197Y}.

Reduced MaNGA data, including reconstructed datacubes, are produced by
the automated MaNGA \drp\ and have been made publicly available since
the thirteenth SDSS data release \citep[DR13;][]{2017ApJS..233...25A}.
Inspired by previous ``value-added catalogs'' (VACs) based on SDSS data,
with the
MPA-JHU\footnote{\url{https://www.sdss.org/dr15/data_access/value-added-catalogs/mpa-jhu-stellar-masses}}
catalog for SDSS-I/II being a prime example, members of the MaNGA team
have also provided publicly available VACs in previous data
releases.\footnote{For example:
\url{https://www.sdss.org/dr14/data_access/value-added-catalogs/}}

The complexity and richness of spatially-resolved data sets, like those
produced by MaNGA, as well as the desire to take on common analysis
tasks that would otherwise be duplicated by large numbers of users has
motivated SDSS-IV to invest in a ``project-led'' MaNGA Data Analysis
Pipeline (\dap).  By providing a uniform set of commonly desired
analysis products, the \dap\ also enables rapid and interactive delivery
methods for these
data products, helping scientists quickly design samples of interest and
make discoveries.  It was also appreciated that while researchers would
likely perform custom measurements of primary interest to their science,
the ability to combine these with readily available \dap\ measurements
that might otherwise be outside their expertise would open up new
opportunities.  In general, a broad-based and robust \dap\ makes MaNGA
data more science-ready.  This strategy aligns with other IFU surveys
including ATLAS3D\footnote{
\url{http://purl.org/atlas3d}}
\citep{Cappellari2011a}, CALIFA\footnote{
\url{http://califa.caha.es}}
\citep{sanchez12}, and SAMI\footnote{
\url{https://sami-survey.org}}
\citep{croom12} that have released high-level data products.

The MaNGA \dap\ has been under development since 2014, evolving through
several versions and a transition from an original {\tt IDL} to {\tt
python} implementation.  Although the \dap\ has primarily been used as a
survey-level pipeline for providing analysis products to the SDSS
collaboration, our development strategy has also emphasized flexibility
to prospective users.  The low-level, core algorithms have been
constructed in a way that is largely independent of their specific use
with the MaNGA data, allowing a user to write new {\tt python} scripts
around \dap\ functions or classes for analysis of more varied data sets.
Additionally, the high-level interface is written such that the detailed
execution of many of the internal algorithms can be modified using a set
of configuration files, allowing a user to tailor how the \dap\ analyses
MaNGA data to better suit their scientific needs.  Although much of the
high-level functionality assumes one is working with MaNGA data, it is
possible to apply the \dap\ to data from different instruments with
modest modification.

The MaNGA science teams have published studies based on \dap\ output
throughout its development, using internal data releases to the SDSS
collaboration that we term MaNGA Product Launches (MPLs).  The \dap\
source code continues to evolve as we improve its fidelity and expand
its functionality.  We encourage community involvement in these efforts
via our public repository on GitHub.\footnote{
\url{https://github.com/sdss/mangadap}}
The GitHub repository also contains some example scripts that use
low-level \dap\ functions to fit a single spectrum, as well as the
higher-level modules (Section \ref{sec:workflow}) that can be used to
fit a single datacube.  Additional example scripts will be provided as
development continues, and we encourage others to submit their own
scripts via a GitHub pull request.

This paper describes the first public release of \dap\ products, as
part of SDSS Data Release 15 (DR15), and includes both the code
(version {\tt 2.2.1}) and output data products. The public \dap\
output is available for download from the SDSS website\footnote{
\url{https://www.sdss.org/dr15/data_access}}
and via \marvin\footnote{\url{https://dr15.sdss.org/marvin}}
\citep{2019AJ....158...74C} --- an interactive web-based interface to
both MaNGA datacubes and \dap\ quantities ({\tt Marvin-Web}), as well
as a {\tt python} package ({\tt Marvin-tools}) that enables seamless
remote and/or local access to these MaNGA data that can be
incorporated in any {\tt python}-based analysis workflow.

The primary output of the \dap\ includes stellar kinematics, fluxes
and kinematics of emission lines, and continuum spectral indices. In
deriving these measurements, the \dap\ makes heavy use of \ppxf\
\citep{2004PASP..116..138C, 2017MNRAS.466..798C} as a workhorse
spectral-fitting routine. Our philosophy has been to focus on
measurements that are made directly on the MaNGA spectra and that do
not require significant model-based assumptions. For example, the
\dap\ fits stellar template mixes plus a polynomial component to the
stellar continuum. This provides an excellent representation of the
data and derived stellar absorption-line kinematics, but is not
necessarily appropriate for accurate stellar-population properties.
For maps of estimated stellar age, metallicity, star-formation
histories, and other model-derived data products, the Firefly
\citep{2017MNRAS.466.4731G} and Pipe3D \citep{2016RMxAA..52...21S,
2016RMxAA..52..171S} VACs\footnote{
\url{https://www.sdss.org/dr15/data_access/value-added-catalogs/}}
are valuable resources. Pipe3D products in particular also include
alternative measurements of kinematics and emission lines.

In this contribution, we present an overview of the MaNGA \dap, its
algorithmic structure, and its output data products.  We begin in
Section \ref{sec:guidance} with a ``quick-start'' guide that provides a
more detailed introduction to the \dap\ output and nomenclature by way
of examples.  Along with many other resources cited therein, we expect
Section \ref{sec:guidance} to be a useful road map to the \dap\ products
and the rest of our paper.  In Section \ref{sec:data}, we describe the
input data required by the \dap, including a general summary of MaNGA
spectroscopy.  Section \ref{sec:workflow} provides an overview of the
\dap\ workflow through its six main analysis modules.  After first
describing the spectral templates we use in DR15 and the method used to
generate them in Section \ref{sec:mileshc}, we dedicate the following
five sections to the detailed description of each of the \dap\ analysis
modules:\footnote{
In these sections, we focus on the specific way that the \dap\ has been
executed for the data provided as part of DR15, not on an exhaustive
description of what the \dap\ {\it can} do.  More exhaustive and
evolving documentation of the code is provided as part of the code
distribution and hosted at
\url{https://sdss-mangadap.readthedocs.io/en/latest/}.}

Section \ref{sec:binning} describes our spatial binning approach and the
importance of including spatial covariance in these calculations.
Section \ref{sec:stellarkin} goes into particular depth with regard to
the assessments of the stellar kinematics. We present results from
several input/output simulations, as well as a statistical comparison of
the results for MaNGA galaxies with multiple observations. In
particular, these repeat observations allow us to provide a detailed
assessment of the errors reported by the \dap. We use Section
\ref{sec:bandpass} to introduce our bandpass-integral formalism that is
used when measuring both non-parameteric emission-line fluxes and
spectral indices. Section \ref{sec:emlfit} describes our emission-line
modeling algorithm, with detailed assessments of the module and usage in
DR15 provided by our companion paper, Belfiore et al.\ {\it accepted}.
Finally, Section \ref{sec:spindex} describes our quantification of
continuum features using spectral indices, and we assess the accuracy
and precision of our measurements using a similar approach used for the
stellar kinematics. Detailed quality assessments of the two
full-spectrum-fitting modules are provided in Section
\ref{sec:stellarkin} (stellar kinematics) and Belfiore et al., {\it
accepted}, (emission-line modeling), and we comment on the overall
performance of the \dap\ for DR15 in Section \ref{sec:performance}. In
particular, we note specific regimes where we find the \dap\ currently
requires further development. Section \ref{sec:output} provides a
detailed discussion of the \dap\ output products and important aspects
of these products that users should keep in mind. Finally, we provide
some brief conclusions in Section \ref{sec:summary}. Appendices
\ref{sec:resolution}, \ref{sec:sigmaerr}, and \ref{sec:datamodel}
provide, respectively, the \dap\ procedure used to match the spectral
resolution of two spectra, an assessment of how instrumental resolution
errors propagate to errors in the stellar velocity dispersion
measurements, and tables describing the \dap\ output data models.

Unless stated otherwise, throughout this paper: (1) We adopt a
$\Lambda$CDM cosmology with $\Omega_m=0.3$, $\Omega_\Lambda=0.7$, and
$H_0 = 100 h$ \kms\ Mpc$^{-1}$. (2) All wavelengths are provided in
vacuum. (3) All flux densities have units of 10$^{-17}$
ergs/cm$^2$/s/\AA/spaxel.

\begin{figure*}
\includegraphics[width=1.0\textwidth]{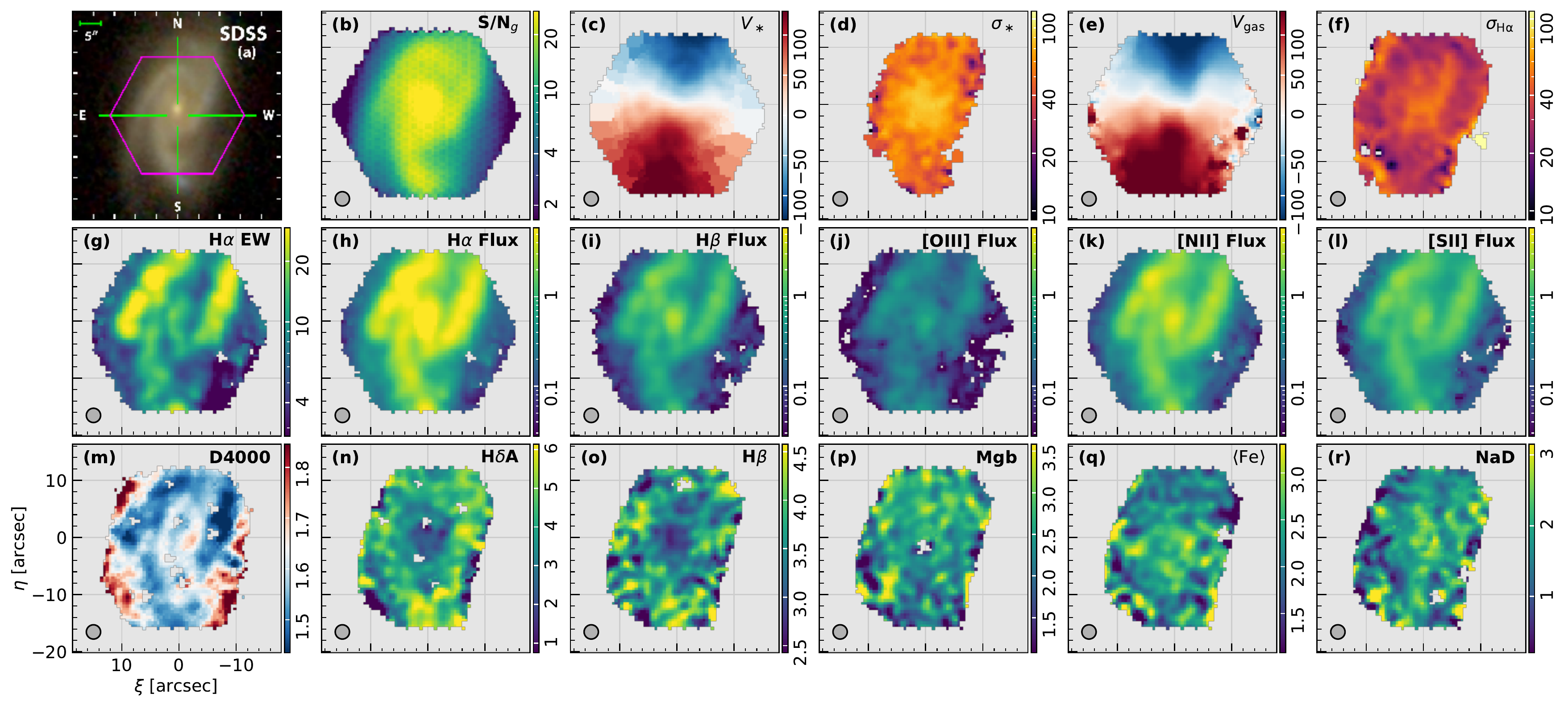}
\caption{A subset of the \dap-derived quantities for datacube {\tt
8439-12703}, the observation of MaNGA galaxy 1-605884 ($z=0.025$;
$\log(\mass_\ast/ {\rm h}^{-2} \mass_\odot) = 10.2$; $\reff =
11\farcs8 \sim 4.2$ h$^{-1}$ kpc; see
\url{https://sas.sdss.org/marvin/galaxy/8439-12703/} for more
information about this galaxy and its properties in the larger
context of the MaNGA sample.) From top-to-bottom and left-to-right:
(a) the SDSS $gri$ composite image with the nominal size of the IFU
outlined in purple; (b) the $g$-band S/N per channel, \snrg; (c) the
stellar line-of-sight (LOS) velocity, $V_\ast$; (d) the stellar
velocity dispersion, $\sigma_\ast$; (e) the ionized-gas LOS velocity,
$V_{\rm gas}$; (f) the velocity dispersion of the H$\alpha$ emission
line, $\sigma_{{\rm H}\alpha}$; (g) the equivalent width (EW) of the
H$\alpha$ emission line; (h) the flux of the H$\alpha$ emission line;
(i) the flux of the H$\beta$ emission line; (j) the total flux in the
[\ion{O}{3}]$\lambda$4959,5007 emission lines; (k) the total flux in
the [\ion{N}{2}]$\lambda$6548,6583 emission lines; (l) the total flux
in the [\ion{S}{2}]$\lambda$6716,6730 emission lines; (m) the D4000
spectral index; (n) the H$\delta$A spectral index; (o) the Mgb
spectral index; (q) the average of the Fe5270 and Fe5335 spectral
indices, $\langle{\rm Fe}\rangle$; (r) and the NaD spectral index.
The gray circle in the bottom-left corner of panels b through r is
the nominal FWHM of MaNGA's spatial resolution element (beam size;
$2\farcs5$). Bins or spaxels with \snrg$<$10 are masked in panel d
($\sigma_\ast$); \snrg$<$3 are masked in panel m (D4000); \snrg$<$5
are masked for the plotted absorption-line indices (panels n through
r); and H$\alpha$ fluxes less than $2.5\times10^{-18}$ erg/s/cm$^2$
are masked in panel f ($\sigma_{{\rm H}\alpha}$).}
\label{fig:showcasemaps}
\end{figure*}

\section{\dap\ Quick-Start Guide}
\label{sec:guidance}

We begin with a ``quick-start guide'' to the MaNGA \dap, jumping
right into example output products that highlight key aspects of the
\dap\ measurements. To be clear from the beginning, {\it all} of the
\dap\ input and output files we discuss here and throughout our paper
are included as part of DR15. The narrative of this Section aims to
help the reader quickly get a sense of those products and navigate
their way to subsequent sections of our paper that are most relevant
to their goals. Although some general guidance can be sufficiently
provided here, other more nuanced advice requires the backdrop of our
assessments of the \dap\ data, performed throughout our paper. In
Section \ref{sec:generalusage}, we highlight specific sections where
the reader can go for that advice, as well as other documentation
provided as part of SDSS DR15. Finally, Section
\ref{sec:known_issues} provides a list of known issues with the \dap\
data provided with DR15 that users should take into account.

In Figure \ref{fig:showcasemaps}, we show the SDSS $gri$ composite image
for the galaxy targeted by integral-field unit (IFU) {\tt 12703} on
plate {\tt 8439} and a sampling of the quantities produced by the \dap\
for this datacube.\footnote{
The python code used to produce this plot and many others in this paper
can be found at
\url{https://github.com/sdss/mangadap/tree/master/docs/papers/Overview/scripts}.}
From top to bottom and left to right, these images roughly follow the
order of the \dap\ workflow (Section \ref{sec:workflow}), from
assessments of the $g$-band signal-to-noise (\snrg\ per spectral pixel;
panel b; Section \ref{sec:binning}); to the measurements of the stellar
kinematics (panels c and d; Section \ref{sec:stellarkin}); to the
emission-line modeling that produces fluxes, equivalent-widths (EWs),
and kinematics (panels e through l; Section \ref{sec:emlfit}; Belfiore
et al.\ {\it accepted}); and finally to the spectral-index measurements
(panels m through r; Section \ref{sec:spindex}).

The images, or maps, plotted in Figure \ref{fig:showcasemaps} are
provided by the primary \dap\ output file, a multi-extension fits
file called the \dapmaps\ file (Section \ref{sec:mapsfile}). The
\dapmaps\ files provide each \dap\ measurement in a two-dimensional
image format, or map, that exactly matches the spatial dimensions of
the \drp\ datacubes. Where appropriate, each mapped quantity has
associated inverse-variance and quality-assessment measurements. Our
quality assessments are provided by a set of bitmasks defined in
Appendix \ref{sec:datamodel}, and their use is a critical aspect of
any workflow incorporating the \dap\ output data. In Figure
\ref{fig:showcasemaps}, empty regions are masked either because they
are outside the hexagonal footprint of MaNGA's dithered field-of-view
or because they do not meet the \dap\ quality-assurance criteria
(Section \ref{sec:prelim}). For Figure \ref{fig:showcasemaps}
specifically, we also mask regions in the maps of the stellar
velocity dispersion ($\sigma_\ast$), H$\alpha$ velocity dispersion
($\sigma_{{\rm H}\alpha}$), and spectral indices at low \snrg\ and
low flux; see the Figure caption.\footnote{
Masking in the \dap\ is largely limited to indicating numerical or
computational issues occurring during the course of the analysis.
Flagging of any given measurement based on its {\it expected} quality
is more limited due to the difficulty of defining criteria that are
generally robust and not overly conservative.}
A user-customized set of \dap\ maps are easily displayed for a
specific galaxy in \marvin, both via the web interface\footnote{
\url{https://dr15.sdss.org/marvin}}
and its core {\tt python} package.\footnote{
\url{https://github.com/sdss/marvin}}

The \dap\ results for {\tt 8439-12703} (the datacube for galaxy
1-605884) were chosen at random for Figure \ref{fig:showcasemaps} and
are representative of the DR15 results as a whole.\footnote{
Because the same galaxy may be observed more than once (see Table
\ref{tab:repeats}), we generally refer to a specific datacube using its
{\tt PLATEIFU} designation throughout this paper (e.g., {\tt 8439-12703}
in this case), as opposed to the unique MaNGA ID associated with each
survey target.}
Using these data as an example, it is important to note that the
spatial pixel, or spaxel, size ($0\farcs5\times0\farcs5$) is
significantly smaller than the full-width at half maximum (FWHM) of
the on-sky point-spread function ($2\farcs5$ diameter) shown as a
gray circle in the bottom-left corner of each panel. This spaxel
sampling was chosen as a compromise between the covariance introduced
in the datacubes by our reconstruction approach and the spatial scale
needed to properly sample the dithered fiber observations
\citep[cf.][]{2019arXiv190606369L}. Even so, the subsampling of the
fiber beam leads to significant covariance between adjacent spaxels
\citep[Section \ref{sec:snr};][]{2016AJ....152...83L} and a few
noteworthy implications. First, accounting for this covariance is
critical to accurately meeting a target S/N threshold when spatially
binning the datacubes (Section \ref{sec:vorbin}). Second, spatial
variations in the mapped measurements driven by random errors in the
observed spectra should be smooth on scales of roughly $5\times5$
spaxels. That is, significant spaxel-to-spaxel variations due to a
random sampling are highly improbable given the well-defined
correlation matrix of the datacube. Instead, a useful rule of thumb
to keep in mind when inspecting \dap\ images is that significant
spaxel-to-spaxel variations are driven by systematic error,\footnote{
The systematic error involved may only yield an increased stochasticity
in the measurements that average out over many spaxels, and does not
necessarily imply a systematic shift of the posterior distribution away
from the true value.}
not astrophysical structure.  On the other hand, structure on scales
similar to the beam size could be astrophysical --- such as the
increased H$\alpha$ EW along the spiral arms of galaxy 1-605884 --- or
driven by noise in the fiber observations --- such as is likely the
cause of the strong, beam-size variations in the gas velocity field
toward the IFU periphery or the high-frequency modulations of its
spectral-index maps.

The maps shown in Figure \ref{fig:showcasemaps} are the result of a
``hybrid'' binning approach (see the introduction to Section
\ref{sec:emlfit} and the algorithm description in Section
\ref{sec:emlhybrid}). In this approach, the stellar kinematics are
measured for spatially binned spectra that meet a minimum of
\snrg$\gtrsim$10 (Section \ref{sec:vorbin}) and the emission-line and
spectral-index measurements are performed for individual spaxels. We
expect these results to be preferable for the majority of users,
providing the benefits of both unbiased stellar kinematics and
unbinned emission-line maps. The selection of these products is made
via the {\tt DAPTYPE}, as we define below.

A core design principle of the \dap\ has been to abstract and
modularize the analysis steps to maintain flexibility. The specific
combination of the settings used for all analysis steps --- e.g., the
specific binning algorithm and the templates used to measure the
stellar kinematics --- adopted for the survey-level execution of the
\dap\ is used to construct a unique keyword called the {\tt DAPTYPE};
see Section \ref{sec:workflow} and Figure \ref{fig:workflow} for more
detail. For DR15, two unique approaches to the analysis were
performed, meaning that each galaxy datacube is analyzed twice and
users must choose which set of products to use (see Section
\ref{sec:output}).\footnote{
For reference, the execution time of the \dap\ strongly depends on
the number of spectra being analyzed, such that larger IFU bundles
require more time. For a single core of our Utah cluster (comparable
to a single core on a laptop), the median execution time to complete
both analysis approaches on a single datacube for DR15 was 2.0, 3.1,
4.6, 6.6, and 8.7 hours for the 19-, 37-, 61-, 91-, and 127-fiber
datacubes, respectively. However, some of the 127-fiber datacubes
required up to 25 hours to complete.}
The fundamental difference between these two {\tt DAPTYPE}s is
whether the output is based on the hybrid binning approach --- the
output we recommend users start with --- or if {\it all} analyses
have been performed using spectra Voronoi-binned to a target
\snrg$\gtrsim$10.


\begin{figure*}
\includegraphics[width=1.0\textwidth]{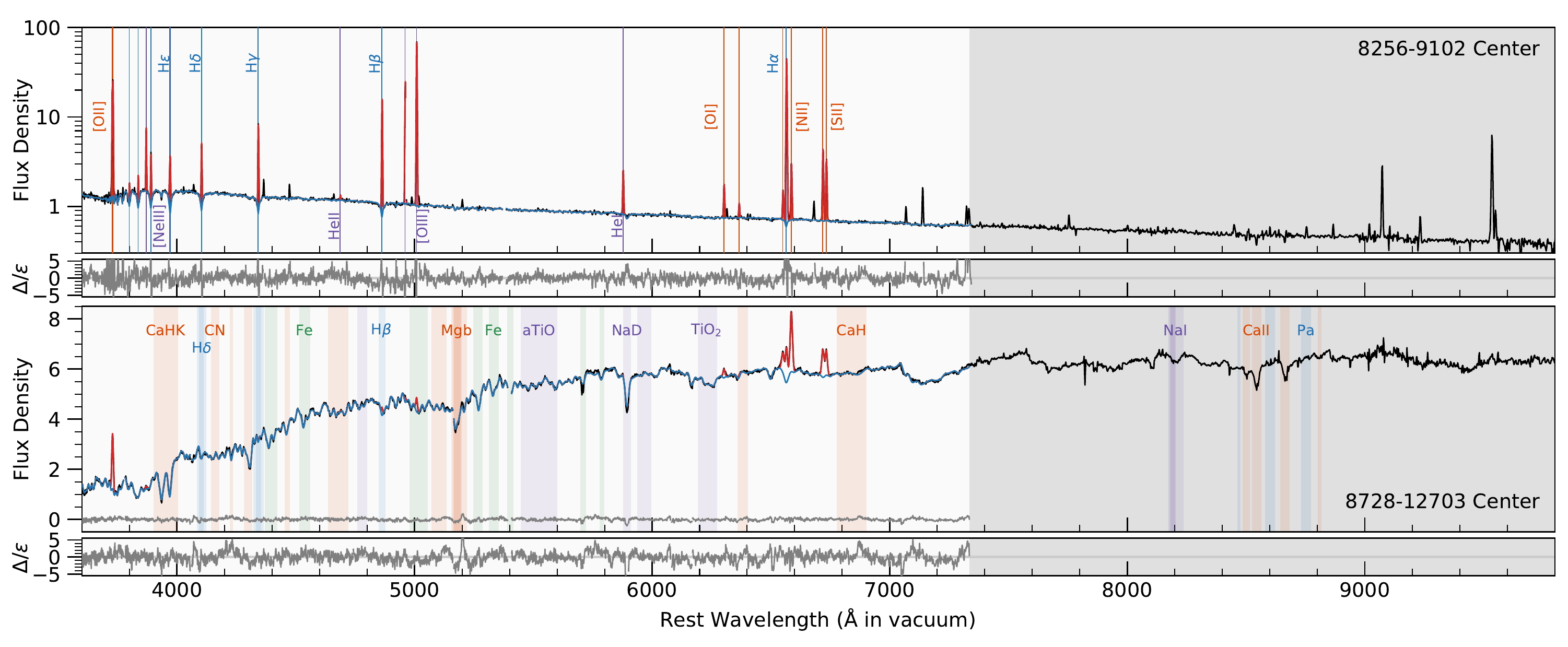}
\caption{The central spaxel of datacubes {\tt 8256-9102} (top) and
{\tt 8728-12703} (bottom) and the best-fitting \dap\ model spectra.
Flux densities are plotted in units of $10^{-17}$
erg/s/cm$^2$/\AA/spaxel. In both panels, the observed spectrum is
shown in black, the best-fitting model spectrum (stellar continuum
plus emission lines) is shown in red, and the stellar-continuum-only
model is shown in blue. The fit residuals are shown directly for {\tt
8728-12703}, and also shown in separate panels for both spectra after
normalizing by the spectral errors, $\Delta/\epsilon$. The gray region
at $\lambda \gtrsim 7340$~\AA\ is not included in any full-spectrum
fit as the stellar spectral templates used for DR15 have no coverage
at these wavelengths. The 22 emission lines fit in DR15 (Table
\ref{tab:emldb}) are labeled and marked in the top panel by vertical
lines colored according to the groups identified in Section 5.3 of
Belfiore et al., {\it accepted}. The primary passbands of the 43
absorption-line indices measured in DR15 (Table \ref{tab:indexdb})
are shown against the {\tt 8728-12703} spectrum: Hydrogen bands are
marked in blue; C, N, Ca, or Mg bands are in orange; Fe bands are in
green; and Na or TiO bands are in purple. A few bands are labeled
according to their name or the element present in their name.}
\label{fig:showcasespec}
\end{figure*}

To minimize the complexity of the output \dap\ data model,
measurements made on binned spectra are mapped to each spaxel in the
bin. This can be seen in the maps of the stellar kinematics shown in
Figure \ref{fig:showcasemaps}, where regions of constant stellar
velocity are the visual result of all the relevant spaxels being
binned into a single spectrum during the fitting process.
Alternatively, the remaining maps all show quantities varying
spaxel-to-spaxel. If the maps in Figure \ref{fig:showcasemaps} were
instead from the analysis that only used the Voronoi-binned spectra
(not the hybrid approach), {\it all} maps would show identical
regions of constant values. Although they are convenient for visual
inspection and for the simplicity of the data model, it is important
to identify and select only the unique measurements for detailed
analysis (see other data-model-specific advice in Section
\ref{sec:output}).

Inevitably, visual inspection of the \dap\ maps will lead one to find
features that are physically counter-intuitive and/or erroneous; we
discuss some of these cases in Section \ref{sec:outliers}. When in
doubt about features seen in the \dap\ maps, there is no substitute
for directly inspecting the MaNGA spectral data \citep[Section
\ref{sec:data};][]{2016AJ....152...83L} and the associated \dap\
model spectra. The latter are provided by the second main \dap\
output file, a multi-extension fits file called the model \dapcube\
file (Section \ref{sec:cubefile}). This file contains all the model
spectra fit to the observations, or provides the information needed
to reconstruct the models (see point 2 in Section
\ref{sec:cubefile}). Our \marvin\ software package and web-based
interface \citep{2019AJ....158...74C} provide particularly useful
tools for this kind of data inspection.

Figure \ref{fig:showcasespec} shows two high-\snrg\ spectra and the
best-fitting \dap\ model spectra.  Note that the full model (stellar
continuum and emission lines) is shown in red with the
stellar-continuum-only model overlaid in blue; except for the emission
features, therefore, the two are identical such that the model appears
blue in most of the Figure.  The model spectra are the result of the two
full-spectrum fits in the \dap, both of which use \ppxf.  The first fit
masks the emission lines and determines the best fit to the stellar
continuum to measure the stellar kinematics (Section
\ref{sec:stellarkin}).  The second fit simultaneously optimizes the
stellar continuum and emission lines while keeping the stellar
kinematics fixed to the result from the first fit (Section
\ref{sec:emlfit}).  The stellar continuum is handled differently between
these two fits and yield slightly different results; the continuum shown
in Figure \ref{fig:showcasespec} is the result of the combined fit of
the second full-spectrum fit (cf.\ Belfiore et al., {\it accepted},
Figure 2).

Note that the wavelength range fit by the \dap\ full-spectrum-fitting
modules is limited to 0.36--0.74 $\mu m$ for DR15 because of the
spectral range of the templates used (Section \ref{sec:mileshc}).  This
has two primary effects.  First, it limits the spectral range over which
we can fit emission lines.  Most notably this excludes modeling of the
near-infrared [\ion{S}{3}] lines.  For this purpose, in particular, we
aim to soon take advantage of our in-house stellar library,
\mastar\footnote{
\url{https://www.sdss.org/surveys/mastar/}}
\citep{2018arXiv181202745Y}, so that our continuum models are fit over
MaNGA's full spectral range.  Second, while we can measure spectral
indices at all wavelengths, we can only calculate the
velocity-dispersion corrections (Section \ref{sec:indexcorr}) for those
measurements in regions with valid model fits.  This means any spectral
index provided in DR15 with a main passband centered at $\lambda >
0.74\mu m$ does not include a velocity-dispersion correction, which can
be critical when, e.g., analyzing absorption-line strengths as a
function of galaxy mass.

The two galaxy spectra in Figure \ref{fig:showcasespec} were selected to
illustrate the features fit by the \dap.  The central spaxel of datacube
{\tt 8256-9102} for the star-forming galaxy 1-255959 has extremely
bright nebular emission with nearly all 22 emission lines measured in
DR15 (Sections \ref{sec:emlfit} and \ref{sec:bandpass}; Table
\ref{tab:emldb}) identifiable by eye.  Indeed, many more emission lines
are visible that are not currently fit by the \dap, largely from H and
He recombination and N, O, S, and Ar forbidden transitions.  We expect
to add to the list of lines included in the fit in future releases of
the \dap, both by extending the spectral range of the continuum models
(cf.\ Belfiore et al.\ {\it accepted}) and controlling for any problems
caused by attempting to fit what are generally much weaker lines.  The
central spaxel of datacube {\tt 8728-12703} of the early-type galaxy
1-51949 has relatively weak emission features but exhibits many of the
absorption features measured by the 46 spectral indices provided in DR15
(Section \ref{sec:spindex} and Table \ref{tab:indexdb}).


\subsection{Usage Guidance}
\label{sec:generalusage}

Anyone planning to use the \dap\ data is strongly encouraged to read
Section \ref{sec:output}. There we describe the main output files
provided by the \dap\ and we highlight a number of aspects of the data
important to their use. Some of these are simple practicalities of the
data model, but others are critical to the proper interpretation of the
data. Basic introductions and usage advice for the \dap\ data products
are also included in the DR15 paper \citep{2019ApJS..240...23A} and
data-release website, \url{https://www.sdss.org/dr15/}. From the latter,
note the following in particular: an introduction to working with MaNGA
data (\url{https://www.sdss.org/dr15/manga/getting-started/}) and some
of the intricacies involved
(\url{https://www.sdss.org/dr15/manga/manga-data/working-with-manga-data/}),
worked tutorials
(\url{https://www.sdss.org/dr15/manga/manga-tutorials/}, a list of known
problems and caveats
(\url{https://www.sdss.org/dr15/manga/manga-caveats/}), and the general
SDSS helpdesk (\url{https://www.sdss.org/dr15/help/}).

In terms of its general success in fitting MaNGA spectra, Section
\ref{sec:performance} provides a useful reference if one encounters a
missing \dap\ product or a counter-intuitive measurement. In particular,
Section \ref{sec:outliers} provides a list of regimes where fit-quality
metrics have been used to identify aberrantly poor spectral fits
produced by the \dap.

Guidance for each of the three primary \dap\ product groups (stellar
kinematics, emission-line properties, and continuum spectral indices)
are provided in, respectively, Section \ref{sec:stellarkin}, Belfiore et
al., {\it accepted}, and Section \ref{sec:spindex}. For the {\bf stellar
kinematics}, we particularly recommend Section \ref{sec:svdusage}, which
provides guidance for how to use our stellar velocity dispersion
measurements. For the {\bf emission-line measurements}, we recommend
that users read Sections 6 and 7 of Belfiore et al., {\it accepted}, at
least, and then follow-up with other Sections of their paper as
relevant. Finally, for the {\bf spectral indices}, we recommend users
read the summary of the assessments we have performed herein, provided
in Section \ref{sec:sisummary}.

\subsection{Known Issues in DR15 \dap\ Products}
\label{sec:known_issues}

For general reference, below we provide a list of known issues with the
DR15 version of the \dap\ software and the \dap-derived data products.
Where relevant, we provide references to subsequent Sections of our
paper with more information; see also
\url{https://www.sdss.org/dr15/manga/manga-caveats/}. More up-to-date
information and documentation of source code changes are included in the
source-code distribution, see
\url{https://github.com/sdss/mangadap/blob/master/CHANGES.md}.
\begin{enumerate}
\item Detailed assessments of the uncertainties provided for the
non-parametric emission-line measurements have not been performed,
meaning their accuracy is not well characterized. We expect they are of
similar quality to the spectral-index uncertainties (Section
\ref{sec:sirepeat}); however, they should be treated with caution.
\item Some Milky Way foreground stars that fall within MaNGA galaxy
bundles have not been properly masked. More generally, the \dap\ does
not correctly handle the presence of multiple objects in the IFU bundle
field-of-view (Section \ref{sec:outliers}).
\item The wings of particularly strong or broad emission lines will not
have been properly masked during the stellar-continuum fits used to
measure the stellar kinematics (Section \ref{sec:stellarkinmask}).
\item The {\tt MASK} extension in the model \dapcube\ files cannot be
used to reconstruct the exact mask resulting from the stellar-kinematics
module.
\item The $\chi^2_\nu$ measurements reported for the stellar-kinematics
module are not correct; they do not exclude pixels that were rejected
during the fit iterations.
\item The highest order Balmer line fit by the \dap\ is H$\theta$, even
though many galaxies show higher-order lines (Figure
\ref{fig:showcasespec}).
\item Measurements for the H$\zeta$ Balmer line are unreliable given its
blending with the nearby \ion{He}{1} line (as reported by Belfiore et
al., {\it accepted}).
\item The velocity-dispersion measurements for the [\ion{O}{2}] line are
improperly masked (also reported by Belfiore et al., {\it accepted}).
\item The \dapall\ file (Section \ref{sec:dapall}) reports spectral
indices within 1 $R_e$ that have {\it not} been corrected for the
observed stellar velocity dispersion (Section \ref{sec:indexcorr}).
\item Velocity dispersion corrections for index measurements will
include velocity effects because all measurements are done using the
single bulk redshift to offset the band definitions (Section
\ref{sec:spindex}).
\end{enumerate}

\section{\dap\ Inputs}
\label{sec:data}

\subsection{MaNGA Spectroscopy}

\citet{2015AJ....149...77D} provide a detailed description of the
MaNGA fiber-feed system, which is composed of 17 integral-field units
(IFUs): two 19-fiber IFUs, four 37-fiber IFUs, four 61-fiber IFUs,
two 91-fiber IFUs, and five 127-fiber IFUs. The plate scale of the
2.5-meter Sloan telescope yields an on-sky fiber diameter of
$2\arcsec$. The combination of the seeing conditions at Apache Point
Observatory (APO), the dithering pattern of the MaNGA observational
strategy \citep{2015AJ....150...19L}, and the method used to
construct the datacubes typically provides a spatial point-spread
function (PSF) with a FWHM of $\sim$$2\farcs5$
\citep{2016AJ....152...83L}. All of the IFUs have their fibers packed
in a hexagonal, regular grid with a field-of-view (FOV) directly
related to the number of fibers. Including the fiber cladding, the
nominal FOV diameters are $12\arcsec$, $17\arcsec$, $22\arcsec$,
$27\arcsec$, and $32\arcsec$ for the 19-, 37-, 61-, 91-, and
127-fiber IFUs, respectively.

The MaNGA fiber-feed systems are coupled to the SDSS-III/BOSS
spectrographs \citep{2013AJ....146...32S}, a pair of spectrographs
with ``blue'' and ``red'' cameras that receive, respectively,
reflected ($\lambda \lesssim 0.63 \mu{\rm m}$) and transmitted
($\lambda \gtrsim 0.59 \mu{\rm m}$) light from a dichroic
beamsplitter. The full spectral range obtained for each fiber
spectrum is 0.36 $\mu$m $\lesssim \lambda \lesssim 1.03 \mu{\rm m}$
after combining the data from both cameras. Each arm of each
spectrograph uses a volume-phase holographic grism yielding spectral
resolutions of $R_\lambda = \lambda/\Delta\lambda \approx 2000$ at
$\lambda = 0.55 \mu{\rm m}$ for the two blue cameras and $R_\lambda
\approx 2500$ at $\lambda = 0.9 \mu{\rm m}$ for the two red cameras
\citep[see][Figure 20]{2016AJ....152..197Y}.

Following the observational strategy outlined by
\citet{2015AJ....150...19L}, each MaNGA plate is observed using a
three-point dither pattern to fill the IFU interstitial regions and
optimize the uniformity of the FOV sampling for all 17 targeted
galaxies on a plate. Depending on the observing conditions, 2--3
hours of total observing time is required to reach the survey-level
constraints on the signal-to-noise (S/N) ratio, as defined by
\citet{2016AJ....152..197Y}.

These data are reduced by the MaNGA \drp, an {\tt IDL}-based software
package, yielding wavelength-, flux-, and astrometrically calibrated
spectra.  The reduction procedures are similar to those used by the
SDSS-III/BOSS pipeline,\footnote{
\url{https://www.sdss.org/dr15/spectro/pipeline/}}
but with significant adjustments as required by the MaNGA observations.
The \drp\ is described in detail by \citet{2016AJ....152...83L}, the
spectrophotometric calibration technique is described by
\citet{2016AJ....151....8Y}, and relevant updates to these procedures
for DR15 are discussed by \citet{2019ApJS..240...23A}.

The spectra from all four cameras are combined into a single spectrum
for each fiber resampled to a common wavelength grid. Spectra are
produced with both linear and log-linear wavelength sampling. All
spectra for a given {\tt PLATEIFU} designation are included in a
single file as a set of row-stacked spectra ({\tt RSS}) and as a
uniformly sampled datacube ({\tt CUBE}). The MaNGA datacubes are
constructed by regridding the flux in each wavelength channel to an
on-sky pixel (spaxel) sampling of $0\farcs5$ on a side following the
method of \citet{Shepard1968}; see \citet[][Section
9]{2016AJ....152...83L} for details \citep[cf.][]{sanchez12,
2019arXiv190606369L}. The interpolating kernel is a two-dimensional
Gaussian with a standard deviation of $0\farcs7$ and a truncation
radius of $1\farcs6$. Although this interpolation process leads to
significant covariance between the spaxels in a given wavelength
channel \citep[][Section 9.3]{2016AJ....152...83L}, the current \dap\
release is primarily focused on working with these resampled
datacubes. Also, the \dap\ currently only analyzes the spectra that
are sampled with a log-linear step in wavelength of
$\Delta\log\lambda = 10^{-4}$ (corresponding to a velocity scale of
$\Delta V = 69$ \kms).

\subsection{Photometric Metadata}
\label{sec:phot}

For convenience, the \dap\ uses measurements of the ellipticity
($\epsilon = 1-b/a$) and position angle ($\phi_0$) of the $r$-band
surface-brightness distribution to calculate the semi-major-axis
elliptical polar coordinates, $R$ and $\theta$, where the radius is
provided in arcseconds as well as in units of the effective (half-light)
radius, $R_e$.  In the limit of a tilted thin disk, these are the
in-plane disk radius and azimuth.  Except for some targets from MaNGA's
ancillary programs, the photometric data are taken from the parent
targeting catalog described by \citet[][Section 2]{2017AJ....154...86W}.
This catalog is an extension of the NASA-Sloan Atlas (NSA)\footnote{M.
Blanton; \url{www.nsatlas.org}} toward higher redshift ($z\lesssim0.15$)
and includes an elliptical Petrosian analysis of the surface-brightness
distributions.  Despite this difference with respect to the NSA catalog
provided by the catalog website, we hereafter simply refer to this
extended catalog as the NSA.  The primary advantage of the NSA is its
reprocessing of the SDSS imaging data to improve the sky-background
subtraction and to limit the ``shredding'' of nearby galaxies into
multiple sources \citep{2011AJ....142...31B}.

\begin{figure*}
\begin{center}
\includegraphics[width=0.9\textwidth]{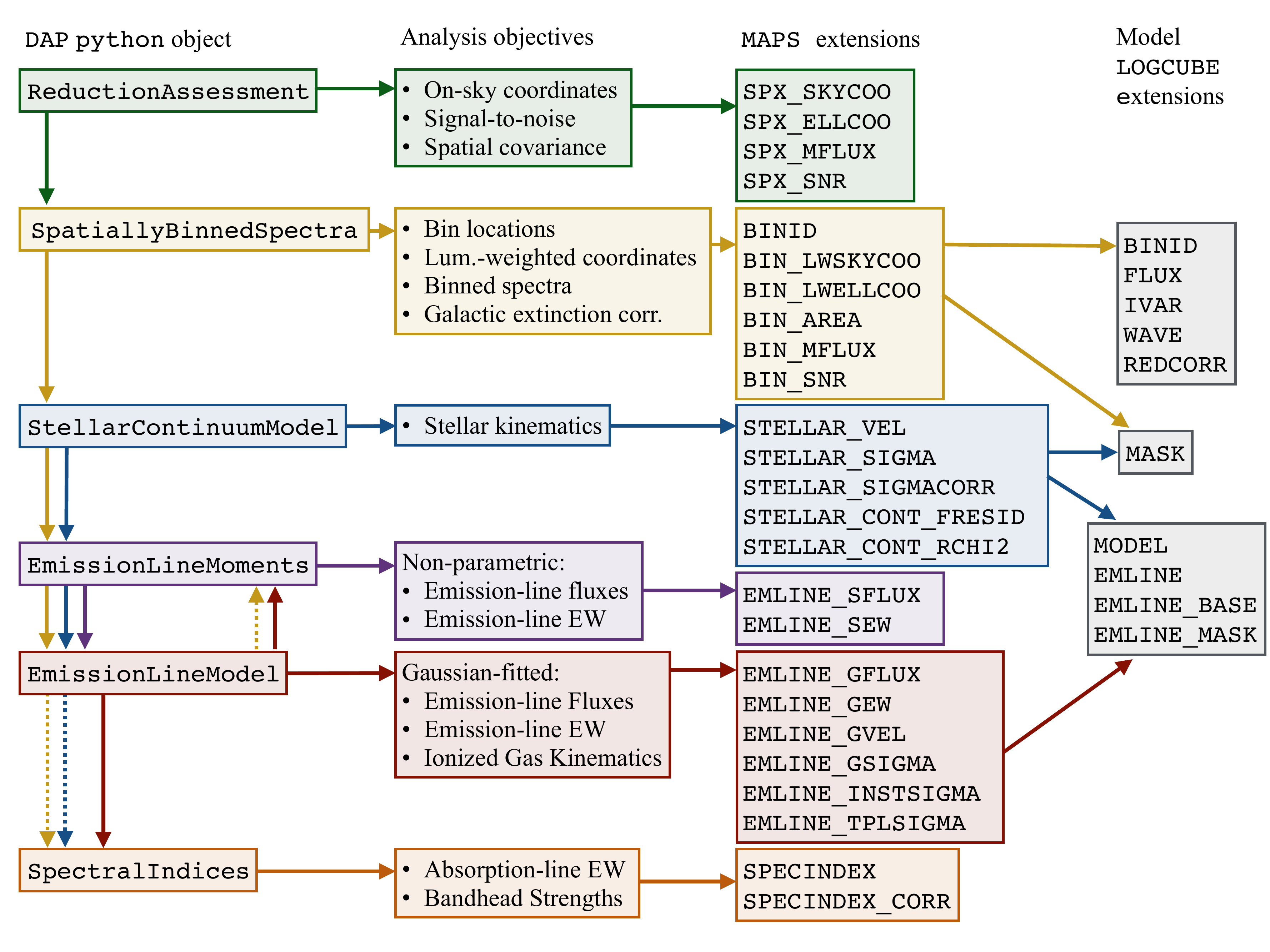}
\end{center}
\caption{Schematic diagram of the \dap\ workflow. From left-to-right,
the schematic provides the relevant {\tt python} modules, the
analysis objectives of each module, and the associated \dapmaps\ and
model \dapcube\ extensions generated by each module, as indicated by
the arrows and colors. The {\tt python} modules, contained within the
named \dap\ {\tt python} objects, are ordered from top to bottom by
their execution order; an exception to this is that the emission-line
moments are computed both before and after the emission-line modeling
(see Sections \ref{sec:emlfit} and \ref{sec:emlmom}), as indicated by
the two sets of arrows pointing toward the {\tt EmissionLineMoments}
module. Arrow directions indicate the execution order and colors
indicate the module dependencies. For example, the {\tt
EmissionLineMoments} object depends on the results of both the {\tt
SpatiallyBinnedSpectra} object and the {\tt StellarContinuumModel}
object. The dashed arrows indicate {\it conditional} dependencies.
For example, the {\tt EmissionLineModel} deconstructs the bins in the
hybrid-binning approach, such that the {\tt SpectralIndices} are
independent of the primary results of the {\tt
SpatiallyBinnedSpectra}. However, there is an explicit dependence of
the {\tt SpectralIndices} on the {\tt SpatiallyBinnedSpectra} when
the hybrid-binning approach is not used.}
\label{fig:workflow}
\end{figure*}

\section{Workflow}
\label{sec:workflow}

At the survey level, the \dap\ is executed once per \drp\ datacube
({\tt PLATEIFU}). The \dap\ will attempt to analyze any datacube
produced by the \drp, as long as it is an observation of a galaxy
target that has an initial estimate of its redshift. Specifically, we
only analyze observations selected from the \drpall\ file
\citep{2016AJ....152...83L} for galaxies in either the main MaNGA
survey or its ancillary programs\footnote{
\url{https://www.sdss.org/dr15/manga/manga-target-selection/ancillary-targets/}}
(respectively, either {\tt mngtarg1} or {\tt mngtarg3} are non-zero)
and with a redshift of $cz > -500$ \kms. The restriction on the
redshift is required because of the $\pm$2000 \kms\ limits we impose
in the fit of the kinematics for each observation (Sections
\ref{sec:stellarkin} and \ref{sec:emlfit}); we allow galaxies with a
small blueshift so that the \dap\ will analyze a few observations of
local targets from ancillary programs. Importantly, the \dap\ will
analyze datacubes that the \drp\ has marked as critical failures
({\tt DRPQUAL} is {\tt CRITICAL} in the \drpall\ file); the
appropriate flag is propagated to the global \dap\ quality bit ({\tt
DAPQUAL}; see Appendix \ref{sec:datamodel}). As noted by
\citep[][Section B.4]{2016AJ....152...83L}, datacubes marked with
{\tt CRITICAL} quality flags should be used with caution or simply
omitted from use. However, the approach to flagging reductions as
{\tt CRITICAL} is purposely conservative, meaning that some of these
reductions may yet be valid. Therefore, we simply include them in our
\dap\ analysis but caution users similarly concerning their use. The
most conservative approach is for users to ignore data marked as {\tt
CRITICAL} by MaNGA quality bits. In total, the \dap\ has analyzed
4731 observations for DR15.

The primary \drp-produced output passed to the \dap\ for analysis are
the MaNGA datacubes that are sampled logarithmically in wavelength
(i.e., the \drp\ {\tt LOGCUBE} files).\footnote{
When needed for covariance calculations (Section \ref{sec:binning}), the
{\tt LOGRSS} files are also used following a computation identical to
what the \drp\ uses to produce the $griz$ covariance matrices provided
in DR15 \citep{2019ApJS..240...23A}.}
The \dap\ uses two additional text files to set its execution
procedures: The first provides the photometric and redshift data for
each target, which are most often drawn from the NSA (Section
\ref{sec:phot}).  The second defines a set of ``analysis plans'' that
are executed in sequence.  We refer to each analysis plan as the {\tt
DAPTYPE} of a given output data set.  An analysis plan is composed of a
set of keywords that select preset configurations of the low-level
parameters that dictate the behavior of each \dap\ module, with one
keyword per primary module.  The six primary modules of the \dap\ have
the following analysis goals: (1) perform basic data-quality assessments
and calculations using the \drp-produced data (Sections \ref{sec:prelim}
and \ref{sec:snr}), (2) spatially bin the \drp\ datacube (Sections
\ref{sec:vorbin} and \ref{sec:stacking}), (3) measure the stellar
kinematics (Section \ref{sec:stellarkin}), (4) use bandpass integrals to
compute non-parametric moments of the emission lines (Section
\ref{sec:emlmom}; Belfiore et al.\ {\it accepted}), (5) fit parametric
models to the emission lines (Section \ref{sec:emlfit}), and (6) use
bandpass integrals to compute a set of absorption-line and bandhead
indices (Section \ref{sec:spindex}).

Figure \ref{fig:workflow} provides a schematic of the \dap\ workflow
through its six modules.  The modules are executed in series, from top
to bottom in the Figure, with each module often depending on the results
of all the preceding modules.  All modules are executed once per {\tt
DAPTYPE}, with the exception of the emission-line moment calculation
(the fourth module), which is run once before and once after the
emission-line model fitting (see Sections \ref{sec:emlfit} and
\ref{sec:bandpass} for more detail).  The analysis objectives of each
module are also listed in the Figure.

The results of each module are saved in its ``reference file''. These
reference files: (1) allow the \dap\ to reuse (as opposed to
re-compute) analysis results common to multiple analysis plans (e.g.,
using the same data-quality assessments from the first module with
different binning schemes), (2) allow the \dap\ to effectively
restart at the appropriate module in case of a failure, and (3)
provide access to a more extensive set of data beyond what is
currently propagated to the two main output files, the \dapmaps\ and
model \dapcube\ files. The reference files are released as part of
DR15 with their data model documented at the DR15 website;\footnote{
\url{https://www.sdss.org/dr15/manga/manga-data/data-model}}
however, we do not expect most users to interact with these files.
Instead, much of the data in the reference files is consolidated into
specific extensions of the two primary \dap\ output files, the
\dapmaps\ and model \dapcube\ files, as indicated in Figure
\ref{fig:workflow}. A complete description of the two main \dap\
output files is provided in Section \ref{sec:output} with the data
models given in Appendix \ref{sec:datamodel}.

Two of the six \dap\ modules, {\tt StellarContinuumModel} and {\tt
EmissionLineModel}, employ a full-spectrum-fitting approach, and both
of these modules use \ppxf\ \citep{2017MNRAS.466..798C}; see Sections
\ref{sec:stellarkin} and \ref{sec:emlfit}. Critical to the \ppxf\
procedure are the spectral templates used. The \dap\ repository
provides a number of spectral-template libraries that we have
collected over the course of \dap\ development (cf.\ Belfiore et al.
{\it accepted}, Section 4.1); however, the results provided for DR15
focus on a distillation of the MILES \citep{Sanchez-Blazquez2006,
FalconBarroso2011miles} stellar-template library using a hierarchical
clustering (HC) technique. We refer to the template library resulting
from this analysis as the \mileshc\ library, and we discuss the
generation of this library in full in Section \ref{sec:mileshc}.
Sections \ref{sec:binning} -- \ref{sec:spindex} discuss the details
of the algorithms used in the six main \dap\ modules.

\section{Hierarchical Clustering of a Spectral-Template Library}
\label{sec:mileshc}

To reduce computation time when using large stellar libraries as
templates, one generally tries to select subsamples of stars that are
representative of the entire library. For example, the execution time
for the \ppxf\ method, which the \dap\ uses to both measure the
stellar kinematics (Section \ref{sec:stellarkin}) and model the
emission lines (Section \ref{sec:emlfit}), is typically slightly
larger than $O(N_{\rm tpl})$ for $N_{\rm tpl}$ templates.
Distillation of the information content of a spectral library into a
minimal number of templates can therefore be critical to meeting the
computational needs of large-scale surveys like MaNGA.

One way to sub-sample a library is to select stars that uniformly
sample a grid in known stellar physical parameters, like effective
temperature ($T_{\rm eff}$), metallicity ([Fe/H]), and surface
gravity ($g$) \citep[e.g.,][]{Shetty2015}. A disadvantage of this
approach is that stellar parameters may not always be available and
are not necessarily a direct proxy of all relevant variation in the
spectral information provided by the library. Alternatively, one
could use a principle-component analysis to isolate the eigenvectors
of the full stellar library \citep[cf.][]{Chen2012pc}. However, one
then loses the ability to enforce positivity constraints on the
weights of the templates when modeling the galaxy spectra, which is a
useful prior for reducing unphysical results at low S/N. Although
more complex methods exist(e.g., Non-Negative Matrix Factorization,
NMF; \citealt{lee1999nmf}; cf.\ \citealt{Blanton2007nmf}), we have
adopted a simple approach that sufficiently avoids the limitations of
the above alternatives and is generally applicable to any
spectral-template library.

The key idea of our method is to apply a clustering algorithm
\citep{jain1999clustering} to the $N_{\rm tpl}$ spectral templates
composed of $M$ spectral channels by treating them as $N$ vectors in an
$M$-dimensional space.\footnote{
One can think of a vast range of practical implementations of this
general idea, given the large number of solutions that were proposed for
the clustering problem; however, our simple approach has proven
reasonable for our purposes, if not necessarily optimal.}
For DR15, we have adopted a hierarchical-clustering approach
\citep{johnson1967hierarchical} due to its simplicity, availability of
robust public software, and the limited number of tuning
parameters.\footnote{
The Python code used for generating the library is available at
\url{https://github.com/micappe/speclus}.}
We have applied our approach to the full MILES stellar library\footnote{
We used MILES V9.1 available at \url{http://miles.iac.es/.}}
\citep{Sanchez-Blazquez2006,FalconBarroso2011miles} of 985 stars by
defining the ``distance'' between two spectra, $S_j$ and $S_k$, as
\begin{equation}
d_{jk} = \frac{2\delta(S_j-S_k)}{\rm \langle S_j\rangle},
\label{eq:hcdistance}
\end{equation}
where $2\delta(S_j-S_k)$ is a robust estimate of the standard
deviation of the residual, computed as one half of the interval
enclosing 95.45\% of the residuals, in a \ppxf\ fit of spectrum $S_k$
using spectrum $S_j$ as the template. For this exercise, we include
an eighth-order additive Legendre polynomial in the \ppxf\ fits to be
consistent with the method used when fitting the stellar kinematics
of the galaxy spectra in the \dap\ (Section \ref{sec:stellarkin}).
The individual elements of $d_{jk}$ from Equation \ref{eq:hcdistance}
are used to construct a distance matrix for input to a
hierarchical-clustering algorithm.\footnote{
Specifically, we use the {\tt scipy} \citep{Scipy2001} function
\texttt{cluster.hierarchy.linkage} with \texttt{method='average'}.  This
implements the nearest-neighbors chain hierarchical-clustering algorithm
described by \citet{mullner2011clustering}.}

\begin{figure}
\begin{center}
\includegraphics[width=1.0\columnwidth]{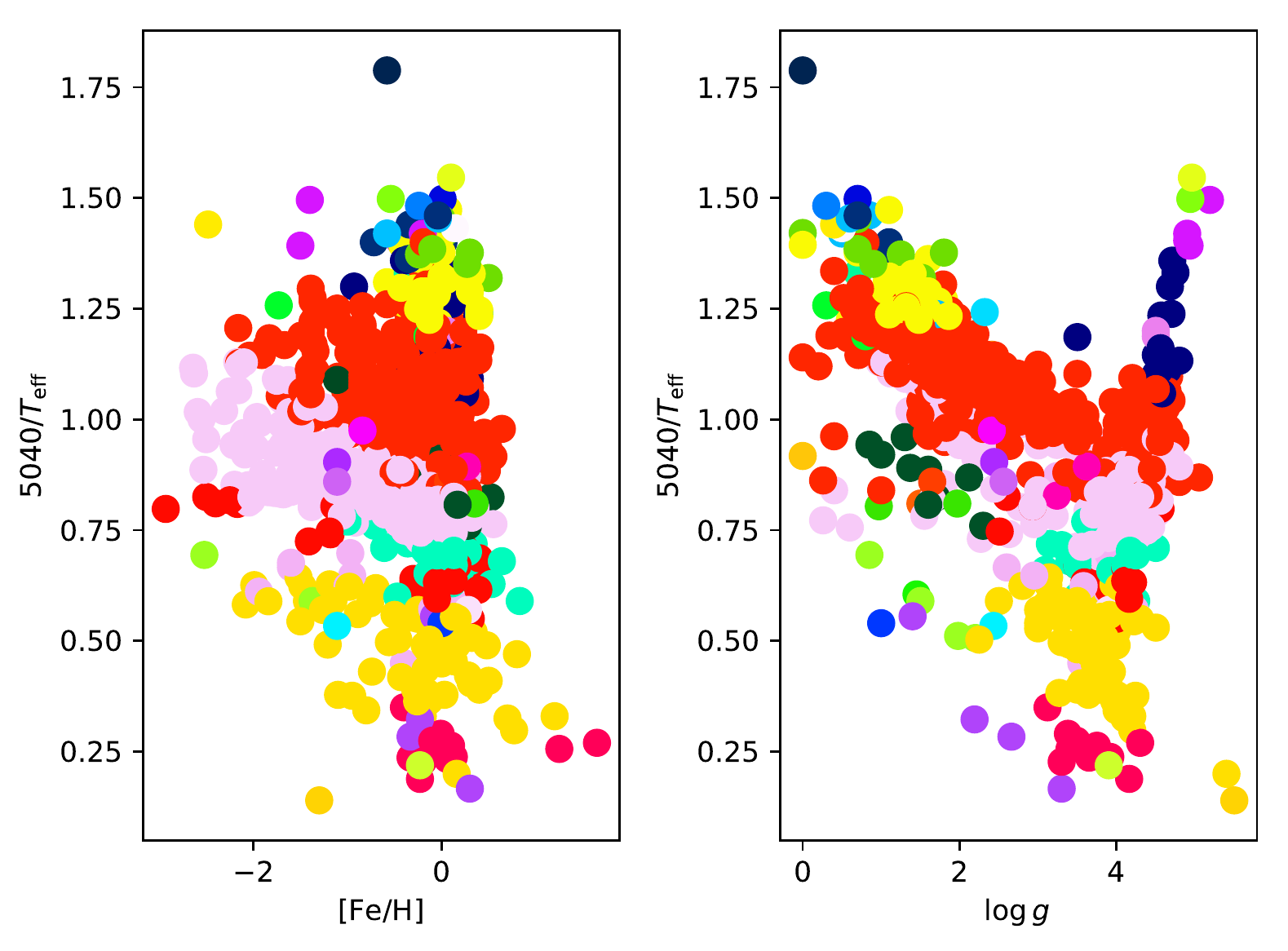}
\end{center}
\caption{Effective temperature $T_{\rm eff}$, metallicity [Fe/H], and
surface gravity $g$ of the stars in the MILES stellar library
\citep{FalconBarroso2011miles}. Each datum is assigned a color based
on its assigned cluster from the hierarchical-clustering algorithm
(Section \ref{sec:mileshc}). Some clusters contain a single star,
whereas others include about a hundred stars. Cluster boundaries
generally do not follow lines of constant stellar parameter due to
the degeneracy between the three physical parameters.}
\label{fig:miles_clusters}
\end{figure}

We form flat clusters such that the cluster constituents have a maximum
distance of $d_{\rm max}$; lower values of $d_{\rm max}$ yield a larger
number of flat clusters.  To construct the spectral templates for the
distilled library, we normalize each MILES spectrum to a mean of unity
and then average all the spectra in each cluster without weighting.  We
refer to the result of our hierarchical clustering of the MILES stellar
library as the \mileshc\ library throughout this paper.  We have
optimized $d_{\rm max}$ by comparing \ppxf\ fits of high-S/N MaNGA
spectra from a few representative young/old galaxies using either the
full set of 985 MILES stars or the \mileshc\ library produced by the
given iteration of $d_{\rm max}$.

\begin{figure*}
\begin{center}
\includegraphics[width=0.9\textwidth]{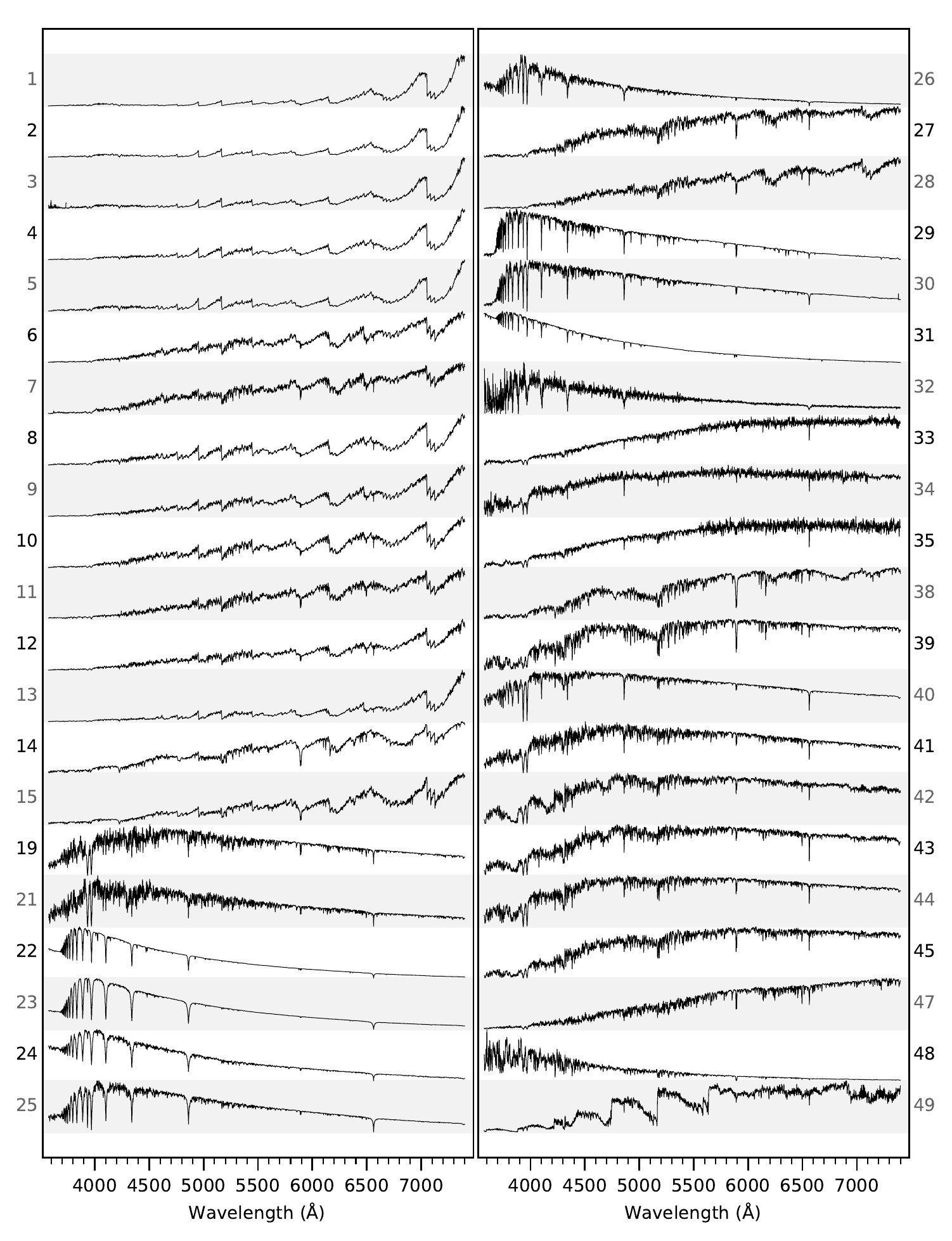}
\end{center}
\caption{ The 42 templates in the \mileshc\ library identified along
the left or right ordinate by their cluster group number. Missing
clusters in the sequence (e.g., cluster 16) were removed because of
low S/N or the presence of emission lines from flaring late-type
stars. The spectra are ordered by their cluster number, which is not
generally related to the mean stellar parameters of the cluster
constituents. }
\label{fig:mileshc}
\end{figure*}

For $d_{\rm max} = 0.05$,\footnote{
Specifically, we use the {\tt scipy} function
\texttt{cluster.hierarchy.fcluster} with \texttt{criterion='distance'}
and a \texttt{threshold} of 5\%.}
our analysis yields 49 clustered spectra from the full MILES stellar
library of 985 spectra.  The number of spectra assigned to each cluster
varies dramatically, from clusters composed of individual spectra to
others that collect hundreds of stars from the MILES library.  However,
as one would expect, the clusters tend to concentrate in regions of
stellar parameter space, as shown in Figure \ref{fig:miles_clusters}.
Although compelling from a perspective of stellar spectroscopy, the
details of the distribution are less important to our application than
whether or not the clustering has successfully captured the information
content of the full MILES spectral library relevant to our full-spectrum
fitting.

From the original set of 49 cluster spectra, we remove templates with
prominent emission lines or relatively low S/N (e.g., from clusters
composed of a single spectrum), leading to a final set of 42 spectra in
the \mileshc\ template library shown in Figure
\ref{fig:mileshc}.\footnote{
These spectra are made available through the \dap\ GitHub repository;
specifically,
\url{https://github.com/sdss/mangadap/tree/master/data/spectral\_templates/miles\_cluster}.}
We compare the stellar kinematics measured using the full MILES and
\mileshc\ libraries in Section \ref{sec:milesvsmileshc}. As one would
expect, the use of \mileshc\ yeilds a moderately worse fit, as
determined by the fit residuals and chi-square statistics; however, the
affect on the resulting kinematics is acceptable, particularly given the
gain of roughly a factor of 25 in execution time. Also, in their Section
4, Belfiore et al., {\it accepted}, compare the emission-line modeling
results when the stellar continuum is fit using the \mileshc\ library
and various simple-stellar-population (SSP) templates. The \mileshc\
library shows specific differences in the continuum shape and Balmer
absorption depths compared to the \citet[][BC03]{2003MNRAS.344.1000B}
library, given its lack of early-type (O) stars.  However, the quality
of the fits to the MaNGA spectra using \mileshc\ are generally no worse
than when using SSP templates.

\section{Spatial Binning}
\label{sec:binning}

Unbiased measurements of stellar kinematics require a minimum S/N
ratio, particularly for the stellar velocity dispersion. It is
therefore generally necessary to bin spectra by averaging neighboring
spaxels to meet a given S/N threshold. To bin for this purpose, we
use the adaptive spatial-binning scheme implemented by the Voronoi
algorithm of \citet{2003MNRAS.342..345C}.\footnote{
In DR15 specifically, we use the Python package {\tt vorbin} version
{\tt 3.1.3} found at \url{https://pypi.org/project/vorbin/}.}
The datacube construction scheme in MaNGA \citep[][Section
9]{2016AJ....152...83L} follows the method of \citet{Shepard1968}
\citep[cf.][]{sanchez12}, leading to significant covariance between
adjacent spaxels that must be accounted for when combining spaxel
data. Indeed, given that the Voronoi-binning algorithm is predicated
on meeting a minimum S/N, the success of the algorithm hinges on an
accurate calculation of the binned S/N. However, calculation of the
full covariance matrix in each datacube is prohibitively expensive,
prompting a few approximations in our approach.

\begin{figure}
\begin{center}
\includegraphics[width=1.0\columnwidth]{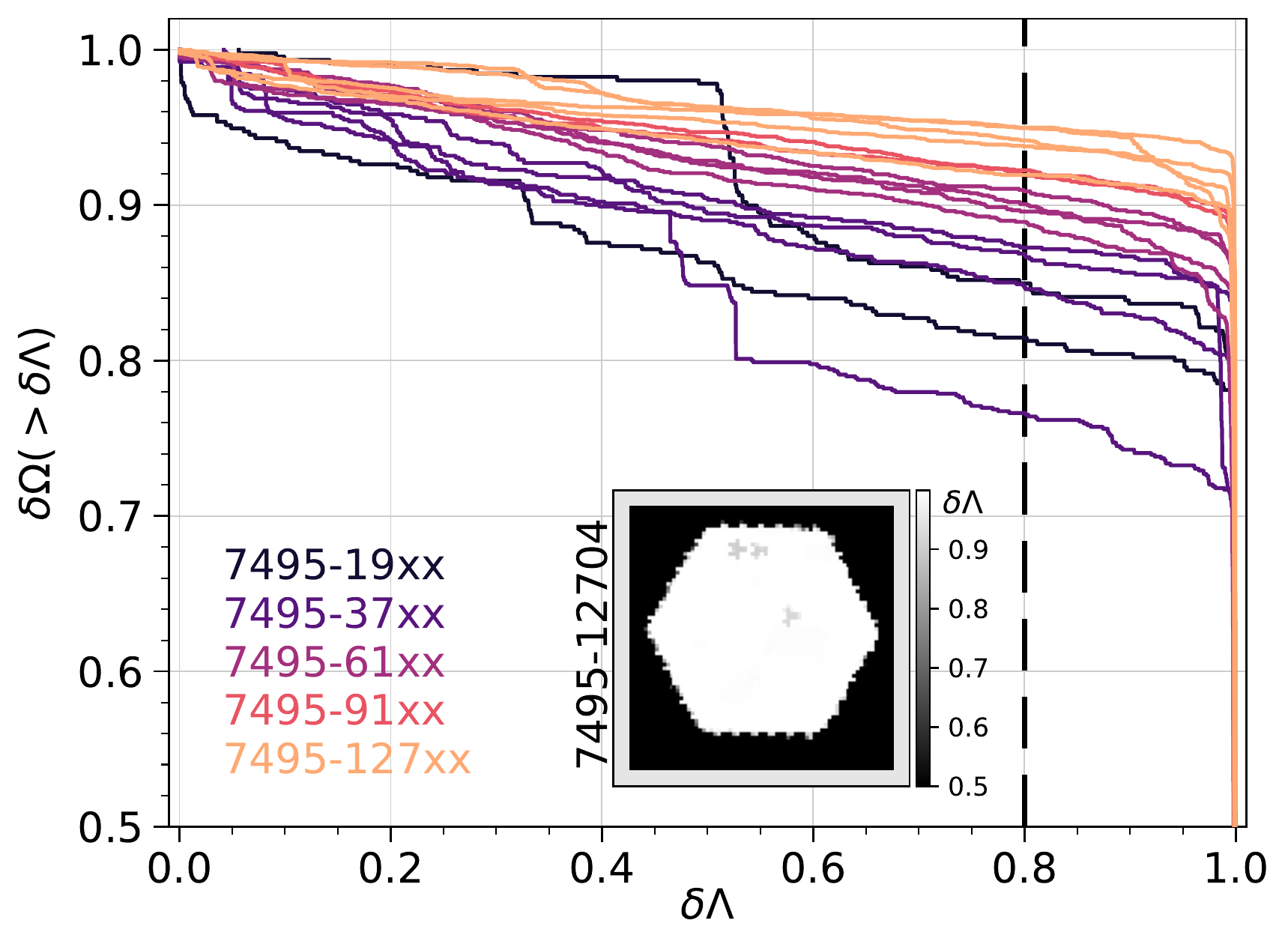}
\end{center}
\caption{The fraction of valid wavelength channels ($\delta\Lambda$)
over the full spectral range versus the fraction of spaxels
($\delta\Omega$; see the definition in Section \ref{sec:prelim}) with
at least $\delta\Lambda$ over the IFU field-of-view. Data are shown
for all 17 observations from plate 7495, colored by the IFU size. The
inset map shows $\delta\Lambda$ in each spaxel of the datacube for
observation 7495-12704: the hexagonal area with non-zero
$\delta\Lambda$ is surrounded by a buffer of spaxels with
$\delta\Lambda=0$ resulting from the datacube construction. The \dap\
only analyzes spaxels with $\delta\Lambda > 0.8$ (vertical dashed
line).}
\label{fig:goodspaxels}
\end{figure}

\begin{figure}
\begin{center}
\includegraphics[width=1.0\columnwidth]{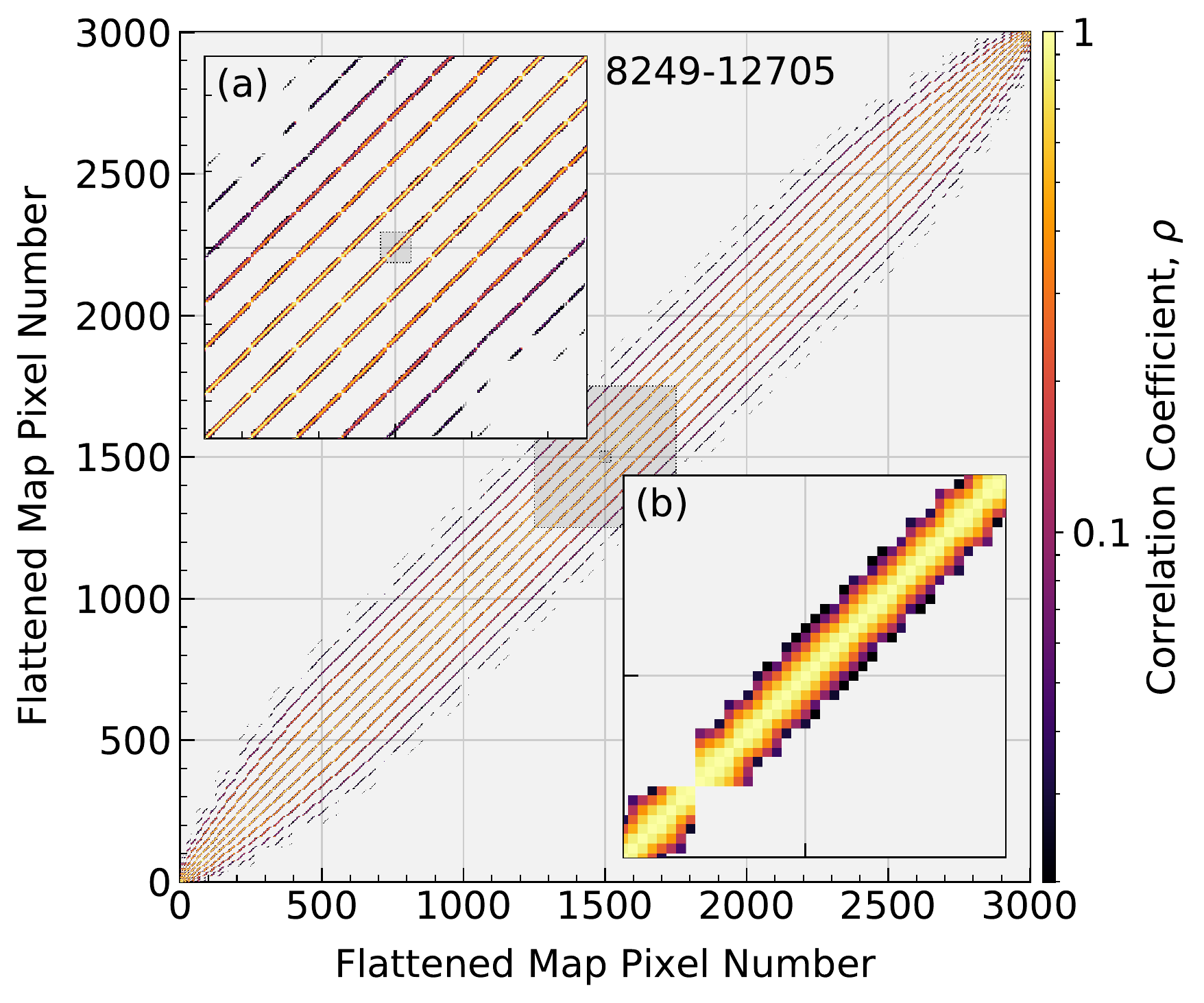}
\end{center}
\caption{
The correlation matrix in channel 1132 ($\lambda=4700$\AA) of datacube
{\tt 8249-12705} for all spaxels with \snrg$>$1.  The correlation
matrix is symmetric and has a correlation coefficient of $\rho = 1$
along the matrix diagonal, by definition.  The majority of the matrix is
empty with $\rho = 0$.  The inset panels provide an expanded view of two
matrix subregions:  Panel (a) shows $\pm$250 pixels around the matrix
center as indicated by the large gray box, and panel (b) shows $\pm$20
pixels around the matrix center as indicated by the small gray box, also
shown in panel (a).  The diagonal banding in panel (a) is an artifact of
the ordering of adjacent pixels in the flattened vector of the datacube
spatial coordinates; adjacent pixels are separated by the width of the
map in one on-sky dimension.  The number of bands in panel (a) roughly
matches the width in pixels along the main diagonal with non-zero $\rho$
in panel (b) as expected by the spatial correlation acting along both
on-sky dimensions.}
\label{fig:correlation}
\end{figure}

The following sections describe the first two modules of the \dap\
workflow (Figure \ref{fig:workflow}) that ultimately yield the binned
spectra used for the determination of the stellar kinematics.  The
distinction between these two modules is that the first is independent
of any specific binning algorithm (Sections \ref{sec:prelim} and
\ref{sec:snr}), whereas the second performs the binning itself (Sections
\ref{sec:vorbin} and \ref{sec:stacking}).  The incorporation of spatial
covariance when aggregating spaxels to meet a minimum S/N (Section
\ref{sec:vorbin}) and when propagating the uncertainties in the binned
spectra (Section \ref{sec:stacking}) are treated separately for
computational expediency.

\subsection{On-Sky Spaxel Coordinates and Datacube Mask}
\label{sec:prelim}

The \drp\ provides a World Coordinate System (WCS) for each datacube,
which the \dap\ uses to calculate the on-sky coordinates of each
spaxel relative to the target center. The target center is provided
in the datacube header with the keywords {\tt OBJRA} and {\tt
OBJDEC}.\footnote{
These are typically, but not always, the same as the pointing center of
the IFU given by the keywords {\tt IFURA} and {\tt IFUDEC}.}
The on-sky coordinates provided by the \dap\ are sky-right in
arcseconds, with positive right-ascension offsets toward the East;
note the abscissae in Figure \ref{fig:showcasemaps} increase from
right to left. The \dap\ then uses the photometric position angle and
ellipticity to calculate the semi-major-axis coordinates, $R$ and
$\theta$. For DR15, these are simply calculated and included in the
output \dapmaps\ file.

The \drp\ also provides detailed masks for each wavelength channel,
which the \dap\ uses to exclude measurements from analysis in any
given module. Large swaths of the full MaNGA spectral range can be
masked by the \drp\ because of broken fibers, known foreground-star
contamination, detector artifacts, or (in the majority of cases)
simply because the spaxel lies outside of the hexagonal IFU
field-of-view (FOV). The \dap\ excludes measurements affected by
these issues by ignoring any measurement flagged as either {\tt
DONOTUSE} or {\tt FORESTAR}. For each spaxel, we calculate the
fraction of the MaNGA spectral range, $\delta\Lambda$, that is viable
for analysis. For DR15, the \dap\ ignores any spaxel with
$\delta\Lambda < 0.8$. As a representative example, Figure
\ref{fig:goodspaxels} shows the viable fraction of spaxels,
$\delta\Omega$, with {\it any} valid flux measurement as a function
of $\delta\Lambda$ for the datacubes observed by plate {\tt 7495}.

\begin{figure}
\begin{center}
\includegraphics[width=1.0\columnwidth]{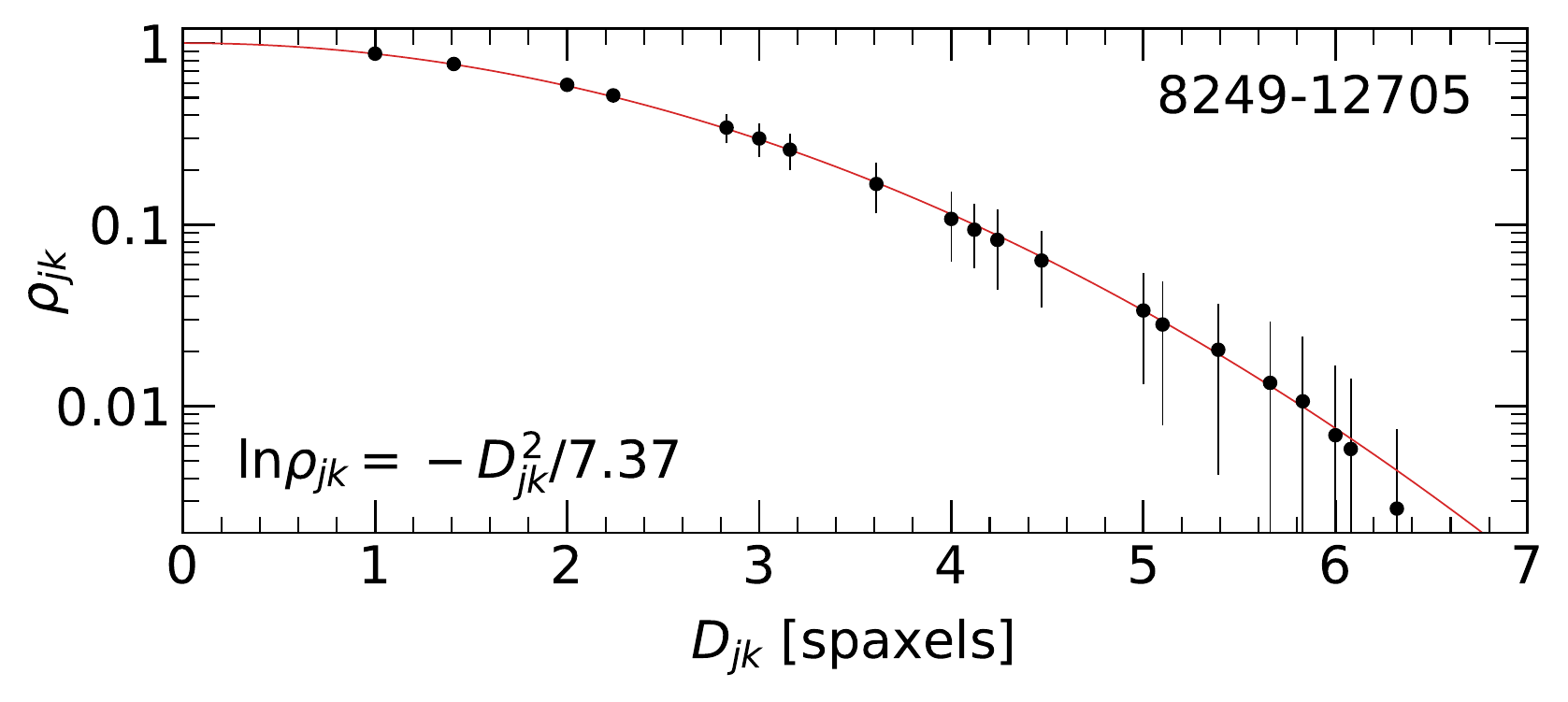}
\end{center}
\caption{The mean (points) and standard deviation (errorbars) of the
correlation coefficient, $\rho_{ij}$, for all spaxels within the
convex hull of the fiber-observation centers at channel 1132
($\lambda=4700$\AA) of datacube {\tt 8249-12705} as a function of the
spaxel separation, $D_{ij}$. The best-fitting Gaussian trend (red)
has a scale parameter of $\sigma = 1.92$ spaxels, leading to the
equuation provided in the bottom-left corner.}
\label{fig:rhogauss}
\end{figure}
    
\begin{figure*}
\begin{center}
\includegraphics[width=0.9\textwidth]{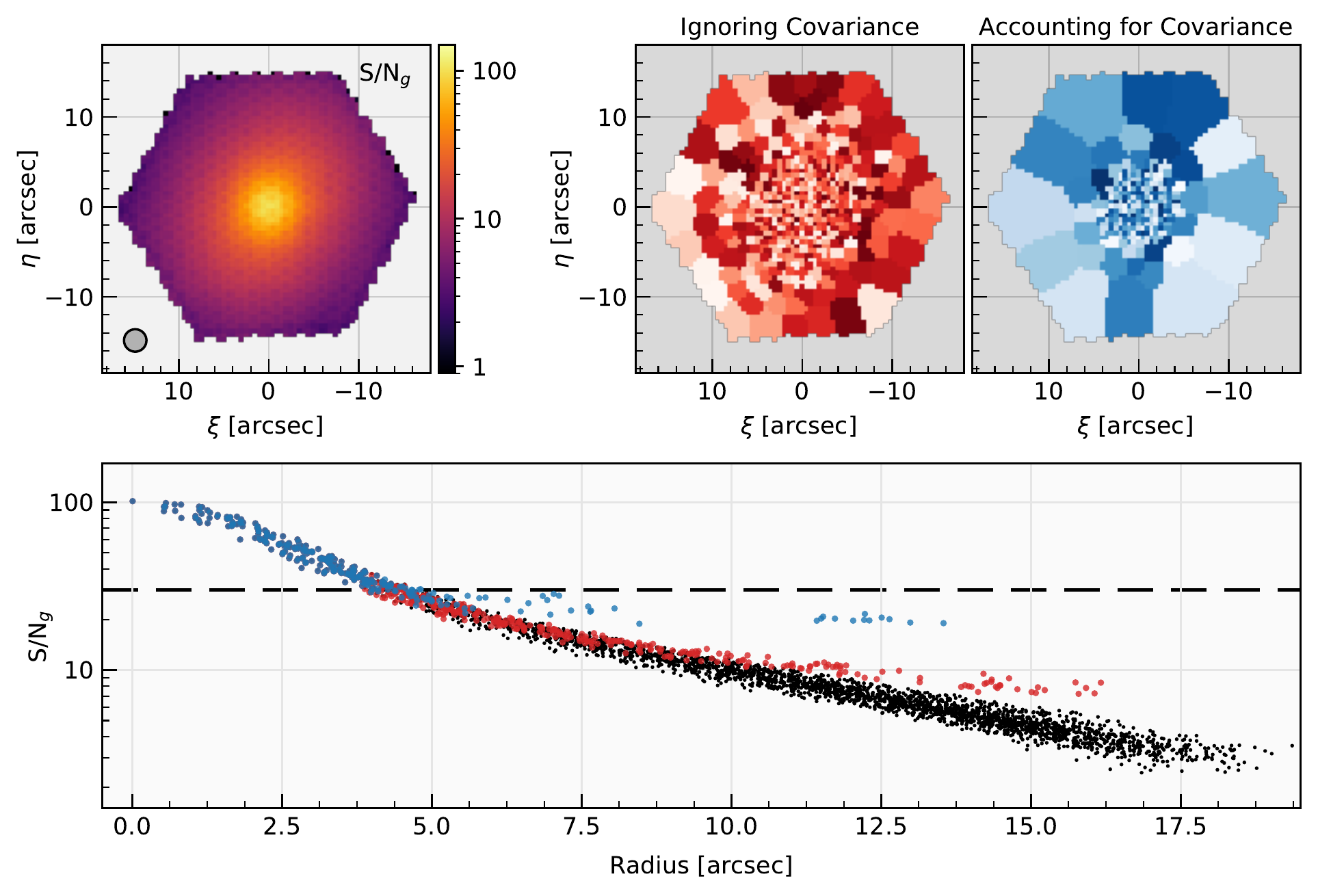}
\end{center}
\caption{Effect of spatial covariance on the result of the Voronoi
binning algorithm. The top-left panel shows the \snrg\ measurements
for datacube {\tt 8249-12705}. We then apply the Voronoi binning
algorithm to these data with a \snrg\ threshold of 30. The resulting
bin distribution that does not include the spatial correlation from
Figure \ref{fig:correlation} is shown in the top-middle panel, and
the bin distribution that does include the correlation is shown in
the top-right panel. The colors in the top-middle and top-right
panels are used to differentiate between spaxels in a given bin. The
bottom panel shows the formally correct \snrg\ as a function of
radius for the individual spaxels (black), the bins derived assuming
no covariance (red), and the bins that include the covariance (blue).
The Voronoi algorithm expect the red data to have \snrg$\sim$30 based
on the \snrg\ calculation that excludes covariances; however, the
formally correct \snrg\ is well below that.}
\label{fig:binning_covar}
\end{figure*}

\subsection{Spectral Signal-to-Noise and Spatial Covariance}
\label{sec:snr}

Both as a basic output product and for binning purposes, the \dap\
calculates a single measurement of S/N for each spaxel. In DR15, this
fiducial S/N --- hereafter referred to as \snrg\ --- is the average
S/N per wavelength channel weighted by the $g$-band response
function.\footnote{
Specifically, we use the response function produced by Jim Gunn in
2001 provided at
\url{https://www.sdss.org/wp-content/uploads/2017/04/filter\_curves.fits}
with the description the SDSS Survey imaging camera at
\url{https://www.sdss.org/instruments/camera/}.}
We calculate \snrg\ for all spaxels, excluding masked channels,
regardless of whether or not they meet our criterion of
$\delta\Lambda>0.8$ (Section \ref{sec:prelim}).

For this fiducial \snrg, we also calculate a single spatial
covariance matrix in two steps: (1) We calculate the spatial
correlation matrix for the wavelength channel at the
response-weighted center of the $g$-band following equation 7 from
\citet[][cf.\ Equation \ref{eq:bincovar}
herein]{2016AJ....152...83L}. We find that the spatial correlation
matrix varies weakly with wavelength over the $g$-band such that, to
first order, we can simply adopt the correlation matrix from this
single wavelength channel. (2) We renormalize the single-channel
correlation matrix by the mean variance in the flux over the $g$-band
to construct a covariance matrix.

Figure \ref{fig:correlation} provides the correlation matrix for
wavelength channel 1132 in datacube {\tt 8249-12705} calculated
following the first step described above, where the correlation
coefficient is defined as $\rho_{jk} = C_{jk}/\sqrt{C_{jj}C_{kk}}$
and $C_{jk}$ is the covariance between spaxels $j$ and $k$. Only
spaxels with \snrg$>$1 are included in the Figure. Critically, note
that the indices $j$ and $k$ are {\it not} the two-dimensional
indices of an individual spaxel on sky, but instead indices for the
spaxels themselves. That is, spaxel $j$ will have on-sky coordinates
($x_j$, $y_j$) and appropriate array indices in the \dap\ map. This
explains the diagonal banding in Figure \ref{fig:correlation} as an
effect of spatially adjacent spaxels being separated by the width of
the map in one dimension in the correlation matrix. Figure
\ref{fig:correlation}b is an expanded view of the $\pm$20 pixels
about the main diagonal and has a width of approximately 10 pixels.
The number of discrete diagonal bands in Figure
\ref{fig:correlation}a and the width of the off-diagonal distribution
in Figure \ref{fig:correlation}b demonstrates that spaxels separated
by fewer than 5 or 6 spaxels have $\rho > 0$, consistent with the
subsampling of the MaNGA $2\farcs5$-diameter fiber beam into
$0\farcs5\times0\farcs5$ spaxels.

We show this explicitly in Figure \ref{fig:rhogauss}, which combines
the correlation data for all spaxels within the convex hull of the
fiber centers used to construct wavelength channel 1132.\footnote{
Spaxels outside the convex hull of the fiber centers naturally have
larger correlation coefficients because fewer fibers contribute to
these spaxels. More generally, this effect will be true of regions in
the IFU field-of-view masked at the fiber level due to, e.g., a
broken or very low-throughput fiber.}
We find in this case, and generally, that $\rho_{jk}$ is well-fit by
a Gaussian distribution in the distance between spaxels, $D_{jk}$.
The optimal fit to this channel is given in the Figure, where the
Gaussian has a scale parameter of $\sigma=1.92$ spaxels
($0\farcs96$).

Finally, as a metric for the degree of covariance in this wavelength
channel, we compute $N^2/\sum_{jk} \rho_{jk} = 142.5$, which provides
a rough estimate of the number of independent measurements. Note that
in the limit of fully independent and fully correlated measurements,
$N < \sum_{jk} \rho_{jk} < N^2$, respectively. As expected, the rough
estimate of independent measurements within the datacube is comparable
to the number of fibers in the relevant IFU (127); however, this is
substantially smaller than the 1905 independent fiber observations
used to construct the datacube. For a more in-depth discussion of
datacube reconstruction and a method that minimizes datacube
covariance, see \citet{2019arXiv190606369L}.

\subsection{Voronoi Binning with Covariance}
\label{sec:vorbin}

As we have stated above, the fidelity of the Voronoi-binning approach
critically depends on a proper treatment of the spatial covariance.
For illustration purposes, we have applied the Voronoi-binning
algorithm to the \snrg\ measurements for datacube {\tt 8249-12705}
both with and without an accounting of the spatial covariance. A map
and radial profile of the \snrg\ measurements are shown in the
top-left and bottom panels of Figure \ref{fig:binning_covar},
respectively, and the correlation matrix used is shown in Figure
\ref{fig:correlation}. Although the threshold used in DR15 is
\snrg$\sim$10, we use a threshold of \snrg$\sim$30 to accentuate the
effect. Application of the algorithm without using the correlation
matrix data results in the bin distribution shown in the upper-middle
plot of Figure \ref{fig:binning_covar}; the distribution resulting
from the formally correct \snrg\ calculation is shown in the
upper-right panel.

The effect of the covariance dramatically increases the number of
spaxels needed to reach the target \snrg, as evidenced by comparing
the size of the bins in the upper-middle and upper-right panels. If
we apply the formal calculation of the \snrg\ to the bins generated
without the covariance (red points in the bottom panel of Figure
\ref{fig:binning_covar}), we show that the actual \snrg\ of these
bins is far below the desired threshold. In this example, we note
that the calculation that includes covariance also falls short of the
target \snrg; however, this is due to the details of Voronoi-binning
algorithm, not an inconsistency in the S/N calculation. In detail,
not every S/N function can partition the FOV into compact bins with
equal S/N. To increase their S/N in this example, the outermost bins
would have to become elongated (e.g., like a circular annulus
following the edge). This is prevented by the roundness criterion of
the Voronoi-binning algorithm and, therefore, limits the \snrg\ of
these bins. However, this example is not representative of a
systematic difference between our target \snrg$\sim$10 and what is
achieved by our use of the Voronoi-binning algorithm (cf.\ Figure
\ref{fig:sibin}).

To minimize the systematic errors at low \snrg\ for the stellar
velocity dispersions, we have chosen a \snrg\ threshold of 10 per
wavelength channel for DR15, which is discussed further in Section
\ref{sec:stellarkin}. This is sufficient for the first two kinematic
moments, but one likely needs an increased threshold for higher order
moments ($h_3$ and $h_4$).

\subsection{Spectral Stacking Calculations}
\label{sec:stacking}

For use in the subsequent modules of the \dap, the procedure used to
stack spaxels must yield the flux density, inverse variance, mask,
and wavelength-dependent spectral resolution of each binned spectrum.
The stacked flux density is a simple masked average of the spectra in
each bin, whereas the computations for the uncertainty and spectral
resolution of the binned spectra are more subtle and discussed in
detail below. These procedures are fundamentally independent of the
specific algorithm that determines which spaxels to include in any
given bin, and our treatment of spatial covariance is slightly
different.

\begin{figure}
\begin{center}
\includegraphics[width=1.0\columnwidth]{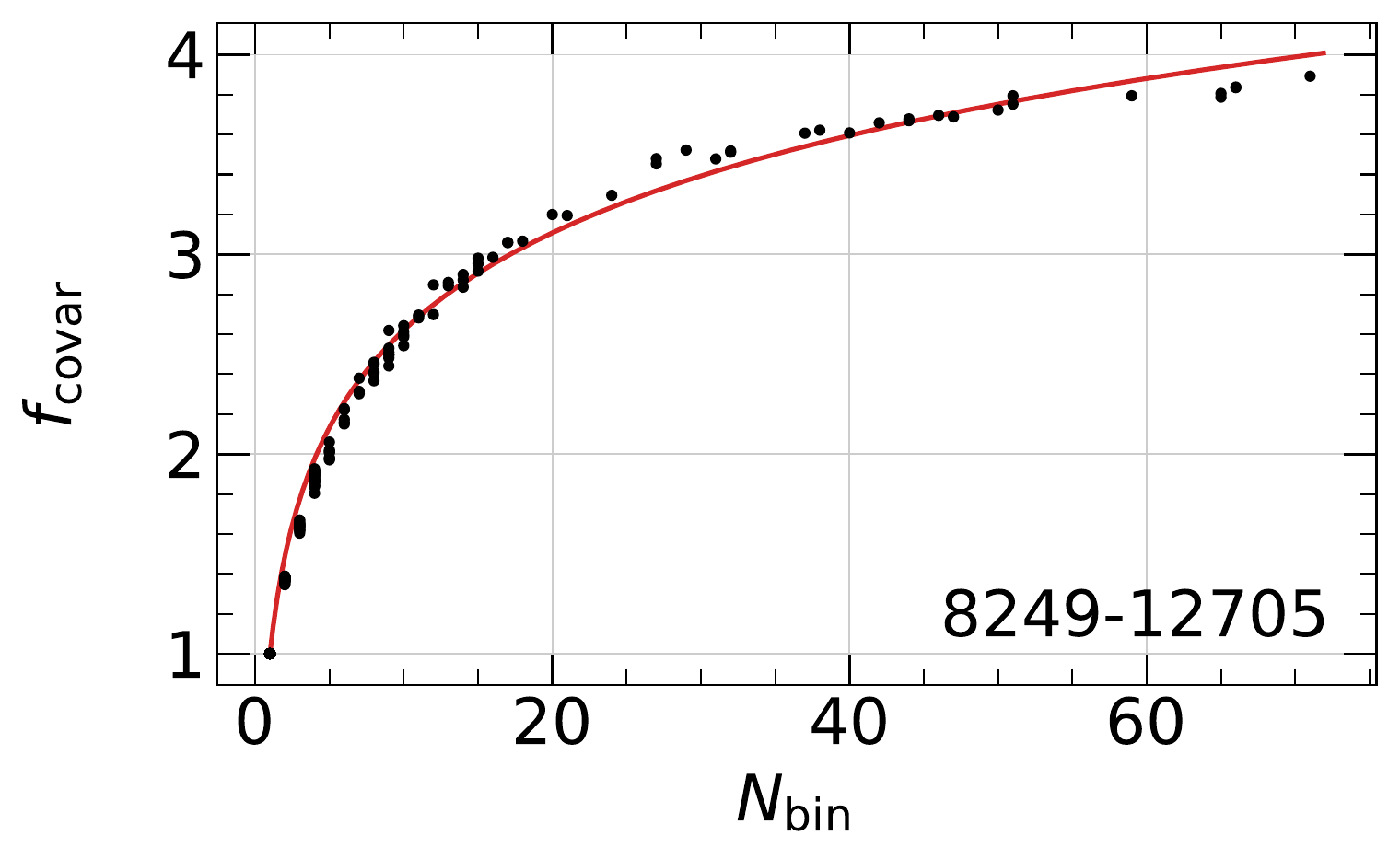}
\end{center}
\caption{The computed factor that properly rescales noise vectors
computed without accounting for covariance to those that do, $f_{\rm
covar}$, as a function of the number of binned spaxels, $N_{\rm
bin}$, for the Voronoi bins constructed for observation 8249-12705
(points). These data are based on the median ratio obtained from the
direct calculation of the covariance matrix in 11 wavelength channels
spanning the full spectral range of the data. For comparison, the
nominal calibration, $f_{\rm covar} = 1+1.62\log(N_{\rm bin})$, from
\citet{2016AJ....152...83L} is shown in red.}
\label{fig:covarcalib}
\end{figure}

The variance in the binned spectra is determined from the covariance
matrix as follows. Similar to the calculation of the covariance in
the datacubes, the covariance in the binned spectra at wavelength
$\lambda$ is
\begin{equation}
{\mathbf C}_{\lambda,{\rm bin}} = {\mathbf T}_{\rm bin} {\mathbf
C}_{\lambda,{\rm spaxel}} {\mathbf T}_{\rm bin}^\top,
\label{eq:bincovar}
\end{equation}
where ${\mathbf C}_{\lambda,{\rm spaxel}}$ is the covariance matrix
for the spaxel data and ${\mathbf T}_{\rm bin}$ is an $N_{\rm bin}
\times N_{\rm spaxel}$ matrix where each row flags the spaxels that
are collected into each bin. To avoid the expensive calculation of
the full datacube covariance matrix, \citet{2016AJ....152...83L} ---
following the original proposal by \citet{2013A&A...549A..87H} ---
recommended the easier propagation of the error that ignores
covariance and provided a simple functional form for a factor,
$f_{\rm covar}$, that nominally recalibrates these error vectors for
the effects of covariance based on the number of binned spaxels. In
the \dap, we instead base $f_{\rm covar}$ on directly calculated
covariance matrices sampled from 11 wavelength channels across the
full spectral range of the data. As an example, Figure
\ref{fig:covarcalib} shows the applied recalibration for the Voronoi
bins in observation 8249-12705 compared to our suggested nominal
calibration provided by \citet{2016AJ....152...83L}.

It is important to note that covariance persists in the rebinned
spectra, even between large, adjacent bins.\footnote{
It is effectively impossible to rebin the MaNGA {\it datacube} in a
way that removes the covariance. One has to restart with the fiber
data in the {\tt RSS} files.}
Therefore, it is important to account for this covariance if users
wish to rebin the binned spectra;\footnote{
If needed, the 11 covariance matrices used to recalibrate the error
in the binned spectra are provided in the \dap\ reference file.}
however, in this case, we recommend simply rebinning the original
datacube.

The spectral resolution in the binned spectrum is determined by a
nominal propagation of the per-spaxel measurements of the line-spread
function (LSF), newly provided with the datacubes released in DR15
\citep{2019ApJS..240...23A}. Similar to how these LSF cubes are
produced by the \drp, we calculate the second moment of the
distribution defined by the sum of the Gaussian LSFs determined for
each spaxel in the bin; i.e.,
\begin{equation}
\sigma^2_{\rm inst,bin}(\lambda) = \frac{1}{N_{\rm bin}} \sum_i^{N_{\rm
bin}-1} \sigma_{{\rm inst},i}^2(\lambda),
\end{equation}
where $\sigma^{-1}_{{\rm inst},i} = R_i \sqrt{8\ln2}/\lambda$ for
each spaxel $i$, with resolution $R_i$, in the bin. A limitation of
this calculation is that the assumption of a Gaussian LSF with a
dispersion of $\sigma^2_{\rm inst,bin}$ for the binned spectrum
becomes less accurate as the range in $\sigma_{{\rm inst},i}$
increases. For MaNGA data, the variation in the LSF between spaxels
in a datacube is generally only a few percent, meaning that this
should not be a concern to first order. However, this may be more of
an issue when binning spectra across observations. Additionally,
deviations of the LSF from a Gaussian {\it will} be a concern for,
e.g., higher-order moments of the stellar LOSVD (e.g., $h_3$, $h_4$).
In DR15, we only provide the first two moments of the stellar LOSVD,
but this could be a concern for future releases and certainly for
those performing their own measurements of $h_3$, etc.

Although never used in the \dap\ directly, we also calculate the
luminosity-weighted coordinates of the binned data in the \dapmaps\
output file. These are simple weighted means of the coordinates of
each spaxel using the mean $g$-band flux (also provided in the
\dapmaps\ file) as the weight.

\subsection{Galactic Extinction Correction}
\label{sec:galext}

After the stacking procedure, all spectra to be fit are corrected for
Galactic extinction using the $E(B-V)$ value from the
\citet{1998ApJ...500..525S} maps provided by the {\tt EBVGAL} keyword in
the header of the \drp\ {\tt LOGCUBE} file.  In DR15, we use the
Galactic extinction law presented by \citet{1994ApJ...422..158O}; the
exact de-reddening vector used for each datacube is provided in the
model \dapcube\ output file (Section \ref{sec:cubefile}).

\subsection{Flagging}

The \snrg\ metrics and binned spectra are flagged according to the
following criteria. Spaxels that are ignored because they do not meet
the \snrg\ or spectral-coverage criteria are masked as {\tt IGNORED}.
Any binned spectrum with individual channels that were masked for
{\it all} spectra in the bin are masked as {\tt FLUXINVALID}, and
pixels with invalid inverse variance values are flagged as {\tt
IVARINVALID}; see Table \ref{tab:dapspecmask}.

\section{Stellar Kinematics}
\label{sec:stellarkin}

The workhorse of our stellar-continuum module (discussed here) and
emission-line fitting module (Section \ref{sec:emlfit}; Belfiore et al.,
{\it accepted}) is the penalized pixel-fitting method\footnote{
We use the {\tt python} package {\tt ppxf} version {\tt 6.7.8} found
here: \url{https://pypi.org/project/ppxf/}.}
(\ppxf) by \citet{2017MNRAS.466..798C}, which is an upgrade to the
original algorithm by \citet{2004PASP..116..138C}.  We refer the reader
to these papers for a detailed description of the method.  In brief,
\ppxf\ assumes that a galaxy spectrum is composed of a mixture of
template spectra, convolved with the line-of-sight velocity distribution
(LOSVD) function of the kinematic component to which each is assigned.
The primary improvement introduced by \citet{2017MNRAS.466..798C} is
that this convolution is now accurate to arbitrarily small velocity
dispersion, achieved by defining the convolution kernel in Fourier space
(cf.\ Section \ref{sec:lineprofile}; Equations \ref{eq:gas_template} and
\ref{eq:gas_template_pixel}).  Currently the \dap\ assigns all
stellar-continuum templates to a single kinematic component while the
emission-line fitting module allows for multiple dynamical components
\citep[cf.][]{Johnston2013, Mitzkus2017}.  The \dap\ also includes
low-order additive and/or multiplicative adjustments to the continuum
via Legendre polynomials, as allowed by \ppxf\ functionality.

The core \ppxf\ algorithm is abstracted and generalized to allow for its
broad application; therefore, it is important to discuss its specific
use in the \dap\ for analyzing MaNGA spectra.  Other applications of the
\ppxf\ method and software  for kinematic measurements in IFS galaxy
surveys include SAURON \citep{Emsellem2004}, ATLAS$^{\rm 3D}$
\citep{Cappellari2011a}, VENGA \citep{Blanc2013venga}, CALIFA
\citep{2017A&A...597A..48F}, and SAMI \citep{2017ApJ...835..104V,
Scott2018sami}.  

Our primary concerns when optimizing our approach are (1) the
selection of the template spectra (see Sections \ref{sec:mileshc},
\ref{sec:milesvsmileshc}, and \ref{sec:tpldownselect}), both in terms
of their pedigree (empirical vs.\ theoretical and individual stars
vs.\ stellar-population synthesis) and their resolution and sampling;
(2) the limitations in the results caused by the instrumental
line-spread function (LSF) and signal-to-noise ratio (S/N); and (3)
the optimization of the parameters provided by the algorithm, such as
the order of the polynomials included in the fit and the penalization
bias applied during fits that include the non-Gaussian moments
($h_3,h_4$) of the LOSVD.

In this Section, we describe our fitting algorithm as it has been
applied to the datacubes provided in DR15 (Section
\ref{sec:scalgorithm}); we briefly compare the stellar kinematics
measured using the full MILES and \mileshc\ libraries to motivate our
use of the latter (Section \ref{sec:milesvsmileshc}) and we justify
some of the nuances of our fitting algorithm (Section
\ref{sec:choices}) based on the data briefly discussed in Section
\ref{sec:testingdata}; we quantify the performance of our algorithm
using both simulated and observed MaNGA data (Section
\ref{sec:scperf}); and we describe the quality flags provided in DR15
(Section \ref{sec:scflags}). Finally, given our particular treatment
and presentation of the stellar velocity-dispersion measurements, we
provide guidance and recommendations for their use in Section
\ref{sec:svdusage}.

\begin{figure*}
\begin{center}
\includegraphics[width=\textwidth]{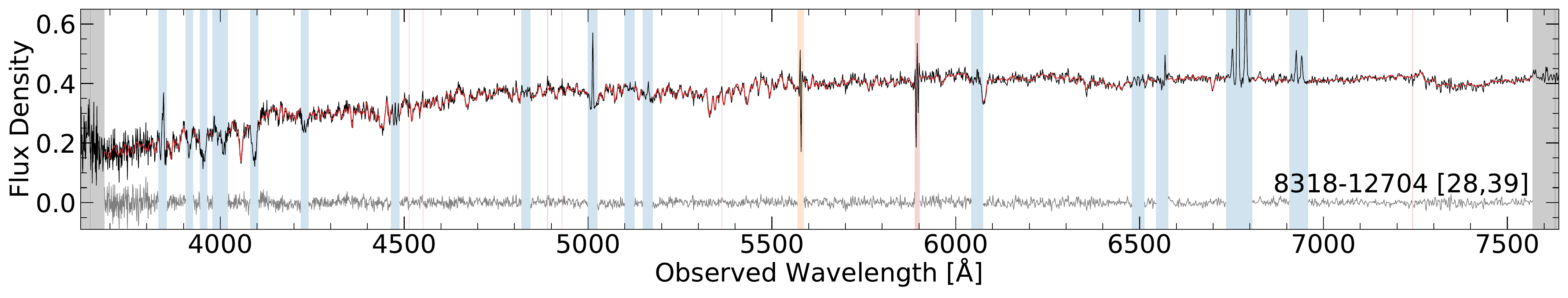}
\end{center}
\caption{An illustration of the spectral masking both input to and
resulting from a typical execution of the stellar-continuum modeling
used to measure stellar kinematics. The observed spectrum ({\it
black}) is from observation {\tt 8318-12704}, specifically the spaxel
at coordinates (x,y) = (28,39). (If read using {\tt astropy.io.fits},
this is the spectrum in the {\tt FLUX} array selected using the slice
{\tt [:,39,28]}.) Before beginning the fit, masks are constructed to
omit regions outside the spectral range of the template spectra
({\it vertical gray bands}), at the expected locations of emission
lines ({\it vertical blue bands}), and at the location of the strong
[\ion{O}{1}] night-sky emission line ({\it vertical orange band}).
After the first iteration, 3-$\sigma$ outliers are identified and
rejected, which in this case omits a few individual channels and an
artifact at $\lambda\sim5900$\AA\ ({\it vertical red bands}). The
resulting model is shown in red, overlayed on the observed spectrum,
and the model residuals are shown in gray.}
\label{fig:ppxfmask}
\end{figure*}

\subsection{Fitting Algorithm}
\label{sec:scalgorithm}

\subsubsection{Template Preparation}
\label{sec:tplprep}

To be flexible to changes in the data and facilitate testing, we have
built methods into the \dap\ that allow us to easily switch between
spectral template libraries in our full-spectrum-fitting modules.  The
template libraries provided with the \dap\ repository\footnote{
\url{https://github.com/sdss/mangadap/tree/master/data/spectral_templates}}
may be linearly or logarithmically sampled in wavelength, can be
provided with a vacuum or air wavelength calibration, have different
spectral resolutions, and adopt different conventions for their overall
flux normalization.  For example, the \mileshc\ spectra have a spectral
resolution of $\Delta\lambda = 2.5$~\AA\ \citep{FalconBarroso2011miles,
2011A&A...531A.109B} and a normalization near unity, whereas the {\tt
M11-STELIB} library spectra --- some of the stellar population models
provided by \citet{2011MNRAS.418.2785M} --- have a spectral resolution
of $\Delta\lambda = 3.4$~\AA\ and a normalization in physical units of
ergs/s/\AA/$\mass_\odot$.  Therefore, we have designed the \dap\ to be
flexible to this variety among the spectral libraries by always
performing a few steps to prepare the templates for use when fitting the
MaNGA spectra.  Here, we focus on the template set used for DR15, the
\mileshc\ library (Section \ref{sec:mileshc}); however, in Belfiore et
al.\ ({\it accepted}), we use many more template libraries to test the
effects of the stellar-continuum modeling on the emission-line
properties.  

The template preparation steps are as follows: (a) the wavelengths
are converted to vacuum, if necessary, to match the MaNGA data; (b;
{\it optional}) we nominally match the template-library resolution to
the MaNGA data by convolving each spectrum with a
wavelength-dependent Gaussian kernel (Appendix \ref{sec:resolution}),
(c) we resample each template to a spectral channel width that is a
fixed fraction of the MaNGA {\tt LOGCUBE} spectral
sampling;\footnote{
This resampling allows for a non-uniform
wavelength step as a function of wavelength. This is primarily to
account for the non-linear conversion from air to vacuum wavelengths
in step (a).}
and (d) we normalize the flux such that the mean flux {\em over all
templates} is unity. In our nominal approach to measuring stellar
kinematics, we skip step (b), leaving the spectral resolution of the
templates unaltered. This offset in spectral resolution between the
template and galaxy spectra is particularly important: the
implications for our fitting algorithm are discussed in Section
\ref{sec:sigmacorr} and the reasoning behind this choice is discussed
in Section \ref{sec:resmatch}. The spectral sampling of the
logarithmically binned MaNGA spectra is fixed to $\delta\log\lambda =
10^{-4}$, which corresponds to a velocity scale of $\Delta
V\approx69$ km s$^{-1}$. We take advantage of the optional behavior
of \ppxf\ in allowing the template spectra to be sampled at some
integer rate higher than the object spectra to avoid undersampling
high-resolution libraries. Therefore, our nominal approach in step
(c) is to sample the \mileshc\ library template spectra four times
per MaNGA spectral channel ($\delta\log\lambda = 2.5\times10^{-5}$).
However, we note that \citet[][Figure 2]{2017MNRAS.466..798C}
demonstrates this oversampling of the template spectra is not
strictly necessary to accurately recover velocity dispersions for an
LOSVD that is undersampled.

\subsubsection{Masking}
\label{sec:stellarkinmask}

A detailed mask is constructed for each spectrum in the MaNGA
datacube before passing the spectrum to \ppxf, as illustrated in
Figure \ref{fig:ppxfmask}: First, \ppxf\ restricts the number of
spectral channels in the template spectra (accounting for any
pixel-scale differences) to be the same or larger than in the object
spectra. However, MaNGA spectra typically have a larger spectral
range than the templates we have included in our testing. In
particular, the \mileshc\ library covers $\sim$3600--7400~\AA, which
is little more than half of the MaNGA spectral range (see Figure
\ref{fig:showcasespec}). The \dap\ therefore censors the MaNGA
spectra based on the expected overlap with the Doppler-shifted
template spectra, given $z_0$ and an assumed maximum velocity range
of $\pm$400 \kms. We mask an additional $\pm$3$\sigma_\ast$ ---
$\sigma_\ast$ is the stellar velocity dispersion --- at the edges of
the spectral range to limit convolution aliasing; instead of
dynamically masking during the fitting, we mask assuming a maximum of
$\sigma_\ast = 400$ \kms. Both of these masks are shown in gray in
Figure \ref{fig:ppxfmask}. Second, we mask any spectrum with
\snrg$<$1, any spectral channels with {\tt MANGA\_DRP3PIXMASK} bits
set to either {\tt DONOTUSE} or {\tt FORESTAR} by the \drp, and the
region from 5570--5586~\AA\ to avoid the near-ubiquitous subtraction
residuals of the strong [\ion{O}{1}] night-sky line; the latter is
shown in orange in Figure \ref{fig:ppxfmask}. Third, we mask a
$\pm$750 \kms\ region around the expected center of each emission
line in Table \ref{tab:emldb}, Doppler shifted to match the expected
recession velocity using $z_0$ (blue in Figure \ref{fig:ppxfmask});
the emission-line mask is applied regardless of whether or not any
emission line is detected. This emission-line mask is generally
sufficient for MaNGA galaxies; however, particularly broad
emission-line cores (e.g., broad-line AGN) or wings (e.g.,
star-formation outflows) are notable exceptions (see our discussion
of performance in Section \ref{sec:performance}).

\subsubsection{Fit Iterations}
\label{sec:stellarfititer}

After preparing the templates and constructing the default mask, the
\dap\ proceeds through two fit iterations. Each fit iteration uses a
common set of templates, an additive eighth-order Legendre
polynomial, and a Gaussian LOSVD; the necessary vetting and
optimization required for fitting a Gauss-Hermite LOSVD has not yet
been done for MaNGA within the \dap. Each iteration runs \ppxf\
twice, once to isolate 3-$\sigma$ outliers in the spectrum and then
with the outlying wavelength channels removed. The 3-$\sigma$
outliers are determined using a 100-channel ($\sim$6900 \kms) boxcar
determination of the local mean and standard deviation in the fit
residuals.

The first iteration fits the masked average of all spectra in the
datacube --- i.e., there is only one spectrum fit during this
iteration --- to isolate the subset of templates allocated non-zero
weight. All spectra, either from a spatial bin or individual spaxel,
are fit in the second iteration. Templates not included in the
non-negative least-squares algorithm used by \ppxf\ to solve for the
template weights in the first iteration are excluded from the second
iteration. This downsampling of the templates used in the second
iteration both expedites that iteration and limits the effect of
noise-driven inclusion of templates in fits to lower \snrg\ spectra.
We discuss the effects of limiting the templates used to fit each
spectrum on the resulting kinematics in Section
\ref{sec:tpldownselect}. The output \dapmaps\ files (Section
\ref{sec:output}; Appendix \ref{sec:datamodel}) provide the
measurements from the second iteration.

\subsubsection{Velocity Definition}
\label{sec:vdef}

The \dap\ does not de-redshift the spectra before executing the fits
used to determine the stellar kinematics or when performing the
emission-line modeling (Section \ref{sec:emlfit}).  However, the
velocities provided in the \dapmaps\ files have been offset to remove
their cosmological redshift with respect to the solar barycentric rest
frame.  This is done in two steps.  First, the velocities returned by
\ppxf\ are (see section 2.3 of \citealt{2017MNRAS.466..798C} for an
explanation)
\begin{eqnarray}
V_{\rm ppxf} & = & {\rm d}V \Delta p \nonumber \\
& = & c \ln(\lambda_{\rm obs}/\lambda_0) \nonumber \\
& = & c \ln(1+z)
\label{eq:vppxf}
\end{eqnarray}
where ${\rm d}V = c\ {\rm d}\ln\lambda$ is the size of the
logarithmically binned pixel in \kms\ and $\Delta p$ is the shift in
pixels found to attain the best fit between the template and galaxy
spectra.  Given that we are not deredshifting the spectra before
executing \ppxf, we must first use equation \ref{eq:vppxf} to calculate
the observed redshift,\footnote{
The value stored in the reference files is actually $cz_{\rm obs}$.}
$z_{\rm obs}$, from $V_{\rm ppxf}$ for each spaxel. Second, for each
galaxy, we remove from these observed redshift measurements the
effect of the {\it input} cosmological redshift of the galaxy, $z_0$,
to bring the stellar velocities to the reference frame of the galaxy.
These are the values that satisfy standard Newtonian laws that one
should use for, e.g., dynamical models (see section 2.4 of
\citealt{2017MNRAS.466..798C} for an explanation), and these are the
velocities \citep[see][]{1999astro.ph..5116H} reported in the output
\dapmaps\ file:
\begin{equation}
V = c\ (z_{\rm obs} - z_0)/(1+z_0).
\label{eq:voffset}
\end{equation}
The cosmological redshift, $z_0$, is most often identical to the
redshift provided by the NSA, except when NSA data is not available for
the galaxy (e.g., it is an ancillary target) or when the NSA redshift
has been corrected based on an improved measurement from the MaNGA data
itself \citep[e.g.,][]{2018MNRAS.477..195T}.

\subsubsection{Velocity-Dispersion Corrections}
\label{sec:sigmacorr}

As discussed above, the approach of the \dap\ is to fit the templates to
the MaNGA spectra at their native resolution (see Section
\ref{sec:resmatch}).  This means that the velocity dispersion returned
by \ppxf\ will be
\begin{equation}
\sigma_{\rm obs}^2 = \sigma_\ast^2 + \delta\sigma_{\rm inst}^2,
\label{eq:sigmaobs}
\end{equation}
where $\sigma_\ast$ is the true astrophysical stellar velocity
dispersion and $\delta\sigma_{\rm inst}$ is an effective difference in
the instrumental dispersion of the template and MaNGA data.  It is
useful to note that, even in the case where one obtains template spectra
from the same instrument as the galaxy data, velocity-dispersion
corrections may still be necessary given the redshift of the galaxy
spectra; see the Appendix of \citet{2011ApJS..193...21W} and additional
discussion below.

\begin{deluxetable}{r r r r r}
\tabletypesize{\scriptsize}
\tablewidth{0pt}
\tablecaption{Repeat observations \label{tab:repeats} }
\tablehead{ &&&& \\ & \multicolumn{4}{c}{{\tt PLATEIFU}} \\ \cline{2-5}
\colhead{MaNGA ID} & \colhead{(1)} & \colhead{(2)} & \colhead{(3)} & \colhead{(4)} }
\startdata
  1-113375 &    7815-9101 &   7972-12704 &      \nodata &      \nodata \\
  1-113379 &    7815-6101 &    7972-3701 &      \nodata &      \nodata \\
  1-113469 &   7815-12702 &   7972-12705 &      \nodata &      \nodata \\
  1-113525 &    7815-1902 &    8618-6103 &    7972-9102 &      \nodata \\
  1-113567 &   7815-12701 &    8618-1902 &      \nodata &      \nodata \\
  1-134760 &   8555-12701 &    8600-9102 &      \nodata &      \nodata \\
  1-137801 &    8247-3702 &    8249-3701 &      \nodata &      \nodata \\
  1-137845 &    8250-9101 &    8249-6104 &      \nodata &      \nodata \\
  1-137853 &    8250-3702 &   8249-12705 &      \nodata &      \nodata \\
  1-149686 &   8997-12701 &   8996-12705 &      \nodata &      \nodata \\
  1-166754 &    8459-3704 &   8461-12703 &      \nodata &      \nodata \\
  1-166919 &    8459-3702 &    8461-3704 &      \nodata &      \nodata \\
  1-166932 &    8459-3701 &    8461-6104 &      \nodata &      \nodata \\
  1-167356 &    8456-6104 &    8454-6103 &      \nodata &      \nodata \\
  1-177236 &    7958-1901 &    9185-1901 &      \nodata &      \nodata \\
  1-177250 &    7958-3703 &    9185-3702 &      \nodata &      \nodata \\
  1-178442 &    7962-6101 &    9085-3703 &      \nodata &      \nodata \\
  1-178443 &    7962-6104 &    9085-3704 &      \nodata &      \nodata \\
  1-178473 &    7962-3701 &    9085-3701 &      \nodata &      \nodata \\
  1-209770 &    9031-6102 &    9036-6104 &      \nodata &      \nodata \\
  1-209772 &    9031-3704 &    9036-3703 &      \nodata &      \nodata \\
  1-209786 &    9031-3701 &    9036-1901 &      \nodata &      \nodata \\
  1-209823 &   9031-12701 &   9036-12703 &      \nodata &      \nodata \\
  1-210186 &   9031-12705 &    9036-6101 &      \nodata &      \nodata \\
  1-210604 &    8600-3702 &   8979-12704 &      \nodata &      \nodata \\
  1-210611 &    8600-1902 &    8979-3703 &      \nodata &      \nodata \\
  1-210700 &   8603-12701 &    8588-3701 &      \nodata &      \nodata \\
  1-211017 &   8312-12703 &    8550-9102 &      \nodata &      \nodata \\
  1-235398 &   8326-12701 &   8325-12703 &      \nodata &      \nodata \\
  1-235530 &    8329-1901 &    8326-3701 &      \nodata &      \nodata \\
  1-255691 &    8256-6102 &    8274-6102 &    8451-3702 &      \nodata \\
  1-255959 &    8256-9102 &    8274-9102 &      \nodata &      \nodata \\
  1-256048 &    8256-6103 &    8274-6103 &    8451-6102 &      \nodata \\
  1-256104 &    8256-9101 &    8274-9101 &    8451-9101 &      \nodata \\
  1-256456 &   8256-12703 &   8274-12703 &   8451-12704 &      \nodata \\
  1-256457 &    8256-1902 &    8274-1902 &    8451-1902 &      \nodata \\
  1-258311 &    8261-1901 &    8262-1901 &      \nodata &      \nodata \\
  1-266074 &    8329-3703 &   8333-12704 &      \nodata &      \nodata \\
  1-277103 &    8256-6101 &    8274-6101 &    8451-6101 &      \nodata \\
  1-277154 &    8256-1901 &    8274-1901 &      \nodata &      \nodata \\
  1-277159 &    8256-3702 &    8274-3702 &      \nodata &      \nodata \\
  1-277161 &    8256-3701 &    8274-3701 &      \nodata &      \nodata \\
  1-277162 &   8256-12702 &   8274-12702 &      \nodata &      \nodata \\
  1-277691 &   8256-12701 &   8274-12701 &      \nodata &      \nodata \\
  1-277858 &    8256-3703 &    8274-3703 &    8451-3703 &      \nodata \\
  1-278485 &    8256-3704 &    8274-3704 &    8451-3704 &      \nodata \\
  1-456757 &    8479-3703 &    8480-3701 &    8953-3702 &    9051-6103 \\
  1-548221 &   8567-12702 &    8239-6104 &      \nodata &      \nodata \\
  1-558910 &    8256-6104 &    8274-6104 &    8451-6103 &      \nodata \\
  1-558912 &   8256-12704 &   8274-12704 &   8451-12701 &      \nodata \\
  1-561017 &   7960-12702 &    9185-3704 &      \nodata &      \nodata \\
  1-569225 &    8329-3701 &   8333-12701 &      \nodata &      \nodata \\
  1-587938 &   8256-12705 &   8274-12705 &   8451-12702 &      \nodata \\
  1-592881 &    8329-3704 &    8333-3702 &      \nodata &      \nodata \\
  1-635503 &   7815-12705 &    8618-6101 &      \nodata &      \nodata \\
   1-93876 &    8555-3704 &    8484-9101 &      \nodata &      \nodata
\enddata
\end{deluxetable}

\begin{deluxetable}{r r r r r}
\tabletypesize{\footnotesize}
\tablewidth{0pt}
\tablecaption{Representative Observations \label{tab:selected}}
\tablehead{\colhead{MaNGA ID} & \colhead{{\tt PLATEIFU}} & \colhead{$M_i$}
           & \colhead{$\log({\mathcal M}_\ast/{\mathcal M}_\odot)$}
           & \colhead{$NUV-r$} }
\startdata
1-113379 &  7815-6101 & -17.5 &  8.7 & 2.2 \\
1-339041 & 8138-12704 & -22.1 & 11.0 & 3.7 \\
1-377176 &  8131-3702 & -20.6 & 10.3 & 5.5 \\
1-113379 &  8131-6102 & -22.3 & 11.1 & 5.3 \\
\enddata
\end{deluxetable}

For DR15, we provide a first-order estimate of $\delta\sigma_{\rm inst}$
based on the average quadrature difference in the instrumental
dispersion of the template and object spectra over the region fit by
\ppxf.  That is, we calculate
\begin{equation}
\delta\sigma_{\rm inst}^2 = \frac{c^2}{8 \log(2) N_{\rm fit}}\
\sum_{i=0}^{N_{\rm fit}-1} R_{{\rm MaNGA},i}^{-2} - R_{{\rm
tpl},i}^{-2},
\label{eq:sigmacorr}
\end{equation}
where $N_{\rm fit}$ is the number of fitted wavelength channels and
$R_{{\rm tpl},i}$ and $R_{{\rm MaNGA},i}$ are, respectively, the
spectral resolution ($R=\lambda/\Delta\lambda$) of the template
library and MaNGA spectrum at the observed wavelength in channel $i$;
i.e., the resolution vector of the template library is appropriately
shifted to the best-fit redshift of the MaNGA spectrum for this
calculation. In particular, we use the estimate of the resolution
that {\it does not} include the integration of the LSF over the
spectral channel, provided by the {\tt PREDISP} extension in the
\drp\ datacubes, as this is most appropriately matched to the \ppxf\
method. Over the spectral region of the \mileshc\ library, the MILES
resolution is $\sim 16^{+7}_{-3}\%$ higher than the mean resolution
of the spaxels in a MaNGA datacube ($R_{{\rm tpl},i} \sim 1.16\
R_{{\rm MaNGA},i}$), such that we should expect $\delta\sigma_{\rm
inst} \sim 34$ \kms, according to Equation \ref{eq:sigmacorr}.
Indeed, the median correction for the spaxels in a MaNGA datacube is
$\delta\sigma_{\rm inst} \sim 32.6^{+9.0}_{-5.7}$ \kms\ (cf.\ Figure
\ref{fig:mastar_corr}). We provide $\sigma_{\rm obs}$ from \ppxf\ and
$\delta\sigma_{\rm inst}$ from Equation \ref{eq:sigmacorr} in the
\dapmaps\ file. It is important to note that {\it we do not provide
the corrected velocity dispersion}, $\sigma_\ast$. We leave it to the
user to use Equation \ref{eq:sigmaobs} to calculate $\sigma_\ast$
themselves. We discuss this decision in detail in Section
\ref{sec:resmatch} and provide some usage guidance in Section
\ref{sec:svdusage}.

We describe the above calculation of $\delta\sigma_{\rm inst}$ as a
first-order approximation because it makes the simplifying assumption
that all spectral regions contribute equally to the determination of
$\sigma_{\rm obs}$. However, we know that the influence of spectral
features on the $\sigma_{\rm obs}$ measurements is roughly
proportional to their equivalent width. That is, even with the
spectral resolution vectors of the template and object spectra, an
accurate measurement of $\delta\sigma_{\rm inst}$ is non-trivial
because of the unknown relative influence of each spectral feature on
the aggregate kinematics. Our recovery simulations, presented in
Section \ref{sec:recovery}, have shown our first-order estimate of
$\delta\sigma_{\rm inst}$ from Equation \ref{eq:sigmacorr}
systematically underestimates the correction by a few percent. This
leads to a systematic bias in $\sigma_\ast$ of $\lesssim$5\% at
$\sigma_\ast\approx 70$ \kms, with an increasing relative bias toward
lower dispersion. However, our test of the accuracy of our
first-order corrections presented toward the end of Section
\ref{sec:resmatch} show no signs of a systematic error that is this
large. We continue to improve the methodology used for the
determination of the velocity dispersions and their corrections
toward low dispersion and will return to this topic in future papers.

\subsection{Optimization and Performance Evaluation Data}
\label{sec:testingdata}

The optimization and performance characterization of the \dap\ stellar
kinematics are primarily based on three data sets:

\smallskip

\noindent {\bf (1) A representative set of MaNGA spectra} selected to
span the full range in preliminary measurements of $D4000$
\citep{1983ApJ...273..105B}, \halpha\ equivalent width (EW), and
stellar velocity dispersion ($\sigma_\ast$). All spectra are sorted
into a three-dimensional grid defined by bin edges at
\begin{eqnarray}
{\rm D4000} & = & \{ 1.2, 1.4, 1.5, 1.6, 1.8, 2.0, 2.2 \}, \nonumber \\
\sigma_{\rm obs}  & = & \{ 25, 50, 75, 100, 150, 200, 250 \},\ {\rm
and} \nonumber \\
{\rm H}\alpha\ {\rm EW} & = & \{ 0, 2, 8, 16, 32 \},\nonumber
\end{eqnarray}
with bins also for data below/above the first/last bin edge, for a
total of 384 bins. When selecting spectra, any datacubes with the
{\tt MANGA\_DRP3QUAL} bit set to {\tt CRITICAL} by the \drp\ are
ignored. We have selected the spectrum with the highest \snrg\ in
each bin (allowing for bins to have no relevant spectra) and then
visually inspected the results to remove spectra with significant
artifacts and with modeling failures; modeling failures in these
cases are usually the result of interloping objects in the field
outside the redshift boundary imposed by \ppxf\ ($\pm$2000 \kms). Our
selection yielded 292 spectra for testing from 100 unique
observations; 192 and 102 of these spectra have \snrg\ larger than 30
and 60, respectively.

\smallskip

\noindent {\bf (2) A representative set of four MaNGA datacubes} for
galaxies that span the mass-color (${\mathcal M}_\ast$, $NUV-r$)
range of the MaNGA sample, listed in Table \ref{tab:selected}. Note
the two ``blue'' galaxies are the same as used by Belfiore et al.,
{\it accepted}; however, the two ``red'' galaxies chosen for Belfiore
et al., {\it accepted} were selected to also have noticeable emission
lines, whereas the observations used here do not.

\smallskip

\noindent {\bf (3) Fifty-six galaxies with multiple MaNGA
observations} provided in DR15 as listed in Table \ref{tab:repeats}.
As part of its ongoing quality control and calibration strategy,
MaNGA has re-observed targets in fully identical plates ({\tt 8256}
and {\tt 8274}), in identical IFUs on different plates (e.g., {\tt
7958-1901} and {\tt 9185-1901}), and with different sized IFUs (e.g.,
{\tt 7960-12702} and {\tt 9185-3704}). Repeat observations provide an
ideal test-bed for empirically characterizing the measurement
uncertainties. We use these data to assess the robustness of our
stellar kinematics here, our emission-line modeling results in
Section 3 of Belfiore et al., {\it accepted}, and our spectral-index
measurements in Section \ref{sec:siqual}.

\subsection{Template Library Comparison: MILES versus \mileshc}
\label{sec:milesvsmileshc}

\begin{figure}
\begin{center}
\includegraphics[width=\columnwidth]{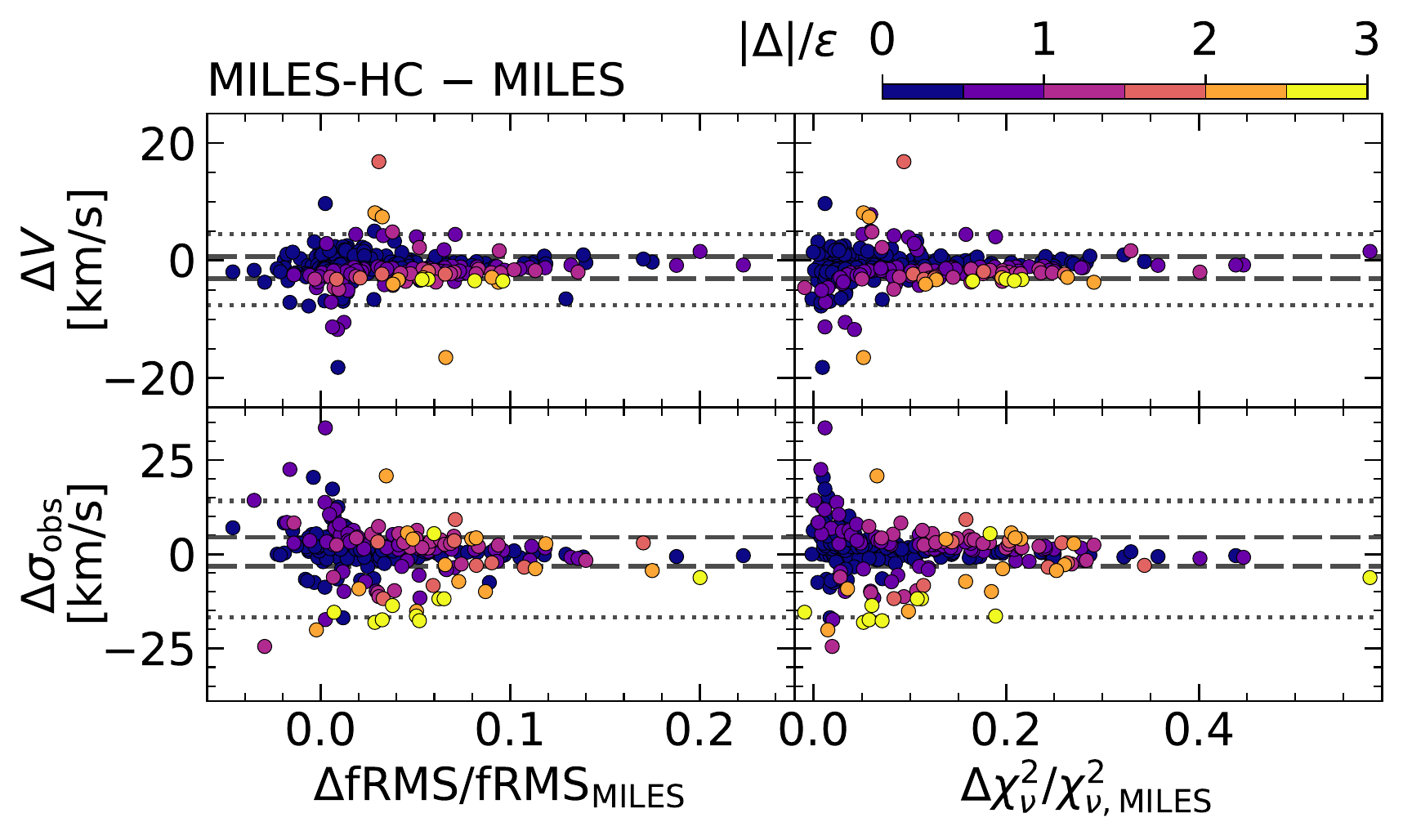}
\end{center}
\caption{Difference between the stellar velocity ({\it top}), $V$,
and the ``observed'' stellar velocity dispersion ({\it bottom}; see
Section \ref{sec:sigmacorr}), $\sigma_{\rm obs}$, measured for the
first dataset listed in Section \ref{sec:testingdata} using our
implementation of \ppxf\ in the \dap\ and either the full MILES or
\mileshc\ template library. Differences are plotted as a function of
the fractional change in the root-mean-square of the fractional
residuals (fRMS; {\it left}) and the reduced chi-square
($\chi^2_\nu$; {\it right}). Positive values mean the \mileshc\
metric or measurement is larger. The difference in the measurement
relative to its error, $|\Delta|/\epsilon$, is represented by the
point color, according to the color bar. The horizontal lines show
the interval enclosing 68\% ({\it dashed}) and 95\% ({\it dotted}) of
all data.}
\label{fig:miles_vs_mileshc}
\end{figure}

\begin{figure*}
\begin{center}
\includegraphics[width=\textwidth]{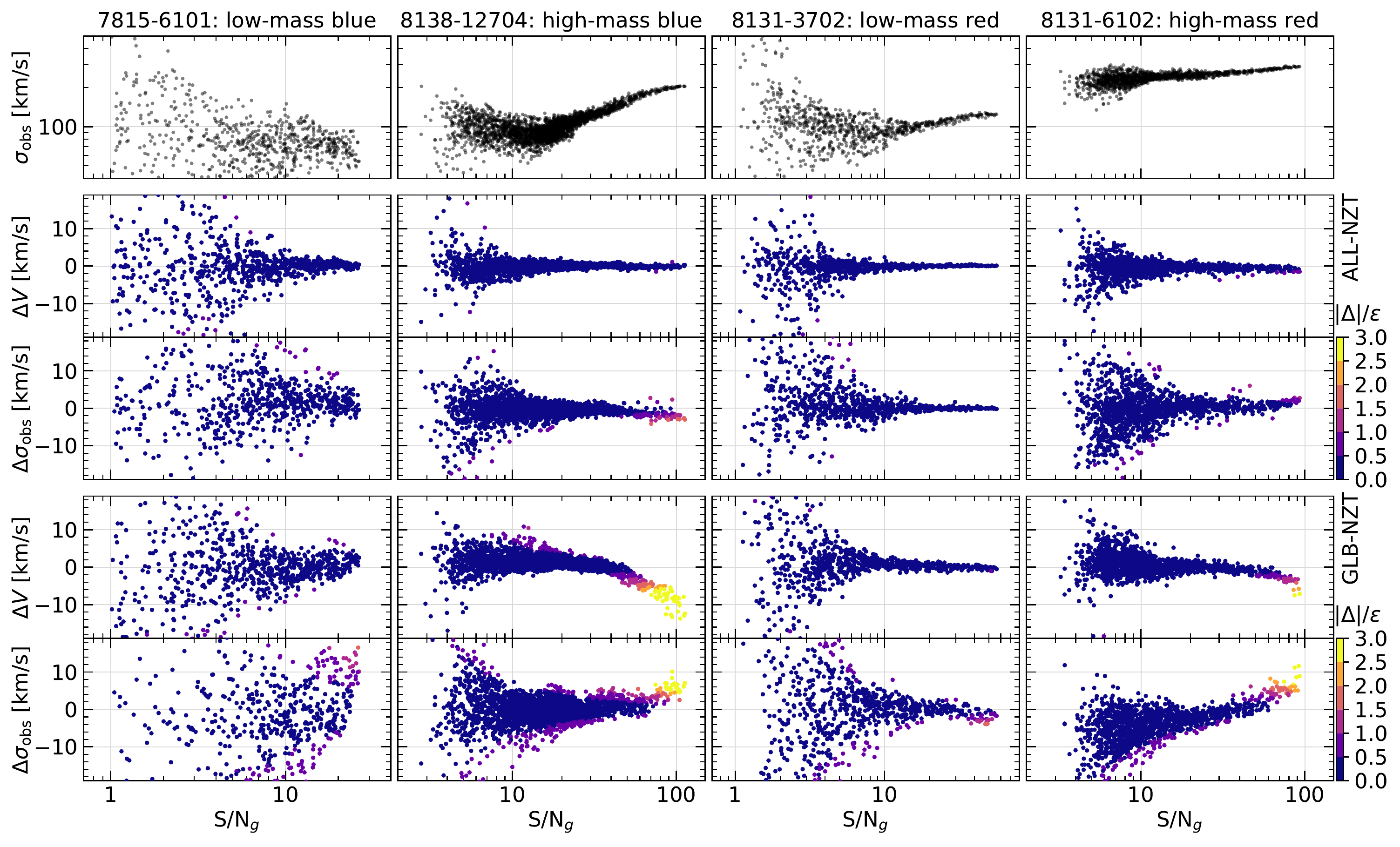}
\end{center}
\caption{Relevance of the template down-selection approach to the
stellar kinematics in four representative MaNGA galaxies (Table
\ref{tab:selected}; Section \ref{sec:testingdata}) as labeled at the
top of each panel column. The top row shows $\sigma_{\rm obs}$ as a
function of \snrg\ for each galaxy resulting from the {\tt NZT}
approach (see Section \ref{sec:tpldownselect}). The second and third
panel rows show, respectively, the velocity difference, $\Delta V$,
and velocity-dispersion difference, $\Delta\sigma_{\rm obs}$, between
the {\tt ALL} and {\tt NZT} approaches. The absolute value of the
difference relative to its error, $|\Delta|/\epsilon$, is given by
the point color. The fourth and fifth panel rows show the differences
when comparing the {\tt GLB} and {\tt NZT} approaches. The difference
between the {\tt GLB} and {\tt NZT} approaches are quite different
for galaxies with significant stellar-population (color) gradients,
like {\tt 8138-12704}, whereas the {\tt ALL} and {\tt NZT} approaches
are more consistent.}
\label{fig:downselect}
\end{figure*}

The motivation for the hierarchical-clustering analysis of the MILES
library (Section \ref{sec:mileshc}) was to limit the templates used
by \ppxf\ to fit the MaNGA spectra while not substantially affecting
the fit quality or the resulting kinematics. Figure
\ref{fig:miles_vs_mileshc} shows the difference in the stellar
kinematics as measured using the MILES and \mileshc\ templates for
the spectra in the first data set described in Section
\ref{sec:testingdata}. Compared to fits using the full MILES library,
use of the \mileshc\ templates when fitting the high-S/N MaNGA test
spectra leads to an increase of the root-mean-square of the
fractional residuals (fRMS) by typically 15\% or less and of the
reduced chi-square ($\chi^2_\nu$) by 30\% or less. Figure
\ref{fig:miles_vs_mileshc} shows that this level of variation leads
to marginal biases in the kinematics
\citep[cf.][Figure~B3]{Emsellem2004}. In detail, we find median,
68\%, and 95\% intervals of $\Delta V =
-1.3~^{+2.0}_{-1.8}~^{+5.8}_{-6.3}$ and $\Delta \sigma_{\rm obs} =
0.9~^{+3.6}_{-4.1}~^{+13.3}_{-17.7}$; these intervals are only
moderately reduced if we restrict the test to the 102 spectra with
\snrg$>$60. However, the difference in execution time is a factor of
25, as expected by the roughly $O(N_{\rm tpl})$ scaling of the \ppxf\
method (i.e., the difference between fitting 985 MILES templates
versus 42 \mileshc\ templates). Given these two results (minimal
effect on the kinematics and dramatically shortened computation
time), we have used the \mileshc\ library for both full-spectrum
fitting algorithms in the DR15 execution of the \dap. Use of the
\mileshc\ library in the continuum modeling needed for the
emission-line measurements are discussed in detail by Belfiore et
al., {\it accepted}.

\subsection{Design Choices}
\label{sec:choices}

We explore and justify three core design choices implemented by our
fitting algorithm described in Section \ref{sec:scalgorithm}. Namely,
we quantify the effect of the algorithmic down-selection of the
templates on the resulting kinematics in Section
\ref{sec:tpldownselect}, we describe tests performed to optimize the
order of the additive Legendre polynomial used in all fits in Section
\ref{sec:polyorder}, and we justify our use of the templates at their
native resolution in Section \ref{sec:resmatch}.

\subsubsection{Algorithmic Down-selection of Templates}
\label{sec:tpldownselect}

In the same vein of reducing its execution time while minimizing the
degradation of the fit quality and stellar kinematics, the \dap\
algorithmically down-selects templates from the larger \mileshc\
library by tuning the templates used to fit the individual spectra of
each galaxy. This is done by first fitting the global spectrum and
only using those templates with non-zero weights in the subsequent
fits (Section \ref{sec:stellarfititer}). This approach reduces the
per-datacube execution time of the stellar-kinematics module by a
factor of 2.5--3.\footnote{
Use of this algorithmic downselection with the full MILES library
leads to execution times that are a factor of $\sim$2 slower than
when used with the \mileshc\ library. Of all the tests we have
performed, we have consistently found that the roughly $O(N_{\rm
tpl})$ execution-time scaling of the \ppxf\ algorithm holds. In this
example, the number of non-zero templates selected from the global
fit using the full MILES library is approximately twice as many as
selected when using the smaller \mileshc\ library.}
How does this further down-selection of the templates change the
resulting stellar kinematics compared to a fit that always uses the
full \mileshc\ library?

Figure \ref{fig:downselect} compares $V$ and $\sigma_{\rm obs}$
(Equation \ref{eq:sigmaobs}) measurements for three different fit
approaches: {\tt ALL} --- a fit that uses all templates from the
\mileshc\ library for all spectra (i.e., no downselection is
performed); {\tt NZT} --- a fit that only uses the templates given a
non-zero weight in a fit to the global spectrum; and {\tt GLB} --- a
fit that only uses a {\it single} template constructed using the
weights determined for the fit to the global spectrum. These fits
were performed using the four example datacubes discussed in point 2
of Section \ref{sec:testingdata} without any spatial binning and
limited to spaxels with \snrg$>$1.

The top row of Figure \ref{fig:downselect} shows \snrg\ versus
$\sigma_{\rm obs}$ resulting from the nominal DR15 approach ({\tt
NZT}), primarily as a reference for the differences seen in the
following panel rows. The next two panel rows show the difference in
$V$ and $\sigma_{\rm obs}$ between the {\tt ALL} and {\tt NZT}
approaches. As given by the point color, the majority of the
differences between the measurements are below 50\% of the error. The
main exceptions to this are the $\Delta \sigma_{\rm obs}$
measurements near the center of the massive blue galaxy. This galaxy
has the strongest broad-band color gradient from its center to its
outskirts, so it is reasonable to find that the templates selected by
a fit to the global spectrum may not capture the templates relevant
to relatively small and/or low-surface-brightness regions. Indeed,
this is much more apparent when only a single template determined by
the fit to the global spectrum is used, as is the case in the bottom
two rows of Figure \ref{fig:downselect}, which compares the {\tt GLB}
and {\tt NZT} approaches. Here, the differences in both $V$ and
$\sigma_{\rm obs}$ for the high-mass blue and high-mass red galaxies
can be more than 3 times the measurement error near their centers.
For this reason, we chose not to adopt the {\tt GLB} approach.

However, the {\tt NZT} and {\tt ALL} approaches are reasonably
consistent. In all cases, $\Delta V$ is small compared to its error.
For the few spaxels where $\Delta \sigma_{\rm obs}$ is up to 1.5
times its error, the differences are small in both an absolute sense
(2--4 \kms) and a relative sense (1--2\%). Tests of the relative
biases in the {\tt NZT} approach are ongoing; however, given the
factor of 2.5--3 decrease in execution time and acceptable level of
change to the kinematics from these few examples, we have adopted the
{\tt NZT} approach for the analysis of all datacubes in DR15.

\begin{figure}
\begin{center}
\includegraphics[width=1.0\columnwidth]{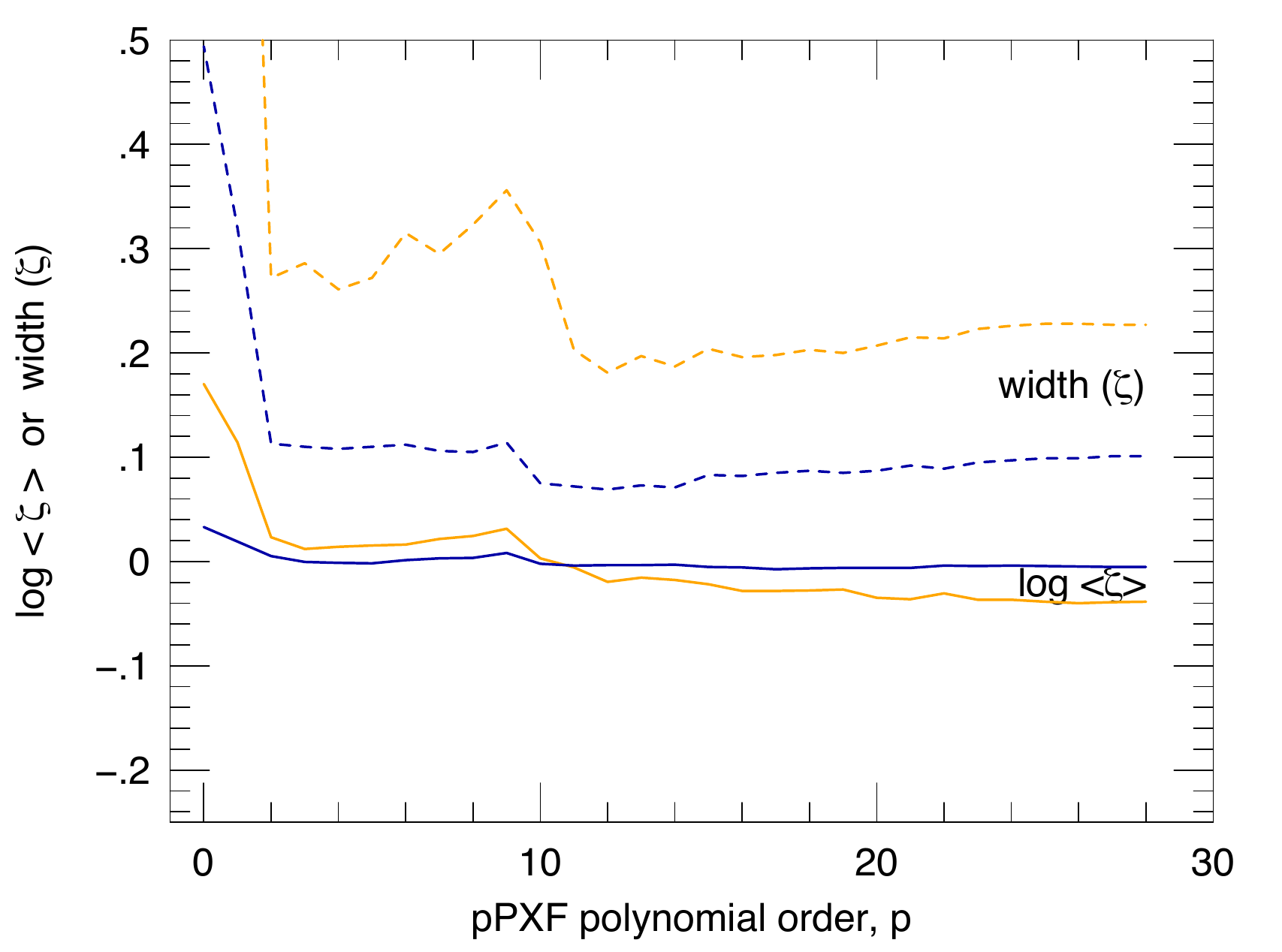}
\end{center}
\caption{Assessment of the influence of the order ($p$) of the
additive Legendre polynomial used during the stellar-kinematics fit
on the best-fitting $\sigma_{\rm obs}$. Statistics of a
representative sample of MaNGA spectra (Section
\ref{sec:testingdata}) are for the normalized velocity dispersion
($\zeta$), and they include the mean and standard deviation (dark
solid and dashed lines, respectively) and the median and median
absolute deviation (light solid and dashed lines, respectively). The
normalization is the mean value of $\sigma_{\rm obs}$ for each
spectrum over all polynomial orders, $p$; i.e., $\zeta = \sigma_{\rm
obs}/ \left\langle\sigma_{\rm obs}\right\rangle_p$. Note the standard
deviation is $\sim$1.5 times larger than the median absolute
deviation for a normal distribution.}
\label{fig:polyorder}
\end{figure}

\subsubsection{Low-order Polynomial Manipulation of the Continuum}
\label{sec:polyorder}

The low-order polynomials included in \ppxf\ model spectra aim to
compensate for any subtle mismatch between the spectral templates and
the science spectra that change slowly with wavelength. Mismatch may
arise due to, e.g., flux-calibration errors, internal or foreground
reddening, and even library incompleteness. The order of the
polynomial should be sufficiently low to avoid impacting individual
line fits except insofar as additive polynomials can modulate the
line equivalent width (but not shape).

However, since polynomials allow for templates to be selected that can
fit detailed lines while relaxing mismatch in their continuum shape,
inclusion of these functions do allow for different solutions.  These
solutions can, in principle, differ at the level of individual
line-profile fits, examples of which are seen clearly in Belfiore et
al., {\it accepted}.  Although the focus of our analysis in this
companion paper is on the strong H and He lines in the blue for young
stellar populations, Figure 14 therein shows differences do extend more
broadly across the spectrum and, by inference therefore, to the weak
metal lines that influence the kinematics solution of the \ppxf\ fits.

Our concern here is then whether the inclusion of polynomials in the
\ppxf\ fitting alters the values of the derived stellar kinematics in
a quantitatively significant fashion. We are primarily concerned here
with velocity dispersion since this (and higher) moments are believed
to be more likely affected by detailed template changes, i.e., the
infamous ``template mismatch'' \citep[e.g., Figure~6
of][]{vanderMarel1994}.

To test for systematic differences in the derived stellar velocity
dispersion, we used the 292 representative spectra that span a range
of parameters (Section \ref{sec:testingdata}). We used \ppxf\ to fit
each spectrum with the \mileshc\ template library but with an
additive Legendre polynomial of order ($p$) ranging from 0 to 30
($p=0$ represents the case for no polynomial term). Of the 292
spectra, 290 were successfully fit for all orders. For {\it each} of
these 290 spectra we determined the mean $\sigma_{\rm obs}$ over all
polynomial orders for the individual spectrum,
$\left\langle\sigma_{\rm obs}\right\rangle_p$, and considered the
statistical trend of $\zeta = \sigma_{\rm obs} /
\left\langle\sigma_{\rm obs}\right\rangle_p$ with polynomial order.

The mean, median, standard deviation, and median absolute deviation of
$\zeta$ for the full sample is shown in Figure \ref{fig:polyorder}.
Strong trends exist in the characteristic value and scatter of $\zeta$
for orders $p<3$.  There is some evidence for other trends when
$p\geq9$, but this is less evident using robust statistics.  Further
exploration reveals that the strength of the trends in the mean and
standard deviation at larger polynomial orders correlate with {\it
decreasing} S/N.  The depth of the MaNGA survey is uniform enough that
S/N is well correlated with surface-brightness over all observations.
Because of this and the correlation of surface-brightness with D4000,
H$\alpha$ EW and $\sigma_\ast$ (younger stellar populations have small
$\sigma_\ast$ and tend to be found in the outskirts of galaxies where
the surface-brightness is low), the strength of the statistical trends
at high polynomial order correlates broadly with many variables
describing the spectral sample.  That said, the systematic changes in
$\sigma_\ast$ with changing polynomial order is small (below a few
percent) for orders $p>3$, and $\sigma_\ast$ for values $p<9$ are
relatively immune to systematics at lower S/N.  A decision to use $p=8$
was made early in the development of the \dap, and this more detailed
analysis demonstrates that there is no compelling reason to revisit that
choice.

\subsubsection{Spectral-Resolution Matching}
\label{sec:resmatch}

The common approach to measuring stellar kinematics is to use
template spectra that have a spectral resolution --- or instrumental
dispersion, $\sigma_{\rm inst}$ --- that is matched to the galaxy
data. In fact, when fitting higher moments of the LOSVD with \ppxf,
the definition of the Gauss-Hermite parametrization
\citep{1993ApJ...407..525V} requires this to be the case. However,
under the simplistic assumption of $\sigma_{\rm inst} \approx 70$
\kms\ for MaNGA, we find that 40\% of all \dap-analyzed spectra in
DR15 have $\sigma_\ast < \sigma_{\rm inst}$, and half of all MaNGA
datacubes show $\sigma_\ast < \sigma_{\rm inst}$ for at least 32\% of
their spectra. Thus, it will be difficult, if not impossible, to
reliably measure the higher-order moments of the LOSVD
\citep[cf.][]{2004PASP..116..138C} for many of the spectra in DR15.
For this reason, we have not measured the Gauss-Hermite moments in
this data release, but have instead restricted our measurements to
the first two velocity moments alone ($V$ and $\sigma_\ast$). Future
improvements of the \dap\ may include fits of the higher-order
moments for a relevant subset of MaNGA spectra.

Free from the {\it requirement} of resolution matching the input
spectra, we explore the difference between fits performed with and
without the matched-resolution spectra.

First, in the limit of a Gaussian LOSVD and Gaussian LSFs in both
spectra, we note that there is no mathematical difference between
first convolving the template spectrum with a Gaussian kernel based
on the resolution difference $\delta\sigma_{\rm inst}$ before
executing the \ppxf\ fit versus subtracting that difference in
quadrature from the \ppxf\ result, as in Equation \ref{eq:sigmacorr}.

Second, because of the Doppler shift between the template and galaxy
spectra, an offset between the spectral resolution of the template
and galaxy data is effectively inevitable even for spectra observed
with identical instrumental resolution. This intrinsic difference
means that, at some level, all velocity-dispersion measurements
require a correction for the detailed, wavelength-dependent
difference of $\sigma_{\rm inst}$ between the template and galaxy
spectra \citep[cf.][]{2011ApJS..193...21W}.

\begin{figure*}
\begin{center}
\includegraphics[width=0.8\textwidth]{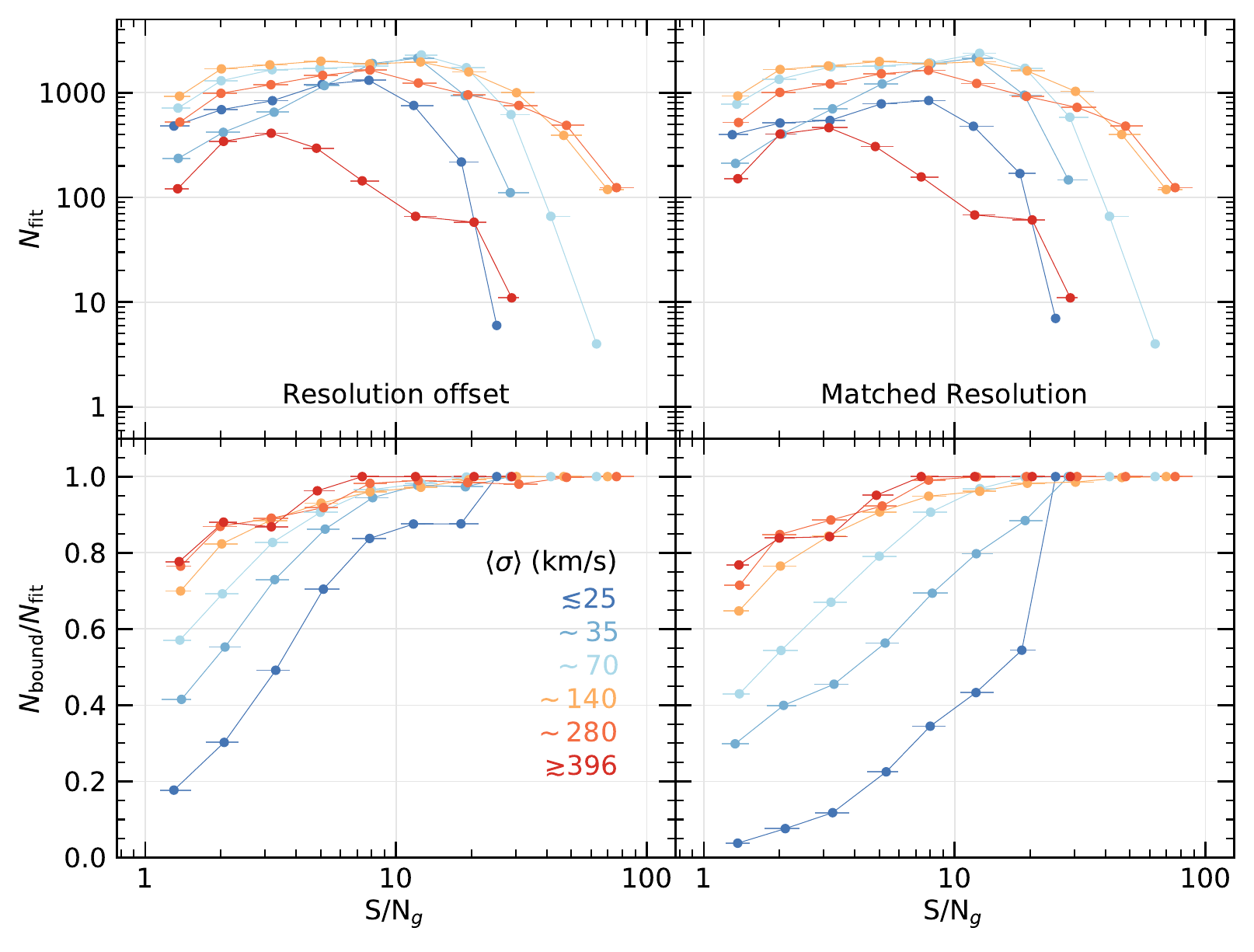}
\end{center}
\caption{For individual spaxels from $\sim$100 datacubes drawn from
Table \ref{tab:repeats}, we show the total number of fitted spectra
($N_{\rm fit}$; {\it top}) and its ratio compared to the number of
fitted spectra with $\sigma_{\rm obs}$ within the \ppxf\ trust-region
boundary ($N_{\rm bound}$; {\it bottom}). The results are shown for
fits using template spectra that have ({\it right}) and have not
({\it left}) had their resolution matched to the MaNGA resolution
(Appendix \ref{sec:resolution}). The results are binned by \snrg\ and
the {\it astrophysical} velocity dispersion, $\sigma_\ast$; the
latter are binned geometrically with the color and geometric center
at the value given by the legend; e.g., the 70 \kms\ bin includes
measurements between $50\lesssim \sigma_\ast\lesssim100$ \kms. The
bottom panels demonstrate that the number of fitted spectra with
viable $\sigma_\ast$ measurements from \ppxf\ is dramatically larger
when a resolution offset exists between the data and the templates,
which allows for improved assessments of the error distribution for
low $\sigma_\ast$.}
\label{fig:sigma_completeness}
\end{figure*}

Third, we therefore expect $\delta\sigma_{\rm inst} > 0$ and, given
that measurements of $\sigma_{\rm inst}$ will always have some
uncertainty, it is useful to understand how those uncertainties for
both the galaxy and template spectra propagate to the uncertainty in
$\sigma_{\rm obs}$. We explore this in Appendix \ref{sec:sigmaerr}
and show that, under some nominal assumptions, it is advantageous in
terms of the error budget to use template spectra with higher
spectral resolution than the galaxy data. Although the resolution
difference between MaNGA and MILES is modest (the median $\sigma_{\rm
inst}$ for MaNGA is $\sim$16\% larger than for MILES over their
common spectral range), our calculations expect a substantial
improvement in the $\sigma_{\rm obs}$ uncertainty, particularly as
$\sigma_\ast$ becomes less than the MaNGA $\sigma_{\rm inst}$.

Fourth, as $\sigma_{\rm obs}$ approaches 0, its error distribution
will become more significantly non-Gaussian, which complicates the
handling of the data both in terms of aggregation and model fitting.
In the limit where measurements of $\sigma_{\rm obs}$ are similar to
fitting a Gaussian function to a set of data, we should expect its
posterior probability to follow an inverse-gamma distribution
\citep[][Section 24.1]{MacKay:itp}. When the ratio of its mean to its
standard deviation is large, the inverse-gamma distribution is
well-approximated by a Gaussian; however, as this ratio decreases,
the inverse-gamma distribution exhibits increasingly significant
positive skew. Thus, in addition to gains in terms of the error
budget (Appendix \ref{sec:sigmaerr}), one benefits from having
$\delta\sigma_{\rm inst} > 0$ in terms of the form of the
$\sigma_{\rm obs}$ error distribution: As $\sigma_\ast$ approaches 0,
$\sigma_{\rm obs}$ approaches the constant $\delta\sigma_{\rm inst}$,
which can be used to limit the ratio of the expectation value of
$\sigma_{\rm obs}$ to its variance at fixed S/N. Again, the
difference in spectral resolution between MaNGA and MILES is rather
modest; however, fitting the MaNGA spectra with the MILES templates
at their native resolution enables us to mitigate some of the issues
with the error distribution of $\sigma_{\rm obs}$ at low
$\sigma_\ast$. In particular, this includes reducing the number of
measurements that hit the $\sigma_{\rm obs}\approx0$ boundary imposed
by \ppxf.\footnote{
To be precise, by default $\delta_v/100 < \sigma_{\rm obs} < 1000$
\kms, where $\delta_v = 10^{-4} \ln(10) c = 69$ \kms\ is the velocity
step per spectral sample of the \drp\ log-linear binned datacubes.}





It is the latter consideration, as motivated below, that was the main
driver of our decision to perform the fit of the first two moments of
the stellar LOSVD without matching the resolution of the templates to
that of the galaxy, instead keeping them at their native resolution.

Before continuing, we note that the public \texttt{python} version of
\ppxf\ employed by the \dap\ uses a novel trust-region implementation
of the Levenberg-Marquardt \citep[see Section~10.3
of][]{nocedal2006numerical} least-squares non-linear optimization
algorithm, which rigorously deals with bound constraints or fixed and
tied parameters.\footnote{
The \texttt{python} version of \ppxf\ has used \texttt{method='capfit'}
in place of \texttt{MPFIT} \citep{Markwardt2009} since version 6.5.}
A best-fit parameter at the boundary of the allowed region does not
necessarily indicate a convergence failure, only that the minimum
$\chi^2$ is at the boundary.

For illustration purposes, we fit the individual spaxels for
$\sim$100 datacubes drawn from the set of repeat observations listed
in Table \ref{tab:repeats}. Each spectrum is fit twice, once with the
\mileshc\ library at its native resolution and once after matching
the template data to the MaNGA spectral resolution (Appendix
\ref{sec:resolution}). We bin the results by their S/N and
$\sigma_\ast$ --- i.e., the direct output from \ppxf\ for the
matched-resolution case and after applying the velocity-dispersion
corrections for the resolution-offset case (Section
\ref{sec:sigmacorr}). The top row of Figure
\ref{fig:sigma_completeness} shows the total number of fitted
spaxels, $N_{\rm fit}$, in each fitting mode. Any difference between
the top two panels of Figure \ref{fig:sigma_completeness} is due to
spaxels being located in different $\sigma_\ast$ bins, mostly seen
for the lowest $\sigma_\ast$ bin. The bottom two panels of Figure
\ref{fig:sigma_completeness} compare $N_{\rm fit}$ to the number of
spaxels that are within the \ppxf\ bounds on $\sigma_{\rm obs}$,
$N_{\rm bound}$.

There is a clear difference in the number of fits with $\sigma_{\rm
obs}$ within the trust-region boundary that result from the treatment of
the spectral-resolution difference, particularly at low $\sigma_\ast$
and low S/N. At virtually infinite S/N and with zero template or LSF
mismatch, \ppxf\ has been shown to measure $\sigma_{\rm obs}$ well below
the instrumental dispersion, as shown by \citet[][Figure
2]{2017MNRAS.466..798C}, meaning that this discrepancy is not an
intrinsic issue with the \ppxf\ algorithm itself. Instead, this is
because spectral-resolution uncertainties --- even at the very modest
level expected of either MILES
\citep[$\sim$2\%;][]{FalconBarroso2011miles, 2011A&A...531A.109B} or
MaNGA ($\sim$3\%) --- and noise can lead to matched-resolution templates
with broader lines than the galaxy spectra and drive the \ppxf\ fit to
the $\sigma_{\rm obs}\approx0$ boundary. The pedestal offset in the
spectral resolution of the MaNGA and \mileshc\ spectra helps avoid this
and yields measurements that more often have nonzero $\sigma_{\rm obs}$,
even with the modest spectral resolution difference between MaNGA and
MILES.


To account for the effect of the resolution offset on the measured
$\sigma_{\rm obs}$ reported by \ppxf, we must calculate a correction
that removes this difference and provides the astrophysical velocity
dispersion of the stars, $\sigma_\ast$ (Section \ref{sec:sigmacorr};
Equation \ref{eq:sigmaobs}). We assess the accuracy of the
velocity-dispersion corrections provided in DR15 in the following two
ways.

\begin{figure}
\begin{center}
\includegraphics[width=1.0\columnwidth]{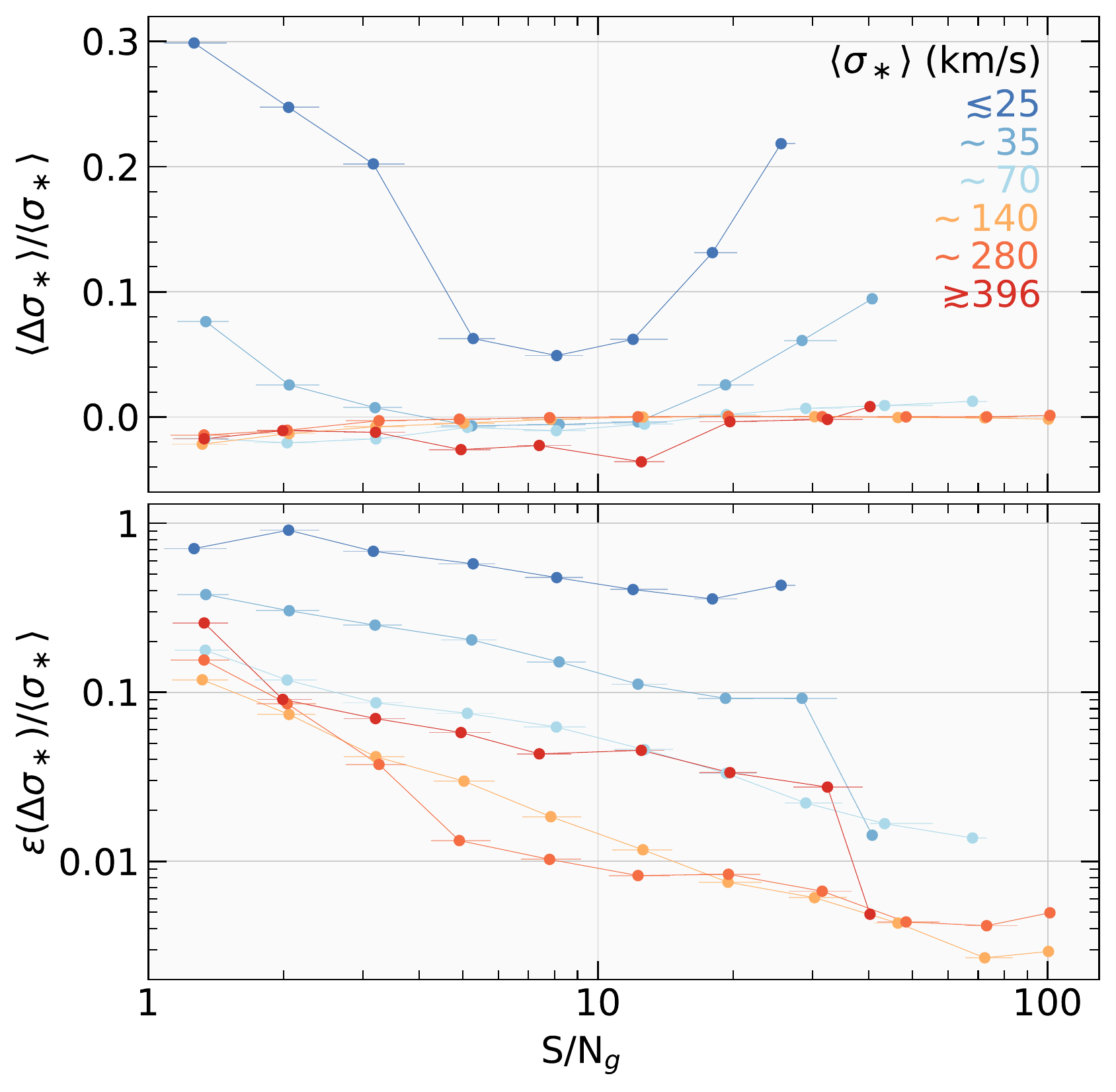}
\end{center}
\caption{The difference between the astrophysical dispersion,
$\sigma_\ast$, obtained for spectra analyzed using templates that
have and have not had their spectral resolution matched to the MaNGA
data (see Section \ref{sec:resmatch}). Data are binned similarly to
Figure \ref{fig:sigma_completeness}. The top panel shows the mean
percentage difference and the bottom panel shows the percent scatter
in the difference; both are plotted as a function of \snrg. The
difference between the two measurements, $\Delta\sigma_\ast$, is
positive if the measurement made with mismatched resolution is larger
than the measurement made with matched resolution. }
\label{fig:matchvoff}
\end{figure}

\begin{figure}
\begin{center}
\includegraphics[width=1.0\columnwidth]{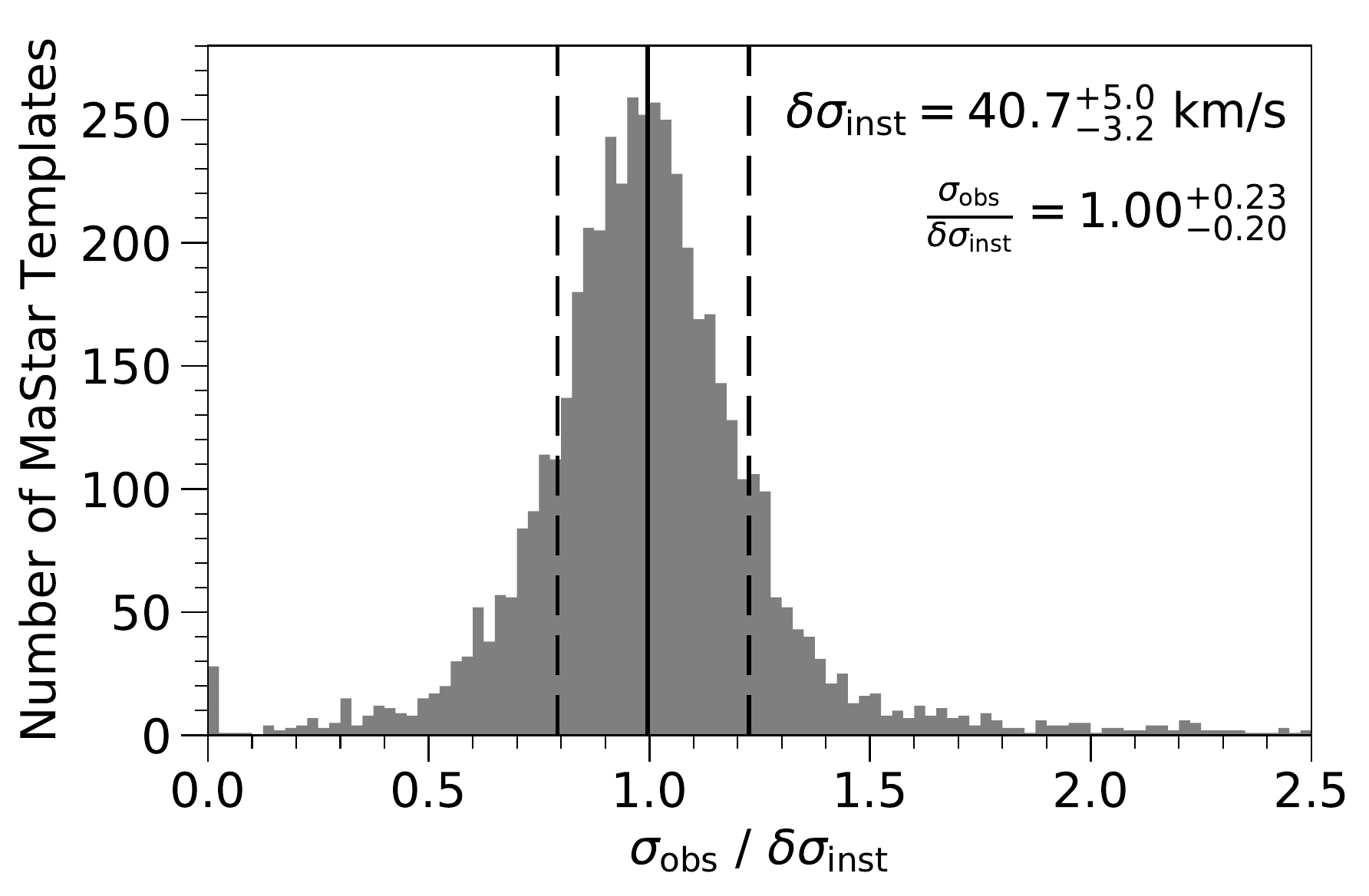}
\end{center}
\caption{A comparison of the first-order velocity-dispersion
correction based on the resolution vectors of a set of $\sim$5000
\mastar\ stellar spectra and the \mileshc\ template library,
$\delta\sigma_{\rm inst}$, and to the effective velocity dispersions
from \ppxf, $\sigma_{\rm obs}$, determined by fitting \mileshc\ to
the \mastar spectra. The plotted histogram of the ratio of these two
measurements has a median of unity with an inner-68\% interval of
$\pm$20\%, as shown in the upper-right corner of the plot. The median
$\delta\sigma_{\rm inst}$ and its inner-68\% interval are also
given.}
\label{fig:mastar_corr}
\end{figure}

First, we compare $\sigma_\ast$ determined with and without matching
the template resolution to the MaNGA data; we exclude any
measurements from the latter with $\sigma_{\rm obs} \leq
\delta\sigma_{\rm inst}$. Figure \ref{fig:matchvoff} illustrates that
the measurements made using either method are consistent to within
1--2\% for $\sigma_\ast \gtrsim 50$ \kms. Systematic differences
become more significant as $\sigma_\ast$ becomes small relative to
$\sigma_{\rm inst} \sim 70$ \kms. There are two considerations that
lead to this discrepancy. First, our current approach (Section
\ref{sec:sigmacorr}) may produce slight underestimates of
$\delta\sigma_{\rm inst}$ that only yield significant differences at
low $\sigma_{\rm obs}$. This underestimation is qualitatively
consistent with results from our idealized simulations and may
motivate a change to the determination of the correction used in
future data releases. Second, at low $\sigma_\ast$, more
measurements from the resolution-offset approach have $\sigma_{\rm
obs} > \delta\sigma_{\rm inst}$ than there are measurements of
$\sigma_{\rm obs}$ within the trust-region boundary in the
matched-resolution approach. This leads to more measurements with
systematically higher $\sigma_{\ast}$ in the resolution-offset case.


Second, we fit a subset of 5000 randomly selected stellar spectra
from the \mastar\ empirical stellar library
\citep{2018arXiv181202745Y} using a similar setup to that used by the
\dap. Assuming negligible stellar rotation or other atmospheric
broadening effects, measurements of $\sigma_{\rm obs}$ determined by
fitting the \mileshc\ templates to \mastar\ spectra provides a direct
measurement of $\delta\sigma_{\rm inst}$ that we can compare to our
first-order calculation (Section \ref{sec:sigmacorr}). Figure
\ref{fig:mastar_corr} shows the measurements of $\sigma_{\rm obs}$
and $\delta\sigma_{\rm inst}$ are very consistent: The median
$\delta\sigma_{\rm inst}$ for the \mastar\ spectra is
$40.7^{+5.0}_{-3.2}$ \kms, and we find that these measurements are
consistent with the directly measured values, $\sigma_{\rm obs}$,
with less than 1\% difference in the median and an inner-68\%
interval of approximately $\pm$20\%.

\begin{figure*}
\begin{center}
\includegraphics[width=0.9\textwidth]{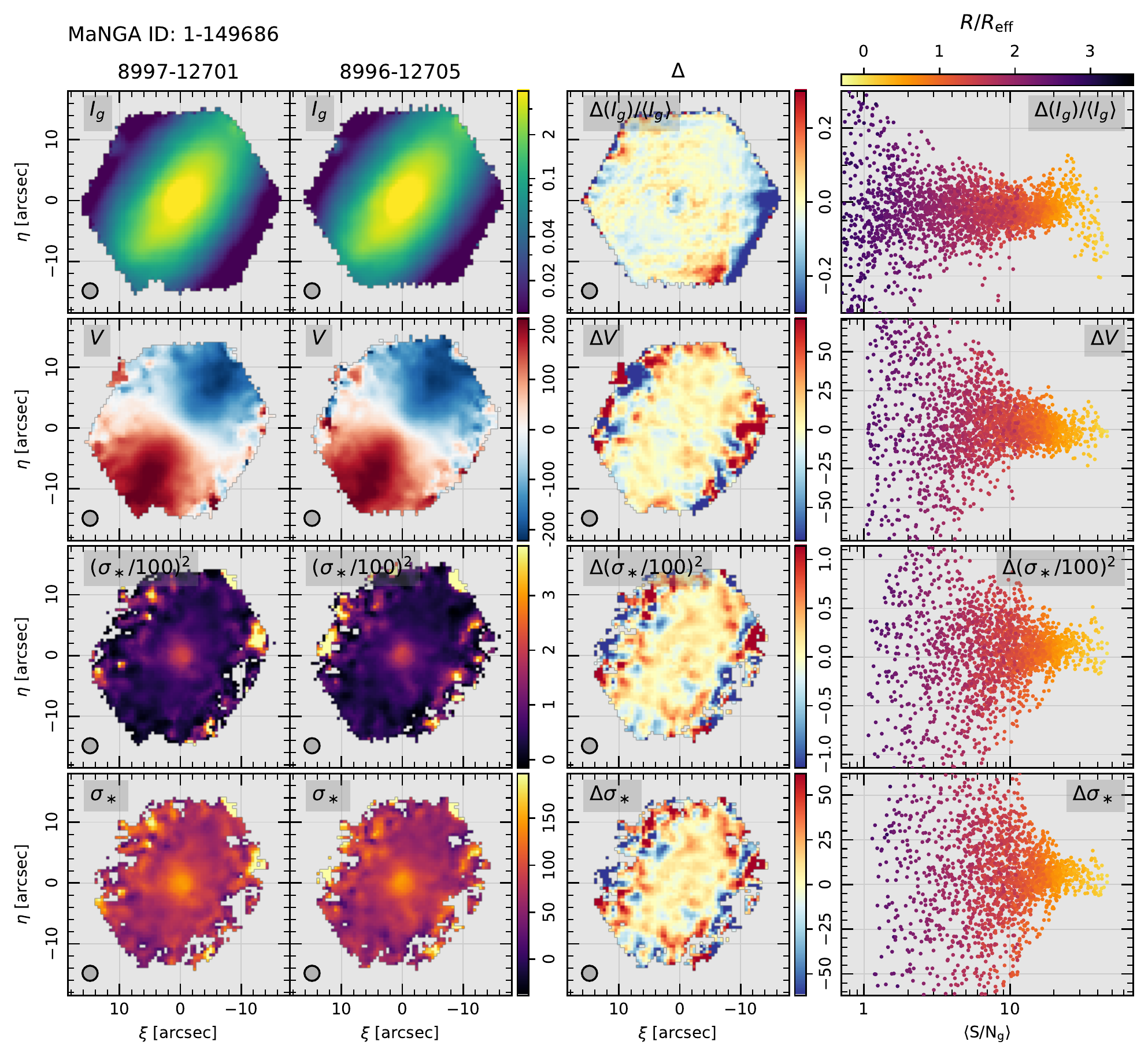}
\end{center}
\caption{The difference in the stellar kinematics of MaNGA galaxy
1-149686 as measured by the \dap\ using observations {\tt 8996-12705}
and {\tt 8997-12705}. From top to bottom, we plot the mean $g$-band
surface-brightness measured from the MaNGA datacubes, the stellar
velocity in \kms, the corrected stellar velocity dispersion squared
(allowing for negative values) in (\kms)$^2$, and the corrected
stellar velocity dispersion, $\sigma_\ast$, in \kms. The left two
columns show the measurements from each observation, the third column
shows their difference, and the right-most column shows the
differences as a function of S/N$_g$, colored by semi-major-axis
radius.}
\label{fig:repeatmaps}
\end{figure*}

Finally, we note that the number of measurements with $\sigma_{\rm
obs} \leq \delta\sigma_{\rm inst}$ for the resolution-offset
measurement is nearly the same as the number of measurements with
$\sigma_{\rm obs} = 0$ for the matched-resolution fits, except at the
very lowest $\sigma_\ast$. One expects this given the consistency of
the comparison shown in Figure \ref{fig:matchvoff}. Therefore, simply
ignoring measurements with $\sigma_{\rm obs} \leq \delta\sigma_{\rm
inst}$ is one possible approach we discuss in Section
\ref{sec:svdusage} in terms of how to use the DR15 data, and this
should lead to results that are consistent with matched-resolution
fits that ignore measurements at the lower trust-region boundary of
$\sigma_{\rm obs}$. However, one should consider alternative
approaches presented in Section \ref{sec:svdusage} that take
advantage of the resolution offset and help mitigate $\sigma_\ast$
biases at low $\sigma_\ast$ and low S/N.

\subsection{Performance}
\label{sec:scperf}

\subsubsection{Empirical Uncertainties from Repeat Observations}
\label{sec:repeats}

We use the first two observations of the 56 galaxies with multiple
MaNGA observations (Table \ref{tab:repeats}) to characterize the
trends of the errors in our stellar kinematics with \snrg\ and test
the accuracy of the formal error estimates returned by
\ppxf (as described toward the end of this Section).
The datacube-reconstruction algorithm generally makes it
straight-forward to compare multiple observations of a single target,
even for IFUs of different size. The target is always centered in the
datacube such that registering the WCS coordinates of MaNGA data from
different IFUs is a simple offset of the arrays to align their
spatial center. In the few cases where this is not true, we align the
data by simply interpolating the relevant values to a common
coordinate grid.

\begin{figure*}
\begin{center}
\includegraphics[width=0.8\textwidth]{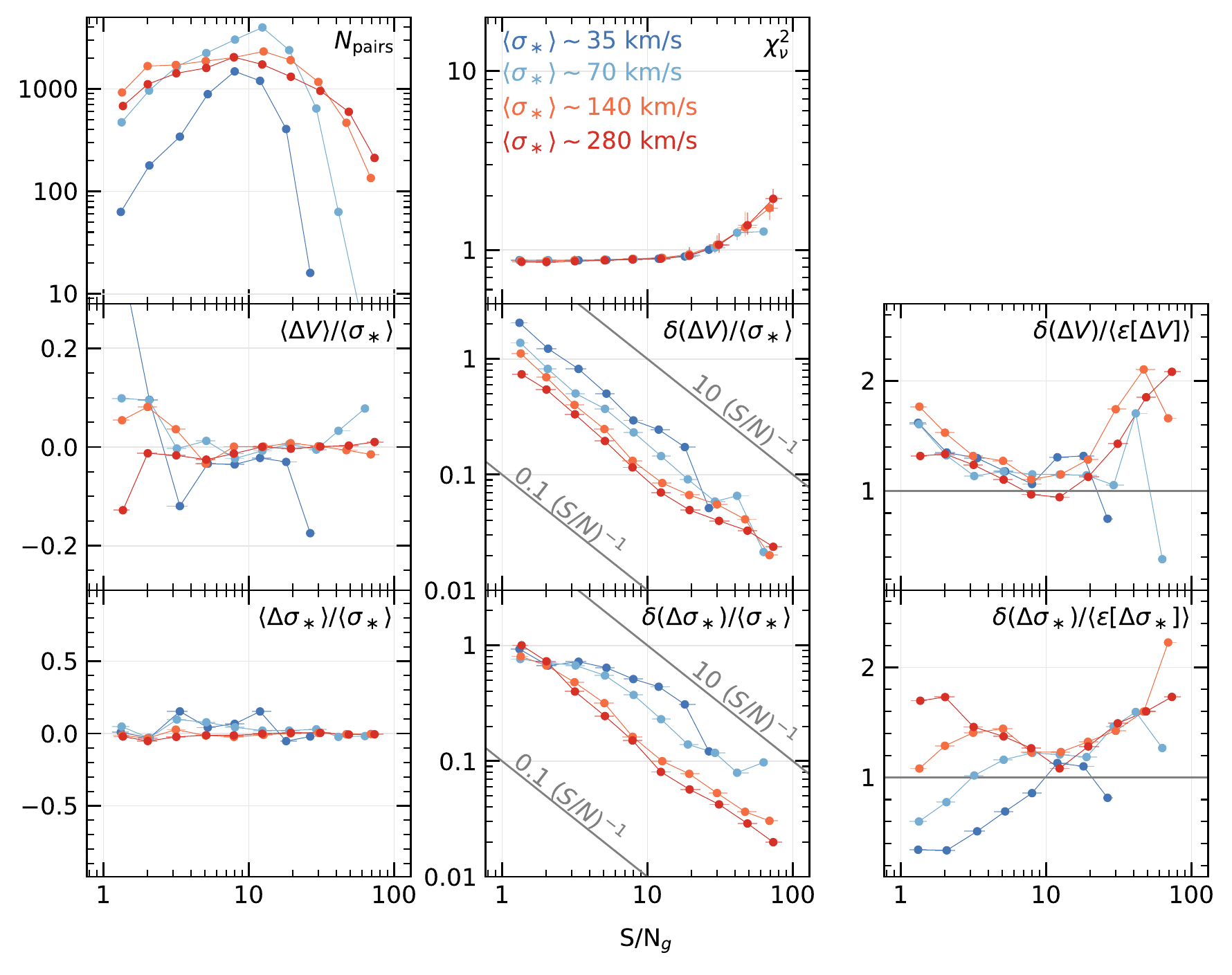}
\end{center}
\caption{A statistical comparison of the spaxel-by-spaxel
measurements of the stellar kinematics from repeat observations of a
set of 56 galaxies (Table \ref{tab:repeats}). Results are binned as a
function of \snrg, according to the abscissa of each panel, and
corrected velocity dispersion ($\sigma_\ast$), according to the color
in the legend. From left-to-right, the top row shows the number of
spaxel pairs compared and the mean $\chi^2_\nu$ for all spaxels in
each \snrg-$\sigma_\ast$ bin. The middle and bottom rows,
respectively, assess the robustness of the velocity, $V$, and
velocity dispersion, $\sigma_\ast$, and their \dap-reported errors
($\epsilon$). For these two rows: The left panels shows the mean
difference between the two kinematic measurements --- $\langle\Delta
V\rangle$, $\langle\Delta \sigma_\ast\rangle$ --- relative to the
mean $\sigma_\ast$, an assessment of systematic error. The middle
panels show the standard deviation in the difference between the two
kinematic measurements --- $\delta(\Delta V)$, $\delta(\Delta
\sigma_\ast)$ --- relative to the mean $\sigma_\ast$, an assessment
of random error. The right panels compare the random error in the
repeat observations to the mean formal error for each kinematic
measurement differences --- $\langle\epsilon[\Delta V]\rangle$,
$\langle\epsilon[\Delta \sigma_\ast]\rangle$, an assessment of the
accuracy of the formal errors.}
\label{fig:repeats}
\end{figure*}

Figure \ref{fig:repeatmaps} compares the two observations of MaNGA
galaxy 1-149686 ({\tt 8996-12705} and {\tt 8997-12705}) as an example
of the differences in the stellar kinematics that we quantify
statistically across all repeat observations in Figure
\ref{fig:repeats}. No registration was required for the observations
in Figure \ref{fig:repeatmaps} because they have identical array
sizes and WCS coordinates. We show the differences in the $g$-band
surface-brightness ($I_g$, calculated directly for each spaxel), the
stellar velocity ($V$), and the corrected stellar velocity dispersion
($\sigma_\ast$); for the latter, we show maps of $\sigma_\ast^2$ that
include negative values resulting from Equation \ref{eq:sigmaobs} and
maps of $\sigma_\ast$ where those spaxels with $\sigma_\ast^2 < 0$
have been removed.

Figure \ref{fig:repeatmaps} shows general consistency between the
mapped quantities and leads one to a sense of the influence of random
errors on the \drp\ and \dap\ products. First, note that the
kinematic residuals do not generally vary on a spaxel-by-spaxel
scale, but more beam-by-beam. This is a natural effect of the
significant spatial covariance (Figure \ref{fig:correlation}) in the
MaNGA data (see Section \ref{sec:guidance}). Second, in the lowest
surface-brightness pixels, the residuals in $\sigma_\ast$ begin to
vary spaxel-by-spaxel, which is driven by systematic error in these
measurements at low S/N and low $\sigma_\ast$.


To empirically assess the random errors in the \dap\ stellar
kinematics in DR15, we similarly register and perform a
spaxel-by-spaxel comparison of the measurements from the first two
observations of all 56 galaxies with multiple MaNGA observations in
Table \ref{tab:repeats}. Here and henceforth, we only focus on our
standard approach for DR15, which is to measure the stellar
kinematics {\it without} resolution matching the template and galaxy
spectra and applying the appropriate correction to correct the \ppxf\
output $\sigma_{\rm obs}$. Spaxel pairs at identical spatial
locations in the repeat observations are binned (geometrically) as a
function of \snrg\ and $\sigma_\ast$. Pairs with a measurement of
$\sigma_\ast^2 < 0$ for either observation are excluded. For each
\snrg-$\sigma_\ast$ bin, we calculate the mean reduced chi-square
($\chi^2_\nu$) of the fit, the mean $\sigma_\ast$
($\langle\sigma_\ast\rangle$), the mean and standard deviation in the
difference in the kinematics (e.g., $\langle\Delta V\rangle$ and
$\delta(\Delta V)$), and the mean of the \dap-reported error in the
difference (e.g., $\langle\epsilon[\Delta V]\rangle$). The results of
these calculations are shown in Figure \ref{fig:repeats}.

From the top middle panel of Figure \ref{fig:repeats}, we note the
increase in $\chi^2_\nu$ toward large \snrg. This is an expected
trend, due to the systematic differences in the \mileshc\ library and
the observed galaxy spectra becoming a more significant fraction of
the random errors in the flux density. These results are consistent
with our findings for the full DR15 sample, as discussed more at
length as part of our general performance assessments of the \dap\ in
Section \ref{sec:performance}.

The bottom two rows of Figure \ref{fig:repeats} are meant to assess,
from left to right, the systematic error, the random error, and the
accuracy of the \dap-reported errors in $V$ and $\sigma_\ast$. The
left panels show that, in the mean, there is little systematic
difference between the $V$ and $\sigma_\ast$ measurements relative to
the scatter in the difference (middle panels). These measurements are
very useful to assess the repeatability of the measurements and the
reliability of the \dap\ formal errors in a relative sense. However,
these data cannot assess the systematic error in an absolute sense,
as we do not know the {\it true} kinematics and are instead comparing
two measurements that may both suffer from an absolute systematic
error. Moreover, by binning the data by the {\it measured}
$\sigma_\ast$, the statistics we have calculated are more of an
assessment of the distribution of the data within the bin instead of
an assessment of the distribution of data about the intrinsic
$\sigma_\ast$.

The expectation found by many authors \citep[e.g.,][]{Jorgensen1995,
2011ApJS..193...21W} is that random errors in both $V$ and
$\sigma_\ast$ should be directly proportional to $\sigma_\ast$
--- at fixed S/N, the centroid uncertainty increases with
the line width --- and inversely proportional to spectral S/N ---
higher S/N spectra provide smaller random errors. The middle column
of the bottom two panel rows in Figure \ref{fig:repeats} demonstrates
that this is generally true for the \dap\ results, with some notable
exceptions. For measurements with $\sigma_\ast\gtrsim100$ \kms\
(i.e., those measurements following the red and orange lines in
Figure \ref{fig:repeats}), we find that the velocity errors are
roughly $\delta(\Delta V) \approx \langle\sigma_\ast\rangle ({\rm
S/N}_g)^{-1}$. That is, the velocity errors are 10\% of $\sigma_\ast$
at \snrg$=$10. The errors in $\sigma_\ast$ are slightly larger than
that, but are also well-approximated by a single proportionality
constant for $\sigma_\ast\gtrsim100$ \kms. For
$\sigma_\ast\lesssim100$ \kms, however, when $\sigma_\ast \sim
\sigma_{\rm inst} \sim 70$ \kms, the proportionality constant changes
such that the errors become a more substantial fraction of
$\sigma_\ast$. In the lowest $\langle\sigma_\ast\rangle$ bin, the $V$
and $\sigma_\ast$ errors are increased to, respectively, $\sim$30\%
and $\sim$60\% of $\sigma_\ast$ at S/N$=$10. We expect this is
because the width of the observed features in the spectrum become
increasingly dominated by $\sigma_{\rm inst}$ such that the
kinematics errors become increasingly independent of $\sigma_\ast$.
Finally, although $\delta(\Delta V) \propto \langle\sigma_\ast\rangle
({\rm S/N}_g)^{-1}$ holds for all $\sigma_\ast$ bins at all \snrg,
the proportionality is lost for $\delta(\Delta \sigma_\ast)$ at low
$\sigma_\ast$ and low \snrg, where $\delta(\Delta \sigma_\ast)$
becomes roughly independent of \snrg. We expect this is because a
larger number of measurements hit the trust-region boundary on
$\sigma_{\rm obs}$ in this regime, due to the large uncertainties.

The right-most panels in Figure \ref{fig:repeats} compare the
uncertainties in the kinematics estimated directly from the repeat
observations, $\delta(\Delta V)$ and $\delta(\Delta \sigma_\ast)$, to
the mean of the errors provided by the formal calculation in \ppxf,
$\langle\epsilon[\Delta V]\rangle$ and $\langle\epsilon[\Delta
\sigma_\ast]\rangle$. Formal uncertainties provided directly by
\ppxf\ are based on the covariance matrix of the fitted parameters,
which can be computed using the Hessian matrix generated as a
byproduct of the Levenberg-Marquardt non-linear least-squares
optimization algorithm; see Section 15.5 of
\citet{NumRecThirdEd}.\footnote{
See also {\tt ppxf.capfit.cov\_err} in the \ppxf\ {\tt python}
package.}
This is a standard approach to measurement errors made under the
(strong) assumptions that the $\chi^2$ space is smooth and unimodal,
that $\chi^2_\nu$ is unity, that the spectral errors are all
independent and Gaussian, and that the covariance between parameters
is negligible (only the diagonal of the covariance matrix is used).
It is worth noting that we have not rescaled the spectrum errors to
artificially yield $\chi^2_\nu = 1$ in our calculation of the
parameter errors we present.

With the exception of the $\sigma_\ast$ uncertainties at low
$\sigma_\ast$ and \snrg, the formal errors tend to underestimate the
empirical errors; however, in the mean, the two measurements are most
often consistent within a factor of two. We return to a discussion of
these results from our repeat observations in the context of the
parameter-recovery simulations presented in the next section.

\subsubsection{Parameter-Recovery Simulations}
\label{sec:recovery}

An industry-standard way to test the performance of fitting algorithms
is to simulate data with known input parameters, apply the fitting
algorithm to those data, and compare the input and output parameters as
a function of S/N \citep[e.g.,][]{Bender1990, Rix1992losvd,
2011ApJS..193...21W}.  Indeed, this is the invaluable first-order check
of the validity of any fitting algorithm.  Such simulations have been
performed multiple times with \ppxf\ over the past $>$15 years,
including simulations specific to MaNGA spectra
\citep{2016MNRAS.462.3955P} as well as in the original \ppxf\ papers
\citep{2004PASP..116..138C, 2017MNRAS.466..798C}.

In the most idealized approach to such simulations, the algorithm used
to generate the model fit to the data is also used to generate the
synthetic data, and the mock spectrum is constructed by one or more
templates contained in the same set used for the fit.  A key feature of
these idealized tests is that any synthetic spectrum can be exactly
reproduced by the provided template library, to within the limits of the
random noise.  Additionally, there is no resolution difference between
the template and synthetic data and the stellar LOSVD is exactly
parameterized (in the case of \ppxf\ by a Gaussian or Gauss-Hermite
function).  In such idealized tests, \ppxf\ has been shown to be robust,
while the method, by design, penalizes the LOSVD towards a Gaussian when
the data do not contain enough information to constrain the higher
moments \citep{2017MNRAS.466..798C}. 

\begin{figure*}
\begin{center}
\includegraphics[width=0.8\textwidth]{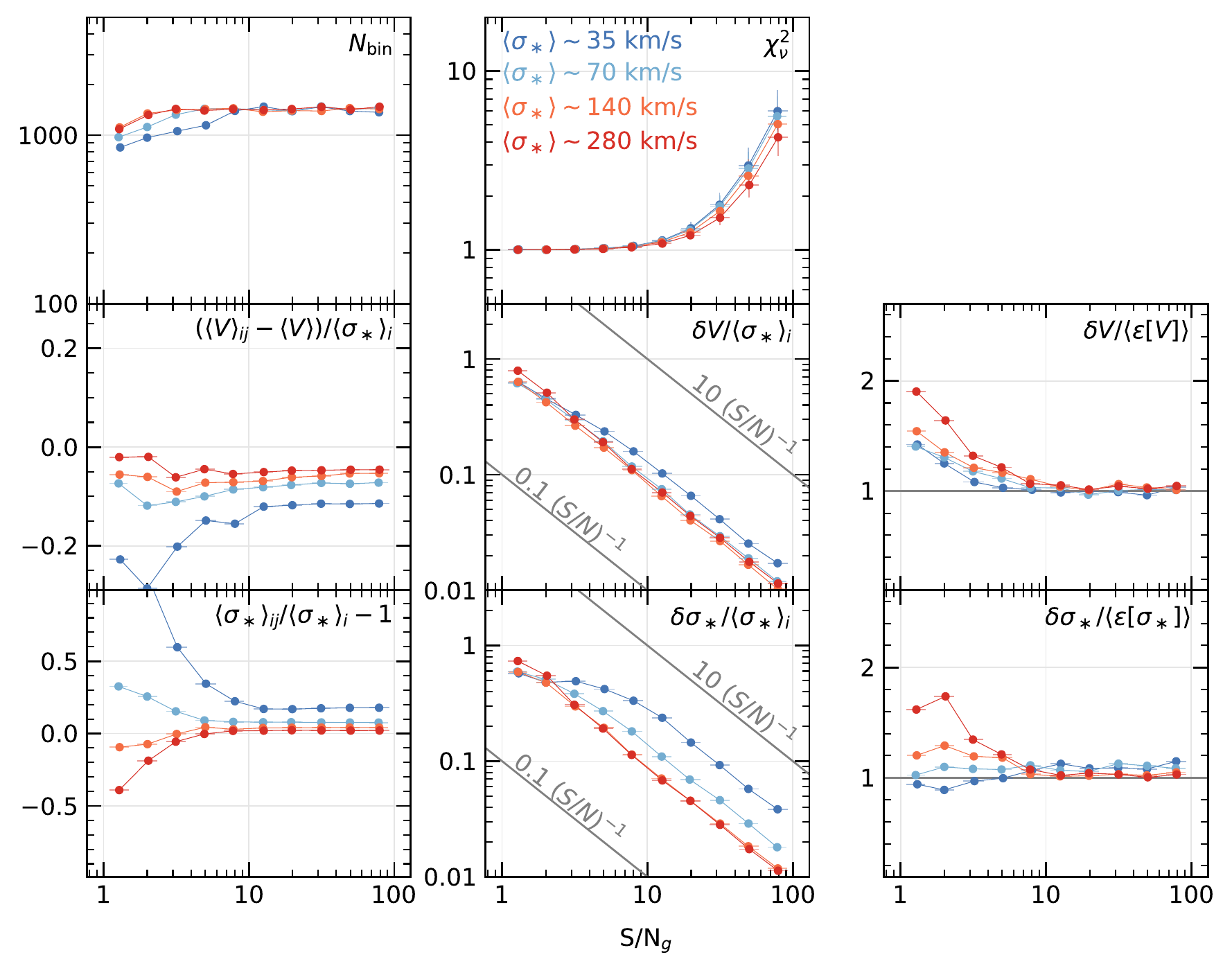}
\end{center}
\caption{Nearly the same as Figure \ref{fig:repeats}, but based on
the idealized simulations discussed in Section \ref{sec:recovery}.
Instead of comparing the results from two fits, we compare the output
kinematic measurement to the known input value used to construct each
synthetic spectrum. The results are {\it binned by the known input
value for $\sigma_\ast$} (cf.\ Figure \ref{fig:sims_out}).}
\label{fig:sims_inp}
\end{figure*}

\begin{figure*}
\begin{center}
\includegraphics[width=0.8\textwidth]{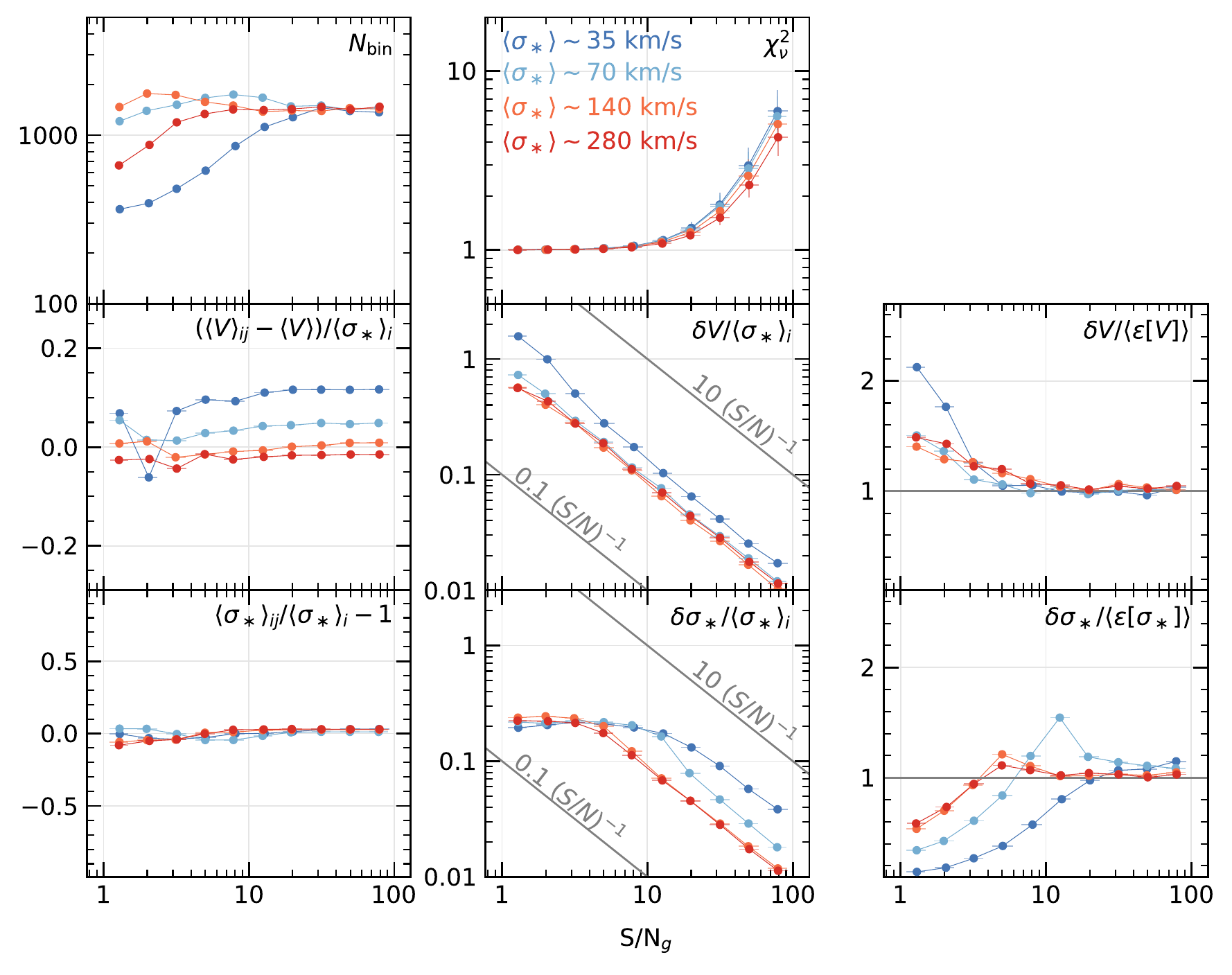}
\end{center}
\caption{The same as Figure \ref{fig:sims_inp}, except with the
simulation results {\it binned by the measured output value for
$\sigma_\ast$}.}
\label{fig:sims_out}
\end{figure*}

This kind of idealized simulation is useful to assess possible intrinsic
limitations of any method, and set limits on what can be achieved, but
are not necessarily representative of what one should expect in real
applications.  In fact, in essentially any practical application of
\ppxf, the template library will be fundamentally limited in its ability
to represent any given galaxy spectrum, the resolution estimation or
resolution matching between the templates and galaxy data will have some
uncertainty, and the stellar LOSVD will not be exactly represented by
parametric forms.  Such model inaccuracies are very difficult to capture
with simulated data in a meaningful way, i.e., in a way that is relevant
to a specific application within the vast parameter space available.

Despite this criticism, parameter-recovery simulations, even if fully
idealized, are still the best way to assess systematic error.  As we
noted in the previous section, repeat observations may provide the most
direct assessment of random error, but they are still limited by the
fact that the true parameter values are unknown and that each
observation could suffer from systematic error in an undetectable way.
Therefore, the goal of our parameter-recovery simulations here is
largely as a comparison to the statistics provided by the repeat
observations and as a check for systematic errors.

In our simulations, we make a small step towards more realistic
synthetic spectra by including the effect of template mismatch and poor
resolution matching, in the sense that our mock solar spectrum is {\em
not} included in the set of 42 \mileshc\ templates we use for the \ppxf\
fit.  Moreover, the resolution of the templates is {\em not} perfectly
matched to that of the mock spectrum.  This implies that, even at
infinite $S/N$, the \ppxf\ fit will never perfectly reproduce the mock
spectrum and some systematic deviations in the recovery should be
expected. 

To construct our synthetic MaNGA spectra, we convolve the BASS 2000
solar spectrum\footnote{
\url{http://bass2000.obspm.fr/solar\_spect.php}}
with an input finely-sampled Gaussian LOSVD, and then convolve again by
a finely-sampled Gaussian kernel to match the spectral resolution to a
fiducial MaNGA resolution vector (Appendix \ref{sec:resolution}).  The
fiducial resolution vector is based on the median of the spectral
resolution vectors for observation {\tt 7815-1902}.  We then integrate
the synthetic spectrum over the pixels to match the spectral sampling of
the MaNGA data, and we add noise to meet a specific S/N level.  We
simulate noise in the spectrum by sampling from a Gaussian distribution
that follows the mean trend of the flux variances with wavelength for
observation {\tt 7815-1902}.  The noise trend is scaled to match the
target S/N for each simulated spectrum.  In a single simulation, one
synthetic spectrum is constructed for each observed spectrum in a
specific MaNGA {\tt RSS} file (e.g., {\tt
manga-7815-1902-LOGRSS.fits.gz}) and each spectrum is fit by the \dap\
using \ppxf\ and the \mileshc\ template library as done for the real
galaxies.

The results of our simulation are shown in Figure \ref{fig:sims_inp},
using a panel layout that is identical to Figure \ref{fig:repeats} for
the repeat observations.  Also similar to our analysis of the repeat
observations, we bin the data by S/N and $\sigma_\ast$, however, in this
case we know and can bin by the {\it true} $\sigma_\ast$ of the
synthetic spectrum, where the number of synthetic spectra falling in
each bin is given by $N_{\rm bin}$.  Of course, instead of comparing
paired results, these simulations compare the input and output
kinematics.

Qualitatively, Figures \ref{fig:repeats} and \ref{fig:sims_inp} have
many common features: $\chi^2_\nu$ similarly increases toward high S/N
reflecting the inability of the \mileshc\ library to exactly produce the
solar spectrum, the random errors are similarly proportional to
$({\rm S/N})^{-1}$, and there is a similar transition of the
random errors in $\sigma_\ast$ to a constant value at low $\sigma_\ast$
and S/N.  Also, the increasing inaccuracy of the formal velocity errors
at S/N$<$10 is qualitatively similar to the behavior of the repeat
observations.  


However, a trend of increasing systematic error emerges at low
$\sigma_\ast$ and ${\rm S/N}$. This is due to the obvious fact that
in that regime the $\sigma_\ast$ uncertainties are so large that the
best-fitting solution must often lie at the physical $\sigma_\ast=0$
positivity boundary. In the limit where $\sigma_\ast=0$ is the true
value, the {\em average} $\sigma_\ast$ is positively biased by an
amount proportional to the $\sigma_\ast$ errors, or inversely to the
${\rm S/N}$. Small inaccuracies in the instrumental resolution of the
template additionally contribute to the trends observed in this
regime, given that even a small mismatch will produce a large effect
on $\sigma_\ast$ at very low $\sigma_{\rm obs}$. The modest bias in
velocity for all $\sigma_\ast$ is likely due to a slight error in the
correction to 0 heliocentric velocity for one or both the BASS2000
spectrum and \mileshc\ templates; this difference of $\lesssim$12
\kms\ is much smaller than the MaNGA pixel scale (70 \kms). Finally,
we note that the formal errors do not exhibit the same trend toward
underestimating the true error toward high S/N seen in the repeat
observations.

For a more direct comparison with Figure \ref{fig:repeats}, it is
useful to rebin the data using the {\it output} $\sigma_\ast$,
instead of the true value, as presented in Figure \ref{fig:sims_out}.
Although it has relatively little effect on the velocity statistics,
the choice of binned quantity substantially changes the $\sigma_\ast$
statistics. This is because now both the abscissa and ordinate are
affected by the positivity boundary bias. In particular, any
indication of systematic error in $\sigma_\ast$ has vanished,
reinforcing our claim that these simulations (i.e., Figure
\ref{fig:sims_inp}) provide key insights into systematic error
despite their other limitations compared to repeat observations.
Also, the transition to a constant fractional error in $\sigma_\ast$
toward low S/N and the spread in $\delta\sigma_\ast /
\langle\epsilon[\sigma_\ast]\rangle$ at low S/N are both more similar
to the empirical behavior in Figure \ref{fig:repeats}.

A distinct component of the repeat-observation statistics in Figure
\ref{fig:repeats} not seen in our simulations is the increase in the
true error relative to the \dap-provided formal error toward high S/N.
Despite the similar increase in $\chi^2_\nu$ toward high S/N, the
simulations show the formal errors are correct to within 10-20\% for
S/N$>$10. We expect this difference between the empirical and simulated
data is because of subtle issues in the repeatability of MaNGA's on-sky
sampling, and propagation of the fiber data to the datacube leads to a
more pronounced difference between the repeat datacubes at high S/N. In
fact, in similar tests with the emission-line properties, discussed in
Section 3.3 of Belfiore et al.\ {\it accepted}, we find that these
differences are fully consistent with the \drp-reported errors in the
astrometric solution of MaNGA's dithered observations.

\subsubsection{Algorithm Implications}

The main algorithmic choice that we have made for DR15 based on the
empirical and simulated performance of the \dap\ presented in this
section is to bin the data to S/N$\sim$10 for our stellar kinematics
measurements.  This limits the number of $\sigma_\ast$ values at the
lowest $\sigma_\ast=0$ boundary and the corresponding bias of the {\em
average} value at lower S/N (Figure \ref{fig:sims_inp}).  We expect this
approach also limits the difference between the true and \dap-reported
errors in kinematics to $\sim$10-20\% at low \snrg.

\subsection{Flagging}
\label{sec:scflags}

Flagging of the \dap\ output stellar kinematics is somewhat minimal
in DR15 with three basic flags applied (see Tables
\ref{tab:dappixmask} and \ref{tab:dapspecmask}): (1) If a spectrum
(binned or otherwise) does not meet either the binning or minimum S/N
criterion for the fit (S/N$>$1), the spaxel is masked and flagged as
having {\tt NOVALUE}. (2) In very rare cases (see Section
\ref{sec:performance}), the fit returns a failure status flag
indicating something went wrong in the fit, which we flag as {\tt
FITFAILED}. (3) We use \ppxf\ to impose boundaries restricting the
viability of the derived kinematics, and set the {\tt NEARBOUND} flag
when the derived kinematics lie at one of these boundaries. For
velocities, {\tt NEARBOUND} is triggered if the velocity is within
1\% of the imposed limits (i.e., $-2000<v<-1980$ or $1980<v<2000$
\kms, with respect to the input redshift). This situation may
indicate that the input redshift was not sufficiently accurate, or
there were problems with the input spectrum. For this reason, the
fits and kinematics should not be trusted. For the velocity
dispersion, {\tt NEARBOUND} is triggered if the velocity dispersion
is within 1\% of the log of the range allowed values (i.e., $0.69 <
\sigma < 0.74$ or $929.8 < \sigma < 1000$ \kms). This flag generally
indicates that the galaxy velocity dispersion is lower than can be
reliably measured. See further comments on the spaxel bit masks in
Sections \ref{sec:performance} and \ref{sec:output}.

\subsection{Usage Guidance}
\label{sec:svdusage}

Given the unusual format of our provided stellar velocity dispersion
measurements, here we summarize and provide some usage guidance.

We emphasize again that the measurements of $\sigma_\ast$ are {\it
not} provided. Instead, we provide a ``raw'' measurement,
$\sigma_{\rm obs}$ --- as determined by \ppxf\ when fitting the MaNGA
spectra with \mileshc\ spectral templates (Section \ref{sec:mileshc})
including the intrinsic spectral-resolution offset --- and a
correction that removes the effects of the MaNGA-MILES resolution
offset, $\delta\sigma_{\rm inst}$. The correction is applied by
solving for $\sigma_\ast$ in Equation \ref{eq:sigmaobs}. With a
typical value of 33 \kms\ (Section \ref{sec:sigmacorr}), the
correction is $\sim$10\% and $\sim$5\% when $\sigma_\ast = 70$ and
100 \kms, respectively. We have calculated spaxel-by-spaxel values of
$\delta\sigma_{\rm inst}$ using the spectral resolution vectors
provided for every MaNGA spaxel, and we recommend they be used
instead of adopting a single correction for all spaxels. Convenience
methods are provided in \marvin\ for this purpose.

Unfortunately, it is difficult to construct definitive criteria for
when one should and should not trust any given determination of
$\sigma_\ast$ in DR15. We first emphasize that Figures
\ref{fig:repeats} and \ref{fig:sims_inp} demonstrate that
uncertainties in $\sigma_\ast$ are a function of {\it both}
$\sigma_\ast$ and \snrg, not just one or the other. Although we have
binned our data to \snrg$\gtrsim$10 to minimize systematic error
$\sigma_\ast$, not all bins will reach \snrg$\sim$10 (cf.\ Figure
\ref{fig:binning_covar}) and not all systematic errors are absent at
\snrg$\gtrsim$10, particularly at low dispersion. Having said that,
Figures \ref{fig:repeats} and \ref{fig:sims_inp} suggest that
measurements of $\sigma_\ast \sim 35$ \kms\ are reasonable at
\snrg$>$20; however, the occurrence of such measurements should be
small, one can expect a systematic error of 10--20\%, and the random
error will be large (30--40\%). A more appropriate limit,
particularly given uncertainties in the LSF width, is likely higher
such that we recommend extreme caution with measurements of
$\sigma_\ast < 50$ \kms\ at S/N$\sim$10.

Based on Figures \ref{fig:sims_inp} and \ref{fig:sims_out}, we expect
the random errors in our kinematics are valid given our binning to
\snrg$\gtrsim$10. However, an important caveat to their accuracy is
the increase in the standard deviation of the results from the repeat
observations in the right-most panels of Figure \ref{fig:repeats}. As
we briefly mention and Belfiore et al., {\it accepted}, discuss more
fully, this increase in the standard deviation is consistent with the
astrometric errors incurred when registering the individual, dithered
MaNGA observations used to reconstruct each datacube. That is, we
expect this increase in error is caused by associating the kinematic
measurement with a particular position on the galaxy, not an
inaccuracy in the reported kinematic error as determined from the
spectrum itself. We expect users will want to associate each
measurement with its spatial location and should therefore include
this increase in the error, assuming they do not instead incorporate
the astrometric errors directly. A rough, by-eye assessment of Figure
\ref{fig:repeats} suggests one should increase the error by a factor
of $\log{\rm S/N}_g$, valid over the range
10$\lesssim$\snrg$\lesssim$100.

Critically, all of the tests performed herein have not considered the
influence of errors in the LSF measurements --- neither in MaNGA nor
in the fitted template spectra --- on the best-fit kinematics. For
example, we have not yet presented a comparison of the MaNGA
$\sigma_\ast$ measurements with those made at higher spectral
resolution \citep[cf.][]{2013MNRAS.428.2980R, 2017A&A...597A..48F}.
Although we continue to improve our measurements of the
wavelength-dependent MaNGA LSF, we currently measure an error of
$\sim$3\%. This leads to $\sim$10\% and $\sim$6\% error in
$\sigma_\ast$ at $\sim$35 and $\sim$50 \kms, respectively.
Particularly at low $\sigma_\ast$, we encourage users to investigate
how this error may or may not influence their analysis and subsequent
inference.


Finally, we comment on the prevailing concern of how to treat
measurements with $\sigma_{\rm obs} < \delta\sigma_{\rm inst}$; i.e.,
when $\sigma^2_\ast$ is negative. First note that, of all
measurements made for DR15, fewer than 2\% show $\sigma_{\rm obs} <
\delta\sigma_{\rm inst}$. In terms of spectra within a datacube, the
median fraction of measurements with $\sigma_{\rm obs} <
\delta\sigma_{\rm inst}$ is 0.3\% and fewer than 5\% of datacubes have
$\sigma_{\rm obs} < \delta\sigma_{\rm inst}$ for more than 14\% of
their measurements.  Even so, importantly, one should {\it not}
consider such measurements equivalent to, nor a valid measurement of,
$\sigma_\ast = 0$. We expect the predominant reason for measurements
of $\sigma_{\rm obs} < \delta\sigma_{\rm inst}$ to be an effect of
the S/N and spectral-resolution errors, as we discuss in detail in
Section \ref{sec:resmatch}.

A minimalist approach is to instead simply ignore these measurements.
As we discuss in Section \ref{sec:resmatch}, the measurements with
$\sigma_{\rm obs} < \delta\sigma_{\rm inst}$ are very similar to the
$\sigma_{\rm obs} = 0$ measurements made by the matched-resolution
fit, with some exceptions at the lowest values of $\sigma_\ast$.
Therefore, ignoring $\sigma_{\rm obs} < \delta\sigma_{\rm inst}$
measurements is nearly equivalent to ignoring measurements of
$\sigma_{\rm obs}$ measurements from \ppxf\ near the lower
trust-region boundary on $\sigma$ in its more common usage.

However, the minimalist approach ignores the possible benefits of
including these $\sigma_{\rm obs} < \delta\sigma_{\rm inst}$
measurements, particularly for aggregation (e.g., the
luminosity-weighted mean $\sigma_\ast$ at fixed radius) and model
fitting. In both cases, results may be biased if one assumes a
Gaussian error distribution when it is instead either truncated at
$\sigma_\ast > 0$ or has a significant positive skew. Given this, we
recommend including the negative $\sigma^2_\ast$ measurements when,
e.g., constructing a mean by computing $(\sum_i \sigma^2_{{\rm
obs},i} - \delta\sigma^2_{{\rm inst},i})^{1/2}$.\footnote{
With the addition of the luminosity weighting, this is the
calculation we perform for the aggregated stellar velocity dispersion
within 1$R_e$ provided in the \dapall\ file; see Section
\ref{sec:dapall}.}
Such a computation can still lead to an imaginary number, which one
can still ignore, but it provides a better accounting of the error
distribution. In terms of model fitting, one can, e.g., reformulate
the merit (likelihood) function to incorporate both $\sigma_{\rm
obs}$ and $\delta\sigma_{\rm inst}$; a reasonable estimate for the
error in $\delta\sigma_{\rm inst}$ is 3\%. The resolution difference
between MaNGA and MILES is small ($\sim$16\%), such that the benefits
of including $\sigma_{\rm obs} < \delta\sigma_{\rm inst}$
measurements may be limited, but still non-negligible.

\section{Bandpass Integrals}
\label{sec:bandpass}

Integrations over spectral regions are general to measurements of
both emission-line moments (Section \ref{sec:emlmom}) and spectral
indices (Section \ref{sec:spindex}), further discussed and defined in
those sections. For clarity, we first introduce the formalism and
nomenclature used when discussing these calculations below.

For a generic vector with values $y_i$ sampled by $i=0...N-1$ pixels
at wavelengths $\lambda_i$, we calculate a bandpass integral via the
discrete sum:
\begin{equation}
S(y) = \sum_i y_i\ {\rm d}p_i\ {\rm d}\lambda_i,
\label{eq:bpintegral}
\end{equation}
where ${\rm d}p_i$ is the fraction of the pixel, $i$, included in a
given passband and ${\rm d}\lambda_i$ is the wavelength step of the
full pixel. For DR15, note that we set ${\rm d}p_i = 0$ for any pixel
within the band that is masked by the \drp, instead of attempting to
replace the errant pixel by interpolation or sampling of a best-fit
model spectrum; spectra with such masked pixels within the relevant
passbands are flagged (Sections \ref{sec:emlmomflags} and
\ref{sec:sindxflags}). Similarly, we calculate the mean of the
spectrum as:
\begin{equation}
\langle y\rangle = \frac{S(y)}{S(1)}.
\label{eq:mean}
\end{equation}
Finally, we calculate the weighted center of a passband as:
\begin{equation}
\langle\lambda\rangle = \frac{S(\lambda y)}{S(y)},
\end{equation}
The \dap\ currently only analyzes the spectra that are log-linearly
sampled in wavelength meaning that $d\lambda_i \propto \lambda_i$.

When required for the measurement, we compute a linear
pseudo-continuum using data in two passbands (sidebands) to either
side (toward shorter --- blue --- and longer --- red --- wavelengths)
of the main feature as follows:
\begin{equation}
C_i = (\langle f\rangle_{\rm red} - \langle f\rangle_{\rm blue})\
\frac{\lambda_i-\langle\lambda\rangle_{\rm
blue}}{\langle\lambda\rangle_{\rm red}-\langle\lambda\rangle_{\rm blue}}
+ \langle f\rangle_{\rm blue},
\label{eq:continuum}
\end{equation}
where $f$ is the flux density and the ``blue'' and ``red'' subscripts
denote calculations in the respective sidebands.

\section{Emission-line Measurements}
\label{sec:emlfit}

The \dap\ provides two sets of emission-line measurements. Of primary
interest are the results of Gaussian line-profile modeling, which is
the focus of this section. However, we also provide non-parametric
measurements based on moment calculations of the continuum-subtracted
data, described in Section \ref{sec:emlmom}. Table \ref{tab:emldb}
provides relevant data for the fitted emission lines with
measurements in DR15; these data are referred to throughout this
Section.


The line-profile modeling performed by the \dap\ is the result of a
second full-spectrum fit using \ppxf\ that simultaneously models all
emission-line features and re-optimizes the weights of the spectral
templates to the underlying stellar continuum. Importantly, this
second full-spectrum fit keeps the stellar kinematics fixed to the
results from the first fit described in Section \ref{sec:stellarkin}.
For completeness, we provide a detailed description of this algorithm
here. However, detailed assessments of our approach and suggested
avenues for improvement are discussed in our companion paper,
Belfiore et al.\ {\it accepted}.

\begin{deluxetable*}{ r l c c c c c c }
\tabletypesize{\footnotesize}
\tablewidth{0pt}
\tablecaption{Emission-Line Parameters \label{tab:emldb} }
\tablehead{ & & \colhead{$\lambda_{\rm rest}$\tablenotemark{a}} & & \colhead{Flux} & \multicolumn{3}{c}{Passbands (\AA)} \\ \cline{6-8}
\colhead{ID} & \colhead{Ion} & \colhead{(\AA)} & \colhead{Ties\tablenotemark{b}} & \colhead{Ratio} & \colhead{Main}
            & \colhead{Blue} & \colhead{Red} }
\startdata
 1 & [\ion{O}{2}]  & 3727.092\phn    & \nodata & \nodata & $3716.3-3738.3$\tablenotemark{c} & $3696.3-3716.3$ & $3738.3-3758.3$ \\ 
 2 & [\ion{O}{2}]  & 3729.875\phn    &      k1 & \nodata & \nodata & \nodata & \nodata \\[2pt]
 3 & H$\theta$     & 3798.9826       & \nodata & \nodata & $3789.0-3809.0$\phm{\tablenotemark{c}} & $3771.5-3791.5$ & $3806.5-3826.5$ \\ 
 4 & H$\eta$       & 3836.4790       & \nodata & \nodata & $3826.5-3846.5$\phm{\tablenotemark{c}} & $3806.5-3826.5$ & $3900.2-3920.2$ \\ 
 5 & [\ion{Ne}{3}] & 3869.86\phn\phn & \nodata & \nodata & $3859.9-3879.9$\phm{\tablenotemark{c}} & $3806.5-3826.5$ & $3900.2-3920.2$ \\ 
 6 & H$\zeta$      & 3890.1576       & \nodata & \nodata & $3880.2-3900.2$\phm{\tablenotemark{c}} & $3806.5-3826.5$ & $3900.2-3920.2$ \\ 
 7 & [\ion{Ne}{3}] & 3968.59\phn\phn & \nodata & \nodata & $3958.6-3978.6$\phm{\tablenotemark{c}} & $3938.6-3958.6$ & $3978.6-3998.6$ \\ 
 8 & H$\epsilon$   & 3971.2020       & \nodata & \nodata & $3961.2-3981.2$\phm{\tablenotemark{c}} & $3941.2-3961.2$ & $3981.2-4001.2$ \\ 
 9 & H$\delta$     & 4102.8991       & \nodata & \nodata & $4092.9-4112.9$\phm{\tablenotemark{c}} & $4072.9-4092.9$ & $4112.9-4132.9$ \\ 
10 & H$\gamma$     & 4341.691\phn    & \nodata & \nodata & $4331.7-4351.7$\phm{\tablenotemark{c}} & $4311.7-4331.7$ & $4351.7-4371.7$ \\ 
11 & [\ion{He}{2}] & 4687.015\phn    & \nodata & \nodata & $4677.0-4697.0$\phm{\tablenotemark{c}} & $4657.0-4677.0$ & $4697.0-4717.0$ \\ 
12 & H$\beta$      & 4862.691\phn    & \nodata & \nodata & $4852.7-4872.7$\phm{\tablenotemark{c}} & $4798.9-4838.9$ & $4885.6-4925.6$ \\ 
13 & [\ion{O}{3}]  & 4960.295\phn    &     a14 &   0.340 & $4950.3-4970.3$\phm{\tablenotemark{c}} & $4930.3-4950.3$ & $4970.3-4990.3$ \\ 
14 & [\ion{O}{3}]  & 5008.240\phn    & \nodata & \nodata & $4998.2-5018.2$\phm{\tablenotemark{c}} & $4978.2-4998.2$ & $5018.2-5038.2$ \\ 
15 & [\ion{He}{1}] & 5877.243\phn    & \nodata & \nodata & $5867.2-5887.2$\phm{\tablenotemark{c}} & $5847.2-5867.2$ & $5887.2-5907.2$ \\ 
16 & [\ion{O}{1}]  & 6302.046\phn    & \nodata & \nodata & $6292.0-6312.0$\phm{\tablenotemark{c}} & $6272.0-6292.0$ & $6312.0-6332.0$ \\ 
17 & [\ion{O}{1}]  & 6365.535\phn    &     a16 &   0.328 & $6355.5-6375.5$\phm{\tablenotemark{c}} & $6335.5-6355.5$ & $6375.5-6395.5$ \\ 
18 & [\ion{N}{2}]  & 6549.86\phn\phn &     a20 &   0.327 & $6542.9-6556.9$\phm{\tablenotemark{c}} & $6483.0-6513.0$ & $6623.0-6653.0$ \\ 
19 & H$\alpha$     & 6564.632\phn    & \nodata & \nodata & $6557.6-6571.6$\phm{\tablenotemark{c}} & $6483.0-6513.0$ & $6623.0-6653.0$ \\ 
20 & [\ion{N}{2}]  & 6585.271\phn    & \nodata & \nodata & $6575.3-6595.3$\phm{\tablenotemark{c}} & $6483.0-6513.0$ & $6623.0-6653.0$ \\ 
21 & [\ion{S}{2}]  & 6718.294\phn    & \nodata & \nodata & $6711.3-6725.3$\phm{\tablenotemark{c}} & $6673.0-6703.0$ & $6748.0-6778.0$ \\ 
22 & [\ion{S}{2}]  & 6732.674\phn    & \nodata & \nodata & $6725.7-6739.7$\phm{\tablenotemark{c}} & $6673.0-6703.0$ & $6748.0-6778.0$ 
\enddata
\tablenotetext{a}{Ritz wavelengths in vacuum from the National Institute
of Standards and Technology (NIST;
\url{http://physics.nist.gov/PhysRefData/ASD/Html/help.html}).}
\tablenotetext{b}{The velocities of {\em all} lines are tied to one
another; `k{\it n}' signifies the line has all its kinematics
($V$,$\sigma$) tied to line with ID {\it n}; `a{\it n}' signifies that
the line has all its parameters tied to line {\it n} with a fixed flux
ratio.}
\tablenotetext{c}{The [\ion{O}{2}] doublet is unresolved.  Our moment
measurements adopt a primary band that brackets both emission lines.
One must sum the Gaussian results for both lines in the doublet when
comparing to the single, zeroth-moment measurement.}
\end{deluxetable*}

Belfiore et al., {\it accepted}, pay particular attention to the
effects of the continuum fit on the resulting emission-line
properties, and assess the advantages of simultaneously fitting both
components (gas and stars). This approach is similar to, and
motivated by the success of, {\tt GANDALF} \citep[based on an early
IDL version of \ppxf]{Sarzi2006}; however, it is worth highlighting
the following differences. Whereas \ppxf\ treats both the gas and
stellar spectra in exactly the same way \citep[as spectral templates;
see][Section 3.6]{2017MNRAS.466..798C} {\tt GANDALF} models the
emission lines as Gaussian functions that are added to the stellar
spectra within the code. The \ppxf\ approach for the gas emission was
made possible by the analytic Fourier convolution
\citep{2017MNRAS.466..798C} introduced in version 6.0 of \ppxf\
package in {\tt python}. This approach is more general and allows for
greater flexibility to the user, but it requires some initial setup.
Therefore, we go into some detail about the construction of the
emission-line templates in Section \ref{sec:emltpl}.

For DR15, the template library used for the stellar component
(\mileshc) is also used during the stellar-kinematics fit. However,
even though we expect the \mileshc\ library to provide the best
measurements of the stellar kinematics, they prove more limiting in
the emission-line modeling. In particular, the spectral range of the
MILES library limits the emission lines that can be modeled (Figure
\ref{fig:showcasespec}), prohibiting models of the [\ion{S}{3}] lines
at 9071.1~\AA\ and 9533.2~\AA. Also, the unrestricted optimization of
the \mileshc\ templates ignores all of the physics that we understand
about stellar mixes relevant to how stars are formed and evolve in
galaxies. Belfiore et al., {\it accepted}, explore the use of
stellar-population-synthesis templates with larger spectral coverage
in a development version of the \dap\ that allows for different
templates to be used for the stellar kinematics and the emission-line
modeling.

\begin{figure*}
\begin{center}
\includegraphics[width=\textwidth]{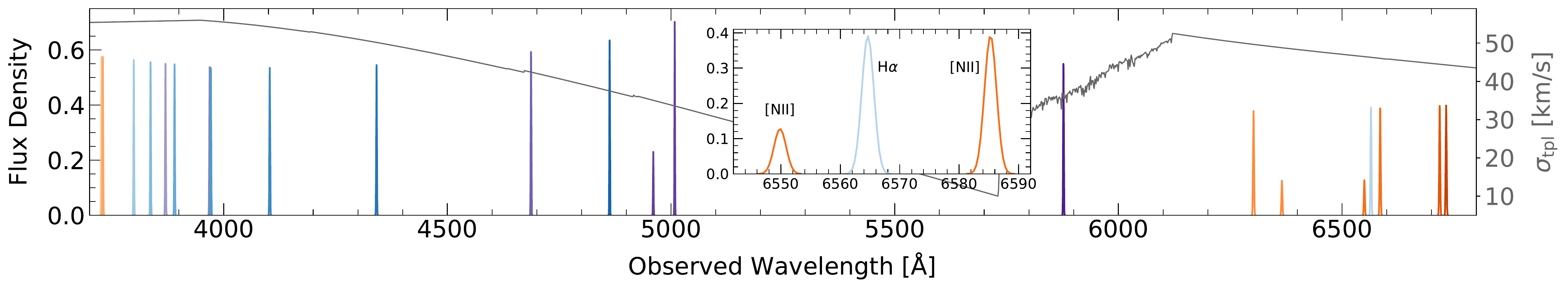}
\end{center}
\caption{Emission-line templates constructed for fitting observation
{\tt 7815-3702}, as described in Section \ref{sec:emltpl}, for the
emission lines listed in Table \ref{tab:emldb}. Spectra associated
with each template are colored according to the line groups defined
by Belfiore et al., {\it accepted}: Balmer lines in blue,
low-ionization lines ([\ion{O}{2}], [\ion{O}{1}], [\ion{N}{2}], and
[\ion{S}{2}]) in orange, and high-ionization lines ([\ion{Ne}{3}],
\ion{He}{2}, [\ion{O}{3}], and \ion{He}{1}) in purple (cf.\ Figure
\ref{fig:showcasespec}). The inset shows the [\ion{N}{2}]$+$H$\alpha$
complex. Note the [\ion{N}{2}] doublet features are components of the
same template spectrum such that the relative kinematics and flux
ratio of the two lines are held fixed. The resolution of the template
features, $\sigma_{\rm tpl}$, is plotted in gray following the right
ordinate.}
\label{fig:etpl}
\end{figure*}

The spatial binning of the data (Section \ref{sec:binning}) for DR15 is
primarily driven by the desire to meet a S/N threshold that minimizes
any systematic errors in the stellar kinematics (Section
\ref{sec:scperf}).  It is reasonable then that the definition of S/N
used for binning the data is based on the broad-band S/N within the SDSS
$g$ filter.  However, this means that the spatial binning applied in
Section \ref{sec:binning} is likely inappropriate for emission-line
focused science.  It is indeed common-place nowadays for emission lines
to be fit with a different binning scheme than used for analyses of the
broad-band continuum.  We have implemented a similar strategy in the
\dap, where the stellar kinematics are performed on the Voronoi-binned
data and the emission-line modeling can be performed on individual
spaxels.  We refer to this as the ``hybrid'' binning output.  However,
it is also useful to have the emission lines fit using the same spectra
as used for the stellar continuum, particularly if the goal is simply to
subtract them for subsequent analysis
\citep[e.g.,][]{2017MNRAS.472.4297W}.  This difference between fitting
the emission lines in the binned spectra or the individual spaxel data
is the primary distinction between the two types of data provided by the
\dap\ for DR15 ({\tt DAPTYPE} is {\tt VOR10-GAU-MILESHC} or {\tt
HYB10-GAU-MILESHC}; see Section \ref{sec:output}).  Sections
\ref{sec:emlhybrid} and \ref{sec:emlbin} separately describe the fit
iterations in these two methods for clarity.

\subsection{Emission-Line Template Construction}
\label{sec:emltpl}

Similar to {\tt GANDALF}, the \dap\ sets how the emission-line templates
are constructed via an input file.  We are free to define (1) the
functional form used to construct the emission line, (2) any lines that
should have a fixed flux ratio, and (3) how line parameters should be
tied together to force all lines to have the same velocity and/or
velocity dispersion.  Table \ref{tab:emldb} provides the fixed flux
ratios adopted for the [\ion{O}{3}], [\ion{O}{1}], and [\ion{N}{2}]
lines in DR15, as well as how lines are tied kinematically.  We force
the velocities of all lines to be identical, and we force the velocity
dispersion of the [\ion{O}{2}], [\ion{O}{3}], [\ion{O}{1}], and
[\ion{N}{2}] doublet features to be the same.  Belfiore et al., {\it
accepted}, explores the effects of different tying strategies on the
emission-line properties, such as grouping lines by ionization potential
(Figure \ref{fig:showcasespec}).

The emission-line templates are constructed once per fitted datacube.
Each emission feature is included in the relevant template with a
known flux (typically unity), line center (at the rest wavelength),
and width. The width of the line is set based on the minimum
instrumental dispersion over all spatial positions in the datacube,
as given by the \drp-provided spectral-resolution measurements.
However, we apply a quadrature offset such that the minimum
line-profile dispersion is 10 \kms. That is, we first compute the
minimum instrumental dispersion for each wavelength, $\lambda$, over
all spatial positions, ($x$,$y$):
\begin{equation}
\sigma_{\rm inst,min} (\lambda) = {\rm min}_{xy}\left[
     \frac{c}{\sqrt{8\ln 2}}\ R^{-1}_{xy}(\lambda) \right],
\end{equation}
where the resolution at a given spatial position, $R_{xy}(\lambda)$,
is sampled at the observed wavelengths, $\lambda = \lambda_{\rm tpl}
(1+z_0)$ ($z_0$ is the cosmological redshift; cf.\ Section
\ref{sec:vdef}). Then we apply the quadrature offset:
\begin{equation}
\sigma^2_{\rm tpl} (\lambda) = \sigma^2_{{\rm inst,min}} (\lambda)
     - {\rm min}_\lambda[\sigma^2_{\rm inst,min}] + 10^2.
\end{equation}
By significantly lowering the width of the lines in the template, we
ensure that the \ppxf\ fit yields a non-zero dispersion measurement
for {\it all} lines, accommodating uncertainties in and variation of
the LSF within the datacube. For example, the emission-line templates
and the relevant $\sigma_{\rm tpl}$ used during the fit of
observation {\tt 7815-3702} are shown in Figure \ref{fig:etpl}.

Lines that are forced to have the same flux ratio can be modeled by
including them in a single template. For example, see the
[\ion{N}{2}] lines in the inset panel of Figure \ref{fig:etpl}. The
fixed flux ratio is then achieved naturally in the linear
optimization of the template weights performed by \ppxf. All other
lines must have their own unique template, such that there are 19
emission-line templates used in DR15. The \dap\ assigns each
emission-line template to a velocity and velocity-dispersion group,
which is used to construct the initial parameter guesses and the
parameter tying structure used by \ppxf. Finally, the emission-line
templates are combined with the stellar templates and assigned
kinematic components, one component for the stars and separate
components as needed for the kinematic parameters to be fit for the
emission lines.

\subsubsection{Rigorous Sub-Pixel Emission-Line Templates}
\label{sec:lineprofile}


As stated previously, the quadrature offset applied to the
emission-line profiles in our emission-line templates is to ensure
that we measure a non-zero velocity dispersion for {\it all} emission
lines when fitting the MaNGA galaxy spectra. However, in making the
template lines narrower, we risk under-sampling the emission
features. Indeed, if the lines are too narrow, use of a Gaussian
function directly to define the line profiles would lead to very
inaccurate results. In general, therefore, we account for this issue
in a mathematically rigorous manner by defining the template
emission-line profile directly in Fourier space
\citep{2017MNRAS.466..798C}:\footnote{
One can compute this profile using the {\tt python} function
\texttt{ppxf.ppxf\_util.gaussian} provided with the \ppxf\ software
package. For the results from Equation \ref{eq:gas_template} or
Equation \ref{eq:gas_template_pixel}, set {\tt pixel=False} or {\tt
pixel=True}, respectively.}
\begin{equation}
\mathcal{L}(x) = \mathcal{F}^{-1}\left[\exp\left(-\frac{\omega^2
\sigma_x^2}{2} - {\rm i}\, \omega\, x_0\right)\right],
\label{eq:gas_template}
\end{equation}
where the expression within the square brackets is the analytic Fourier
transform of a Gaussian with dispersion $\sigma_x$ and center $x_0$ in
the same logarithmically spaced spectral pixels as for the stellar
templates.  Here, $\mathcal{F}^{-1}$ is the inverse of the {\em
discrete} Fourier transform of real input.  Importantly, the calculation
of the inverse Fourier transform is required only once per emission
line, meaning that the overhead compared to sampling a Gaussian function
is entirely negligible.

With this definition, the convolution of the emission-line template
within \ppxf\ produces, by construction, a numerically accurate Gaussian
regardless of the width of the emission line.  Indeed, this approach
even allows one to use mathematically correct Dirac delta functions as
emission-line profiles in \ppxf\ by simply setting $\sigma_x=0$ in
Equation \ref{eq:gas_template}.  When the emission line is well sampled,
namely when $\sigma_x\gtrsim1$, Equation \ref{eq:gas_template} reduces
to a normalized Gaussian function that {\it sums} to unity.

In general, for maximum accuracy, one will want to include the
integration of the emission lines over the spectral pixels before
fitting the observed spectrum.  Ignoring this effect is equivalent to
adopting an instrumental dispersion $\approx3$\% lower than the ``true''
value one would measure with a well-sampled line-spread function
\citep[Figure~4 of][]{2017MNRAS.466..798C}.  In this case, the
emission-line profiles becomes
\begin{equation}
\mathcal{L}(x) = \mathcal{F}^{-1}\left[\exp\left(-\frac{\omega^2
\sigma_x^2}{2} - {\rm i}\, \omega\, x_0\right)\, {\rm
sinc}\left(\frac{\omega}{2\pi}\right)\right],
\label{eq:gas_template_pixel}
\end{equation}
where the expression in square brackets now represents the analytic
Fourier transform of a Gaussian convolved with an unitary pixel, with
the sinc function being the Fourier transform of a unitary box function.
Like before, this line profile produces a mathematically accurate
Gaussian integrated within the pixels when convolved with the
ionized-gas LOSVD in the \ppxf\ fit.

\subsubsection{Usage Notes}

We make the following two notes regarding the practical use of this
method of constructing the emission-line profiles in the \dap.

First, at least for the data in DR15, the emission-line templates are
constructed on the same wavelength grid as the stellar templates,
meaning that the pixel sampling is a factor of four more frequent
than the MaNGA galaxy spectra (Section \ref{sec:tplprep}; i.e., the
pixel width is 17.3 \kms). Even so, the definition of the Gaussian
profiles for the emission-lines directly in Fourier space will be
critical to robust measurements for lines falling in spectral regions
at our minimum $\sigma_{\rm tpl}$ of 10 \kms.

Second, although not relevant to the example templates shown in
Figure \ref{fig:etpl} given the form of $\sigma_{\rm tpl}$, the line
profiles generated using equation \ref{eq:gas_template_pixel} will
show characteristic Fourier ringing when $\sigma_{\rm tpl}$ is much
smaller than the size of the pixel. Although germane to the goal of a
mathematically rigorous convolution, these features may remain in the
final fit if the best-fit, convolved profile is significantly
under-sampled. We know of no specific examples of this in the \dap\
results provided in DR15; however, we raise the issue here and
continue to assess its influence, if any, in general use of the \dap.

\subsection{Fit Iterations: Remapping from Binned Spectra to Individual
Spaxels, the Hybrid Scheme}
\label{sec:emlhybrid}

Our ``hybrid'' approach to emission-line fitting refers to the
combination of having the stellar kinematics determined using the
binned spectra, whereas the combined
stellar-continuum$+$emission-line fit is performed for individual
spaxels. In this approach, the emission-line fitter runs through
three fit iterations:

\smallskip

\noindent {\bf Step (1):} We fit the binned spectra assuming the
ionized gas is part of a single kinematic component; i.e., all lines
have the same velocity and velocity dispersion. Only multiplicative
polynomials are used to adjust the stellar-continuum model such that
the absorption-line equivalent-width distribution from the stellar
library is preserved. The multiplicative polynomial is only applied
to the stellar templates, not the emission-line templates, to
maintain any flux ratio between line doublets. The spectra are fit
twice, once on the input binned spectra and then again after removing
3-$\sigma$ outliers. When fitting each binned spectrum, the weights
of {\it all} templates (\mileshc\ for DR15) are optimized; i.e., the
emission-line module does not implement the {\tt NZT} approach used
by the stellar-kinematics module (Sections \ref{sec:stellarfititer}
and \ref{sec:tpldownselect}). Instead, the results of this fit are
used to set stellar-continuum template weights and initial guess
kinematics that are {\it unique to each binned spectrum} and applied
to the spaxels associated with that bin. Spaxels are associated to
the binned spectra by simple on-sky proximity.\footnote{
This approach is a development hold-over. Of course, we know exactly
which spaxels were combined into each bin, meaning that this step is
unnecessary and can actually lead to incorrectly matching a spaxel to
the relevant bin. In recent updates to the \dap\ (MPL-8; {\tt version
2.3.0}), we associate each spaxel directly to the correct parent
bin.}
The stellar continuum used in the fit to the individual spaxels is
determine by re-weighting the optimal template determined for the
associated binned spectrum. That is, in all remaining iterations, the
templates used by \ppxf\ to model the spectra consist of a single,
optimized template for the stellar continuum and the full set of
emission-line templates.

\smallskip

\noindent {\bf Step (2):} We fit the individual spaxels, again
assuming the ionized gas is in a single kinematic component. The fit
includes a 3-$\sigma$ rejection iteration and the initial starting
guess for the next iteration are updated from the previous fit.

\smallskip

\noindent {\bf Step (3):} Finally, the individual spaxels are fit
again, however, this iteration assigns the kinematic components as
dictated by the input file; see Section \ref{sec:emltpl} and Table
\ref{tab:emldb}. This iteration {\it does not} include a rejection
iteration.

\subsection{Fit Iterations: Binned Spectra, No Remapping}
\label{sec:emlbin}

When the emission-line fits are provided for the binned spectra and not
remapped to the individual spaxels, the emission-line fitter runs
through two fit iterations.  These iterations are virtually identical to
the last two iterations in Section \ref{sec:emlhybrid} with the
additional step of obtaining the optimal stellar-continuum template in
the first iteration.  In the interest of completeness, the two iteration
steps are:

\smallskip

\noindent {\bf Step (1):} We fit the binned spectra assuming the
ionized gas is part of a single kinematic component and including a
3-$\sigma$ rejection iteration. This iteration determines the optimal
stellar-continuum template and initial starting guesses for the
second iteration (cf. Step 1 in Section \ref{sec:emlhybrid}).

\smallskip

\noindent {\bf Step (2):} The binned spectra are fit again, however,
this iteration assigns the kinematic components as dictated by the
input file; see Section \ref{sec:emltpl} and Table \ref{tab:emldb}.
This iteration {\it does not} include a rejection iteration (cf. Step
3 in Section \ref{sec:emlhybrid}).

\subsection{Emission-Line Modeling Results}
\label{sec:emlmeasurements}

Unlike the stellar-continuum fit, there is a non-trivial step in parsing
the results of the \ppxf\ fit into the relevant quantities for the
emission-line modeling.  The nuances of this parsing are discussed here,
and particularly important usage implications are further discussed in
Section \ref{sec:output}.

Our general approach is to provide flux, equivalent-width, velocity
(see Section \ref{sec:vdef}), and velocity-dispersion measurements
for {\it all} lines listed in Table \ref{tab:emldb}, even for
quantities that are constrained by multiple lines. For fluxes, our
convention is to construct the emission-line templates such that the
{\it integral} of the line is unity, except for those lines with a
fixed flux ratio (Section \ref{sec:emltpl}). However, the convolution
performed by \ppxf\ conserves the {\it sum} of the template pixel
values, not its integral. Therefore, to calculation the best-fitting
flux for each emission line based on the input flux of line $l$,
$F_{{\rm tpl},l}$, and the optimized weight of the associated
template $j$, $w_j$, we account for the effect of the convolution
convention on the line {\it integral} following:
\begin{equation}
F_l = w_j F_{{\rm tpl},l} (1 + z_j);
\label{eq:lineflux}
\end{equation}
note that $F_{{\rm tpl},l} = 1$ except where given otherwise in Table
\ref{tab:emldb}. We emphasize that the need for Equation
\ref{eq:lineflux} is an artifact of the template construction and the
\ppxf\ convolution.

Critically, note that all emission-line fluxes have been corrected
for Galactic foreground extinction because the spectra are
appropriately de-reddened (Section \ref{sec:galext}) before being
modeled. However, we have not attempted to correct any of the line
fluxes for attenuation by the dust content of the target galaxy
itself.

Emission-line equivalent widths based on the line-profile modeling
results are calculated as:
\begin{equation}
{\rm EW}_l = \frac{1}{1+z_l}\ \frac{F_l}{C_l},
\label{eq:lineew}
\end{equation}
where $F_l$ is from Equation \ref{eq:lineflux} and $C_l$ is the
pseudo-continuum of the observed spectrum (see Equation
\ref{eq:continuum}) interpolated at the line center, $\lambda =
(1+z_l)\lambda_{\rm rest}$, using the definition of the sidebands in
Table \ref{tab:emldb} (see the discussion of the passbands used here
and in the non-parametric calculations in Section \ref{sec:emlmom}).
Note that, in a slight modification to Equation \ref{eq:continuum},
the computation of $C_l$ in Equation \ref{eq:lineew} does not use the
passband-integrated mean flux in each sideband (e.g., $\langle
f\rangle_{\rm red}$), but instead uses a simple median of the flux
within the passband. Also note that, although Equation
\ref{eq:lineew} is defined such that emission and absorption yield,
respectively, positive and negative EW, the non-negative constraint
on the template weights imposed by \ppxf\ means that the model-based
equivalent widths {\it cannot} be negative. Finally, note that the
factor of $(1+z_l)$ in Equation \ref{eq:lineew} is needed to convert
our {\it observed-frame} equivalent-width measurement to the {\it
rest-frame}.

The velocity and velocity dispersion of each line is determined by
its associated kinematic component. The velocity measurements are
offset by the cosmological redshift of the galaxy in a manner
identical to the method used for the stellar velocities described in
Section \ref{sec:vdef}. Also similar to the stellar kinematics, the
gas velocity dispersions are provided as one would measure directly
from the spectrum and must be corrected by the user using the
provided instrumental dispersion (see Section
\ref{sec:gassigmacorr}).

Errors are also provided for each line flux, equivalent-width,
velocity, and velocity dispersion regardless of whether or not the
parameter was fit independently. This has important implications for
the propagation of errors in averaged quantities. For example, {\it
all} the velocity errors for the lines in a given spectrum are the
same because the velocities of all lines are tied during the fit,
meaning that one cannot combine the velocities of multiple lines in a
nominal way to improve the velocity precision.

\subsection{Velocity Dispersion Corrections}
\label{sec:gassigmacorr}

For the ionized-gas velocity dispersions, the instrumental resolution
$\sigma_{{\rm inst},i}$ at the wavelength of each line $i$ is
provided (Section \ref{sec:output}; Appendix \ref{sec:datamodel})
such that the corrected velocity dispersion of each line is
\begin{equation}
\sigma^2_{l,i} = \sigma^2_{{\rm obs},i} - \sigma^2_{{\rm inst},i},
\label{eq:sigmaeml}
\end{equation}
where $\sigma^2_{\rm obs}$ is the velocity dispersion of the line in
the observed spectrum; the \dap\ \dapmaps\ files provide
$\sigma^2_{\rm obs}$ and $\sigma^2_{\rm inst}$.

\subsection{Flagging of Modeling Results}

Flagging of the emission-line model fitting results (see Tables
\ref{tab:dappixmask} and \ref{tab:dapspecmask}) are similar to those
used for the stellar kinematics, including the caveat that they are
currently rather limited.  The {\tt NOVALUE} and {\tt NEARBOUND} flags
have the same meaning; the criteria used to flag data as {\tt NEARBOUND}
are the same as used for the stellar kinematics.  In addition to a core
failure of the fitting algorithm, the {\tt FITFAILED} flag is also used
to signify an error in the computation of the formal errors in the
best-fit parameters.

\subsection{Non-parametric Emission-Line Measurements}
\label{sec:emlmom}

Although more precise measurements of the emission-line properties
are determined by our Gaussian modeling, it is useful to perform
direct, non-parametric measurements of the emission features. In
particular, these measurements provide initial estimates for the
model optimization, and lead to valuable assessments of the model
results that can identify catastrophic errors and non-Gaussianity of
the line profiles. Therefore, we compute zeroth-, first-, and
second-order moments of the emission-line profiles using
continuum-subtracted spectra, and we combine the zeroth moment and
measurements of the local continuum to provide non-parametric
emission-line equivalent widths.


In the calculation of the emission-line moments, the model continuum is
subtracted from the data to produce emission-line-only spectra.  Even
with this subtraction, however, coherent deviations of the baseline
local to each emission line may remain.  Therefore, we also subtract a
linear baseline below each emission feature following from Equation
\ref{eq:continuum} and two sidebands to either side of the emission
feature (see below).  The definitions of the main passband over which
the moments are calculated, as well as the two sideband definitions, are
provided in Table \ref{tab:emldb}.  These same passbands are used in the
equivalent-width calculation (see Section \ref{sec:emlmeasurements}).

The definition of the emission-line passbands are taken from
\citet[][Table 3]{2006ApJ...648..281Y} for the majority of the strong
lines.  For the additional lines in Table \ref{tab:emldb}, we adopt
20~\AA\ passbands and place the sidebands directly to either side of
main passband.  The exceptions to this are when the sidebands overlap
with the main passband of other lines we intend to measure.  In those
cases, the sideband limits are adjusted as necessary to regions that
should be free from emission.  Other specific changes relative to the
definitions from \citet{2006ApJ...648..281Y} are that we apply a slight
shift to the main passband for H$\beta$ to center the passband on the
line and we use a narrower H$\alpha$ main passband to avoid overlap with
the [\ion{N}{2}] lines.  Given their small separation, a number of line
groups use the same sidebands; these groups are (1) H$\eta$, H$\zeta$,
and [\ion{Ne}{3}]; (2) H$\alpha$ and [\ion{N}{2}]; and (3) the two
[\ion{S}{2}] lines.  Also note that only one main passband is used for
the [\ion{O}{2}] doublet because these lines are unresolved by the BOSS
spectrographs.\footnote{
Space is actually allocated for two non-parametric measurements of the
[\ion{O}{2}] doublet in the \dapmaps\ file; however, this is simply to
establish a symmetry between the data format of the non-parametric and
Gaussian-fit results.  More detail is provided in Section
\ref{sec:output} and Appendix \ref{sec:datamodel}.}
All passbands are defined at rest wavelengths and the passbands are
appropriately redshifted such that all measurements are performed on the
observed spectra.

Following the definition in Equation \ref{eq:bpintegral}, we calculate
the zeroth moment of the each line in flux units as:
\begin{equation}
\mu_{l,0} = S(f_l - B),
\end{equation}
where $f_l$ is the continuum-subtracted, emission-line-only spectrum and
$B$ is the linear baseline.  The linear baseline is determined using
Equation \ref{eq:continuum}, except that we use the unweighted center of
each of the ``blue'' and ``red'' passbands, instead of the
spectrum-weighted center, because the passband-integrated flux of $f_l$
is nearly zero.  We then calculate the first and second moments in units
of \kms\ as:
\begin{equation}
\mu_{l,1} = \frac{c}{\mu_{l,0}}\
S\left[\left(\frac{\lambda}{\lambda_{\rm rest}}-1\right) (f_l -
B)\right],
\end{equation}
and
\begin{equation}
\mu_{l,1}^2 + \mu_{l,2}^2 = \frac{c^2}{\mu_{l,0}}\
S\left[\left(\frac{\lambda}{\lambda_{\rm rest}}-1\right)^2 (f_l -
B)\right],
\end{equation}
respectively; $\mu_{l,1}$ is equivalent to a non-parametric Doppler
shift of the line and $\mu_{l,2}$ is the dispersion of the line
profile about that Doppler shift. The calculation of the
non-parametric equivalent width is identical to that used for by the
line-profile modeling, except that we set $F_l = \mu_{l,0}$ in
Equation \ref{eq:lineew}. Errors are provided for the non-parametric
measurements; however, they have not been as well vetted as the
model-based results (see Belfiore et al., {\it accepted}). We expect
their accuracy to be similar to what we find for the spectral
indices (see Sections \ref{sec:sisim} and \ref{sec:sirepeat}).
However, we have not calibrated their accuracy, as done for the
spectral indices, such that the values provided in DR15 should be
treated with caution or simply ignored.

The non-parametric emission-line measurements are performed twice
(Figure \ref{fig:workflow}). The measurements are first performed
before the emission-line modeling using $z_0$ as the redshift {\it
for all spectra in the datacube} and the best-fitting model from the
stellar-kinematics module as the continuum. The first moment of the
H$\alpha$ line from these measurements is used as the initial guess
for the velocity in the emission-line modeling of each spectrum. The
measurements are performed a second time after the emission-line
modeling and they serve two purposes: (1) The emission-line modeling
includes a simultaneous adjustment of the stellar continuum, such
that the continuum from the first full-spectrum fit (Section
\ref{sec:stellarkin}) can be different from the continuum used by the
emission-line modeling. Redoing the measurements after the
emission-line modeling is finished allows us to force the continuum
used in both modules to be identical.\footnote{
The baseline, $B$, is always included, meaning the {\it local}
continuum can still be different between the non-parametric and
Gaussian-fit results.}
(2) Instead of a single redshift for all spectra, we use the
best-fitting emission-line velocities (tied for all lines in a given
spectrum in DR15) as the Doppler shift, again allowing us to minimize
the systematic differences in the non-parametric and Gaussian-fit
measurements.  Only the second set of measurements are provided in the
main \dap\ output files (Section \ref{sec:output}; Appendix
\ref{sec:datamodel}).

\subsubsection{Flagging}
\label{sec:emlmomflags}

Through these non-parametric calculations, we set a number of flags (see
Table \ref{tab:dappixmask}): (1) If any of the passbands are empty, no
measurement is made and the {\tt NOVALUE} maskbit is set.  (2) When
constructing the emission-line-only spectrum, discontinuities in the
continuum will occur in the transition between regions that are and are
not fit by the relevant full-spectrum-fitting module.  If such a
discontinuity lands within or between passbands, no measurement is made
and the {\tt FITFAILED} maskbit is set.  This is not really a concern
for DR15, but listed here for completeness.  (3) If the measurement is
made in a region without the continuum subtracted, the measurement is
provided, but the {\tt NOCORRECTION} maskbit is set. (4) If there are
masked pixels within any of the passbands, the measurement is provided,
but the {\tt UNRELIABLE} maskbit is set.  (5) In rare cases, the
calculation of the moments or equivalent widths requires a division by
zero; these cases are flagged as {\tt MATHERROR}.

\section{Spectral Indices}
\label{sec:spindex}

\begin{deluxetable*}{ r l c c c r c c r r }
\tabletypesize{\scriptsize}
\tablewidth{0pt}
\tablecaption{Spectral-Index Parameters \label{tab:indexdb} }
\tablehead{ &&&&&&& \\ & & \multicolumn{3}{c}{Passbands (\AA)} & & & \\ \cline{3-5}
\colhead{ID} & \colhead{Index} & \colhead{Main} & \colhead{Blue} & \colhead{Red} & \colhead{Medium} & \colhead{Units} & \colhead{Ref.\tablenotemark{a}} & \colhead{$\varepsilon$} & \colhead{$\delta\epsilon$}}
\startdata
  1 & CN1     &  $4142.125-4177.125$ & $4080.125-4117.625$ & $4244.125-4284.125$ &    Air &     mag & 1 & -0.54 & 1.3 \\
  2 & CN2     &  $4142.125-4177.125$ & $4083.875-4096.375$ & $4244.125-4284.125$ &    Air &     mag & 1 & -0.45 & 1.6 \\
  3 & Ca4227  &  $4222.250-4234.750$ & $4211.000-4219.750$ & $4241.000-4251.000$ &    Air &     \AA & 1 &  0.79 & 1.4 \\
  4 & G4300   &  $4281.375-4316.375$ & $4266.375-4282.625$ & $4318.875-4335.125$ &    Air &     \AA & 1 &  1.01 & 1.4\\
  5 & Fe4383  &  $4369.125-4420.375$ & $4359.125-4370.375$ & $4442.875-4455.375$ &    Air &     \AA & 1 &  1.13 & 1.9 \\
  6 & Ca4455  &  $4452.125-4474.625$ & $4445.875-4454.625$ & $4477.125-4492.125$ &    Air &     \AA & 1 &  0.83 & 1.4 \\
  7 & Fe4531  &  $4514.250-4559.250$ & $4504.250-4514.250$ & $4560.500-4579.250$ &    Air &     \AA & 1 &  1.03 & 1.7 \\
  8 & C24668  &  $4634.000-4720.250$ & $4611.500-4630.250$ & $4742.750-4756.500$ &    Air &     \AA & 1 &  1.18 & 1.9 \\
  9 & H$\beta$ &  $4847.875-4876.625$ & $4827.875-4847.875$ & $4876.625-4891.625$ &    Air &     \AA & 1 & 0.76 & 1.2 \\
 10 & Fe5015  &  $4977.750-5054.000$ & $4946.500-4977.750$ & $5054.000-5065.250$ &    Air &     \AA & 1 &  1.13 & 1.9 \\
 11 & Mg1     &  $5069.125-5134.125$ & $4895.125-4957.625$ & $5301.125-5366.125$ &    Air &     mag & 1 & -0.82 & 1.4 \\
 12 & Mg2     &  $5154.125-5196.625$ & $4895.125-4957.625$ & $5301.125-5366.125$ &    Air &     mag & 1 & -0.73 & 1.3 \\
 13 & Mgb     &  $5160.125-5192.625$ & $5142.625-5161.375$ & $5191.375-5206.375$ &    Air &     \AA & 1 &  0.81 & 1.4 \\
 14 & Fe5270  &  $5245.650-5285.650$ & $5233.150-5248.150$ & $5285.650-5318.150$ &    Air &     \AA & 1 &  0.85 & 1.4 \\
 15 & Fe5335  &  $5312.125-5352.125$ & $5304.625-5315.875$ & $5353.375-5363.375$ &    Air &     \AA & 1 &  0.96 & 1.7 \\
 16 & Fe5406  &  $5387.500-5415.000$ & $5376.250-5387.500$ & $5415.000-5425.000$ &    Air &     \AA & 1 &  0.87 & 1.7 \\
 17 & Fe5709  &  $5696.625-5720.375$ & $5672.875-5696.625$ & $5722.875-5736.625$ &    Air &     \AA & 1 &  0.87 & 1.6 \\
 18 & Fe5782  &  $5776.625-5796.625$ & $5765.375-5775.375$ & $5797.875-5811.625$ &    Air &     \AA & 1 &  0.84 & 1.5 \\
 19 & NaD     &  $5876.875-5909.375$ & $5860.625-5875.625$ & $5922.125-5948.125$ &    Air &     \AA & 1 &  0.90 & 1.5 \\
 20 & TiO1    &  $5936.625-5994.125$ & $5816.625-5849.125$ & $6038.625-6103.625$ &    Air &     mag & 1 & -0.69 & 1.7 \\
 21 & TiO2    &  $6189.625-6272.125$ & $6066.625-6141.625$ & $6372.625-6415.125$ &    Air &     mag & 1 & -0.80 & 1.8 \\
 22 & H$\delta_{\rm A}$ &  $4083.500-4122.250$ & $4041.600-4079.750$ & $4128.500-4161.000$ &    Air &     \AA & 2 & 1.02 & 1.2 \\
 23 & H$\gamma_{\rm A}$ &  $4319.750-4363.500$ & $4283.500-4319.750$ & $4367.250-4419.750$ &    Air &     \AA & 2 & 1.00 & 1.3 \\
 24 & H$\delta_{\rm F}$ &  $4091.000-4112.250$ & $4057.250-4088.500$ & $4114.750-4137.250$ &    Air &     \AA & 2 & 0.83 & 1.1 \\
 25 & H$\gamma_{\rm F}$ &  $4331.250-4352.250$ & $4283.500-4319.750$ & $4354.750-4384.750$ &    Air &     \AA & 2 & 0.75 & 1.1 \\
 26 & CaHK    &  \phn\phn$3899.5-4003.5$\phn\phn & \phn\phn$3806.5-3833.8$\phn\phn & \phn\phn$4020.7-4052.4$\phn\phn &    Air &     \AA & 3 & 1.43 & 1.5 \\
 27 & CaII1\tablenotemark{b,e}   &  \phn\phn$8484.0-8513.0$\phn\phn & \phn\phn$8474.0-8484.0$\phn\phn & \phn\phn$8563.0-8577.0$\phn\phn &    Air &     \AA & 4 & \nodata & \nodata \\
 28 & CaII2\tablenotemark{b,e}   &  \phn\phn$8522.0-8562.0$\phn\phn & \phn\phn$8474.0-8484.0$\phn\phn & \phn\phn$8563.0-8577.0$\phn\phn &    Air &     \AA & 4 & \nodata & \nodata \\
 29 & CaII3\tablenotemark{b,e}   &  \phn\phn$8642.0-8682.0$\phn\phn & \phn\phn$8619.0-8642.0$\phn\phn & \phn\phn$8700.0-8725.0$\phn\phn &    Air &     \AA & 4 & \nodata & \nodata \\
 30 & Pa17\tablenotemark{b,e}    &  \phn\phn$8461.0-8474.0$\phn\phn & \phn\phn$8474.0-8484.0$\phn\phn & \phn\phn$8563.0-8577.0$\phn\phn &    Air &     \AA & 4 & \nodata & \nodata \\
 31 & Pa14\tablenotemark{b,e}    &  \phn\phn$8577.0-8619.0$\phn\phn & \phn\phn$8563.0-8577.0$\phn\phn & \phn\phn$8619.0-8642.0$\phn\phn &    Air &     \AA & 4 & \nodata & \nodata \\
 32 & Pa12\tablenotemark{b,e}    &  \phn\phn$8730.0-8772.0$\phn\phn & \phn\phn$8700.0-8725.0$\phn\phn & \phn\phn$8776.0-8792.0$\phn\phn &    Air &     \AA & 4 & \nodata & \nodata \\
 33 & MgICvD  &  \phn\phn$5165.0-5220.0$\phn\phn & \phn\phn$5125.0-5165.0$\phn\phn & \phn\phn$5220.0-5260.0$\phn\phn & Vacuum &     \AA & 5 & 0.94 & 1.4 \\
 34 & NaICvD\tablenotemark{e}  &  \phn\phn$8177.0-8205.0$\phn\phn & \phn\phn$8170.0-8177.0$\phn\phn & \phn\phn$8205.0-8215.0$\phn\phn & Vacuum &     \AA & 5 & \nodata & \nodata \\
 35 & MgIIR\tablenotemark{e}   &  \phn\phn$8801.9-8816.9$\phn\phn & \phn\phn$8777.4-8789.4$\phn\phn & \phn\phn$8847.4-8857.4$\phn\phn & Vacuum &     \AA & 5 & \nodata & \nodata \\
 36 & FeHCvD\tablenotemark{e}  &  \phn\phn$9905.0-9935.0$\phn\phn & \phn\phn$9855.0-9880.0$\phn\phn & \phn\phn$9940.0-9970.0$\phn\phn & Vacuum &     \AA & 5 & \nodata & \nodata \\
 37 & NaI\tablenotemark{e}     &  $8168.500-8234.125$ & $8150.000-8168.400$ & $8235.250-8250.000$ &    Air &     \AA & 6 & \nodata & \nodata \\
 38 & bTiO    &  $4758.500-4800.000$ & $4742.750-4756.500$ & $4827.875-4847.875$ &    Air &     mag & 7 & -0.64 & 1.8 \\
 39 & aTiO    &  $5445.000-5600.000$ & $5420.000-5442.000$ & $5630.000-5655.000$ &    Air &     mag & 7 & -0.74 & 2.5 \\
 40 & CaH1    &  $6357.500-6401.750$ & $6342.125-6356.500$ & $6408.500-6429.750$ &    Air &     mag & 7 & -0.68 & 1.9 \\
 41 & CaH2    &  $6775.000-6900.000$ & $6510.000-6539.250$ & $7017.000-7064.000$ &    Air &     mag & 7 & -0.74 & 2.8 \\
 42 & NaISDSS\tablenotemark{e} &  \phn\phn$8180.0-8200.0$\phn\phn & \phn\phn$8143.0-8153.0$\phn\phn & \phn\phn$8233.0-8244.0$\phn\phn &   Air &  \AA & 8 & \nodata & \nodata \\
 43 & TiO2SDSS &  $6189.625-6272.125$ & $6066.625-6141.625$ & \phn\phn$6422.0-6455.0$\phn\phn & Air &  mag  & 8 & -0.79 & 2.0 \\
 44 & D4000\tablenotemark{c}   &              \nodata & $3750.000-3950.000$ & $4050.000-4250.000$ &    Air & \nodata & 9 & -0.33 & 1.4 \\
 45 & Dn4000\tablenotemark{c}  &              \nodata & $3850.000-3950.000$ & $4000.000-4100.000$ &    Air & \nodata & 10 & -0.33 & 1.1 \\
 46 & TiOCvD\tablenotemark{d,e}  &              \nodata & $8835.000-8855.000$ & $8870.000-8890.000$ & Vacuum & \nodata & 5 & \nodata & \nodata 
\enddata
\tablenotetext{a}{References: (1) \citet{1998ApJS..116....1T}; (2)
\citet{1997ApJS..111..377W}; (3) \citet{2005ApJ...627..754S}; (4)
\citet{2001MNRAS.326..959C}; (5) \citet{2012ApJ...747...69C}; (6)
\citet{2012ApJ...753L..32S}; (7) \citet{2014MNRAS.438.1483S}; (8)
\citet{2013MNRAS.433.3017L}; (9) \citet{1983ApJ...273..105B}; (10)
\citet{1999ApJ...527...54B}. }
\tablenotetext{b}{The CaII triplet and Paschen bands are each meant to
be measured as a group using a set of five interspersed continuum
passbands.  See \citet[][Section 4.3]{2001MNRAS.326..959C} for details.
However, these indices are currently treated identically to other
indices, i.e., independently measured and with two associated sidebands.  Future
improvements of the \dap\ will provide functionality for these more
complex index definitions.}
\tablenotetext{c}{Bandpass integration is performed over $F_\nu$, not
$F_\lambda$. The index is defined as ratio of the red bandpass
integral divided by the blue bandpass integral. See Equation
\ref{eq:d4000}.}
\tablenotetext{d}{Index defined as ratio of the blue bandpass integral
divided by the red bandpass integral.  See Equation \ref{eq:tio}.}
\tablenotetext{e}{Due to the limited spectral range of the \mileshc\
library, index does not have a velocity-dispersion correction. These
data should be used with caution or ignored.}
\end{deluxetable*}

We use the term spectral index generally to refer to a measurement of
a specific continuum feature in a galaxy spectrum made manifest by
its stellar population. Because the spectral indices are meant to
quantify features on the continua of galaxy spectra, all of our
spectral-index measurements are made after first subtracting the
best-fit emission-line model from the observed spectrum. The spectral
indices provided by the \dap\ fall into the following two groups.

Absorption-line indices --- like those defined in the Lick Index system
\citep{1998ApJS..116....1T} --- are measured similarly to emission-line
equivalent widths.  These indices are defined to measure the strength of
absorption features predominantly associated with one, or a small
number, of atoms or molecules present in stellar atmospheres.  Following
Equations 2 and 3 from \citet{1994ApJS...95..107W} and Equation
\ref{eq:bpintegral} above, we compute absorption-line indices as:
\begin{equation}
{\mathcal I}_a = \left\{
\begin{array}{ll}
\frac{1}{1+z}\ S(1 - f_c/C), & \mbox{for \AA\ units} \\[3pt]
-2.5\log\left[S(f_c/C)\right], & \mbox{for magnitude units}
\end{array}\right.,
\label{eq:absindex}
\end{equation}
where $f_c$ is the continuum-only spectrum determined by subtracting
the best-fitting emission-line model and $C$ is calculated using
Equation \ref{eq:continuum} and the continuum-only spectrum. As
indicated by Equation \ref{eq:absindex}, absorption-line indices can
be measured in magnitude units or \AA\ and are constructed such that
features seen in absorption relative to the pseudo-continuum yield
positive index values. Despite some subtle difference in the
definition, $EW_l \approx -{\mathcal I}_a$ when ${\mathcal I}_a$ has
\AA\ units and both use the same passbands. The first 43 rows in
Table \ref{tab:indexdb} provide the three passbands used in measuring
${\mathcal I}_a$ by the \dap\ in DR15; the Table includes the units
of each index and whether the passbands were defined for air or
vacuum wavelengths.

Bandhead, or color, indices are measurements that simply compare the
flux in two passbands, usually placed to either side of significant
continuum-break features due to the atomic or molecular composition
in stellar atmospheres. As listed in Table \ref{tab:indexdb}, there
are three such indices provided in DR15, the ubiquitous D4000 and
Dn4000 and a TiO bandhead. Recall that the D4000 and Dn4000 indices
both quantify the strength of the 4000~\AA\ break, but with slighly
different definitions for their two sidebands. These three indices
have subtly different definitions,\footnote{
The $\lambda^2$ terms in Equation \ref{eq:d4000} are due to the
D(n)4000 indices being defined as an integration in wavelength over
flux per unit frequency ($f_\nu$) instead of over $f_\lambda$.}
as noted in the Table; we provide them here for completeness (cf.\
Equation \ref{eq:continuum}):
\begin{eqnarray}
{\rm D(n)4000} & = & \frac{\langle \lambda^2 f_c \rangle_{\rm
red}}{\langle \lambda^2 f_c \rangle_{\rm blue}}, \label{eq:d4000} \\
{\rm TiO} & = & \frac{\langle f_c \rangle_{\rm blue}}{\langle f_c
\rangle_{\rm red}}. \label{eq:tio}
\end{eqnarray}

As with the definition and use of the emission-line passbands, all
spectral-index passbands are defined at rest and then redshifted
appropriately for measurements on the {\it observed} spectra. This leads
to the $(1+z)$ factor in Equation \ref{eq:absindex}, such that the
indices are always provided in the rest frame. In DR15, we adopt a
single redshift, $z_0$ (see Section \ref{sec:vdef}), for all
spectral-index measurements within a given datacube.

\subsection{Velocity Dispersion Corrections}
\label{sec:indexcorr}

The numerous, blended absorption features in the continuum spectra of
galaxies lead to a dependence of the determination of the
pseudo-continuum and the main bandpass integral on the effective
resolution of the data.  By effective resolution, we mean the
convolution of the intrinsic spectrum with the combined kernel made up
of the instrumental resolution and the astrophysical Doppler broadening.
For comparison of the index measurements from the MaNGA spectra with
model grids \citep[e.g.,][]{2011MNRAS.412.2183T}, it is important that
both are performed at the same effective resolution.  This is often done
by matching the instrumental resolution of the data to a fiducial
resolution --- e.g., the original Lick resolution of 8.4~\AA\ --- and
correcting the measurements to a fiducial astrophysical velocity
dispersion.

Instead of degrading the resolution of the MaNGA data, we make the
measurements directly at their native resolution.  We then calculate a
velocity-dispersion correction for each index that converts the
measurement to one that would be made if the galaxy had $\sigma_\ast =
0$.  This correction is constructed by calculating the index in both the
best-fitting continuum model, ${\mathcal I}_{a,m}$, (from the combined
emission-line and stellar continuum fit; see Section \ref{sec:emlfit})
and the same model constructed with $\sigma_\ast = 0$, ${\mathcal
I}_{a,m,0}$.  The correction depends on the index units such that:
\begin{equation}
\delta{\mathcal I}_a = \left\{
\begin{array}{ll}
{\mathcal I}_{a,m,0}/{\mathcal I}_{a,m}, & \mbox{for \AA\ units} \\[3pt]
{\mathcal I}_{a,m,0}-{\mathcal I}_{a,m}, & \mbox{for magnitude units}
\end{array}\right.,
\end{equation}
which can be applied to the data to get the corrected index, ${\mathcal
I}_a^c$, as follows:
\begin{equation}
{\mathcal I}_a^c = \left\{
\begin{array}{ll}
{\mathcal I}_a \delta{\mathcal I}_a, & \mbox{for \AA\ units} \\[3pt]
{\mathcal I}_a + \delta{\mathcal I}_a, & \mbox{for magnitude units}
\end{array}\right..
\end{equation}

Recalling from our discussion of the velocity-dispersion corrections for
the stellar kinematics in Section \ref{sec:sigmacorr}, the ``raw''
velocity dispersion measurements we make, $\sigma_{\rm obs}$, include
both the astrophysical stellar velocity dispersion and the difference
between the MaNGA and MILES spectral resolution (see Equation
\ref{eq:sigmaobs}).  We use this to our advantage in the determination
of the velocity-dispersion correction for the spectral indices: By
calculating ${\mathcal I}_{a,m}$ based on the best-fit model and
${\mathcal I}_{a,m,0}$ using the templates at their native resolution,
we account for both astrophysical dispersion and the difference in
resolution between MaNGA and MILES.  Therefore, the corrected indices
from the \dap\ can then be compared to model grids made for indices
measured with $\sigma_\ast = 0$ at the MILES spectral resolution
\citep[e.g.,][]{2011MNRAS.412.2183T}.

In DR15, we provide ${\mathcal I}_a$ and $\delta{\mathcal I}_a$ such
that users can calculate ${\mathcal I}_a^c$, or derive and apply
their own corrections. Just as with the stellar velocity dispersion
correction, {\it the user must apply the spectral index corrections
themselves}. Importantly, although provided as part of the release,
the spectral indices outside of the MILES spectral range should only
be used as a rough guide because no velocity-dispersion corrections
are available. The velocity-dispersion corrections are critical to
both the comparison of the relative absorption-line strengths within
and among galaxies, as well as for the inference of stellar
population properties when compared to model values. Spectral indices
outside of the MILES spectral range in DR15 will therefore have
limited use, and should only be used, even in a relative sense, with
spectra of similar velocity dispersion.

\subsection{Possible Model-Driven Biases}

Before presenting our quality assessments of the \dap\ spectral-index
measurements, we first comment on two possible sources of systematic
error that we do {\it not} explore in detail here.

First, any systematic error in the emission-line model will propagate
to biases in the spectral indices measured with passbands that span
emission-line regions because the emission-line model is subtracted
from the data before the spectral indices are measured. It is,
therefore, important to consult Belfiore et al., {\it accepted}, for
an in-depth investigation of the systematic errors in the
emission-line modeling. Although other sources are possible (e.g.,
non-Gaussianity in the line profiles), Belfiore et al., {\it
accepted}, find that the dominant source of systematic error, of
those they explored, is the choice of the stellar-continuum
templates. Importantly, there is little power in the fit-quality
figures-of-merit to discriminate between continuum models based on
the \mileshc\ library and three different simple stellar-population
model template sets. However, based on their comparison between the
\citet[][BC03]{2003MNRAS.344.1000B} and \mileshc\ templates, Belfiore
et al., {\it accepted}, show it is possible that the latter yields
systematically shallow Balmer absorption-line depths. In the context
of the spectral-index measurements, this would imply a systematic
under-subtraction of the Balmer emission lines and a systematically
small absorption-line equivalent width for Balmer-line indices (e.g.,
H$\beta$ and H$\delta_{\rm A}$).

Second, although the raw spectral-index measurements are largely
model-independent (apart from the emission-line subtraction), the
velocity-dispersion corrections are constructed solely from the
stellar-continuum model. Systematic errors due to the
velocity-dispersion correction can manifest in two ways: (1) biases
due to a propagated bias in the velocity-dispersion measurement
itself and (2) biases due to local inaccuracies in the ability of the
model to reproduce the observed spectrum. Effects due to the former
are explored in the experiments performed in Section \ref{sec:sisim}.
In the context of the latter, we note discrepancies that Belfiore et
al., {\it accepted}, find between the Mgb and NaD absorption features
between the BC03 and \mileshc\ templates (see Figure 8 and Section
4.2 of Belfiore et al., and see discrepancies in the model fit to the
central spectrum of observation {\tt 8728-12703} in our Figure
\ref{fig:showcasespec}).\footnote{
We note that some of this discrepancy in the NaD absorption depth may
be to due interstellar absorption, as opposed to a limitation in
stellar parameter coverage.}
In fact, Figure \ref{fig:miles_clusters}
shows that the range of stellar parameters consolidated into a single
\mileshc\ template can be quite broad in [Fe/H] and $\log g$. It is
therefore likely that the \mileshc\ library cannot, in detail, mimic
the full range of features seen in the full MILES library, which
compounds the ability of the entirety of the MILES library to match
the detailed features in the MaNGA galaxy spectra due to its limited
coverage of stellar parameter space. Nonetheless, the
velocity-dispersion corrections are typically a few percent, meaning
that the influence of such systematic errors on the corrected
spectral index will be of the same order at most.

\subsection{Quality Assessments}
\label{sec:siqual}

Similar to the tests of the stellar kinematics in Section
\ref{sec:stellarkin}, we assess the quality of the spectral-index
measurements using both idealized simulations (Section \ref{sec:sisim})
and repeat observations (Section \ref{sec:sirepeat}). Although more
limited in scope than our stellar kinematics assessments, these tests
provide guidance in terms of the limitations of the data provided by the
\dap. Specifically, we note that in both binning cases provided in DR15,
bins and/or spaxels may be below the S/N needed to meet the goals for a
given analysis of the spectral indices. In Section \ref{sec:sibin}, we
discuss the procedure one can use to combine index measurements from
multiple bins to improve their precision. A comparison of the results
provided by the \dap\ and the Firefly Value-Added Catalog\footnote{
\url{https://www.sdss.org/dr15/manga/manga-data/manga-firefly-value-added-catalog/}}
was performed by \citet{GoddardPHD} for the data available as of SDSS
DR14 \citep{2018ApJS..235...42A}.\footnote{
Specifically, see Figure 3.20 at
\url{https://researchportal.port.ac.uk/portal/files/11006420/Daniel_Stephen_Goddard_Thesis_ICG_Portsmouth.pdf}.}

\subsubsection{Idealized Simulations}
\label{sec:sisim}

\begin{figure}
\begin{center}
\includegraphics[width=1.0\columnwidth]{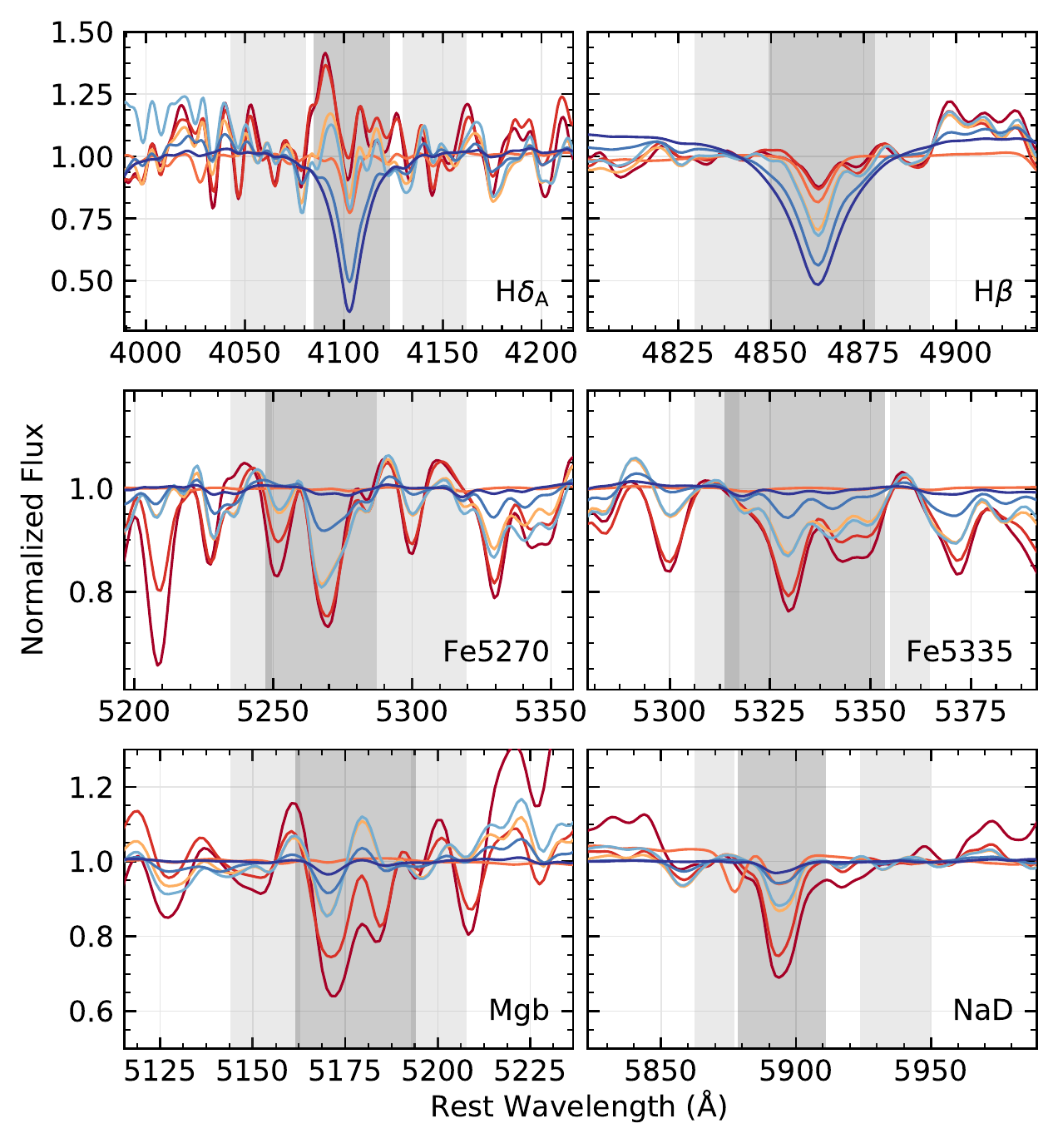}
\end{center}
\caption{
A subset of seven template spectra from the \mileshc\ library in six
spectral windows near absorption-line indices measured by the \dap.
The template spectra have been broadened by $\sigma_\ast = 140$ \kms,
and their line color is ordered from blue to red by the measured
$H\beta$ equivalent width; see Table \ref{tab:tplindex}. From
top-to-bottom, left-to-right, the spectral regions are near the
H$\delta_{\rm A}$, H$\beta$, Fe5270, Fe5335, Mgb, and NaD absorption
indices, as marked in the lower right corner of each panel. The dark
gray region highlights the main passband of the index, with lighter
gray regions showing the two sidebands; note that the sidebands can
overlap with the main band (e.g., Fe5335).}
\label{fig:tplindices}
\end{figure}

We limit the scope of our idealized simulations to consider
measurements of the following subset of indices (see Table
\ref{tab:indexdb}): H$\delta_{\rm A}$, H$\beta$, Mgb, Fe5270, Fe5335,
NaD, and D4000. We also limit the exploration of the systematic and
random errors in these indices to their measurement for a subset of
single spectra chosen from the \mileshc\ templates: template 21, 23,
25, 28, 31, 41, and 45 (see Figure \ref{fig:mileshc}). These
templates were chosen to roughly span the range of values for the
selected indices relevant to the galaxy data. The spectral region of
each spectral index (except D4000) are shown in Figure
\ref{fig:tplindices} for the subset of seven \mileshc\ templates.

We convolve each of the seven selected \mileshc\ templates with a
wavelength-independent Gaussian with velocity dispersions of 35, 70,
140, 280, and 396 \kms. No velocity shift is applied, and we also
include the original spectrum (i.e., a $\sigma_\ast = 0$ \kms\
spectrum) such that there are six noise-free spectra used for each of
the seven \mileshc\ templates. Noise is added to each spectrum to
meet a $g$-band S/N of $2^i$ for $i=0,...,7$ with a wavelength
dependence that matches the median wavelength dependence of the noise
vectors measured for an example MaNGA datacube ({\tt 7495-12704} in
this case). We measure the spectral indices for both the noise-free
synthetic spectra and 1000 noise realization for each \snrg.

\begin{figure*}
\begin{center}
\includegraphics[width=\textwidth]{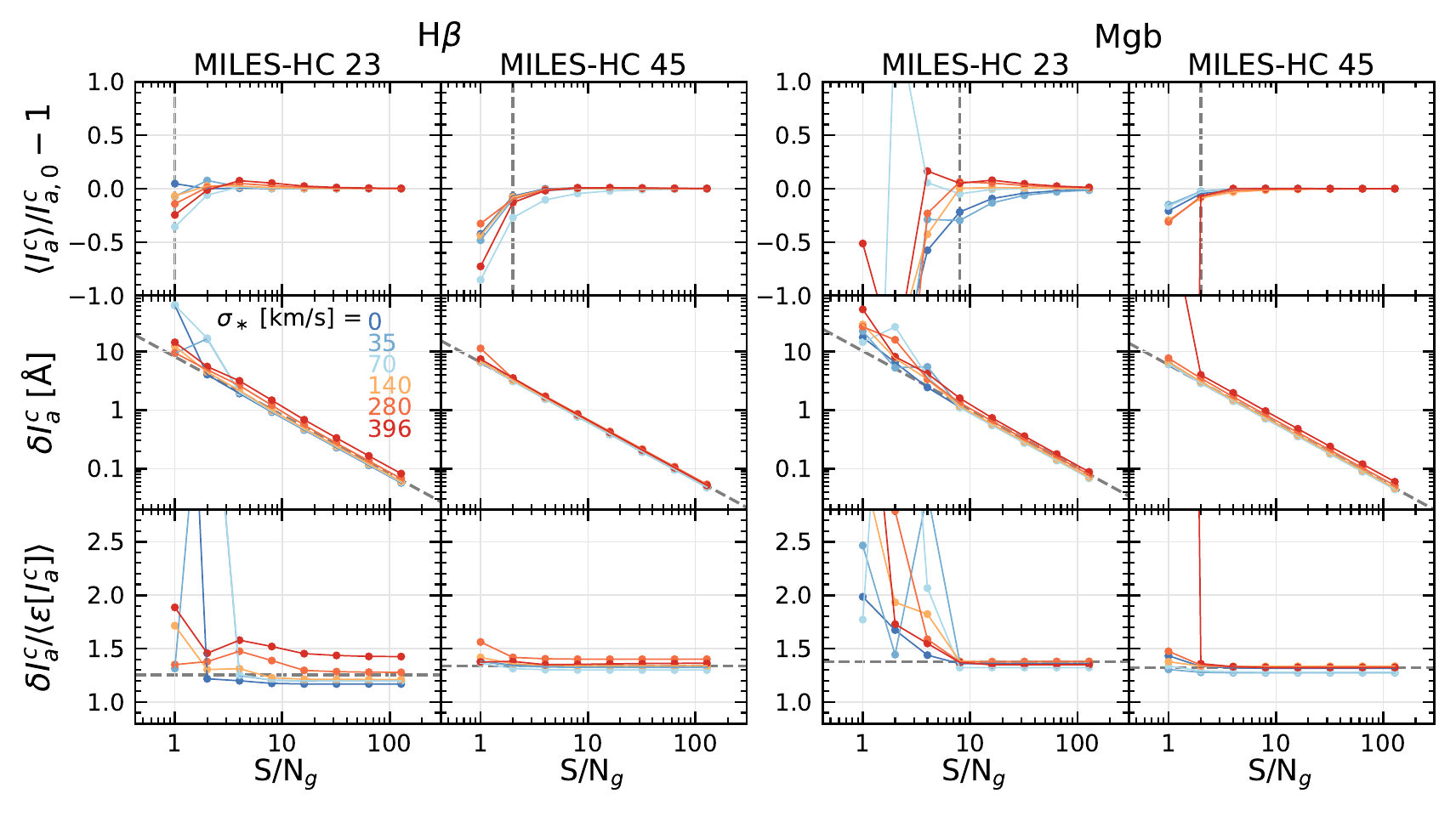}
\end{center}
\caption{Results for idealized simulations testing the recovery of
the H$\beta$ and Mgb absorption-line indices as a function of \snrg\
and spectral template. See Section \ref{sec:sisim} for a description
of the simulations. The results for two \mileshc\ templates (template
23 and 45; Figure \ref{fig:mileshc}) are shown for each index as
marked. Line colors in all panels indicate the stellar velocity
dispersion, $\sigma_\ast$, of the LOSVD kernel. Each column of panels
shows the fractional recovery of the input index (top), the empirical
estimate of the random error (middle), and the ratio of the empirical
estimate and formal calculation of the index error. The vertical
dashed line in the top panels indicate the \snrg\ above which the
median and full range of the binned data are less than 30\% and 50\%,
respectively. The dashed line in the middle panels show the optimal
inverse proportionality between the index error and \snrg. The
dashed line in the bottom panels show the median ratio between the
empirical estimate and formal calculation of the error for all
\snrg\ and $\sigma_\ast$ bins.}
\label{fig:sisim}
\end{figure*}

The ``true'' velocity-dispersion corrections for each index,
unaffected by noise, are calculated using the noise-free spectra and
the known input dispersion. However, for each noise realization, the
corrections are calculated similarly to the observed galaxy spectra:
We first fit the stellar kinematics of each synthetic spectrum using
the same code used to fit the galaxy data --- including the full
\mileshc\ library in the fit --- to obtain a stellar velocity
dispersion that is subject to relevant random and systematic error;
this velocity-dispersion measurement is used to construct the
correction. With both the true and measured velocity-dispersion
corrections, we have explored the accuracy and precision of the
velocity-dispersion corrections themselves, as well as their relative
influence on the uncertainties in the corrected indices.

For each of the seven spectral indices measured in the idealized
simulation, we construct plots like the examples provided in Figure
\ref{fig:sisim}. In each column, the top panel shows the accuracy of
the mean {\it corrected} spectral index measured from the simulated
spectra, $\langle\cindx\rangle$ compared to the known value
from the noise-free spectrum, ${\mathcal I}_{a,0}^c$. The middle panel
provides the standard deviation in the measured index (i.e., the
empirical error), $\delta\cindx$, and the bottom panel shows
the ratio of $\delta\cindx$ to the mean of the error provided
by the formal error-propagation calculations performed by the \dap.

\begin{deluxetable*}{ l r c c c r c  r c c c c r c  r c c c c r c }
\tabletypesize{\footnotesize}
\tablewidth{0pt}
\tablecaption{Example \mileshc\ Spectral Indices \label{tab:tplindex} }
\tablehead{\\[-5pt] & \multicolumn{6}{c}{H$\delta_{\rm A}$} && \multicolumn{6}{c}{H$\beta$}
            && \multicolumn{6}{c}{Mgb} \\
            \cline{2-7} \cline{9-14} \cline{16-21}
            \colhead{ID} & \colhead{[\AA]} & \colhead{corr} & \colhead{$\varepsilon$} & \colhead{\snrg} & \colhead{$\gamma$} & \colhead{$\delta\epsilon$}
            && \colhead{[\AA]} & \colhead{corr} & \colhead{$\varepsilon$} & \colhead{\snrg} & \colhead{$\gamma$} & \colhead{$\delta\epsilon$}
            && \colhead{[\AA]} & \colhead{corr} & \colhead{$\varepsilon$} & \colhead{\snrg} & \colhead{$\gamma$} & \colhead{$\delta\epsilon$} }
\startdata
28 &  -5.0 & 1.049 &  1.66 &   16 &  0.05 &   1.4 &&   1.2 & 1.106 &  0.81 &    1 &  0.08 &   1.3 &&   4.9 & 0.982 &  0.73 &    1 &  0.08 &   1.3 \\
45 &  -6.1 & 1.030 &  1.39 &    4 &  0.05 &   1.4 &&   1.2 & 1.070 &  0.82 &    2 &  0.05 &   1.3 &&   3.4 & 0.990 &  0.77 &    2 &  0.02 &   1.3 \\
31 &   2.0 & 1.015 &  0.81 &    1 &  0.05 &   1.2 &&   1.6 & 0.999 &  0.92 &    4 &  0.02 &   1.4 &&  -0.2 & 1.137 &  1.15 &   64 &  0.05 &   1.6 \\
41 &  -1.1 & 1.058 &  1.16 &   16 &  0.03 &   1.3 &&   3.3 & 1.038 &  0.82 &    1 &  0.02 &   1.3 &&   0.3 & 0.759 &  0.84 &    8 &  0.07 &   1.4 \\
21 &   0.6 & 0.925 &  0.95 &   16 &  0.51 &   1.3 &&   3.5 & 1.032 &  0.89 &    1 &  0.04 &   1.3 &&   0.2 & 0.497 &  0.92 &   32 &  0.18 &   1.5 \\
25 &   8.0 & 1.002 &  0.89 &    1 &  0.03 &   1.2 &&   6.3 & 1.013 &  0.87 &    1 &  0.17 &   1.2 &&   0.6 & 1.046 &  0.92 &    8 &  0.03 &   1.4 \\
23 &  11.1 & 1.003 &  0.80 &    1 &  0.12 &   1.2 &&   8.2 & 1.016 &  0.91 &    1 &  0.30 &   1.3 &&   0.3 & 1.014 &  1.01 &    8 &  0.04 &   1.4 \\
\hline\\[-5pt]
    & \multicolumn{6}{c}{Fe5270} && \multicolumn{6}{c}{Fe5335} && \multicolumn{6}{c}{NaD} \\
            \cline{2-7} \cline{9-14} \cline{16-21} \colhead{ID} & \colhead{[\AA]} 
            & \colhead{corr} & \colhead{$\varepsilon$} & \colhead{\snrg} & \colhead{$\gamma$} & \colhead{$\delta\epsilon$}
            && \colhead{[\AA]} & \colhead{corr} & \colhead{$\varepsilon$} & \colhead{\snrg} & \colhead{$\gamma$} & \colhead{$\delta\epsilon$}
            && \colhead{[\AA]} & \colhead{corr} & \colhead{$\varepsilon$} & \colhead{\snrg} & \colhead{$\gamma$} & \colhead{$\delta\epsilon$} \\
\hline
28 &   4.2 & 1.087 &  0.70 &    1 &  0.03 &   1.3 &&   4.5 & 1.103 &  0.68 &    1 &  0.02 &   1.5 &&   4.8 & 1.032 &  0.74 &    1 &  0.03 &   1.2 \\
45 &   3.8 & 1.084 &  0.80 &    1 &  0.01 &   1.4 &&   3.7 & 1.114 &  0.80 &    1 &  0.03 &   1.5 &&   3.3 & 1.045 &  0.86 &    1 &  0.04 &   1.2 \\
31 &   0.1 & 0.912 &  1.08 &  128 &  0.08 &   1.4 &&   0.0 & 1.223 &  1.46 &  128 &  0.05 &   1.9 &&   0.6 & 0.830 &  1.37 &   16 &  0.09 &   1.3 \\
41 &   2.9 & 1.075 &  0.91 &    1 &  0.07 &   1.4 &&   2.6 & 1.147 &  0.94 &    2 &  0.02 &   1.6 &&   1.5 & 1.062 &  1.06 &    2 &  0.01 &   1.2 \\
21 &   2.8 & 1.073 &  0.96 &    2 &  0.05 &   1.3 &&   2.8 & 1.142 &  1.04 &    2 &  0.05 &   1.5 &&   1.3 & 1.059 &  1.19 &    8 &  0.04 &   1.2 \\
25 &   1.1 & 1.044 &  1.00 &    2 &  0.06 &   1.4 &&   1.0 & 1.150 &  1.09 &    8 &  0.02 &   1.6 &&   0.7 & 1.040 &  1.23 &    8 &  0.05 &   1.3 \\
23 &   0.1 & 0.959 &  1.08 &   16 &  0.07 &   1.4 &&   0.2 & 1.147 &  1.33 &   16 &  0.06 &   1.6 &&   0.4 & 0.991 &  1.35 &   16 &  0.07 &   1.3
\enddata
\tablecomments{Six columns are provided for each index: (1) The
uncorrected index value for $\sigma_\ast = 140$ \kms, (2) the index
correction for $\sigma_\ast=140$ \kms, (3) the normalization of the
inverse correlation between the empirical error and \snrg, (4) the
\snrg\ below which fractional systematic error is significant, (5)
the maximum ratio between the systematic and random error for all
\snrg\ and $\sigma_\ast$ bins, and (6) the mean ratio between the
empirical error and the \dap-calculated formal error. See the text of
Section \ref{sec:sisim} for additional description.}
\end{deluxetable*}

The specific indices and templates chosen for Figure \ref{fig:sisim}
are illustrative of the general behavior across the seven indices
explored: The accuracy of the measured index depends both on the
\snrg\ of the spectrum and the intrinsic equivalent width of the
feature. Indeed, one should expect that S/N requirements to meet a
given {\it fractional} accuracy are more stringent for weak (shallow)
features than for strong (deep) features. Table \ref{tab:tplindex}
provides each index measured for each noise-free spectrum and shows
that the corrected H$\beta$ index for template 23 (8.3\AA) is much
larger than for template 45 (1.3\AA), whereas the opposite is true of
the Mgb index. Thus, we find that the \snrg\ requirements for a fixed
fractional accuracy of H$\beta$ are lower for an early-type (roughly
A-type) star than for a late-type (roughly K-type) star (see Figure
\ref{fig:mileshc}), and vice versa for the Mgb index. We crudely
quantify this effect by determining the \snrg\ above which the median
and full-spread in index recovery for all $\sigma_\ast$ bins are
better than 30\% and 50\%, respectively. This ``minimum'' \snrg\ is
marked as a vertical dashed line in the top panels of Figure
\ref{fig:sisim}, and provided for all templates and indices used for
this test in Table \ref{tab:tplindex}. We note that, although it is
generally true that the \snrg\ limit is higher for weak features, the
detailed correlation is non-trivial.

The middle panel of each column in Figure \ref{fig:sisim} shows the
tight inverse correlation between the random error in each {\it
corrected} spectral index and the \snrg. This correlation has
relatively weak secondary dependencies on $\sigma_\ast$ and the value
of the spectral index. We characterize the primary relation as an
inverse proportionality and determine the optimal proportionality
constant such that $\delta{\mathcal I}^c_a = 10^\varepsilon ({\rm
S/N}_g)^{-1}$. We note that this relation is generally well-posed and
robust, whereas a similar relation use to quantify the {\it
fractional} error (i.e., $\delta{\mathcal I}^c_a/\langle {\mathcal
I}^c_a\rangle$) is not, particularly at low equivalent width. The
proportionality constants, $10^\varepsilon$, are used to construct the
dashed lines in the middle panels of Figure \ref{fig:sisim}, and
$\varepsilon$ is provided for all spectral indices and templates in
Table \ref{tab:tplindex}. For example, the coefficients in Table
\ref{tab:tplindex} yield a typical random error in the corrected
H$\beta$ index at \snrg=10 of 0.65--0.83 \AA, depending on the
spectrum.

For small values of ${\mathcal I}_{a,0}^c$, the fractional systematic
error is not well posed and it is more informative to consider the
ratio of the systematic and random error, $\gamma = (\langle
\cindx\rangle - {\mathcal I}_{a,0}^c)/\delta\cindx$. In {\it all}
cases (independent of index, template, velocity dispersion, or
\snrg), the systematic error from these idealized simulations is less
than the random error, $\gamma < 1$, most often substantially so.
Also, we do not find any significant trends in this ratio with \snrg\
or $\sigma_\ast$. For reference, we provide the {\it maximum} value
of $\gamma$ for any (\snrg, $\sigma_\ast$) bin for all indices and
templates in Table \ref{tab:tplindex}. We note that the largest
$\gamma$ values occur at high \snrg, where the random error is small
(e.g., the H$\delta_{\rm A}$ index for template 21).


The contribution of the velocity-dispersion correction to the total
error in $\cindx$ is relatively small, despite what can be
substantial error in the correction itself toward low \snrg\ (e.g.,
5\% $\pm$ 5\%). The primary reason is that the corrections are
generally only a few percent such that the fractional error in the
correction itself (e.g., $0.05/1.05$), and therefore its contribution
to the total error, is small. This becomes less true towards low
\snrg, where the corrections can suffer from substantial systematic
error (e.g., as seen in the Mgb index for template 23 and
\snrg$\lesssim$10). We also note a correlation between the error in
the index correction and $\sigma_\ast$, which leads to the increase
in $\delta\cindx$ with $\sigma_\ast$ for H$\beta$ in template 23 seen
in Figure \ref{fig:sisim}; however, we do not provide a general usage
recommendation because this effect has widely varying influence on
any given index for any given underlying spectrum.

The bottom panel of each column in Figure \ref{fig:sisim} shows the
ratio between the empirically measured spectral-index error,
$\delta\cindx$, and the mean of the formally propagated errors
provided by the \dap, $\langle\epsilon[\cindx]\rangle$. Generally
speaking, the true random errors in the spectral indices are larger
than those provided by the \dap. We note, in particular, that the
formal calculations in the \dap\ do not include any error from the
velocity-dispersion correction (no error is calculated on the
correction). Although there can be systematic differences in the
accuracy of the formal errors between spectra of different
$\sigma_\ast$ and at different \snrg\ (e.g., the Mgb index in
\mileshc\ template 23), there is little motivation to try to capture
these variations here and we simply determine the median value of the
ratio, provided as $\delta\epsilon$ in Table \ref{tab:tplindex}. We return to
the accuracy of the \dap-provided formal errors using the repeat
observations.

\begin{figure*}
\begin{center}
\includegraphics[width=\textwidth]{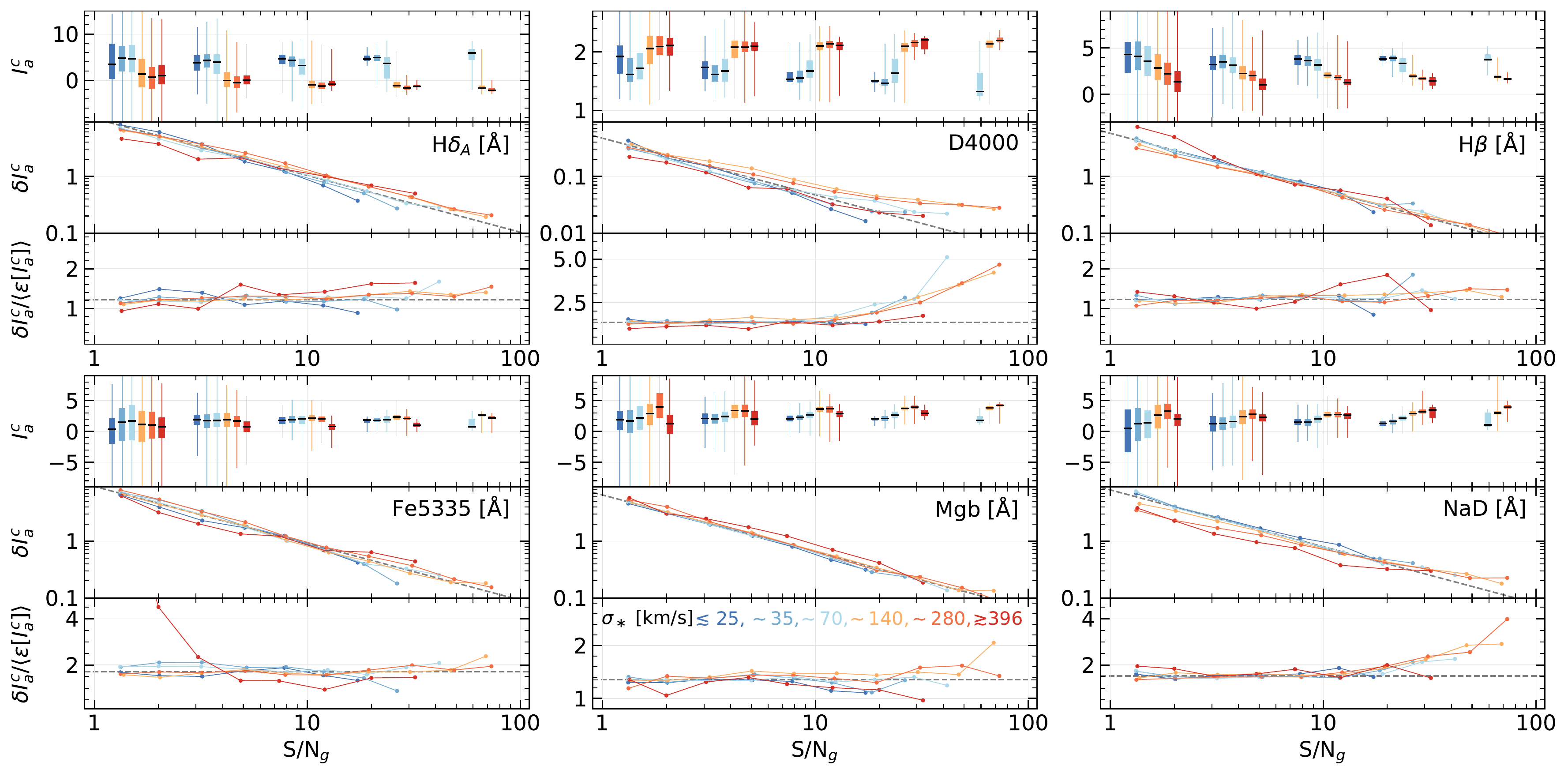}
\end{center}
\caption{Assessments of the random errors in six spectral indices
determined from repeat observations of 56 galaxies. Three panels are
provided for each index: The top panel gives box-and-whisker
representations of the distribution of index measurements in bins of
\snrg\ and $\sigma_\ast$; the $\sigma_\ast$ bins are colored as
labeled in the bottom panel of the Mgb panel group. The box spans the
inner two quartiles of the distribution, and the whiskers span its
full extent. The middle panel in each group provides the empirical
estimate of the error in the index value for each (\snrg,
$\sigma_\ast$) bin. The dashed gray line is the optimized trend for
all (\snrg, $\sigma_\ast$) bins assuming an inverse proportionality
between the error and \snrg. The bottom panel shows the ratio of the
empirically estimated error to the mean error provided by the formal
calculations in the \dap. The dashed gray line shows the median ratio
across all (\snrg, $\sigma_\ast$) bins.}
\label{fig:sirepeat}
\end{figure*}

We have excluded the results for D4000 from Table \ref{tab:tplindex},
summarizing them here instead. Contrary to the results from the
absorption-line indices, the formally propagated errors for D4000 are
accurate such that we always find $\delta\epsilon = 1$. Due to the difference
between the calculation of the bandhead and absorption-line indices,
we infer part of the inadequacy of the formal error calculations for
the absorption-line indices is due to an insufficient propagation of
the error introduced by the continuum calculation (see Equation
\ref{eq:continuum}; nor does it properly propagate the error in the
index correction, as we mentioned above). Although we calculate
velocity-dispersion corrections for both the bandhead and
absorption-line indices, these corrections are less important for the
former; the velocity-dispersion correction for D4000 measured for all
templates used in our idealized simulations are less than 0.1\% for
$\sigma_\ast = 140$ \kms. Of the seven templates included, only the
late-type templates 28 and 45 show any systematic error in D4000;
these extremely red spectra (D4000$\gtrsim$3.5) have a ``minimum''
\snrg$\sim$4. Finally, the proportionality constants for the trend of
random error with \snrg\ in D4000 are generally smaller than for the
absorption-line indices, with errors of a few hundredths at \snrg=10
for early-type spectra; for late-type spectra, these errors reach a
few tenths at the same \snrg.



\subsubsection{Repeat Observations}
\label{sec:sirepeat}

Using the repeat observations of 56 galaxies (Table
\ref{tab:repeats}), we assess the random uncertainties in the
spectral-index measurements and the accuracy of the \dap\ error
calculations. The spatial registration of the repeated observations
is discussed in Section \ref{sec:repeats}.

Figure \ref{fig:sirepeat} shows the results for six of the measured
indices provided in DR15: from left-to-right and top-to-bottom,
H$\delta_{\rm A}$, D4000, H$\beta$, Fe5335, Mgb, and NaD. The mean
and difference in the spectral indices measured for paired spaxels
between repeated observations are calculated and binned by their
\snrg\ and stellar velocity dispersion, $\sigma_\ast$.

Figure \ref{fig:sirepeat} provides a set of three panels for each of
the six spectral indices. The top panel of each group shows the
distribution of the index measurements in (\snrg, $\sigma_\ast$) bins
using a box-and-whisker plot. For each bin, the central box spans the
inner two quartiles of the spectral-index distribution and the lines
span its full extent. These distributions are a result of the
convolution of the error distribution with the intrinsic,
astrophysical distribution. The latter is evident in, e.g., the
higher D4000 measurements for galaxy regions with larger velocity
dispersion. The middle panel shows the empirical estimates of the
random error computed by taking half of the 68\% confidence interval
and dividing by $\sqrt{2}$. Finally, the bottom panel gives the ratio
of the empirical error to the median of the \dap-provided error.

Similar to the idealized simulations in the previous section, we find
that an empirical determination of the error is well-characterized by
an inverse proportionality with \snrg. The trends with \snrg\ show
stronger deviation from the optimized relation when compared to
Figure \ref{fig:sisim} for both H$\beta$ and Mgb; however, this is
expected given the variation in Table \ref{tab:tplindex} of
$\varepsilon$ of a few tenths between different templates. That is, we
have not differentiated between the spectra included in the (\snrg,
$\sigma_\ast$) bin (e.g., by also binning by index value) such that
this varation in $\varepsilon$ is reflective of the astrophysical
variation in the spectra falling into each bin.

We also find that the errors reported by the \dap\ tend to
underestimate the empirically estimated errors by a ratio that is
roughly constant at \snrg$\lesssim$15. At higher \snrg, D4000 and
NaD, in particular, show an increase in the difference between the
empirical and formal estimates of the error, similar to the behavior
seen for the stellar kinematics in Figure \ref{fig:repeats} (and the
emission-line properties explored by Belfiore et al., {\it
accepted}). These difference in the repeated observations are due to
astrometric errors in the dithered MaNGA observations. We expect this
increase is {\it not} seen in the empirical error of the other
indices in Figure \ref{fig:sirepeat} because the random errors in the
measurement of the index itself dominate over the contribution from
the astrometric errors.

In Table \ref{tab:tplindex}, we provide measurements of both
$\varepsilon$ and $\delta\epsilon$ for all indices measured within
the MILES spectral range; i.e., those indices where we have been able
to calculate velocity-dispersion corrections. For $\delta\epsilon$,
we only consider data with \snrg$<$15. We find that the errors are
typically $\sim$0.5-1 \AA\ or $\lesssim$0.03 dex at \snrg$\sim$10 for
indices with units of \AA\ or magnitudes, respectively. Finally, we
find that the errors reported by the \dap\ are smaller than the
empirically estimated errors by $\sim$30-100\%.

\begin{figure*}
\begin{center}
\includegraphics[width=\textwidth]{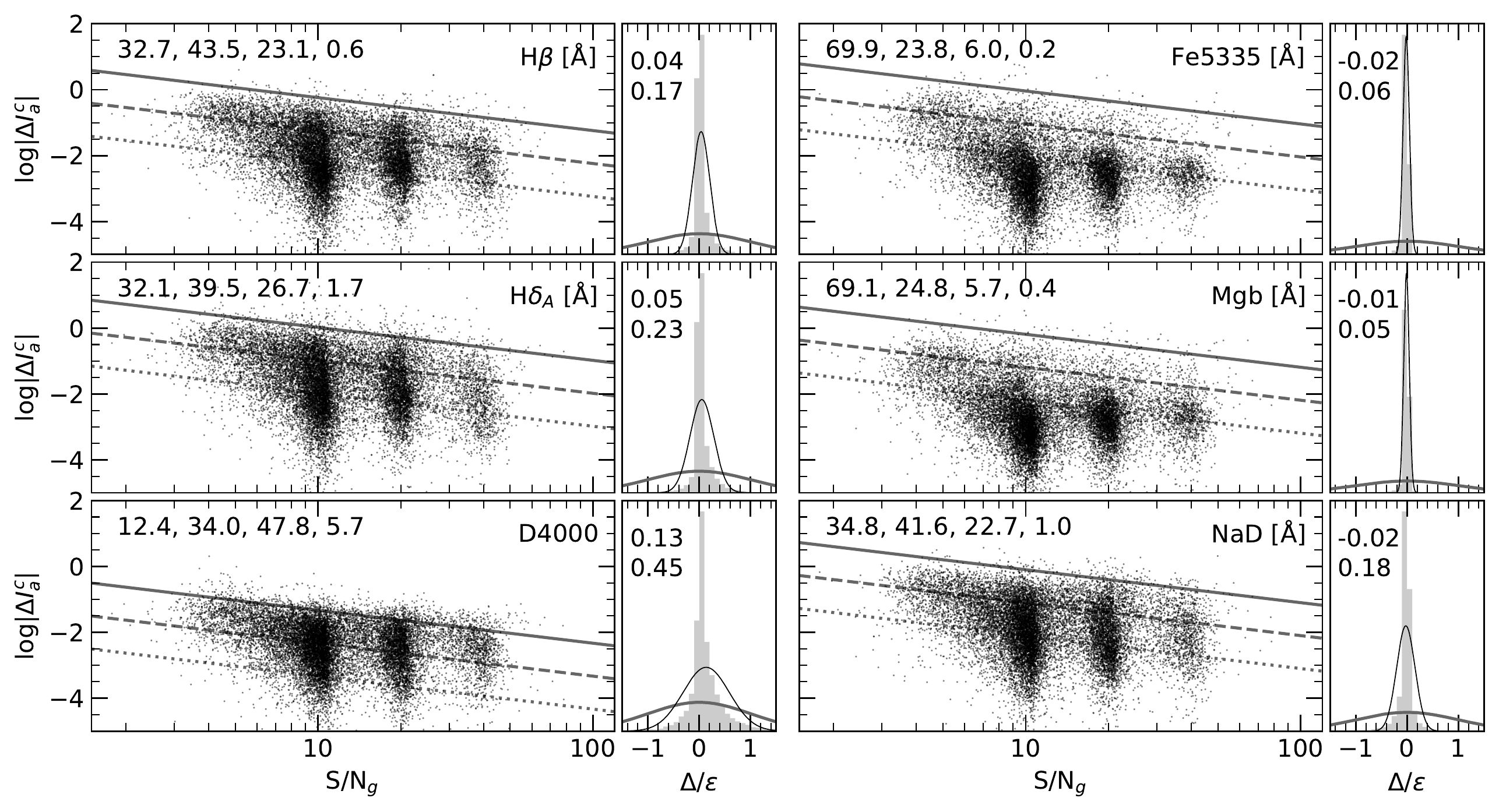}
\end{center}
\caption{A comparison of six spectral indices determined by direct
measurements on a binned spectrum to the result of combining
measuments from individual spectra in the bin. See the description in
Section \ref{sec:sibin}. For each index, we show the difference in
the two measurements, $\Delta {\mathcal I}_a^c$, as a function of
\snrg, and we show the distribution of ratio of the difference to the
error, $\Delta/\varepsilon$, over all \snrg. Gray lines underlying the
distribution of $\Delta {\mathcal I}_a^c$ are based on the expected
error relation derived in Section \ref{sec:sirepeat}; the solid line
is the nominal relation, whereas the dashed and dotted lines show,
respectively, a factor of 10 and 100 below the nominal error. In
these panels, the relevant index is given in the upper-right corner,
and the numbers in the upper left provide, from left-to-right, the
percentage of measurements below the dotted line, between the dotted
and dashed lines, between the dashed and solid lines, and above the
solid line. For example, 43.5\% of the H$\beta$ measurements show
differences that are between 1\% and 10\% of the nominal error in a
single measurement. The distributions of $\Delta/\varepsilon$ are shown
in gray, with the Gaussian distribution based on the nominal error
relation show in dark gray. In the upper-left corner, we provide,
from top-to-bottom, the mean and standard deviation of the
distribution after clipping 10$\sigma$ outliers; the relevant
Gaussian distribution is plotted as a thin black line.}
\label{fig:sibin}
\end{figure*}

\subsubsection{Binning Spectral Indices}
\label{sec:sibin}

In DR15, we provide spectral indices measured both for individual
spaxels, as a product of the hybrid binning approach, and for the
spectra resulting from Voronoi-binning the data to \snrg$\gtrsim$10.
However, the precision of the spectral-index measurements for a given
science pursuit can require measurements at significantly higher S/N
\citep[e.g.][]{2019MNRAS.483.3420P}. Instead of providing additional
\dap\ output resulting from binning to a higher \snrg\ threshold, we
explore and test the accuracy of a method of combining the indices
directly.

Given the use of a linear continuum (Equation \ref{eq:continuum}),
the explicit calculations of the absorption-line indices from Equation
\ref{eq:absindex} are, to good approximation,
\begin{equation}
{\mathcal I}_a \approx \left\{
\begin{array}{ll}
\frac{1}{1+z}[1 - S(f_c)/S(C)], & \mbox{for \AA\ units} \\[3pt]
-2.5\log\left[S(f_c)/S(C)\right], & \mbox{for magnitude units}
\end{array}\right. .
\label{eq:absindexapprox}
\end{equation}
For \AA\ units, these definitions can be used to derive
\begin{equation}
{\mathcal I}^\prime_a [{\rm \AA}] = \frac{1}{1+z}\left[1
    - \frac{\sum_i S(f_{c,i})}{\sum_i S(C_i)}\right]
    = \frac{\sum_i S(C_i) {\mathcal I}_{a,i}}{\sum_i S(C_i)}, 
\label{eq:combindex}
\end{equation}
where ${\mathcal I}^\prime_a$ is the spectral index measured for a
spectrum constructed as the sum of $i=0..N-1$ spectra with spectral
indices ${\mathcal I}_{a,i}$. That is, the combined spectral index is
the weighted sum of the indices from the individual spectra, where
the weights are the value of the continuum integrated over the main
passband.  Equivalently for magnitude units, we find
\begin{eqnarray}
{\mathcal I}^\prime_a [{\rm mag}]
    & = & -2.5\log\left[\frac{\sum_i S(f_{c,i})}{\sum_i S(C_i)}\right]
            \nonumber \\
    & = & -2.5\log\left[\frac{\sum_i S(C_i)
        10^{-0.4 {\mathcal I}_{a,i}}}{\sum_i S(C_i)}\right].
\end{eqnarray}

In principle, these derivations suggest that a simple linear
combination of the spectral indices can be used to construct the
measurement in a summed spectrum. We test this result directly as
follows. Selecting the 112 {\tt PLATEIFU}s from Table
\ref{tab:repeats} that compose the first two observations of each
target, we run the \dap\ four times, once using the hybrid-binning
approach and three times using a Voronoi-binning approach that adopts
thresholds of \snrg\ = 10, 20, and 40, respectively. For spectra
composed of more than one spaxel and directly analyzed in the three
Voronoi-binning cases, we construct the combined index using Equation
\ref{eq:combindex} using the individual spaxel measurements from the
hybrid-binning approach and compare those to the measurements made
directly using the binned spectrum. In this experiment, we apply
Equation \ref{eq:combindex} {\it after} correcting the spectral
indices for the velocity dispersion, and we replace the integral of
the continuum over the main passband by the mean $g$-band flux. The
latter is to make this experiment most directly applicable for users
of the DR15 data; the calculation of $S(C)$ is not provided in DR15,
whereas the mean $g$-band flux is provided. Calculations of $S(C)$
will be provided in future releases to ensure a more accurate
calculation. The results are shown for six spectral indices ---
H$\beta$, H$\delta_{\rm A}$, D4000, Fe5335, Mgb, and NaD --- in Figure
\ref{fig:sibin}.

Figure \ref{fig:sibin} provides two panels for each of the six
spectral indices. The left panel shows the distribution of the
difference in the {\it corrected} spectral indices as a function of
the \snrg in the binned spectrum. We overplot the calibrated error
relations from Table \ref{tab:indexdb}, as well as lines representing
factors of 10 and 100 below this relation. We calculate, and provide
in the Figure, the percentage of points with differences that are
larger than the expected error, between each of the rescaled
relations, and below 1\% of the expected error. The second panel
provides the distribution of the difference normalized by the
expected error and marginalized over all \snrg. We calculate, and
provide in the Figure, the mean and standard deviation of the
distribution. With the exception of D4000, the difference between a
direct spectral-index measurement using a binned spectrum is
consistent with the results of equation \ref{eq:combindex} to better
than 10\% of the expected error for the majority of the calculations
and with a systematic shift of at most a few percent of the expected
error.

Depending on the accuracy and precision needed, this approach of
combining indices could be used directly for scientific inquiry, or
at least provide guidance for follow-up measurements. We emphasize
that this is a {\it statistical} statement; i.e., systematic errors
in the combined index compare to a binned spectrum will largely
average out of a study over many galaxies. Conversely, studies
focused on small regions of individual galaxies should always measure
the indices directly on the binned spectra to ensure the measurement
accuracy.

\subsubsection{Summary}
\label{sec:sisummary}

In our assessments of the quality of the spectral indices provided as
part of DR15, we find:
\begin{itemize}
\item Our idealized simulations show that systematic errors can be
significant with respect to the value of the index, particularly at
low S/N and low EW. However, the systematic error is {\it always} less
than, and typically less than 10\% of, the random error at any \snrg.
For the subset of spectral indices investigated, we provide the
maximum expected systematic error as a fraction of the random error
($\gamma$) in Table \ref{tab:tplindex}.
\item Random errors, determined using both idealized simulations and
repeat observations, are well behaved to very low S/N
(\snrg$\gtrsim$2), following a simple inverse proportionality with
\snrg. Using repeat observations, we calibrate this relation for all
spectral indices provided in DR15 such that users can determine the
\snrg\ required to meet a desired spectral-index error using the data
($\varepsilon$) in Table \ref{tab:indexdb}.
\item The random errors reported by the \dap\ are underestimated, as
determined for both the idealized simulations and the repeat
observations. For all the indices in DR15, we provide a simple
scaling of the reported errors ($\delta\epsilon$ in Table
\ref{tab:indexdb}) to match the results from the repeat observations.
However, in spectra with \snrg$\gtrsim$15, the spectral-index
measurements can be affected by the astrometric errors in the
registration of the dithered observations, as evidenced by stronger
discrepancies between repeat observations.
\item We provide a method of combining index measurements from
multiple spectra that avoids having to recompute the index on the
binned spectra themselves. The method yields results that are
typically consistent with a direct measurement to better than 10\% of
the calibrated error ($\varepsilon$ from Table \ref{tab:indexdb}).
However, this is a {\it statistical} statement that should be treated
with caution, or ignored, when applied to a limited number of spatial
regions.
\end{itemize}

\subsection{Flagging}
\label{sec:sindxflags}

The flags used for the spectral-index measurements are virtually
identical to those used for the emission-line moments (Section
\ref{sec:emlmomflags}; see Table \ref{tab:dappixmask}).  The {\tt
NOVALUE}, {\tt UNRELIABLE}, and {\tt MATHERROR} flags have the same
meaning.  However, any regions without an emission-line model
subtraction are not flagged as {\tt NOCORRECTION} in a synonymous way to
the model subtraction performed for the emission-line moments.  Instead,
measurements are flagged as {\tt NOCORRECTION} if there was an error in
the calculation of the velocity dispersion correction, or if one could
not be calculated because of the spectral range of the fitted models.
The latter is a critical consideration for the spectral indices provided
in DR15 with passbands at $\lambda\gtrsim 7400$~\AA.

\section{Performance}
\label{sec:performance}

In this section, we assess the {\it overall} performance of the \dap.
The performance specifically with regard to the stellar kinematics
are discussed in Section \ref{sec:scperf}, with regard to the
emission-line modeling in our companion paper, Belfiore et al.\ {\it
accepted}, and with regard to the spectral indices in Section
\ref{sec:siqual}. Here, we start with basic statements concerning the
success rate of the \dap\ (Section \ref{sec:success}) and then
provide a more detailed look at the statistical performance of the
two full-spectrum-fitting modules (Section \ref{sec:fsf}). In the
latter, we note particular regimes where the \dap\ performs poorly
(Section \ref{sec:outliers}), which will become a focus for future
development.

\subsection{Success Rate}
\label{sec:success}

Although not necessarily relevant to the quality of the data it
provides, the \dap\ executes successfully for the vast majority of
the datacubes provided in DR15. There are still some corner cases
where the \dap\ ends in error for reasons that we are still
investigating. For the 4731 datacubes that the \dap\ attempted to
analyze in DR15 (Section \ref{sec:workflow}) using two different
methods ({\tt DAPTYPE}s; Sections \ref{sec:workflow} and
\ref{sec:output}), 22 of the 9462 (0.2\%) executions failed. The
observations with \dap\ failures are: 7443-3703, 8140-6101,
8146-3702, 8158-3703, 8309-3703, 8312-6101, 8481-6103, 8549-12703,
8993-1901, 9025-12702, 9507-12702, 9677-12703, 9888-9102.
Observations 8481-6103, 8549-12703, 8993-1901, and 9507-12702 were
successful for the {\tt VOR10-GAU-MILESHC} method but failed the {\tt
HYB10-GAU-MILESHC} approach. In these cases, the successful {\tt
VOR10-GAU-MILESHC} results are provided, and one can select those
galaxies that were successfully analyzed using {\it both} approaches
using the {\tt DAPDONE} and {\tt DAPTYPE} columns in the \dapall\
catalog (Table \ref{tab:dapall}).

\begin{figure*}
\begin{center}
\includegraphics[width=0.8\textwidth]{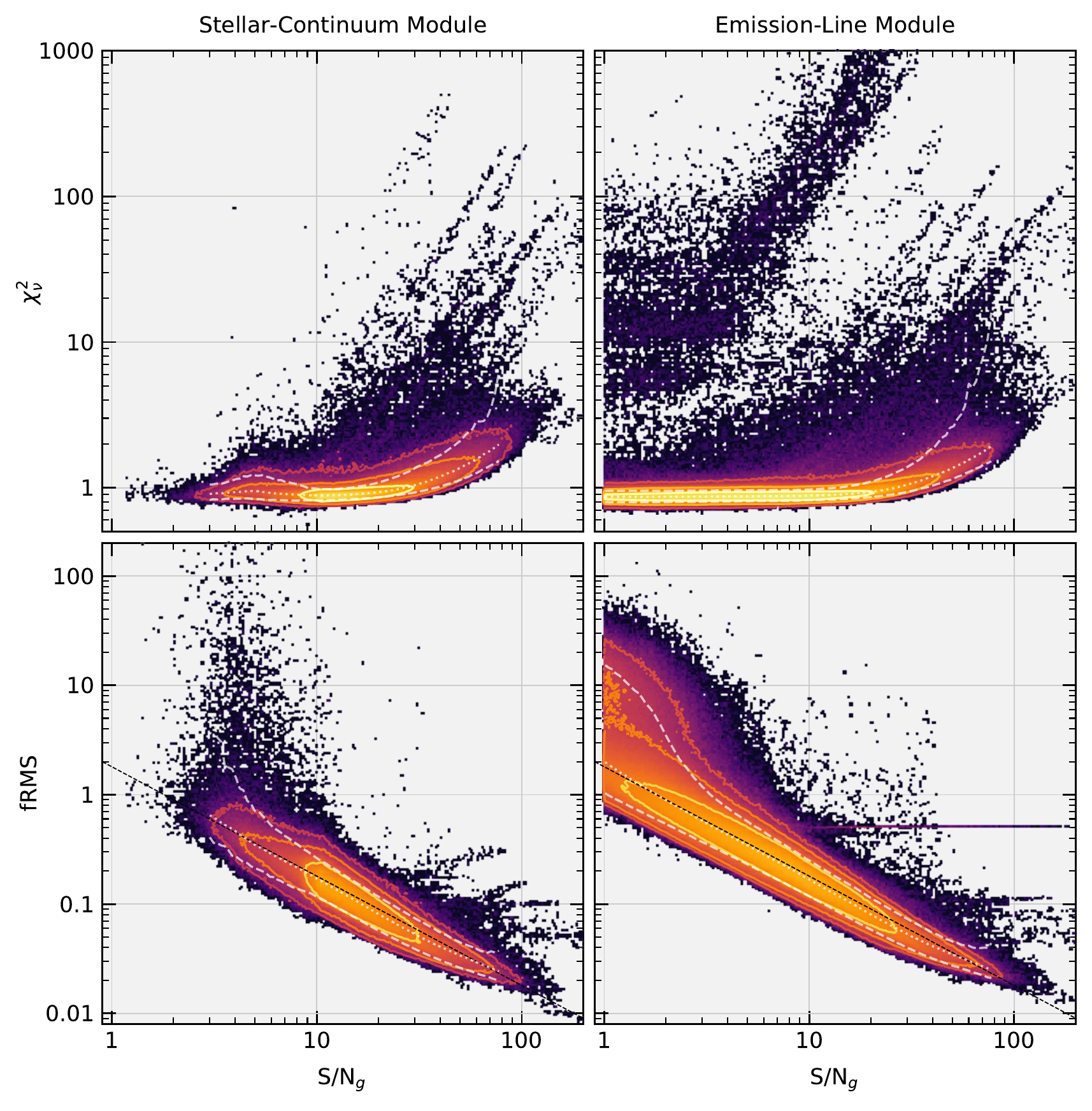}
\end{center}
\caption{The distribution of $\chi^2_\nu$ ({\it top}) and fRMS ({\it
bottom}) as a function of \snrg\ for all spectra fit in DR15 using
the {\tt HYB10-GAU-MILESHC} approach, excluding datacubes marked as
{\it CRITICAL} by the \drp\ or results masked by the relevant \dap\
module. The difference in the \snrg\ distributions for the results of
the stellar-continuum module ({\it left}) and the emission-line
module ({\it right}) are because, for the hybrid approach, the
stellar-continuum module analyzes the Voronoi-binned spectra, whereas
the emission-line module uses the individual spaxels. Note that the
\snrg\ distribution in the left panels show there are quite a few
binned spectra that do not meet the \snrg$\gtrsim$10 threshold. The
density of spectra at each location is indicated by the color, where
density increases from darker to lighter colors. The three colored
contours in each panel enclose 68\%, 95\%, and 99\% of the fitted
spectra. That is, populated regions outside the largest contour
represent 1\% of all MaNGA spectra fit in DR15. The dotted and dashed
white lines show, respectively, the median and 95\% interval at fixed
\snrg.}
\label{fig:snr_fom}
\end{figure*}

\subsection{Full-Spectrum Fitting}
\label{sec:fsf}

Much of the data provided by the \dap\ are the result of its two
full-spectrum-fitting modules described in Sections
\ref{sec:stellarkin} and \ref{sec:emlfit}. The question we address
here is: {\it How well does the \dap\ model each MaNGA spectrum?}

\subsubsection{Fit Quality}

For each model, $m_i$, fit to each MaNGA flux measurement, $f_i$, the
\dap\ calculates the absolute value of the residual ($|\Delta_i| =
|f_i-m_i|$), the fractional residual ($|\Delta_i|/m_i$), and the
error-normalized residual ($|\Delta_i|/\epsilon_i$) for each spectral
channel, $i$. We consider the growth of these quantities over each
fitted spectrum and the following reduced metrics over the full
spectrum: (1) the root-mean-square (RMS) of the fit residuals, (2)
the RMS of the fractional residuals (fRMS), and (3) the $\chi^2$
statistic (the sum of the square of the error-normalized residuals).
For assessments of the latter, we use the reduced $\chi^2$,
$\chi^2_\nu = \chi^2/(N-\nu)$, where $N$ is the number of fitted
spectral pixels and $\nu$ is the number of fitted parameters. For the
stellar-continuum module, $\nu$ is the sum of the number of kinematic
parameters (2), the order of the additive polynomial (8), and the
number of templates with non-zero weight. For the emission-line
module, $\nu$ is the sum of the order of the multiplicative
polynomial (8) and the combination of the number of templates with
non-zero weight and the relevant number of free kinematic parameters
associated with those templates (i.e., if a line is not given any
weight, the kinematics parameters associated with only that line are
not included in $\nu$). These metrics are calculated for (i) the
stellar-continuum fit used to determine the stellar kinematics
(Section \ref{sec:stellarkin}), (ii) the combined emission-line and
stellar-continuum fit used to determine the emission-line properties
(Section \ref{sec:emlfit}), and (iii), except for the growth metrics,
in 15-pixel regions around each emission line (cf.\ Belfiore et al.\
{\it accepted}).\footnote{
The metrics used in this paper have been recalculated post DR15 given
some minor errors in their calculation in \dap\ version 2.2.1. Code
that can be used to recalculate these metrics given the data provided
in DR15 is given in the \dap\ github repository found at
\url{https://github.com/sdss/mangadap}.}

For models that are well fit to the data and assuming robust
\drp-provided flux errors, (1) RMS should be proportional to the
noise in each spectrum, (2) fRMS should be inversely proportional to
its \snrg, and (3) $\chi^2_\nu$ should be very close to unity (i.e.,
the mean value of $|\Delta|/\epsilon$ should be nearly unity). Figure
\ref{fig:snr_fom} demonstrates that these expectations are well met
for both fRMS and $\chi^2_\nu$ resulting from both the
stellar-continuum and emission-line fitting modules.

Toward low \snrg, the median $\chi^2_\nu$ is $\sim$0.9, meaning that
either the model is slightly over-fitting the data (e.g., by an
error-driven selection of templates that are a marginally better fit the
data) or, more likely, that the errors in the data are slightly
overestimated ($\sim$5\%). For \snrg$\lesssim 5$, there are a number of
fits with quite large fRMS that are the result of fits to spectra with
very low flux levels. This indicates that there may be a systematic
underestimation of the continuum level in these \snrg\ regimes. Toward
high \snrg, the sharp lower limit in $\chi^2_\nu$ seen at all \snrg\
increases from $\chi^2_\nu\sim0.7$ at \snrg$\sim$1 to $\sim$2 by
\snrg$\sim$100. This is an expected and is a result of the systematic
model errors gradually beginning to dominate over the random errors in
the observations (cf.\ Section \ref{sec:scperf}; Belfiore et al.\ {\it
accepted}). In the median, $\chi^2_\nu\sim2$ at \snrg$\sim$80; however,
the width of the $\chi^2_\nu$ distribution is dramatically increased at
high \snrg. This is caused by a combination of the paucity of spectra at
such high \snrg\ and the tendency of spectra that are poorly handled by
the \dap\ (see below) to have high \snrg.

\subsubsection{Figure-of-Merit Outliers}
\label{sec:outliers}

Figure \ref{fig:snr_fom} shows contours that enclose 68\%, 95\%, and
99\% of the fitted spectra for the purpose of highlighting that
strong outliers, particularly for the $\chi^2_\nu$ distribution, or
poor fits represent fewer than 1\% of all the spectra fit by the
\dap. Coherent structures exist for both $\chi^2_\nu$ and fRMS, such
as the data groupings with roughly linear correlations between \snrg\
and $\chi^2_\nu$ and roughly constant values of fRMS. Most of these
groupings come from a small number of individual datacubes and
typically fall into one of the following categories:

\smallskip

\noindent{\bf (1) Unmasked foreground stars}: Although many of the
foreground stars that land in the field-of-view (FOV) of each IFU are
masked by the \drp, this masking is incomplete.  Masks for the
foreground stars were initially constructed thanks to a by-eye
inspection performed by L.~Lin and K.~Masters.  Recently, we have
crowd-sourced this inspection using the Galaxy Zoo: 3D
interface;\footnote{
A citizen science project at
\url{https://www.zooniverse.org/projects/klmasters/galaxy-zoo-3d}.}
objects identified as stars by at least 10 people in that project are
masked at present.  The incompleteness of the masking is partly by
design in order to avoid accidentally flagging point-like components of
the targeted galaxy (e.g., a HII regions).  Obviously, this results in
some stars being missed.  In other cases, the foreground star is masked,
but the masked area is too small to capture the wings of the stellar
PSF.  Since the \dap\ assumes every spectrum in each datacube is of the
primary target, these interloping foreground stars will be poorly fit.
Because of their varying luminosity, these outliers can occur over a
large range in \snrg, but they typically identified as significant
$\chi^2_\nu$ outliers at fixed \snrg. One should expect these to be
outliers in both the stellar-continuum and emission-line modules. Most
of the outliers with roughly constant fRMS at high \snrg, or roughly
follow $\chi_\nu\propto$ \snrg, are due to these unmasked stellar
spectra.

\smallskip

\noindent{\bf (2) Non-targeted Galaxies in the Field-of-View}: Many
MaNGA observations include multiple objects in the FOV, which may or may
not have been previously recognized as objects superimposed along the
line-of-sight. There are two ways these objects can lead to poor \dap\
fits. First, if the objects overlap in a given spectrum, the \dap\ will
tend to optimize the fit to the more luminous component and a poor fit
is likely, depending on the surface-brightness ratio and velocity
separation of the two objects. These cases can be difficult to identify
because the effect on $\chi^2_\nu$ can be subtle. However, in the second
case, strong deviations in $\chi^2_\nu$ will occur for regions dominated
by the interloper when it is outside of the redshift range allowed by
each fit ($\pm2000$ \kms\ from the input redshift, $z_0$, typically from
the NSA; Section \ref{sec:stellarkin}). Such poor fits occur at all
\snrg\ and can typically be identified by the deviation from the
$\chi^2_\nu$ distribution of good fits at similar \snrg.

\smallskip

\noindent{\bf (3) Bright/Broad Emission Lines}: The \dap\ currently
assumes that all emission lines are single Gaussian components.
However, particularly for very bright emission lines, a second broad
component is very apparent in the data but cannot be reproduced by
the nominal \dap\ model. Recall that the stellar-continuum fit uses a
fixed $\pm750$ \kms\ mask for the emission lines offset by the input
redshift ($z_0$; Figure \ref{fig:ppxfmask}). This is
typically sufficient to mask the relevant velocity range of the
emission features, but it is not sufficient for sources with
broad-line regions (e.g., AGN). Both effects can lead to dramatic
$\chi^2_\nu$ outliers from both the stellar-continuum and
emission-line fitting modules. Because the latter optimizes the
combined continuum$+$emission-line spectrum, the fit and resulting
$\chi^2_\nu$ are generally better than for the stellar-continuum
module; however, the poor quality persists for the brightest/broadest
spectra. {\it Any AGN-focused studies should be very careful with the
data provided by the \dap. It is likely they will require other
analysis products that better handle broad, multi-component emission
lines.} In principle, including additional emission-line components
in the \dap\ is straight-forward, with minimal code development;
however, the validation and stability of the approach will likely
require a significant investment.

\smallskip

\noindent{\bf (4) Unmasked Cosmic Rays}: The \drp\ removes the vast
majority of the cosmic rays from each MaNGA exposure; however, some
are still missed for a variety of reasons. Most often the affected
spectral channels are easily identified as 3-$\sigma$ outliers and
rejected during the fit iterations for both the stellar-continuum and
emission-line modules. However, in the emission-line module, these
cosmic rays may not be rejected if they fall close enough to emission
lines: To avoid rejecting emission-line flux for particularly strong
lines, we do not allow the rejection iteration to remove pixels near
a fitted emission-line.\footnote{
Specifically, pixels are excluded from rejection if the best-fit
emission-line model has a flux density of $>10^{-6}$
erg/s/cm$^2$/\AA/spaxel.}
Lingering cosmic-ray artifacts generally show up as ``beam-sized''
regions with large values in maps of $\chi^2_\nu$. We also note that,
for the subset of spectra that we have inspected directly, spectra
that populate the upper-left of the upper-right panel in Figure
\ref{fig:snr_fom} are in fact due to cosmic rays near emission-lines.
It is therefore reasonable that a similar distribution is not seen
for the results of the stellar-continuum module. Because the cosmic
rays affect such a small portion of the spectrum, the model may
actually be perfectly reasonable for these cases. Unfortunately, this
needs to be assessed on a case-by-case basis. In particular, we have
found cases (e.g., {\tt 8319-3704}) where cosmic rays have
dramatically affected the ionized-gas velocity by pulling all lines
so that a single line can better fit the cosmic ray.

\smallskip

Beyond the four cases above, there are other more subtle limitations
of the full-spectrum-fitting modules when one isolates fits that are
only just outside the main $\chi^2_\nu$ distribution. Some
interesting examples include galaxies with significantly asymmetric
emission-line profiles and star-forming galaxies with easily
identifiable emission lines that are not currently in the list of
lines fit by the \dap\ (see Table \ref{tab:emldb} and Figure
\ref{fig:showcasespec}). We continue to identify these
astrophysically interesting phenomena that stretch beyond the
standard \dap\ assumptions and work toward improvements that can
properly handle the large variety of MaNGA spectra.

\section{Output Products}
\label{sec:output}

We have touched on the output products provided by the \dap\
throughout our paper, particularly when introducing some salient
details about the data in Section \ref{sec:guidance} and when
discussing the \dap\ workflow in Section \ref{sec:workflow}. In the
latter, we noted that each of the six main \dap\ modules produces a
reference file, which includes all the data produced by the module
and can be used to reconstruct the state of the relevant {\tt python}
object (see Figure \ref{fig:workflow}) to minimize redundant analysis
steps. The final step of the \dap\ is to consolidate and reformat the
data in these reference files into the two main files meant for
general use, the \dapmaps\ file (Section \ref{sec:mapsfile}) and
model \dapcube\ file (Section \ref{sec:cubefile}).

Reformatting the data is a key component of this final step. Most of
the core functionality of the \dap\ treats each spectrum
independently, regardless of whether it is from a bin or an
individual spaxel. The format of the reference files matches this
structure, with spectra organized along rows of 2D arrays and derived
quantities organized in data tables with one row per spectrum. To
ease its use, however, we provide the data in a spatial format that
exactly matches the \drp-produced datacubes. These details are
largely irrelevant to anyone who uses the \dapmaps\ and model
\dapcube\ files, except to emphasize that users must be careful when
interpreting the Voronoi-binned maps and spectra in these files. We
provide guidance in this regard specific to the \dapmaps\ and model
\dapcube\ files in Sections \ref{sec:mapsfile} and
\ref{sec:cubefile}, respectively.

The \dap\ produces a \dapmaps\ and a model \dapcube\ file for each
analysis approach, or {\tt DAPTYPE} (Section \ref{sec:workflow}),
meaning there are two \dapmaps\ and model \dapcube\ files for each
datacube successfully analyzed for DR15 (see Section
\ref{sec:success}). The detailed data models for the \dap\ output
files are provided in Appendix \ref{sec:datamodel} and via the DR15
website.\footnote{
Specifically, see the description at
\url{https://www.sdss.org/dr15/manga/manga-data/data-model/} and the
detailed datamodel at
\url{https://data.sdss.org/datamodel/files/MANGA\_SPECTRO\_ANALYSIS/}.
A brief introduction for how to read the latter can be found at
\url{https://data.sdss.org/datamodel}.}

Once the \dap\ has been executed on the individual datacubes, a final
post-processing step is executed to construct a summary catalog
called the \dapall\ catalog (Section \ref{sec:dapall}). Currently,
the primary intent of this catalog is to aid sample selection. We
continue to improve the quantities provided by the \dapall\ catalog,
but we currently do not recommend these data be used for science
without a detailed understanding of the data quality. In particular,
note that the \dapall\ catalog does not provide {\it any} measurement
uncertainties and only very simple methods are used to perform each
measurement (e.g., the star-formation rate does not account for
attenuation). The \dapall\ summary catalog can be queried using both
CASJobs\footnote{\url{https://skyserver.sdss.org/casjobs/}} and
\marvin. The full list of columns provided in the \dapall\ catalog is
provided in Appendix \ref{sec:datamodel} and via the DR15
website.\footnote{
The \dapall\ catalog is now included in the list of MaNGA Catalogs at
\url{https://www.sdss.org/dr15/manga/manga-data/catalogs/}, with the
detailed datamodel at
\url{https://data.sdss.org/datamodel/files/MANGA\_SPECTRO\_ANALYSIS/DRPVER/DAPVER/dapall.html}.}

\subsection{\dapmaps\ Files}
\label{sec:mapsfile}

The \dapmaps\ file is the primary output file that provides the
spaxel-by-spaxel quantities derived by the \dap.  The measurements are
organized in a series of extensions (Table \ref{tab:mapsfile}) that
contain images, or maps, with a format identical to the spatial
dimensions of the \drp\ datacube.  Extensions may contain a single map,
like the measured stellar velocity, or a series of maps organized in
``channels'', like the fluxes derived for each emission line.  When an
extension contains more than one map, the channels are identified in the
header.  For example, the header of the extension containing the
emission-line fluxes contains the header keyword and value {\tt
C19~=~'Ha-6564'}, indicating that the nineteenth channel contains the
flux of the H$\alpha$ line.  Most extensions with \dap\ measurements
have companion extensions with the inverse variance of the measurements
and a quality mask (see Table \ref{tab:mapsfile}).

Beyond this basic description of the data format, there are a few
critical components of the \dapmaps\ files that users should keep in
mind:

\smallskip

\noindent {\bf (1) Quantities are provided that a user must correct
using the provided corrections.}  In particular, the {\bf stellar
velocity dispersions} are provided as measure by \ppxf\ including the
offset in spectral resolution between MaNGA and the \mileshc\ template
library; see Section \ref{sec:sigmacorr}.  Similarly, the {\bf
emission-line velocity dispersions} are provided as would be determined
by fitting a Gaussian directly to the emission line and must be
corrected for the instrumental resolution; see Section
\ref{sec:gassigmacorr}.  Finally, the {\bf spectral indices} are
provided as measured directly from the spectra and must be corrected for
the effects of the velocity dispersion on the measurement; see Section
\ref{sec:indexcorr}.  

\smallskip

\noindent {\bf (2) Basic quality assessments of the data are provided
via bitmasks and should be used.} The \dap\ performs a number of
quality checks during the measurement process. The mask bits
triggered by the \dap\ modules are consolidated and incorporated into
the bitmasks provided with the \dapmaps\ files (see Table
\ref{tab:dappixmask}). Any non-zero value of the bitmask indicates
that the measurement should be either treated with care or ignored.
Both the \dap\ source code and \marvin\ provide convenience {\tt
python} classes that facilitate the use of the bitmasks\footnote{
\url{https://www.sdss.org/dr15/algorithms/bitmasks/}}
to appropriately flag the \dap\ data.

\smallskip

\noindent {\bf (3) Results for a binned spectrum are repeated for each
spaxel in the bin.}  When using {\it any} data from the {\tt
VOR10-GAU-MILESHC} files or results from the first three modules (Figure
\ref{fig:workflow}) of the {\tt HYB10-GAU-MILESHC} files, the results in
every spaxel do not necessarily represent unique measurements.  This is
critical to consider when, e.g., fitting the data with a model or
binning the data as a function of radius.  The primary use of the {\tt
BINID} extension in the \dapmaps\ file (Table \ref{tab:mapsfile}) is to
allow users to select the unique measurements made for each mapped
quantity.

\smallskip

\noindent {\bf (4) Flux units are per spaxel.}  The units can be
converted to surface brightness by multiplying by the pixel scale (i.e.,
four spaxels per arcsec$^2$).  Integrations of the flux over map
apertures can be done by summing spaxel values; however, be aware of the
previous point about measurements being repeated for binned spaxels.
The spectral stacking procedure is a simple average of the spaxels in
each bin (Section \ref{sec:stacking}), meaning that the units are
correct; however, one should avoid apertures that do not enclose the
full bin.

\smallskip

\noindent {\bf (5) Velocities are offset by the input bulk redshift.} As
discussed in Section \ref{sec:vdef}, the velocities reported in the
\dapmaps\ files have been offset by the input bulk redshift, $z_0$.
These redshifts are most often provided by the NSA, and the value used for
the bulk redshift is saved in \kms\ ($cz_0$) in the header keyword {\tt
SCINPVEL}.  Because these bulk redshifts are not directly determined
from the MaNGA data, they may not accurately offset the velocity to 0
\kms\ at the galaxy center.  One can recover the redshift measured for
each spaxel or binned spectrum, $z_{\rm obs}$, using Equation
\ref{eq:voffset} in Section \ref{sec:vdef}.

\smallskip

\noindent {\bf (6) Some velocity dispersion measurements are below the
MaNGA instrumental resolution.}  As mentioned above, both the stellar
and emission-line velocity dispersions must be corrected; the former is
corrected for the intentional offset between the template resolution and
the galaxy data (Section \ref{sec:resmatch}) and the other is for the
instrumental resolution of the data.  It is possible to find
measurements that are smaller than the quadrature correction due to the
error distribution in the determination of either.  The reason we have
left it to the users to decide how to treat these measurements is
because that treatment may depend on the science goals.  We discuss
these measurements, in particular, and provide advice for their
treatment in Section \ref{sec:svdusage}.

\smallskip

\noindent {\bf (7) Modeled properties of each emission line in a
spectrum are not necessarily independent.} As executed for DR15, the
\dap\ ties all emission-line velocities and ties the fluxes and
velocity dispersion for many of the doublets; see Section
\ref{sec:emltpl} and Table \ref{tab:emldb}. However, as discussed in
Section \ref{sec:emlmeasurements}, the relevant properties and errors
are provided for each line, even if those properties are not
independent parameters in the fit. Unfortunately, there is no way to
determine which parameters are tied based solely on the provided
output files; users must consult Table \ref{tab:emldb}.

\smallskip

\noindent {\bf (8) Emission-line fluxes are corrected for Galactic
foreground extinction but not dust within a given galaxy.}  See point
(4) in the next section.

\smallskip

\noindent {\bf (9) The non-parametric emission-line fluxes are
provided largely as a check on the Gaussian modeling results.} The
\dapmaps\ files provide both an {\tt EMLINE\_GFLUX} and {\tt
EMLINE\_SFLUX} extension with the results from, respectively, the
Gaussian model-fit results (Section \ref{sec:emlmeasurements}) and
the zeroth-moment measurements (Section \ref{sec:emlmom}). In
general, the more precise measurements from the Gaussian modeling
should be used. The two measurements only significantly differ at low
flux levels (see Belfiore et al., {\it accepted}, Figure 4) when
the Gaussian fit can be driven by noise. The rule of thumb is then to
use the Gaussian results when it is similar to the non-parametric
result; otherwise, do not trust either.

\smallskip

\noindent {\bf (10) Errors are based on formal calculations.} All
errors provided by the \dap\ are currently based on either a formal
calculation (see, e.g., the description in the last paragraph of
Section \ref{sec:repeats}) or simple error propagation from the DRP.
In general, we find these errors to be within a factor of 2 of the
true error as determined by both idealized parameter-recovery
simulations and empirical measurements of the random error using
repeat observations. This is true of the stellar kinematics (Section
\ref{sec:scperf}), the model-fit emission-line properties (Belfiore
et al., {\it accepted}), and the spectral indices (Section
\ref{sec:spindex}). In the latter two cases, we have provided simple
prescriptions to recalibrate the provided formal errors to match our
simulation results and/or empirical estimates; however, the user must
apply these calibrations themselves.

\smallskip

\noindent {\bf (11) The mapped properties are covariant.} As we
discuss throughout our paper (Sections \ref{sec:guidance},
\ref{sec:snr}, \ref{sec:repeats}; Figures \ref{fig:correlation},
\ref{fig:rhogauss}, \ref{fig:binning_covar}, and
\ref{fig:covarcalib}), the MaNGA datacubes exhibit significant
spatial covariance given the subsampling of the MaNGA
$2\farcs5$-diameter fiber beam into $0\farcs5\times0\farcs5$ spaxels.
This covariance, of course, propagates to the derived parameters;
however, we have not provided covariance matrices for the \dapmaps\
data in DR15. Initial simulations suggest that, to first order, the
{\it correlation} matrix of the flux in a representative wavelength
channel (e.g., Figure \ref{fig:correlation}) is the same as for the
derived parameters. In particular, a first-order correlation matrix
for the derived quantities can be constructed assuming $\ln \rho_{jk}
= - D^2_{jk}/7.37$, where $j$ and $k$ are the indices of two spaxels
separated by a distance of $D_{jk}$ in spaxels. Although the
significant covariance between spaxels complicates the analysis of the
data, it also allows one to quickly determine the \snrg\ level at
which \dap\ results for individual spaxels may suffer from systematic
error: Any results provide by the \dap\ that do not smoothly vary
between adjacent spaxels are driven by systematic error in the
associated algorithm.

\smallskip

\noindent {\bf (12) Visual inspection of the data can be critical.}
The MaNGA data set is incredibly varied. One may find interesting
outliers or trends when looking for them; however, possibly not for
the expected reason. For example, in searching for datacubes with the
strongest gradient in D4000, one is led to find observations with
more than one target in the IFU field-of-view, not individual
galaxies with the strongest stellar population gradients. Both in the
sense of looking at the mapped properties and the fitted spectra,
conclusions should always be evaluated in the context of the original
source data. \marvin\ is particularly useful for quick visual
assessments of the
data.\footnote{\url{https://dr15.sdss.org/marvin/}}

\begin{figure}
\begin{center}
\includegraphics[width=\columnwidth]{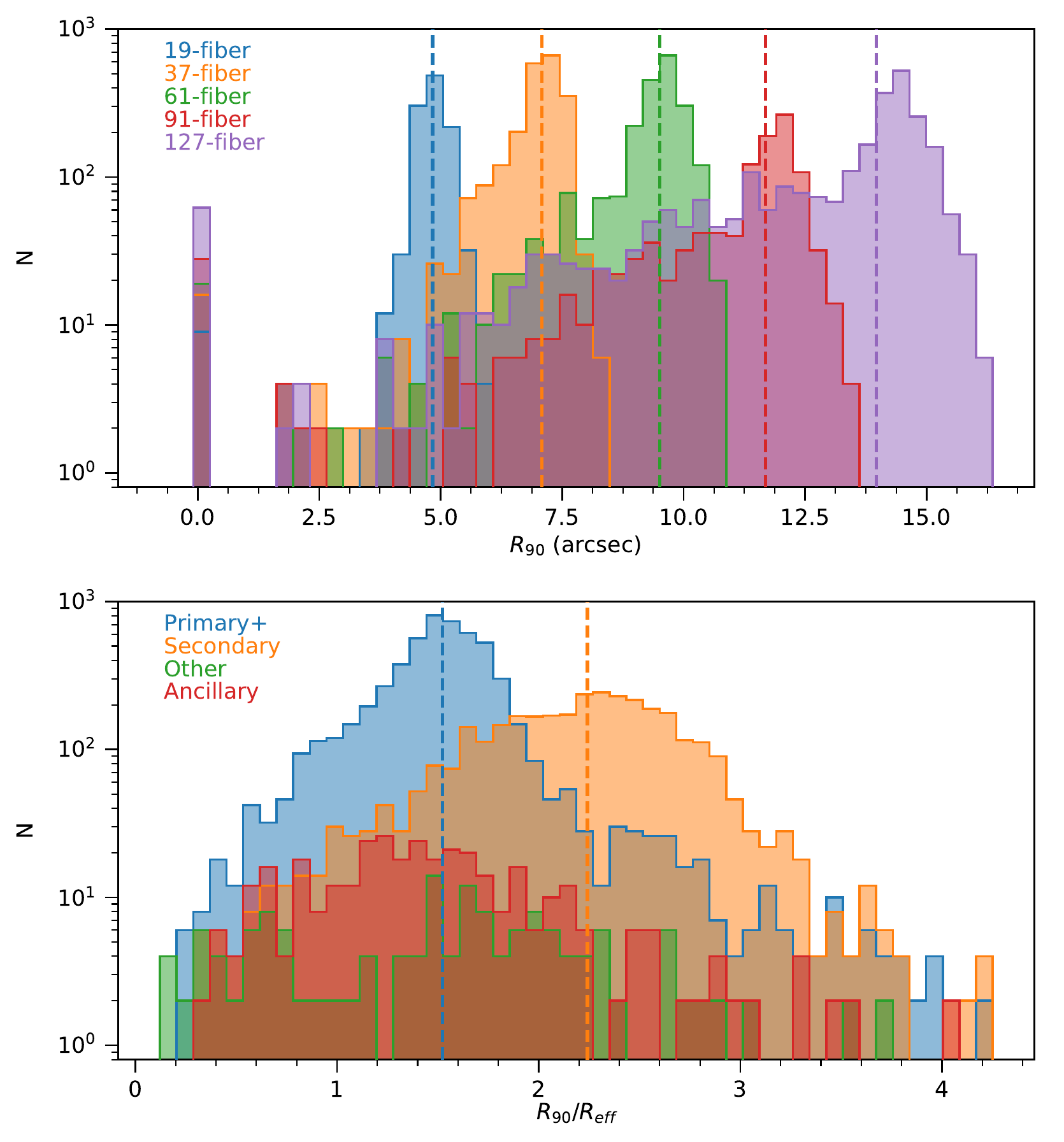}
\end{center}
\caption{The radius to which at least 90\% of a $2\farcs5$ elliptical
annulus is covered by spaxels analyzed by the \dap, $R_{90}$. The
criteria selecting spaxels to be analyzed by the \dap\ is discussed
in Section \ref{sec:prelim}. The top panel shows the distribution of
$R_{90}$ in arcsec for observations taken with each bundle, colored
by the bundle size. The bottom panel shows the distribution of
$R_{90}$ normalized by the elliptical-Petrosian half-light radius,
$R_{\rm eff}$, for galaxies belonging to the Primary$+$ and Secondary
samples, as well as observations of ancillary or filler targets.}
\label{fig:r90}
\end{figure}

\begin{figure}
\begin{center}
\includegraphics[width=\columnwidth]{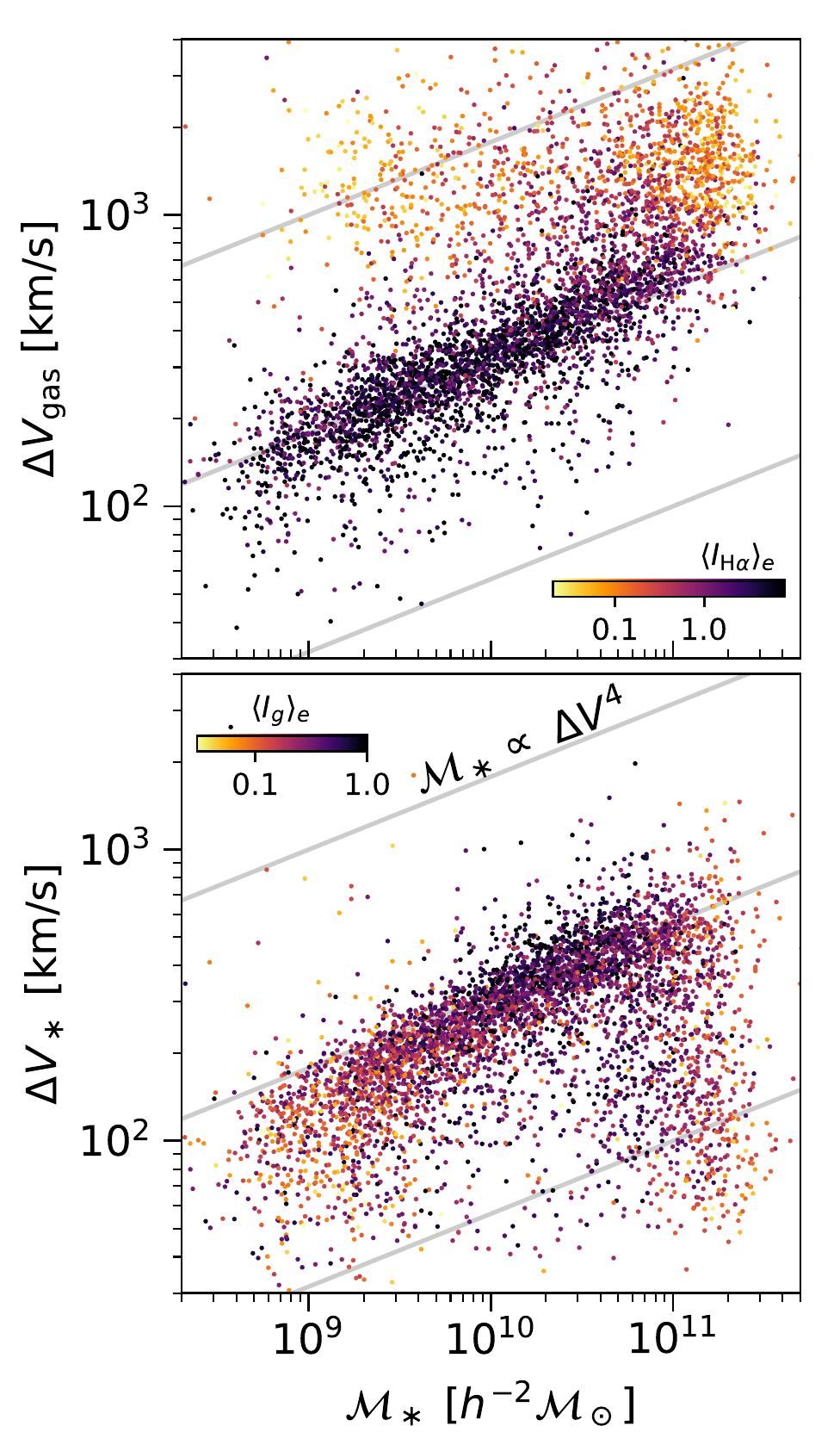}
\end{center}
\caption{NSA stellar mass versus the velocity gradient --- defined as
$\Delta V = (V_{\rm hi} - V_{\rm lo})/(1-(b/a)^2)^{1/2}$, where
$V_{\rm hi}$ and $V_{\rm lo}$ are provided by the \dapall\ file ---
of the emission-line (top) and stellar (bottom) kinematics. Points
are colored according to the mean surface brightness within 1 $R_e$.}
\label{fig:massvel}
\end{figure}

\subsection{Model \dapcube\ Files}
\label{sec:cubefile}

The primary purpose of the model \dapcube\ file is to allow users to
check the results of the two full-spectrum-fitting modules against the
data, particularly when the data in the \dapmaps\ file appear
unreasonable.  The file is made up primarily of extensions that contain
datacubes in the same format as the \drp\ {\tt LOGCUBE} file that the
\dap\ has analyzed.  The full list of extensions and their content is
provided in Table \ref{tab:logcubefile}.

Points (2), (3), and (4) from the previous section on the \dapmaps\ file
also apply when using the model \dapcube\ files.  For the description of
the bitmasks used in the model \dapcube\ file, see Table
\ref{tab:dapspecmask}.  In addition:

\smallskip

\noindent {\bf (1) The flux array provided in the \dap\ model \dapcube\
files is different from the flux array in the \drp\ datacube.}  Although
this is apparent from the data model of each of the relevant files, the
naming convention of the two files can lead to confusion.  For clarity,
the model \dapcube\ file always provides the {\it binned} spectra, and
the name of the model \dapcube\ file always includes the {\tt DAPTYPE}.

\smallskip

\noindent {\bf (2) The best-fit stellar continuum used to determine
the stellar kinematics is not provided directly and must be
constructed.} The models provided in the {\tt MODEL} extension are
the result of the combined continuum$+$emission-line fits performed
by the emission-line module. To construct the best-fitting spectrum
from the stellar kinematics module, one has to remove the emission
lines (in extension {\tt EMLINE}) and the difference between the
stellar continuum determined between the two full-spectrum-fitting
modules (in extension {\tt EMLINE\_BASE}). That is, the stellar
continuum is computed as {\tt MODEL} - {\tt EMLINE} - {\tt
EMLINE\_BASE}.\footnote{
This convention is true for DR15, but the datamodel of the model
\dapcube\ files will change in future releases.}

\smallskip

\noindent {\bf (3) The models provided for the hybrid binning scheme
should be compared to the \drp\ datacube.}  The model \dapcube\ file
always provides the {\it binned} spectra.  However, in the hybrid
binning scheme, the models provided have been fit to the individual
spaxels because they are the result of the emission-line module.
Although there are ``binned spectra'' composed of single spaxels, in
general this means that the {\tt MODEL} extension of the model \dapcube\
file for the hybrid binning scheme ({\tt DAPTYPE} = {\tt
HYB10-GAU-MILESHC}) must be compared to the \drp\ datacube, not its own
{\tt FLUX} array.  The same is {\it not} true for the {\tt DAPTYPE} =
{\tt VOR10-GAU-MILESHC} files.

\smallskip

\noindent {\bf (4) The spectra include Galactic extinction.} Section
\ref{sec:galext} notes that, once the spectra are binned, the
Galactic extinction is removed from the data and all spectral
modeling and measurements provided by the last four modules of the
\dap\ (see Figure \ref{fig:workflow}) use extinction-corrected
spectra. However, to facilitate the comparison of the models with the
\drp-produced datacubes (particularly given the previous point), the
extinction curve is reapplied to the data before being written to the
model \dapcube\ file; the exact reddening correction applied is
provided in the {\tt REDCORR} extension.

\begin{figure}
\begin{center}
\includegraphics[width=1.0\columnwidth]{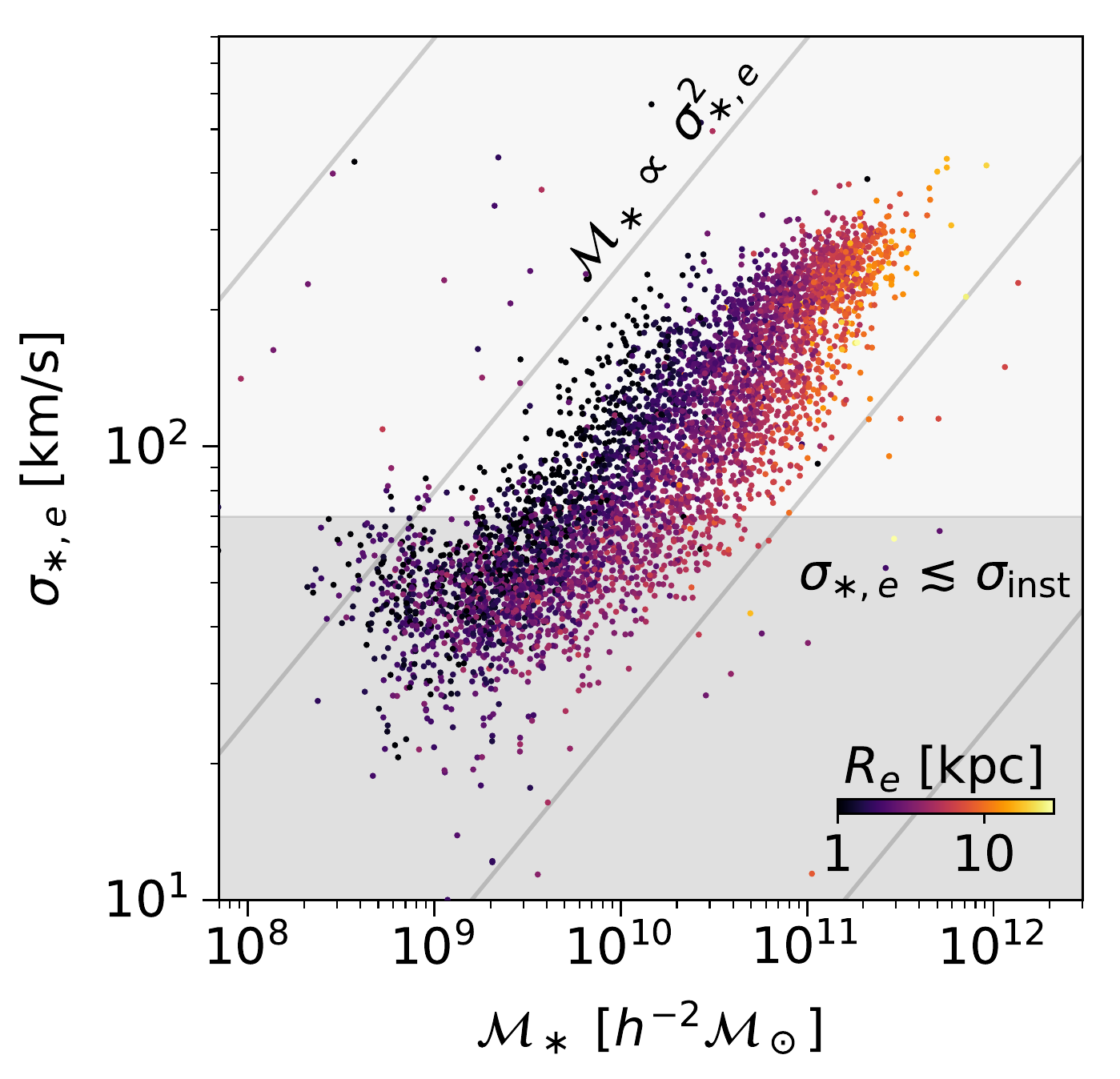}
\end{center}
\caption{NSA stellar mass versus light-weighted stellar velocity
dispersion within 1 $R_e$ from the \dapall\ file.}
\label{fig:masssigma}
\end{figure}

\begin{figure}
\begin{center}
\includegraphics[width=1.0\columnwidth]{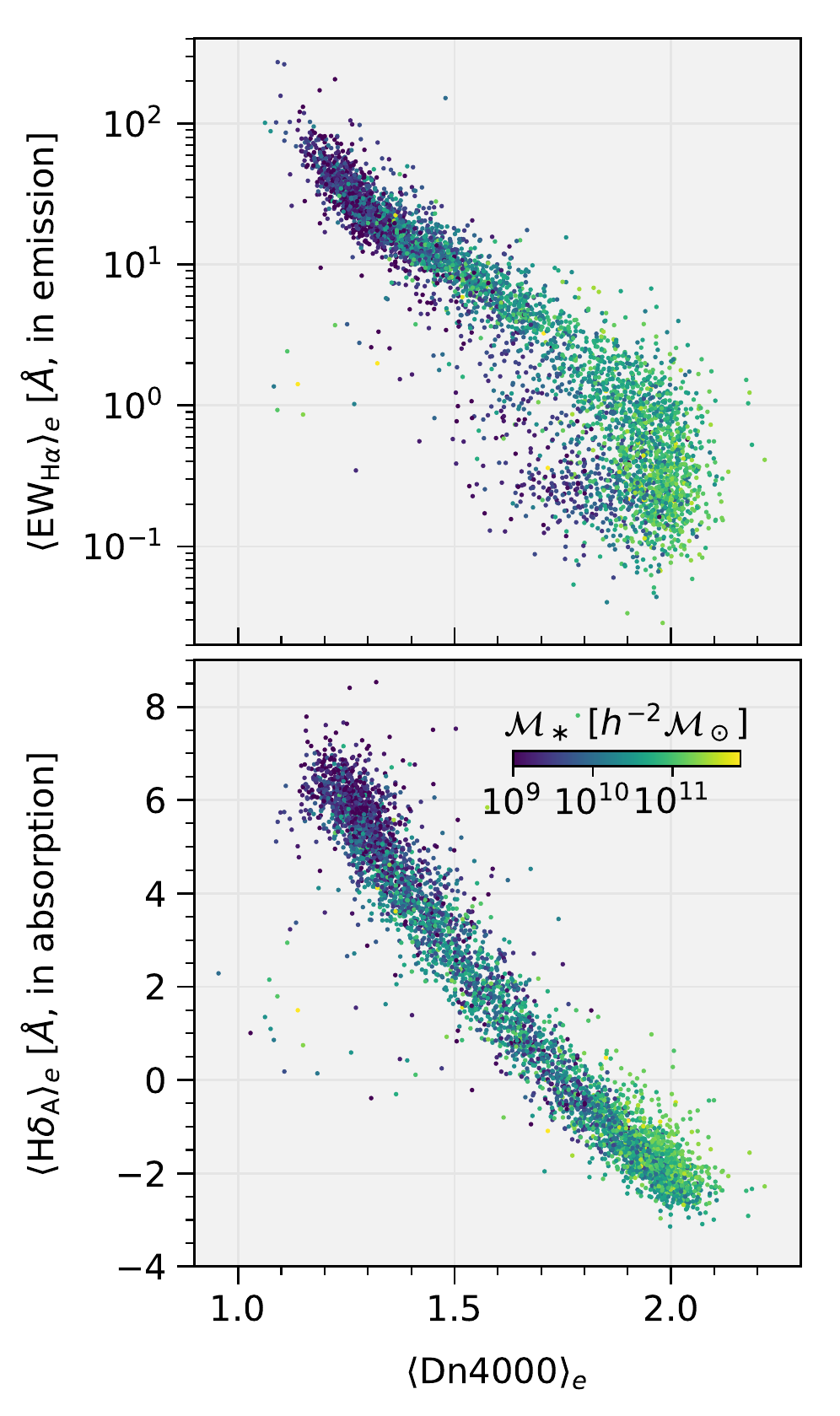}
\end{center}
\caption{Dn4000 versus the H$\alpha$ equivalent width {\it in
emission} (top) and the H$\delta_{\rm A}$ index (equivalent width in
{\it absorption}) after subtracting the best-fitting emission-line
model. Points are colored by NSA stellar mass.}
\label{fig:ew_d4000}
\end{figure}

\begin{figure}
\begin{center}
\includegraphics[width=1.0\columnwidth]{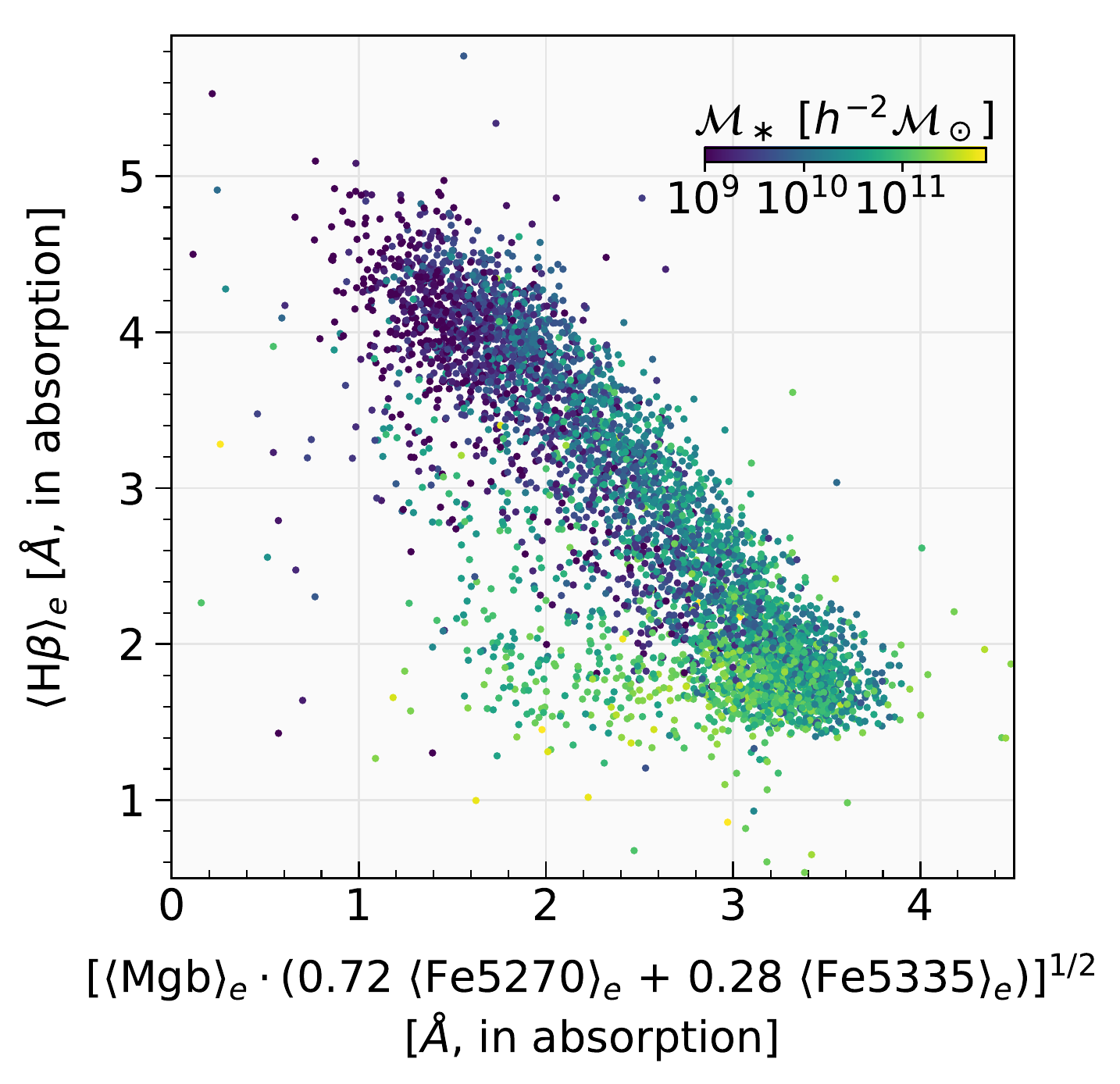}
\end{center}
\caption{Typical stellar-population age-metallicity diagnostic using
the Mgb, Fe5270, Fe5335, and H$\beta$ absorption indices. Points are
colored by NSA stellar mass.}
\label{fig:pops}
\end{figure}

\begin{figure}
\begin{center}
\includegraphics[width=1.0\columnwidth]{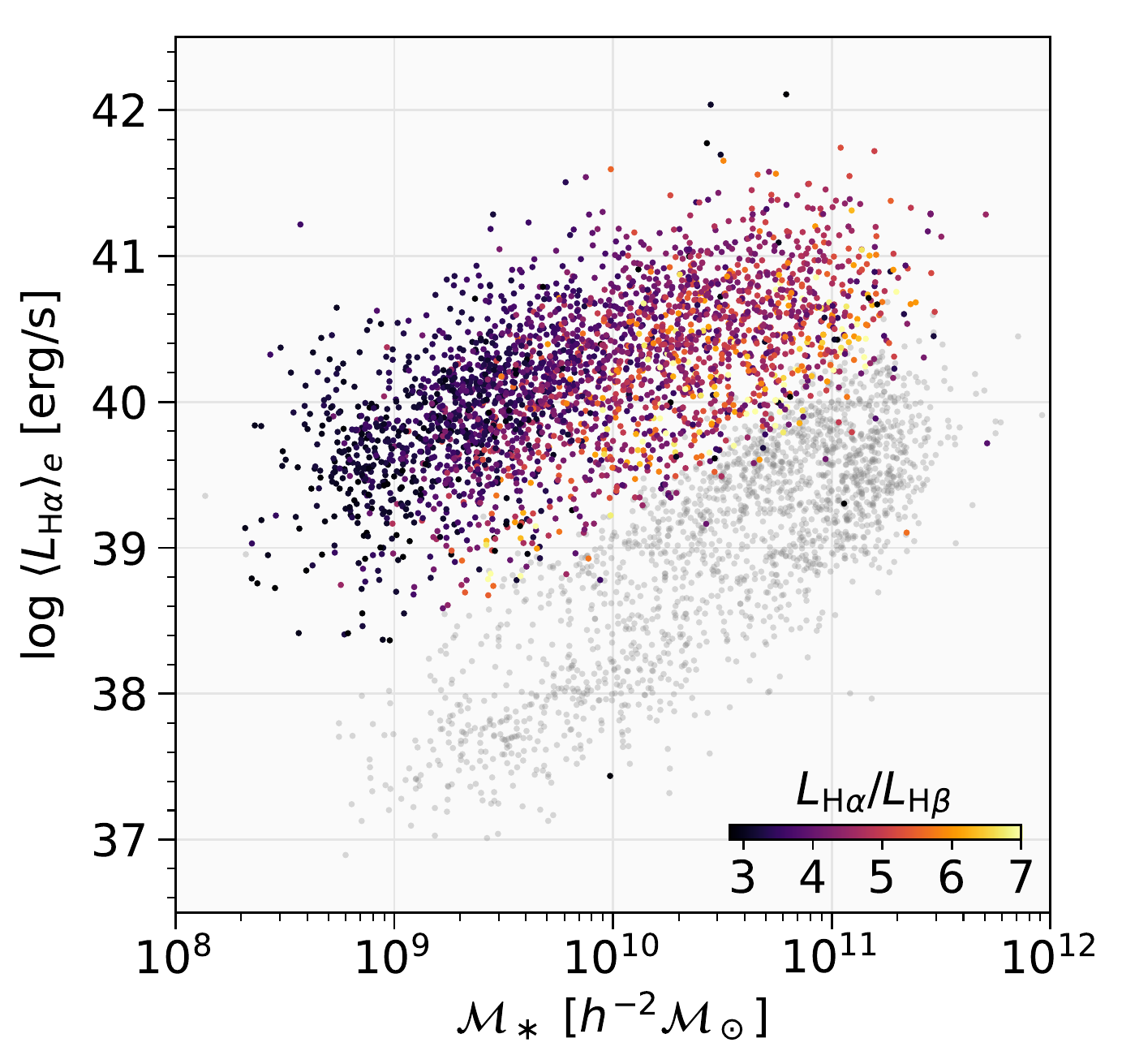}
\end{center}
\caption{NSA stellar mass versus the absolute luminosity in
H$\alpha$. Points with H$\alpha$ EW greater than 2 \AA\ are colored
by the H$\alpha$-to-H$\beta$ luminosity ratio; others are set to
gray.}
\label{fig:sfr}
\end{figure}

\subsection{The \dapall\ Summary Catalog}
\label{sec:dapall}

Similar to the MaNGA \drpall\ file, we provide a summary \dapall\
catalog that collates global information pulled or derived from the
primary output files of the \dap. The \dapall\ catalog contains one
row per {\tt PLATEIFU} and {\tt DAPTYPE} combination. As discussed in
Section \ref{sec:workflow}, the \dap\ analyzed 4731 observations
using two analysis approaches for DR15, meaning the \dapall\ file has
9462 rows. However, a small fraction of those analysis attempts
failed (see Section \ref{sec:success}); the failures are indicated by
the {\tt DAPDONE} column in the \dapall\ catalog. The most basic
selection of rows from the \dapall\ catalog would then select
observations that were successfully analyzed ({\tt DAPDONE == 1}) for
a given analysis approach (e.g., {\tt DAPTYPE == HYB10-GAU-MILESHC}).
For convenience when querying properties in both the \drpall\ and
\dapall\ catalogs, we also provide the row index in the \drpall\
database, {\tt DRPALLINDX}, matched to the same observation ({\tt
PLATEIFU}).

We emphasize again (see the beginning of this Section) that the
current \dapall\ catalog is primarily provided as a convenience to
aid sample selection. Roughly half of the \dapall\ columns contain
either metadata pulled from the \dapmaps\ file headers that are
relevant to the methods used in the analysis or metadata repeated
from the \drpall\ catalog and provided for convenience. The other
half are derived directly from the \dapmaps\ data with the aim of
providing relevant quantities for queries based on the spatial
coverage, signal-to-noise, redshift, internal kinematics, and
composition of each galaxy. The methods used to construct these data
are simple, sometimes at the expense of performing the nuanced
analysis needed for direct scientific use. In particular, no
uncertainties are currently calculated for the properties unique to
the \dapall\ catalog. Of course, any sample selection based on these
quantities should be tempered by an understanding of the sample
biases that may result \citep{2017AJ....154...86W}, as well as the
limitations in the measurement construction and return to the source
\dapmaps\ data as necessary for a more nuanced analysis. Below we
briefly highlight some of these quantities and the details of their
calculation.

The MaNGA galaxy survey is designed with nominal radial coverage and
S/N requirements \citep{2016AJ....152..197Y, 2017AJ....154...86W},
and the \dapall\ file provides assessments of these quantities for
each observation. For example, Figure \ref{fig:r90} shows the
distribution of the radial coverage for each observation in
arcseconds and normalized by $R_e$. We define the radial coverage of
each galaxy as the limiting radius to which at least 90\% of the area
of a $2\farcs5$ elliptical annulus is observed by MaNGA spaxels. The
distribution is as expected with median values that illustrate the
on-sky size of the IFU and the designed 1.5-$R_e$ and 2.5-$R_e$
radial coverage of, respectively, the Primary$+$ and Secondary
samples of the main galaxy survey. The S/N metrics in the \dapall\
file provide, e.g., the median S/N between $1-1.5 R_e$ in the $griz$
bands and the median $g$-band S/N for spectra between $0.0-1.0$,
$0.5-1.5$, and $1.5-2.5 R_e$.

For global kinematic properties, the \dapall\ file provides simple
measurements of the bulk redshift, velocity gradient, and velocity
dispersion within 1$R_e$ for the stellar and ionized gas tracers.
Figure \ref{fig:massvel} shows our assessment of the velocity
gradient for the gas and stars against the NSA stellar mass. This a
crude version of the \citet{Tully1977} (T-F) relation using ionized
gas and stellar mass. We define the velocity gradient as the
difference between the minimum and maximum measured velocity after
removing 3-sigma outliers, and we apply a rough correction for the
projection of the motions along the line-of-sight using the
photometric ellipticity. That is, $\Delta V = (V_{\rm hi}-V_{\rm lo})
(1-(b/a)^2)^{-1/2}$, where $V_{\rm hi}$ and $V_{\rm low}$ are
provided in the \dapall\ catalog and $b/a$ is provided by the \drpall
catalog. Although this is a very basic assessment of the velocity
field, we do find a correlation between the velocity gradient and
stellar mass for galaxies with relatively high \halpha\ and/or
$g$-band surface brightness. The observed trend has roughly the
expected form $\Delta V\propto M_{\ast}^4$ of the T-F relation.

Figure \ref{fig:masssigma} shows the luminosity-weighted mean stellar
velocity dispersion within 1$R_e$ against the NSA stellar mass. In
detail, we calculate $\sum_i I_{g,i} (\sigma^2_{{\rm obs},i} -
\delta\sigma^2_{{\rm inst},i})^{1/2}/\sum_i I_{g,i}$, where $I_{g,i}$
is the $g$-band-weighted mean flux for spaxel $i$, and the sum is
over all $i$ spaxels with luminosity-weighted bin (or individual
spaxel) centers within 1$R_e$. This is effectively stellar-mass view
of the \citet{Faber1976} relation, which is a projection of the
$(\log M_\ast,\log\sigma_{\ast,e},\log R_e)$ mass plane \cite[see
sec.~4 of][for a review]{Cappellari2016} but it includes all
morphological types instead of ETGs alone. As expected, the primary
correlation between stellar and dynamical mass is evident. Above the
instrumental resolution, the upper boundary envelope roughly follows
the trend $M_\ast\propto\sigma_{\ast,e}^2$ observed for much smaller
samples, with the expected flattening at larger masses
($M_\ast\sim3\times 10^{10}$ ) \citep[cf.\ Figure~20
of][]{Cappellari2016}. But the trend persists at lower $\sigma_\ast$
(with a slight upturn at the lowest masses), confirming our ability
to measure $\sigma_\ast$ well below the MaNGA instrumental
dispersion. A different projection of the MaNGA mass plane, for the
galaxies in the DR14 was presented in \citet{Li2018manga}.

The \dapall\ file provides the unweighted median of all emission-line
fluxes and equivalent widths, both from the Gaussian and
non-parametric fits, and spectral indices for spectra with
luminosity-weighted bin (or individual spaxel) centers within 1$R_e$.
These can be used as quick emission-line and stellar-population
diagnostics, as demonstrated in Figures \ref{fig:ew_d4000} and
\ref{fig:pops}. Using the \halpha\ flux measured within each IFU, we
also provide a very rough estimate of the star-formation rate: We
calculate the absolute luminosity in \halpha, using the luminosity
distance --- see the plot of $L_{{\rm H}\alpha}$ against stellar mass
in Figure \ref{fig:sfr} --- and adopt $\log {\rm SFR} = \log L_{{\rm
H}\alpha} - 41.27$ \citep[Kroupa IMF; from the literature compilation
provided by][]{KennicuttEvansARAA}.

\section{Conclusions}
\label{sec:summary}

We have presented a description of the MaNGA Data Analysis Pipeline
(\dap), its output data products, and its performance. We recommend
readers who intend to use the provided data closely read our
``quick-start guide'' in Section \ref{sec:guidance} and the
description of the output products in Section \ref{sec:output} for
particularly useful summary-level information and usage notes. In
particular, Section \ref{sec:guidance} also serves as a guide to
sections throughout this paper with detailed information regard each
data product.

In its automated measurements of stellar kinematics, nebular
emission-line properties, and spectral indices, the \dap\ is highly
successful. In Section \ref{sec:stellarkin}, we demonstrate via
repeat observation and simulation that the \dap\ provides accurate
stellar kinematics to \snrg$\sim$10 to a minimum velocity dispersion
of $\sim$50 \kms. The formally calculated errors in the stellar
kinematics are very consistent with the direct estimates from repeat
observations. In Belfiore et al., {\it accepted}, we show similar
performance for our emission-line fitting module presented in Section
\ref{sec:emlfit}. These two full-spectrum-fitting modules are shown
to robustly fit the MaNGA spectra across its hugely varied dataset,
owing much to the accuracy and fidelity of our data reduction and
flux calibration techniques \citep{2016AJ....151....8Y,
2016AJ....152...83L}. The few exceptions to this, as enumerated in
Section \ref{sec:outliers}, are an area of ongoing improvements being
made to the \dap.

A current drawback of the \dap\ products provided in DR15 is the
limited wavelength range over which measurements are made (0.36-0.74
$\mu m$) owing to the wavelength coverage of the MILES stellar
templates used in our full-spectrum fits. An exciting near-term
development goal for the \dap\ will be the adoption of stellar
templates from the \mastar\ Stellar Library
\citep{2018arXiv181202745Y}, which samples a larger number of stars
across a wider range of stellar parameters and over the full MaNGA
wavelength range. While the \mastar\ library is not appropriate for
stellar kinematics near the MaNGA instrumental resolution (since
\mastar\ stars are \emph{observed} with MaNGA itself), it will enable
measurements of stellar- and gas-phase spectral features as red as
1.0 $\mu m$ and provide the basis for new
stellar-population-synthesis models (Maraston et al.\ {\it in prep}).

The \dap\ design philosophy has included a focus on measurements that
can be made directly from individual MaNGA spectra. These
measurements are generic to galaxy spectra at similar wavelengths and
spectral resolution. Indeed, while the goals of an automated pipeline
require fine-tuning to the MaNGA data set and data format, the \dap\
sub-routines have been written with generality in mind so that they
could be adapted for other data sets.

Future versions or extensions to the \dap\ may incorporate estimates
of higher-level ``model-derived'' quantities. Natural extensions
include continuum fitting in order to derive stellar population
properties (e.g., stellar age and metallicity) and multi-line
analysis of gas-phase emission lines for estimates of ionization and
gas-phase metallicity.

Future extensions in the context of MaNGA or other IFU data might
also move beyond the independent treatment of spectra from each
spatial bin and fit models that attempt to capture the spatial
information in each galaxy's datacube. These could include dynamical
models of \dap-derived kinematic maps, as well as forward models of
spatially-dependent stellar or gas-phase galaxy components. The \dap\
currently inherits structural information (e.g., the galaxy's
effective radius) from extant photometric catalogs. Spatially
dependent modeling might instead be iterative, making use of the
structural information present in the \dap\ output maps themselves.

\acknowledgements

We sincerely appreciate the effort of our referee in providing a
thoughtful, constructive, and thorough report that was of great
benefit to our paper. MC acknowledges support from a Royal Society
University Research Fellowship. MAB acknowledges NSF Award
AST-1517006. CAT acknowledges NSF Award AST-1554877. ZZ is supported
by the National Natural Science Foundation of China No.\ 11703036. MY
gratefully acknowledges the financial support from China Scholarship
Council (CSC). This research made use of Astropy, a
community-developed core {\tt python} package for Astronomy
\citep{2013A&A...558A..33A}; {\tt numpy} \citep{Numpy2007}; {\tt
scipy} \citep{Scipy2001}; and {\tt matplotlib}
\citep{matplotlib2007}.

Funding for the Sloan Digital Sky Survey IV has been provided by the
Alfred P. Sloan Foundation, the U.S. Department of Energy Office of
Science, and the Participating Institutions. SDSS-IV acknowledges
support and resources from the Center for High-Performance Computing at
the University of Utah. The SDSS web site is www.sdss.org.

SDSS-IV is managed by the Astrophysical Research Consortium for the
Participating Institutions of the SDSS Collaboration including the
Brazilian Participation Group, the Carnegie Institution for Science,
Carnegie Mellon University, the Chilean Participation Group, the French
Participation Group, Harvard-Smithsonian Center for Astrophysics,
Instituto de Astrof\'isica de Canarias, The Johns Hopkins University,
Kavli Institute for the Physics and Mathematics of the Universe (IPMU) /
University of Tokyo, Lawrence Berkeley National Laboratory, Leibniz
Institut f\"ur Astrophysik Potsdam (AIP),  Max-Planck-Institut f\"ur
Astronomie (MPIA Heidelberg), Max-Planck-Institut f\"ur Astrophysik (MPA
Garching), Max-Planck-Institut f\"ur Extraterrestrische Physik (MPE),
National Astronomical Observatories of China, New Mexico State
University, New York University, University of Notre Dame,
Observat\'ario Nacional / MCTI, The Ohio State University, Pennsylvania
State University, Shanghai Astronomical Observatory, United Kingdom
Participation Group, Universidad Nacional Aut\'onoma de M\'exico,
University of Arizona, University of Colorado Boulder, University of
Oxford, University of Portsmouth, University of Utah, University of
Virginia, University of Washington, University of Wisconsin, Vanderbilt
University, and Yale University.

\appendix

\section{Spectral-Resolution Matching}
\label{sec:resolution}

To match the resolution of the template library to that of the
\drp-produced spectra, we convolve the discretely sampled flux density,
$f(\lambda)$ in units of \mangafd, with a Gaussian kernel, $g(\lambda)$,
where the standard deviation of the kernel is a function of the
wavelength, $\sigma_\lambda(\lambda)$.  In general, the convolution is
defined as:
\begin{eqnarray}
(f\ast g)(\lambda) & = & \int_{-\infty}^{\infty} f(\Lambda)\
g(\lambda-\Lambda,\sigma_g)\ d\Lambda \nonumber \\
& = & \int_{-\infty}^{\infty} \frac{f(\Lambda)}{\sqrt{2\pi}\
\sigma_g(\Lambda)}\ \exp\left(-\frac{(\lambda-\Lambda)^2}{2\
\sigma_g(\Lambda)^2}\right) d\Lambda .
\label{eq:contconv}
\end{eqnarray}
It is important to note that the integral of the kernel is normalized to
unity.  In practice, application of equation \ref{eq:contconv} should
account for the discrete sampling and censoring of the data over the
observed spectral range.  We do so by normalizing the convolution by the
integral over the kernel, which is significantly different from unity
only near the edges of the observed spectral range.  We additionally
simplify equation \ref{eq:contconv} by performing the convolution in
pixel space, converting $\sigma_\lambda$ to $\sigma_p$.  Therefore, the
convolved spectrum at wavelength $\lambda_i$ becomes the kernel-weighted
mean of the spectrum over all pixels, with the kernel centered at
$\Lambda_j$:
\begin{equation}
(f\ast g)(\lambda_i) = \frac{\sum_j f(\Lambda_j)
g(\lambda_i-\Lambda_j,\sigma_j)}{\sum_j g(\lambda_i-\Lambda_j,\sigma_j)}
.
\label{eq:disconv}
\end{equation}

Adopting a Gaussian line-spread-function (LSF) for both the spectral
templates and the MaNGA data, we determine the kernel parameters,
$\sigma_p(\lambda)$, as follows.  We define the spectral resolution as
$R = \lambda/\Delta\lambda$, where $\Delta\lambda$ is the full-width at
half maximum (FWHM) of the spectral resolution element.  The standard
deviation of the resolution element in angstroms is then $\sigma_\lambda
= \frac{\lambda}{n_\sigma R}$, where $n_\sigma = {\rm \Delta
\lambda}/\sigma_\lambda \sim 2.35$ for a Gaussian LSF.

For two spectra with spectral resolutions $R_1 \geq R_2$, the defining
parameters of the Gaussian LSFs can be related by
\begin{equation}
\sigma^2_{\lambda,2} = \sigma^2_{\lambda,1} + \sigma^2_{\lambda,d},
\label{eq:sigres}
\end{equation}
where we define
\begin{equation}
\sigma^2_{\lambda,d} \equiv \left(\frac{\lambda}{f}\right)^2 (R^{-2}_2 -
R^{-2}_1).
\label{eq:sigdiffl}
\end{equation}

For the application of equation \ref{eq:disconv}, we convert the units
of $\sigma^2_{\lambda,d}$ from wavelength to pixels.  Performing the
convolution in pixel units has the added advantage that it allows for
similar application of equation \ref{eq:disconv} to spectra that are
either sampled linearly or geometrically --- sampled in linear steps of
$\log_b\lambda$ --- in wavelength.  If linearly sampled,
\begin{equation}
\sigma^2_{p,d} \equiv \left(\frac{\lambda}{f\ \delta\lambda}\right)^2
(R^{-2}_2 - R^{-2}_1).
\label{eq:sigdiffplin}
\end{equation}
where $\delta\lambda$ is the pixel scale in angstroms.  If geometrically
sampled, the pixel size is converted to velocity,
\begin{equation}
\delta v = c\ \ln b\ (\delta\log_b\lambda),
\label{eq:dv}
\end{equation}
such that
\begin{eqnarray}
\sigma^2_{p,d} & = & \sigma^2_{v,d} (\delta v)^{-2} \\ \label{eq:sigv2p}
& = & \left(\frac{c}{\lambda}\right)^2 \sigma^2_{\lambda,d} (\delta
v)^{-2} \nonumber \\ & = & \left(\frac{c}{ f\ \delta v}\right)^2
(R^{-2}_2 - R^{-2}_1),
\label{eq:sigdiffplog}
\end{eqnarray}
where $c$ is the speed of light in \kms.

As stated above, equation \ref{eq:sigres} assumes $R_1 \geq R_2$.
However, in practice, some spectral libraries may not have resolutions
that are larger than the MaNGA data over the full spectral range.  In
our resolution matching algorithm, we define a minimum value of
$\sigma_{p,d}$, $\epsilon_\sigma$, below which we approximate the
Gaussian kernel as a Kronecker delta function.  Therefore, as long as
\begin{equation}
\sigma_{p,d} \equiv \sigma^2_{p,d}/\sqrt{|\sigma^2_{p,d}|} \geq
-\epsilon_\sigma\ ,
\label{eq:kron}
\end{equation}
the behavior of the convolution should not be affected.

However, we typically set $\epsilon_\sigma=0$ and need to robustly
handle regions where $R_1 < R_2$.  For such spectral regions, we
highlight three approaches:
\begin{enumerate}
\item trim the spectral range to only those spectral regions where the
existing resolution is better than the target resolution,
\item match the existing resolution to the target resolution up to some
constant offset that must be accounted for in subsequent analyses, or
\item allow for a wavelength dependent difference in the spectral
resolution that must be accounted for in subsequent analyses.
\end{enumerate}
Our code allows for selection of the first or second approach.  Our
standard practice is currently to adopt the first approach; our code
does not allow for the third option.

In the first approach, pixels with $\sigma_{p,d} < -\epsilon_\sigma$ are
masked from subsequent analysis, and the convolution algorithm does not
alter the spectral resolution of these pixels.

In the second approach, we define
\begin{equation}
\sigma^2_{v,o} = -{\rm min}(\sigma^2_{v,d}) - {\rm max}(\epsilon_\sigma
\delta v)^2
\label{eq:sigoff}
\end{equation}
where $\delta v$ is constant for the geometrically binned spectrum and
is wavelength dependent, $\delta v = c\ \delta\lambda/\lambda$,  for the
linearly binned spectra.  If $\sigma^2_{v,o} > 0.0$, it must be that
${\rm min}(\sigma^2_{v,d}) < -{\rm max}(\epsilon_\sigma \delta v)^2$
such that an offset should be applied.  In that case, the returned
kernel parameters are
\begin{equation}
\sigma^\prime_{v,d} = \sqrt{\sigma^2_{v,d} + \sigma^2_{v,o}}
\label{eq:sigprime}
\end{equation}
with the units converted to pixels using equation \ref{eq:sigv2p}.  In
this approach, not pixels are masked and $\sqrt{\sigma^2_{v,o}}$ is
returned for use in subsequent analysis.  Otherwise (i.e., equation
\ref{eq:sigoff} yields $\sigma^2_{v,o} \leq 0.0$), the returned offset
is set to zero.

It should be noted that the offset, $\sigma_{v,o}$, is always kept in
units of \kms, regardless of the spectral sampling.  This facilitates
later adjustment of the offset by a constant Gaussian velocity
dispersion.  This is useful for imposing a single offset for spectral
templates at different resolutions (impose the maximum $\sigma^2_{v,o}$
on all spectral templates), or to apply a constant velocity dispersion
offset for kinematic reasons.

\section{Propagation of spectral-resolution errors in the
         $\sigma_\ast$ error budget}
\label{sec:sigmaerr}

Assuming all instrumental line-spread functions (LSFs) and the
stellar line-of-sight velocity distribution (LOSVD) are Gaussian, we
can write the \ppxf-measured stellar velocity dispersion as (cf.\
Equation \ref{eq:sigmaobs}):
\begin{equation}
\sigo^2 = \sigs^2 + \sigg^2 - \sigt^2 \\
\label{eq:sigo}
\end{equation}
where $\dsigi = \sigg^2 - \sigt^2$ is the quadrature difference in
the instrumental resolution of the galaxy spectra, $\sigg$, and the
template spectra, $\sigt$. Here, we explore how the errors in $\sigg$
and $\sigt$ propagate to the error in $\sigo$ at fixed $\sigs$, with
the aim of minimizing the influence of instrumental-dispersion
uncertainties.

Assuming all errors are Gaussian and one can accurately perform
nominal error propagation, we can write
\begin{equation}
\left(\frac{\epsilon[\dsigi^2]}{\dsigi^2}\right)^2 =
\left(\frac{\epsilon[\sigg^2]}{\sigg^2 - \sigt^2}\right)^2 +
\left(\frac{\epsilon[\sigt^2]}{\sigg^2 - \sigt^2}\right)^2,
\label{eq:sigierr0}
\end{equation}
where $\epsilon[x]$ is the formal error in $x$. Adopting $\xi =
\sigg/\sigt$ and
\begin{equation}
\frac{\epsilon[\sigi]}{\sigi} \equiv \frac{\epsilon[\sigg]}{\sigg} \sim
\frac{\epsilon[\sigt]}{\sigt},
\end{equation} 
we can simplify Equation \ref{eq:sigierr0} to
\begin{equation}
\frac{\epsilon[\dsigi]}{\dsigi} = \frac{\epsilon[\sigi]}{\sigi}\
 \frac{\sqrt{\xi^4 + 1}}{\xi^2 - 1},
\label{eq:sigierr}
\end{equation}
where $\xi > 1$. From Equation \ref{eq:sigmaobs}, we can derive the
error in $\sigo$ at fixed $\sigs$,
\begin{equation}
\frac{\epsilon[\sigo]}{\sigo} = \frac{\dsigi^2}{\sigo^2}\
\frac{\epsilon[\dsigi]}{\dsigi}.
\label{eq:sigoerr0}
\end{equation}
Finally, substituting Equation \ref{eq:sigierr} and rewriting in
terms of $\sigs$, $\sigg$, and $\xi$, we find:
\begin{equation}
\frac{\epsilon[\sigo]}{\sigo} = \frac{\epsilon[\sigi]}{\sigi}\ 
 \frac{\sqrt{\xi^4 + 1}}{\xi^2(\sigs^2/\sigg^2 + 1) - 1}.
\label{eq:sigoerr}
\end{equation}

\begin{figure}
\begin{center}
\includegraphics[width=0.5\columnwidth]{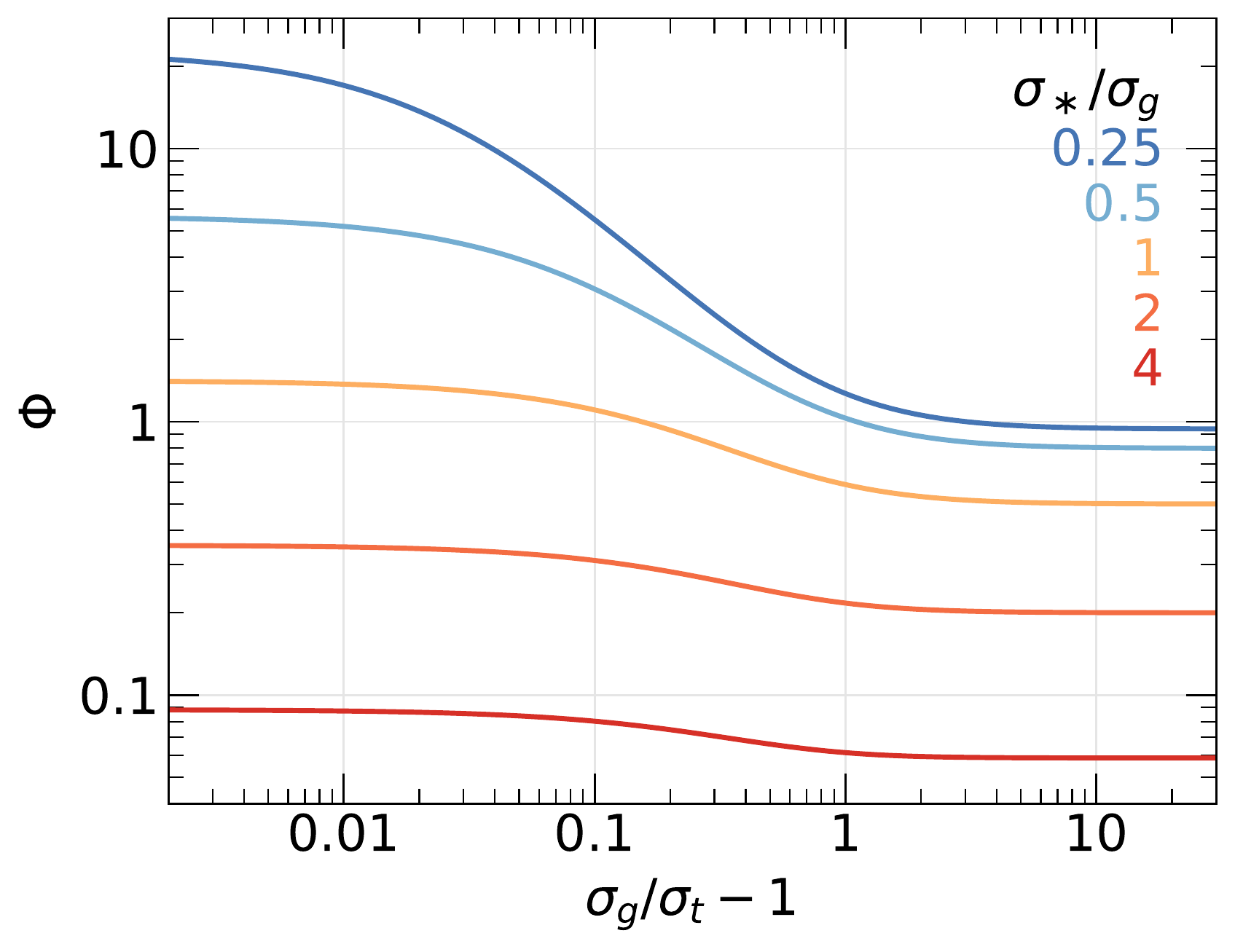}
\end{center}
\caption{Ratio of fractional error in $\sigo$ to the fractional error
in $\sigi$ (i.e., $\Phi$; Equation \ref{eq:phi}) at fixed $\sigs$ as
a function of the ratio between the instrumental resolution of the
galaxy spectra $\sigg$ and the template spectra $\sigt$. Note that
one {\it always} improves the fractional error in $\sigo$ by using
higher resolution templates to measure the stellar kinematics.}
\label{fig:reserrprop}
\end{figure}

Defining
\begin{equation}
\Phi \equiv \frac{\sigi}{\sigo}\
\frac{\epsilon[\sigo]}{\epsilon[\sigi]} = \frac{\sqrt{\xi^4 +
1}}{\xi^2(\sigs^2/\sigg^2 + 1) - 1},
\label{eq:phi}
\end{equation}
we show $\Phi$ as a function of $\xi\equiv\sigg/\sigt$ for five
discrete ratios of the target stellar velocity dispersion to the
instrumental resolution of the galaxy observations in Figure
\ref{fig:reserrprop}. In accordance with intuition, Figure
\ref{fig:reserrprop} shows that one can limit the influence of
$\sigi$ errors on $\sigo$ by performing observations at higher
spectral resolution; i.e., $\Phi$ decreases as $\sigs/\sigg$
increases. Interestingly, Figure \ref{fig:reserrprop} also shows that
one can {\it always} reduce the influence of instrumental resolution
errors by fitting the galaxy spectra with intrinsically higher
resolution templates. Moreover, it shows that the relevant reduction
of $\Phi$ increases as $\sigs/\sigg$ decreases, i.e., as one attempts
to measure $\sigs$ toward and below the instrumental resolution of
the galaxy data.

The limiting values of $\Phi$ when $\xi = 1$ and as $\xi \to \infty$
are, respectively,
\begin{equation}
\Phi_{\xi=1} = \frac{\sqrt{2}}{\sigs^2/\sigg^2}
\end{equation}
and
\begin{equation}
\Phi_{\xi\to\infty} = \frac{1}{\sigs^2/\sigg^2 + 1},
\end{equation}
such that
\begin{equation}
\frac{\Phi_{\xi=1}}{\Phi_{\xi\to\infty}} = \sqrt{2} +
\frac{1}{\sigs^2/\sigg^2}.
\end{equation}
Therefore, improvements in $\Phi$ can be more than a factor of 2 when
$\sigs \sim \sigg$ simply by using higher resolution templates.

Finally, we note from Figure \ref{fig:reserrprop} that $\xi$ need not
be too large to make significant gains. Even with the modest 16\%
difference in resolution between MaNGA and MILES, Equation
\ref{eq:sigoerr} yields $\Phi_{\xi = 1}/\Phi = 1.43$ when $\sigs =
\sigg$. In fact, the asymptotic behavior of $\Phi$ is such that one
expects diminishing returns for $\xi\equiv\sigg/\sigt$ larger than a
factor of $\sim$2--3.

\section{MaNGA \dap\ Data Model}
\label{sec:datamodel}

This appendix provides the \dap\ data model via a series of tables:
\begin{itemize}
\item The two primary output of the \dap\ (Section \ref{sec:output}) are
the \dapmaps\ and model \dapcube\ files.  The names and content of each
extension in these files are provided in Tables \ref{tab:mapsfile} and
\ref{tab:logcubefile}, respectively.
\item Both files include bitmask extensions.\footnote{
For an introduction to SDSS bitmasks and further documentation of the
\dap\ mask bits, see
\url{https://www.sdss.org/dr15/algorithms/bitmasks/}.}
The relevant bitmask types for the \dapmaps\ and model \dapcube\ files
are {\tt MANGA\symbol{95}DAPPIXMASK} and {\tt
MANGA\symbol{95}DAPSPECMASK}, respectively.  The bit values, names, and
descriptions are provided in Tables \ref{tab:dappixmask} and
\ref{tab:dapspecmask}.  The global quality assessment of a \dap\ file is
provided by the {\tt MANGA\symbol{95}DAPQUAL} bitmask type (given as the
{\tt DAPQUAL} header keyword in the primary extension of each file) with
bit values, names, and descriptions provided in Table \ref{tab:dapqual}.
\item Finally, the \dapall\ file (Section \ref{sec:dapall}) is a summary
catalog of the \dap\ parameters and global quantities based on the
output data.  The tabulated data it provides is listed in Table
\ref{tab:dapall}.
\end{itemize}

\startlongtable
\begin{deluxetable*}{ r l c r p{3in} }
\tabletypesize{\footnotesize}
\tablewidth{0pt}
\tablecaption{\dap\ \dapmaps\ File Extensions \label{tab:mapsfile} }
\tablehead{\colhead{Index} & \colhead{Name} & \colhead{Channels} & \colhead{Units} & \colhead{Description} }
\startdata
  0 & {\tt PRIMARY} & 0 & \nodata & Empty extension with primary header information. \\
\hline
\multicolumn{5}{c}{Coordinates and Binning} \\
\hline
  1 & {\tt SPX\symbol{95}SKYCOO}                    &  2 &                             arcsec & Sky-right offsets -- +x toward +RA and +y toward +DEC -- of each spaxel from the galaxy center \\
  2 & {\tt SPX\symbol{95}ELLCOO}                    &  3 &              arcsec, unitless, deg & Elliptical polar coordinates of each spaxel from the galaxy center: $R$, $R/R_e$, $\theta$.  In the limit of tilted thin disk, these are the in-plane disk radius and azimuth; the second channel is the radius normalized by the elliptical Petrosian effective radius from the NSA. \\
  3 & {\tt SPX\symbol{95}MFLUX}                     &  1 & $10^{-17}$ erg/s/cm$^2$/\AA/spaxel & $g$-band-weighted mean flux, {\it not} corrected for Galactic extinction or internal attenuation \\
  4 & {\tt SPX\symbol{95}MFLUX\symbol{95}IVAR}      &  1 &                                    & Inverse variance of $g$-band-weighted mean flux \\
  5 & {\tt SPX\symbol{95}SNR}                       &  1 &                                    & Mean $g$-band weighted signal-to-noise ratio per pixel \\
  6 & {\tt BINID}                                   &  5 &                                    & Numerical ID for spatial bins for the binned spectra, stellar-continuum results, emission-line moment results, emission-line model results, and spectral-index results. \\
  7 & {\tt BIN\symbol{95}LWSKYCOO}                  &  2 &                             arcsec & Light-weighted sky-right offsets -- +x toward +RA and +y toward +DEC -- of each bin from the galaxy center. \\
  8 & {\tt BIN\symbol{95}LWELLCOO}                  &  3 &                arcsec,unitless,deg & Light-weighted elliptical polar coordinates of each bin from the galaxy center: $R$, $R/R_e$, $\theta$.  In the limit of tilted thin disk, these are the in-plane disk radius and azimuth; the second channel is the radius normalized by the elliptical Petrosian effective radius from the NSA. \\
  9 & {\tt BIN\symbol{95}AREA}                      &  1 &                         arcsec$^2$ & Area of each bin \\
 10 & {\tt BIN\symbol{95}FAREA}                     &  1 &                                    & Fractional area that the bin covers for the expected bin shape (only relevant for radial binning) \\
 11 & {\tt BIN\symbol{95}MFLUX}                     &  1 & $10^{-17}$ erg/s/cm$^2$/\AA/spaxel & $g$-band-weighted mean flux for the binned spectra, {\it not} corrected for Galactic extinction or internal attenuation \\
 12 & {\tt BIN\symbol{95}MFLUX\symbol{95}IVAR}      &  1 &                                    & Inverse variance of $g$-band-weighted mean flux for the binned spectra \\
 13 & {\tt BIN\symbol{95}MFLUX\symbol{95}MASK}      &  1 &                                    & Bit mask for the $g$-band-weighted mean flux per bin \\
 14 & {\tt BIN\symbol{95}SNR}                       &  1 &                                    & Mean $g$-band-weighted signal-to-noise ratio per pixel in the binned spectra \\
\hline
\multicolumn{5}{c}{Stellar Kinematics} \\
\hline
 15 & {\tt STELLAR\symbol{95}VEL}                   &  1 &                               \kms & Line-of-sight stellar velocity, relative to the input guess redshift (given as cz in SCINPVEL PRIMARY header keyword and most often identical to the NSA redshift) \\
 16 & {\tt STELLAR\symbol{95}VEL\symbol{95}IVAR}    &  1 &                                    & Inverse variance of stellar velocity measurements \\
 17 & {\tt STELLAR\symbol{95}VEL\symbol{95}MASK}    &  1 &                                    & Data quality mask for stellar velocity measurements \\
 18 & {\tt STELLAR\symbol{95}SIGMA}                 &  1 &                               \kms & Raw line-of-sight stellar velocity dispersion (must be corrected using {\tt STELLAR\symbol{95}SIGMACORR} to obtain the astrophysical dispersion) \\
 19 & {\tt STELLAR\symbol{95}SIGMA\symbol{95}IVAR}  &  1 &                                    & Inverse variance of stellar velocity dispersion \\
 20 & {\tt STELLAR\symbol{95}SIGMA\symbol{95}MASK}  &  1 &                                    & Data quality mask for stellar velocity dispersion \\
 21 & {\tt STELLAR\symbol{95}SIGMACORR}             &  1 &                               \kms & Quadrature correction for {\tt STELLAR\symbol{95}SIGMA} to obtain the astrophysical velocity dispersion. \\
 22 & {\tt STELLAR\symbol{95}CONT\symbol{95}FRESID} &  2 &                                    & 68\% and 99\% growth of the fractional residuals between the model and data \\
 23 & {\tt STELLAR\symbol{95}CONT\symbol{95}RCHI2}  &  1 &                                    & Reduced chi-square of the stellar continuum fit \\
\hline
\multicolumn{5}{c}{Emission-Line Properties} \\
\hline
 24 & {\tt EMLINE\symbol{95}SFLUX}                  & 22 &     $10^{-17}$ erg/s/cm$^2$/spaxel & Non-parametric summed flux {\it after subtracting the stellar-continuum model}.  The emission-line fluxes account for Galactic reddening using the $E(B-V)$ value (copied to the \dap\ primary headers, see {\tt EBVGAL}) provided by the \drp\ header and assuming the reddening law provided by \citet{1994ApJ...422..158O}; however, no attenuation correction is applied due to dust internal to the galaxy. \\
 25 & {\tt EMLINE\symbol{95}SFLUX\symbol{95}IVAR}   & 22 &                                    & Inverse variance for summed flux measurements \\
 26 & {\tt EMLINE\symbol{95}SFLUX\symbol{95}MASK}   & 22 &                                    & Data quality mask for summed flux measurements \\
 27 & {\tt EMLINE\symbol{95}SEW}                    & 22 &                                \AA & Non-parametric equivalent widths measurements (based one {\tt EMLINE\symbol{95}SFLUX}) \\
 28 & {\tt EMLINE\symbol{95}SEW\symbol{95}IVAR}     & 22 &                                    & Inverse variance for non-parametric equivalent width measurements \\
 29 & {\tt EMLINE\symbol{95}SEW\symbol{95}MASK}     & 22 &                                    & Data quality mask for non-parametric equivalent width measurements \\
 30 & {\tt EMLINE\symbol{95}GFLUX}                  & 22 &     $10^{-17}$ erg/s/cm$^2$/spaxel & Gaussian profile integrated flux {\it from a combined continuum+emission-line fit}.  {\it The flux ratio of the [OIII], [OI], and [NII] lines are fixed and cannot be treated as independent measurements.}  The emission-line fluxes account for Galactic reddening using the E(B-V) (copied to the \dap\ primary headers, see {\tt EBVGAL}) value provided by the \drp\ header and assuming the reddening law provided by \citet{1994ApJ...422..158O}; however, no attenuation correction is applied due to dust internal to the galaxy. \\
 31 & {\tt EMLINE\symbol{95}GFLUX\symbol{95}IVAR}   & 22 &                                    & Inverse variance for Gaussian flux measurements \\
 32 & {\tt EMLINE\symbol{95}GFLUX\symbol{95}MASK}   & 22 &                                    & Data quality mask for Gaussian flux measurements \\
 33 & {\tt EMLINE\symbol{95}GEW}                    & 22 &                                \AA & Gaussian-fitted equivalent widths measurements (based on {\tt EMLINE\symbol{95}GFLUX}) \\
 34 & {\tt EMLINE\symbol{95}GEW\symbol{95}IVAR}     & 22 &                                    & Inverse variance of the above \\
 35 & {\tt EMLINE\symbol{95}GEW\symbol{95}MASK}     & 22 &                                    & Data quality mask of the above \\
 36 & {\tt EMLINE\symbol{95}GVEL}                   & 22 &                               \kms & Line-of-sight emission-line velocity, relative to the input guess redshift (given as $cz$ in the {\tt PRIMARY} header keyword {\tt SCINPVEL} and most often identical to the NSA redshift).  A velocity is provided for each line, {\it but the velocities are identical for all lines} because the parameters are tied during the fitting process. \\
 37 & {\tt EMLINE\symbol{95}GVEL\symbol{95}IVAR}    & 22 &                                    & Inverse variance for Gaussian-fitted velocity measurements, which are {\it the same for all lines and should not be combined as if independent measurements}. \\
 38 & {\tt EMLINE\symbol{95}GVEL\symbol{95}MASK}    & 22 &                                    & Data quality mask for Gaussian-fitted velocity measurements \\
 39 & {\tt EMLINE\symbol{95}GSIGMA}                 & 22 &                               \kms & Gaussian profile velocity dispersion as would be measured from a direct Gaussian fit (must be corrected using {\tt EMLINE\symbol{95}INSTSIGMA} to obtain the astrophysical dispersion).  {\it The velocity dispersions of the [OII], [OIII], [OI], and [NII] lines are tied and cannot be treated as independent measurements} \\
 40 & {\tt EMLINE\symbol{95}GSIGMA\symbol{95}IVAR}  & 22 &                                    & Inverse variance for Gaussian profile velocity dispersion \\
 41 & {\tt EMLINE\symbol{95}GSIGMA\symbol{95}MASK}  & 22 &                                    & Data quality mask for Gaussian profile velocity dispersion \\
 42 & {\tt EMLINE\symbol{95}INSTSIGMA}              & 22 &                               \kms & The instrumental dispersion at the fitted center of each emission line  \\
 43 & {\tt EMLINE\symbol{95}TPLSIGMA}               & 22 &                               \kms & The dispersion of each emission line used in the template spectra \\
\hline
\multicolumn{5}{c}{Spectral Indices} \\
\hline
 44 & {\tt SPECINDEX}                               & 46 &                            \AA,mag & Spectral-index measurements \\
 45 & {\tt SPECINDEX\symbol{95}IVAR}                & 46 &                                    & Inverse variance for spectral index maps \\
 46 & {\tt SPECINDEX\symbol{95}MASK}                & 46 &                                    & Data quality mask for spectral index maps \\
 47 & {\tt SPECINDEX\symbol{95}CORR}                & 46 &                       unitless,mag & Corrections to apply to account for the velocity dispersion and effectively determine the index without Doppler broadening \\
\enddata
\end{deluxetable*}

\begin{deluxetable*}{ r l r p{3in} }
\tabletypesize{\footnotesize}
\tablewidth{0pt}
\tablecaption{\dap\ Model \dapcube\ File Extensions \label{tab:logcubefile} }
\tablehead{\colhead{Index} & \colhead{Name} & \colhead{Units} & \colhead{Description} }
\startdata
  0 & {\tt PRIMARY}               &                            \nodata & Empty extension with primary header information. \\
  1 & {\tt FLUX}                  & 10$^{-17}$ erg/s/cm$^2$/\AA/spaxel & Flux of the {\it binned} spectra \\
  2 & {\tt IVAR}                  &                            \nodata & Inverse variance in the binned spectra \\
  3 & {\tt MASK}                  &                            \nodata & Bitmask for the binned and model spectra \\
  4 & {\tt WAVE}                  &                                \AA & Vacuum-wavelength vector \\
  5 & {\tt REDCORR}               &                            \nodata & Reddening correction applied during the fitting procedures; calculate the dereddened flux as {\tt FLUX * REDCORR} \\
  6 & {\tt MODEL}                 & 10$^{-17}$ erg/s/cm$^2$/\AA/spaxel &  The best fitting model spectra (sum of the fitted continuum and emission-line models) \\
  7 & {\tt EMLINE}                & 10$^{-17}$ erg/s/cm$^2$/\AA/spaxel &  The model spectrum with {\it only} the emission lines \\
  8 & {\tt EMLINE\symbol{95}BASE} & 10$^{-17}$ erg/s/cm$^2$/\AA/spaxel &  The adjustment to the stellar continuum made during the combined continuum$+$emission-line fit. \\
  9 & {\tt EMLINE\symbol{95}MASK} &                            \nodata &  The bitmask that only applies to the emission-line modeling. \\
 10 & {\tt BINID}                 &                            \nodata &  Numerical ID for spatial bins in 5 channels: (1) binned spectra, (2) stellar-continuum results, (3) empty, (4) emission-line model results, and (5) empty; i.e., channels 1, 2, and 4 are the same as the BINID extension in the \dapmaps\ files and channels 3 and 5 are empty
\enddata
\end{deluxetable*}

\begin{deluxetable*}{ r l p{4.0in} }
\tabletypesize{\small}
\tablewidth{0pt}
\tablecaption{\dap\ Mapped Quantity Mask Bits ({\tt MANGA\symbol{95}DAPPIXMASK}) \label{tab:dappixmask} }
\tablehead{ \colhead{Bit $(\log_2)$} & \colhead{Name} &
            \colhead{Description} }
\startdata
  0 & {\tt NOCOV} & No coverage in this spaxel \\[2pt]
  1 & {\tt LOWCOV} & Low coverage in this spaxel \\[2pt]
  2 & {\tt DEADFIBER} & Major contributing fiber is dead \\[2pt]
  3 & {\tt FORESTAR} & A foreground star influences the flux in this
  spaxel \\[2pt]
  4 & {\tt NOVALUE} & Spaxel ignored by the \dap\ \\[2pt]
  5 & {\tt UNRELIABLE} & Value is deemed unreliable \\[2pt]
  6 & {\tt MATHERROR} & A mathematical error occurred when computing the
  value \\[2pt]
  7 & {\tt FITFAILED} & Fit to this spaxel failed \\[2pt]
  8 & {\tt NEARBOUND} & Fitted value is too near an imposed boundary \\[2pt]
  9 & {\tt NOCORRECTION} & Appropriate correction is not available \\[2pt]
 10 & {\tt MULTICOMP} & A multi-component velocity feature has been
 detected \\[2pt]
 30 & {\tt DONOTUSE} & Do not use this spaxel for science\tablenotemark{a}
\enddata
\tablenotetext{a}{This bit is a consolidation of all spaxels flagged as
{\tt NOCOV}, {\tt LOWCOV}, {\tt DEADFIBER}, {\tt FORESTAR}, {\tt
NOVALUE}, {\tt MATHERROR}, {\tt FITFAILED}, or {\tt NEARBOUND}.}
\end{deluxetable*}

\begin{deluxetable*}{ r l p{4.0in} }
\tabletypesize{\small}
\tablewidth{0pt}
\tablecaption{\dap\ Model \dapcube\ Mask Bits ({\tt
MANGA\symbol{95}DAPSPECMASK}) \label{tab:dapspecmask} }
\tablehead{ \colhead{Bit $(\log_2)$} & \colhead{Name} &
            \colhead{Description} }
\startdata
 0 & {\tt IGNORED} & Pixel ignored \\[2pt]
 1 & {\tt FORESTAR} & A foreground star influences the flux in this
 spaxel \\[2pt]
 2 & {\tt FLUXINVALID} & Invalid flux measurements in pixel \\[2pt]
 3 & {\tt IVARINVALID} & Invalid inverse variance in pixel ($\leq 0$)
 \\[2pt]
 4 & {\tt ARTIFACT} & Flux measurements affected by a designated
 artifact \\[2pt]
 5 & {\tt FITIGNORED} & Pixel not included in the relevant fit \\[2pt]
 6 & {\tt FITFAILED} & Fit to spectral region failed \\[2pt]
 7 & {\tt ELIGNORED} & Pixel ignored during emission-line fit \\[2pt]
 8 & {\tt ELFAILED} & Fit to emission-line in this spectral region
 failed
\enddata
\end{deluxetable*}

\begin{deluxetable*}{ r l p{4.0in} }
\tabletypesize{\small}
\tablewidth{0pt}
\tablecaption{\dap\ Quality Mask Bits ({\tt MANGA\symbol{95}DAPQUAL}) \label{tab:dapqual} }
\tablehead{\colhead{Bit $(\log_2)$} & \colhead{Name} &
            \colhead{Description} }
\startdata
 0 & {\tt FORESTAR} & A foreground star is present within the data cube
 field-of-view \\[2pt]
 1 & {\tt BADZ} & Mismatch between redshifts derived from MaNGA
 observations and provided by the NASA-Sloan Atlas \\[2pt]
 2 & {\tt LINELESS} & No significant nebular emission detected \\[2pt]
 3 & {\tt PPXFFAIL} & pPXF fails to fit this object \\[2pt]
 4 & {\tt SINGLEBIN} & Voronoi binning algorithm forced all spectra into
 a single bin\\[2pt]
 5 & {\tt BADGEOM} & Invalid input geometry; elliptical coordinates and
 effective radius are meaningless\\[2pt]
28 & {\tt DRPCRIT} & Critical failure in \drp\ \\[2pt]
29 & {\tt DAPCRIT} & Critical failure in \dap\ \\[2pt]
30 & {\tt CRITICAL} & Critical failure in \drp\ or \dap\
\enddata
\end{deluxetable*}

\startlongtable
\begin{deluxetable*}{ l r p{3.5in} }
\tabletypesize{\footnotesize}
\tablewidth{0pt}
\tablecaption{\dapall\ Table Data \label{tab:dapall} }
\tablehead{\colhead{Column} & \colhead{Units} & \colhead{Description} }
\startdata
{\tt PLATE                 } &                              \nodata & Plate number \\
{\tt IFUDESIGN             } &                              \nodata & IFU design number \\
{\tt PLATEIFU              } &                              \nodata & String combination of {\tt PLATE-IFU} to ease searching \\
{\tt MANGAID               } &                              \nodata & MaNGA ID string \\
{\tt DRPALLINDX            } &                              \nodata & Row index of the observation in the \drpall\ file \\
{\tt MODE                  } &                              \nodata & 3D mode of the \drp\ file ({\tt CUBE} or {\tt RSS}) \\
{\tt DAPTYPE               } &                              \nodata & Keyword of the analysis approach used (e.g., {\tt HYB10-GAU-MILESHC}) \\
{\tt DAPDONE               } &                              \nodata & Flag that \dapmaps\ file successfully produced \\
{\tt OBJRA                 } &                                  deg & RA of the galaxy center \\
{\tt OBJDEC                } &                                  deg & Declination of the galaxy center \\
{\tt IFURA                 } &                                  deg & RA of the IFU pointing center (generally the same as {\tt OBJRA}) \\
{\tt IFUDEC                } &                                  deg & Declination of the IFU pointing center (generally the same as {\tt OBJDEC}) \\
{\tt MNGTARG1              } &                              \nodata & Main survey targeting bit \\
{\tt MNGTARG2              } &                              \nodata & Non-galaxy targeting bit \\
{\tt MNGTARG3              } &                              \nodata & Ancillary targeting bit \\
{\tt Z                     } &                              \nodata & Redshift used to set initial guess velocity (typically identical to {\tt NSA\symbol{95}Z}) \\
{\tt LDIST\symbol{95}Z               } &                         h$^{-1}$ Mpc & Luminosity distance, $D_L$, based on {\tt Z} and a standard cosmology\tablenotemark{a} \\
{\tt ADIST\symbol{95}Z               } &                         h$^{-1}$ Mpc & Angular-diameter distance, $D_A$, based on {\tt Z} and a standard cosmology\tablenotemark{a} \\
{\tt NSA\symbol{95}Z                 } &                              \nodata & Redshift from the NASA-Sloan Atlas \\
{\tt NSA\symbol{95}ZDIST             } &                              \nodata & NSA distance estimate using pecular velocity model of \citet{1997ApJS..109..333W}; multiply by $c/H_0$ for Mpc. \\
{\tt LDIST\symbol{95}NSA\symbol{95}Z           } &                         h$^{-1}$ Mpc & Luminosity distance based on {\tt NSA\symbol{95}Z} and a standard cosmology\tablenotemark{a} \\
{\tt ADIST\symbol{95}NSA\symbol{95}Z           } &                         h$^{-1}$ Mpc & Angular-diameter distance based on {\tt NSA\symbol{95}Z} and a standard cosmology\tablenotemark{a} \\
{\tt NSA\symbol{95}ELPETRO\symbol{95}BA        } &                              \nodata & NSA isophotal axial ratio from an elliptical Petrosian analysis of the $r$-band image \\
{\tt NSA\symbol{95}ELPETRO\symbol{95}PHI       } &                                  deg & NSA isophotal position angle from an elliptical Petrosian analysis of the $r$-band image \\
{\tt NSA\symbol{95}ELPETRO\symbol{95}TH50\symbol{95}R    } &                               arcsec & Half-light radius provided by the NSA from an elliptical Petrosian analysis of the $r$-band image; this is the same as $R_e$ below. \\
{\tt NSA\symbol{95}SERSIC\symbol{95}BA         } &                              \nodata & NSA isophotal axial ratio from Sersic fit to the $r$-band image \\
{\tt NSA\symbol{95}SERSIC\symbol{95}PHI        } &                                  deg & NSA isophotal position angle from Sersic fit to the $r$-band image \\
{\tt NSA\symbol{95}SERSIC\symbol{95}TH50       } &                               arcsec & NSA effective radius from the Sersic fit to the $r$-band image \\
{\tt NSA\symbol{95}SERSIC\symbol{95}N          } &                              \nodata & NSA Sersic index from the Sersic fit to the $r$-band image \\
{\tt VERSDRP2              } &                              \nodata & Version of \drp\ used for 2d reductions \\
{\tt VERSDRP3              } &                              \nodata & Version of \drp\ used for 3d reductions \\
{\tt VERSCORE              } &                              \nodata & Version of mangacore used by the \dap\ \\
{\tt VERSUTIL              } &                              \nodata & Version of idlutils used by the \dap\ \\
{\tt VERSDAP               } &                              \nodata & Version of mangadap \\
{\tt DRP3QUAL              } &                              \nodata & \drp\ 3D quality bit \\
{\tt DAPQUAL               } &                              \nodata & \dap\ quality bit \\
{\tt RDXQAKEY              } &                              \nodata & Configuration keyword for the method used to assess the reduced data \\
{\tt BINKEY                } &                              \nodata & Configuration keyword for the spatial binning method \\
{\tt SCKEY                 } &                              \nodata & Configuration keyword for the method used to model the stellar-continuum \\
{\tt ELMKEY                } &                              \nodata & Configuration keyword that defines the emission-line moment measurement method \\
{\tt ELFKEY                } &                              \nodata & Configuration keyword that defines the emission-line modeling method \\
{\tt SIKEY                 } &                              \nodata & Configuration keyword that defines the spectral-index measurement method \\
{\tt BINTYPE               } &                              \nodata & Type of binning used \\
{\tt BINSNR                } &                              \nodata & Target for bin S/N, if Voronoi binning \\
{\tt TPLKEY                } &                              \nodata & The identifier of the template library, e.g., MILES. \\
{\tt DATEDAP               } &                              \nodata & Date the \dap\ file was created and/or last modified. \\
{\tt DAPBINS               } &                              \nodata & The number of "binned" spectra analyzed by the \dap. \\
{\tt RCOV90                } &                               arcsec & Semi-major axis radius (R) below which spaxels cover at least 90\% of elliptical annuli with width $R\pm2\farcs5$.  This should be independent of the {\tt DAPTYPE}. \\
{\tt SNR\symbol{95}MED               } &                              \nodata & Median S/N per pixel in the ''griz'' bands within 1.0-1.5 $R_e$.  This should be independent of the {\tt DAPTYPE}. \\
{\tt SNR\symbol{95}RING              } &                              \nodata & S/N in the ''griz'' bands when binning all spaxels within 1.0-1.5 $R_e$.  This should be independent of the {\tt DAPTYPE}. \\
{\tt SB\symbol{95}1RE                } & 10$^{-17}$ erg/s/cm$^{2}$/\AA/spaxel & Mean $g$-band surface brightness of valid spaxels within 1 $R_e$.  This should be independent of the {\tt DAPTYPE}. \\
{\tt BIN\symbol{95}RMAX              } &                                $R_e$ & Maximum $g$-band luminosity-weighted semi-major radius of any "valid" binned spectrum. \\
{\tt BIN\symbol{95}R\symbol{95}N               } &                              \nodata & Number of binned spectra with $g$-band luminosity-weighted centers within 0-1, 0.5-1.5, and 1.5-2.5 $R_e$. \\
{\tt BIN\symbol{95}R\symbol{95}SNR             } &                              \nodata & Median $g$-band S/N of all binned spectra with luminosity-weighted centers within 0-1, 0.5-1.5, and 1.5-2.5 $R_e$. \\
{\tt STELLAR\symbol{95}Z             } &                              \nodata & Flux-weighted mean redshift of the stellar component within a 2.5 arcsec aperture at the galaxy center. \\
{\tt STELLAR\symbol{95}VEL\symbol{95}LO        } &                                 \kms & Stellar velocity at 2.5\% growth of all valid spaxels. \\
{\tt STELLAR\symbol{95}VEL\symbol{95}HI        } &                                 \kms & Stellar velocity at 97.5\% growth of all valid spaxels. \\
{\tt STELLAR\symbol{95}VEL\symbol{95}LO\symbol{95}CLIP   } &                                 \kms & Stellar velocity at 2.5\% growth after iteratively clipping 3-sigma outliers. \\
{\tt STELLAR\symbol{95}VEL\symbol{95}HI\symbol{95}CLIP   } &                                 \kms & Stellar velocity at 97.5\% growth after iteratively clipping 3-sigma outliers. \\
{\tt STELLAR\symbol{95}SIGMA\symbol{95}1RE     } &                                 \kms & Flux-weighted mean stellar velocity dispersion of all spaxels within 1 $R_e$. \\
{\tt STELLAR\symbol{95}CONT\symbol{95}RCHI2\symbol{95}1RE} &                              \nodata & Median reduced chi$^{2}$ of the stellar-continuum fit within 1 $R_e$. \\
{\tt HA\symbol{95}Z                  } &                              \nodata & Flux-weighted mean redshift of the H$\alpha$ line within a 2.5 arcsec aperture at the galaxy center. \\
{\tt HA\symbol{95}GVEL\symbol{95}LO            } &                                 \kms & Gaussian-fitted velocity of the H$\alpha$ line at 2.5\% growth of all valid spaxels. \\
{\tt HA\symbol{95}GVEL\symbol{95}HI            } &                                 \kms & Gaussian-fitted velocity of the H$\alpha$ line at 97.5\% growth of all valid spaxels. \\
{\tt HA\symbol{95}GVEL\symbol{95}LO\symbol{95}CLIP       } &                                 \kms & Gaussian-fitted velocity of the H$\alpha$ line at 2.5\% growth after iteratively clipping 3-sigma outliers. \\
{\tt HA\symbol{95}GVEL\symbol{95}HI\symbol{95}CLIP       } &                                 \kms & Gaussian-fitted velocity of the H$\alpha$ line at 97.5\% growth after iteratively clipping 3-sigma outliers. \\
{\tt HA\symbol{95}GSIGMA\symbol{95}1RE         } &                                 \kms & Flux-weighted H$\alpha$ velocity dispersion (from Gaussian fit) of all spaxels within 1 $R_e$. \\
{\tt HA\symbol{95}GSIGMA\symbol{95}HI          } &                                 \kms & H$\alpha$ velocity dispersion (from Gaussian fit) at 97.5\% growth of all valid spaxels. \\
{\tt HA\symbol{95}GSIGMA\symbol{95}HI\symbol{95}CLIP     } &                                 \kms & H$\alpha$ velocity dispersion (from Gaussian fit) at 97.5\% growth after iteratively clipping 3-sigma outliers. \\
{\tt EMLINE\symbol{95}SFLUX\symbol{95}CEN      } &            10$^{-17}$ erg/s/cm$^{2}$ & Summed emission-line flux integrated within a 2.5 arcsec aperture at the galaxy center. \\
{\tt EMLINE\symbol{95}SFLUX\symbol{95}1RE      } &            10$^{-17}$ erg/s/cm$^{2}$ & Summed emission-line flux integrated within 1 effective-radius aperture at the galaxy. \\
{\tt EMLINE\symbol{95}SFLUX\symbol{95}TOT      } &            10$^{-17}$ erg/s/cm$^{2}$ & Total integrated flux of each summed emission measurement within the full MaNGA field-of-view. \\
{\tt EMLINE\symbol{95}SSB\symbol{95}1RE        } &     10$^{-17}$ erg/s/cm$^{2}$/spaxel & Mean emission-line surface-brightness from the summed flux measurements within 1 $R_e$. \\
{\tt EMLINE\symbol{95}SSB\symbol{95}PEAK       } &     10$^{-17}$ erg/s/cm$^{2}$/spaxel & Peak summed-flux emission-line surface brightness. \\
{\tt EMLINE\symbol{95}SEW\symbol{95}1RE        } &                                  \AA & Mean emission-line equivalent width from the summed flux measurements within 1 $R_e$. \\
{\tt EMLINE\symbol{95}SEW\symbol{95}PEAK       } &                                  \AA & Peak emission-line equivalent width from the summed flux measurements. \\
{\tt EMLINE\symbol{95}GFLUX\symbol{95}CEN      } &            10$^{-17}$ erg/s/cm$^{2}$ & Gaussian-fitted emission-line flux integrated within a 2.5 arcsec aperture at the galaxy center. \\
{\tt EMLINE\symbol{95}GFLUX\symbol{95}1RE      } &            10$^{-17}$ erg/s/cm$^{2}$ & Gaussian-fitted emission-line flux integrated within 1 effective-radius aperture at the galaxy. \\
{\tt EMLINE\symbol{95}GFLUX\symbol{95}TOT      } &            10$^{-17}$ erg/s/cm$^{2}$ & Total integrated flux of the Gaussian fit to each emission line within the full MaNGA field-of-view. \\
{\tt EMLINE\symbol{95}GSB\symbol{95}1RE        } &     10$^{-17}$ erg/s/cm$^{2}$/spaxel & Mean emission-line surface-brightness from the Gaussian-fitted flux measurements within 1 $R_e$. \\
{\tt EMLINE\symbol{95}GSB\symbol{95}PEAK       } &     10$^{-17}$ erg/s/cm$^{2}$/spaxel & Peak Gaussian-fitted emission-line surface brightness. \\
{\tt EMLINE\symbol{95}GEW\symbol{95}1RE        } &                                  \AA & Mean emission-line equivalent width from the Gaussian-fitted flux measurements within 1 $R_e$. \\
{\tt EMLINE\symbol{95}GEW\symbol{95}PEAK       } &                                  \AA & Peak emission-line equivalent width from the Gaussian-fitted flux measurements. \\
{\tt SPECINDEX\symbol{95}LO          } &                              \AA,mag & Spectral index at 2.5\% growth of all valid spaxels. \\
{\tt SPECINDEX\symbol{95}HI          } &                              \AA,mag & Spectral index at 97.5\% growth of all valid spaxels. \\
{\tt SPECINDEX\symbol{95}LO\symbol{95}CLIP     } &                              \AA,mag & Spectral index at 2.5\% growth after iteratively clipping 3-sigma outliers. \\
{\tt SPECINDEX\symbol{95}HI\symbol{95}CLIP     } &                              \AA,mag & Spectral index at 97.5\% growth after iteratively clipping 3-sigma outliers. \\
{\tt SPECINDEX\symbol{95}1RE         } &                              \AA,mag & Median spectral index within 1 effective radius. \\
{\tt SFR\symbol{95}1RE               } &      h$^{-2}$ $\mathcal{M}_\odot$/yr & Simple estimate of the star-formation rate within 1 effective radius based on the Gaussian-fitted H$\alpha$ flux; $\log{\rm SFR} = \log L_{{\rm H}\alpha} - 41.27$ \citep[Kroupa IMF;][]{2011ApJ...737...67M,2011ApJ...741..124H,KennicuttEvansARAA}, where $\log L_{{\rm H}\alpha} = 4\pi F_{{\rm H}\alpha,1R_e} D_L^2$ and ''no'' attentuation correction has been applied. \\
{\tt SFR\symbol{95}TOT               } &      h$^{-2}$ $\mathcal{M}_\odot$/yr & Simple estimate of the star-formation rate within the IFU field-of-view based on the Gaussian-fitted H$\alpha$ flux; $\log{\rm SFR} = \log L_{{\rm H}\alpha} - 41.27$ \citep[Kroupa IMF;][]{2011ApJ...737...67M,2011ApJ...741..124H,KennicuttEvansARAA}, where $\log L_{{\rm H}\alpha} = 4\pi F_{{\rm H}\alpha,1R_e} D_L^2$ and ''no'' attentuation correction has been applied.
\enddata
\tablenotetext{a}{Calculated assuming h=1, $\Omega_{\mathcal{M}}=0.3$,
and $\Omega_{\Lambda}=0.7$.}
\end{deluxetable*}

\onecolumngrid

\bibliography{master}

\end{document}